%% file: main.tex
\documentclass[aps,prx,twocolumn,amsmath,amssymb,floatfix,superscriptaddress,longbibliography]{revtex4-2}

\usepackage[normalem]{ulem}
\usepackage{graphicx}
\usepackage{bm}
\usepackage[dvipsnames]{xcolor}

\usepackage[colorlinks=true,linkcolor=RoyalBlue,citecolor=RoyalBlue,urlcolor=RoyalBlue,hypertexnames=false]{hyperref}
\usepackage{braket,bbold,color,enumerate,longtable,pifont,multirow,nicematrix}

\def\ba#1\ea{\begin{align}#1\end{align}}
\def\bg#1\eg{\begin{gather}#1\end{gather}}
\def\bpm{\begin{pmatrix}}
\def\epm{\end{pmatrix}}
\newcommand{\bs}[1]{\boldsymbol{#1}}
\newcommand{\nn}{\nonumber \\ }
\newcommand{\bb}[1]{{\mathbf #1}}
\newcommand{\trm}[1]{\textrm{#1}}

\newcommand{\mc}[1]{\mathcal{#1}}

\newcommand{\Z}{\mathbb{Z}}
\newcommand{\N}{\mathbb{N}}

\begin{document}

\title{Stable Real-Space Invariants and Topology Beyond Symmetry Indicators}

\author{Yoonseok Hwang}
\affiliation{Department of Physics, University of Illinois Urbana-Champaign, Urbana IL 61801, USA}
\affiliation{Anthony J. Leggett Institute for Condensed Matter Theory, University of Illinois Urbana-Champaign, Urbana IL 61801, USA}
\affiliation{Blackett Laboratory, Imperial College London, London SW7 2AZ, United Kingdom}

\author{Vaibhav Gupta}
\affiliation{Department of Physics, University of Illinois Urbana-Champaign, Urbana IL 61801, USA}
\affiliation{Anthony J. Leggett Institute for Condensed Matter Theory, University of Illinois Urbana-Champaign, Urbana IL 61801, USA}

\author{Frank Schindler}
\affiliation{Blackett Laboratory, Imperial College London, London SW7 2AZ, United Kingdom}

\author{Luis Elcoro}
\affiliation{Department of Physics, University of the Basque Country UPV/EHU, Apartado 644, 48080 Bilbao, Spain}
\affiliation{EHU Quantum Center, University of the Basque Country UPV/EHU, Apartado 644, 48080 Bilbao, Spain}

\author{Zhida Song}
\affiliation{International Center for Quantum Materials, School of Physics, Peking University, Beijing 100871, China}

\author{B. Andrei Bernevig}
\affiliation{Department of Physics, Princeton University, Princeton, NJ 08544, USA}
\affiliation{Donostia International Physics Center, P. Manuel de Lardizabal 4, 20018 Donostia-San Sebastian, Spain}
\affiliation{IKERBASQUE, Basque Foundation for Science, Bilbao, Spain}

\author{Barry Bradlyn}
\thanks{Corresponding Author: bbradlyn@illinois.edu}
\affiliation{Department of Physics, University of Illinois Urbana-Champaign, Urbana IL 61801, USA}
\affiliation{Anthony J. Leggett Institute for Condensed Matter Theory, University of Illinois Urbana-Champaign, Urbana IL 61801, USA}

\begin{abstract}
We introduce stable real-space invariants (SRSIs), topological invariants defined from adiabatic deformations between Wannier states, generalizing previously discovered local and composite real-space invariants.
SRSIs are $\Z$- and $\Z_n$-valued ($n=2,4$) linear combinations of Wannier state multiplicities characterizing the stable equivalence of atomic insulators.
We enumerate all SRSIs in nonmagnetic space groups with and without spin-orbit coupling.
$\Z$SRSIs are in one-to-one correspondence with momentum-space symmetry data and thus determine symmetry indicators of topology (SIs). $\Z_n$SRSIs capture real-space information beyond momentum-space symmetry data and SIs.
Applying SRSIs to split elementary band representations (EBRs) whose symmetry data decomposes into positive sums of other EBR symmetry data, we diagnose the topology of all 211 cases across 51 space groups except for 8 exceptions in 5 space groups.
Our results solidify Topological Quantum Chemistry beyond SIs and momentum-space symmetry data.
Finally, we use SRSIs to diagnose an obstructed atomic insulator in a realistic material.
\end{abstract}

\maketitle

\section*{Introduction}
\label{sec:intro}
In topological quantum chemistry (TQC)~\cite{Bradlyn17,jenbarryreview21,zaktqcprecursor88,Vergniory17,Cano17-2,elcoro2017,evarestov1997site,michel1999connectivity,bacry1988symmetrya}, bands that cannot be expressed as a positive-integer linear combination of atomic limits, also known as elementary band representations (EBRs), are considered topological.
Such bands are often characterized by nontrivial symmetry indicators (SIs) calculated from momentum-space symmetry data, which include the representation of bands at high-symmetry momenta in the Brillouin zone (BZ)~\cite{FuKane2007,Hughes11,TurnerSymInds12,fang2012,Alexandradinata14,Slager17,Khalaf17,Po_2017, Song_2018,Watanabe18Unified,horeview20,Zhang_2021}.
The combination of TQC and SIs with first-principles calculations has enabled the discovery of numerous topological material candidates~\cite{Vergniory_2019,Zhang_2019,Tang_2019,wang2019two,Xu20Magnetic,vergniory22,Wieder2022}.
One corollary of TQC is that when an EBR splits into disconnected bands, some of these bands must be topological.
Refs.~\cite{AshvinFragile,Cano17,Bouhon18Fragile,BarryFragile} highlighted that, in specific cases, one group of bands in a split EBR may appear trivial -- sharing identical SIs with an atomic limit -- while the other group of bands is indicated as topological, both by TQC and by SIs, as its momentum-space symmetry data is inconsistent with that of sums of atomic limits.
While TQC provides a conceptually more general real-space framework, many known examples suggest that SIs often reproduce the same diagnosis in practice, including but not limited to the case of split EBRs.
This raises the question of whether TQC and SIs yield identical results, and more importantly, when TQC can, in principle, distinguish a broader class of topological phases beyond SIs alone.
It is known, for instance, that SIs do not capture all aspects of band topology, especially in cases involving fragile or non-symmetry-indicated stable topological phases.
Moreover, SIs do not distinguish between distinct trivial atomic insulators that cannot be deformed into each other, despite their topological inequivalence~\cite{bacry1988symmetrya,Cano22Invisible}.

In this work, we introduce stable real-space invariants (SRSIs) to address these limitations.
By generalizing the previously discovered real-space invariants (RSIs)~\cite{ZhidaFragileTwist1,ZhidaFragileTwist2,xu2021threedimensional}, SRSIs provide a comprehensive classification that accounts for the (stable) adiabatic equivalence of bands.
SRSIs enable the identification and characterization of both trivial and topological phases beyond the information contained in momentum-space symmetry data or SIs.
We systematically derive and tabulate all SRSIs, which are of two types; $\Z$- and $\Z_n$-valued SRSIs ($\Z$SRSIs and $\Z_n$SRSIs, where $n=2,4$) in the 230 space groups (SGs) with and without spin-orbit coupling (SOC).
Notably, $\Z$SRSIs determine the full set of SIs, and $\Z_n$SRSIs provide additional information about band structures not captured by momentum-space symmetry data and SIs.
Hence the formalism of TQC with SRSIs contains, but goes beyond, that of SIs.

After introducing the general framework, we demonstrate the application of SRSIs to band topology.
For atomic insulators, we show that SRSIs provide both a classification and a diagnostic of stable equivalence.
Specifically, we prove that if two atomic insulators have the same values for all $\Z$ and $\Z_n$SRSIs, then they can be adiabatically deformed into each other in the presence of auxiliary trivial bands.
For topological insulators, we demonstrate that $\Z$SRSIs can be assigned based on the mapping between momentum-space symmetry data and $\Z$SRSIs, with their fractional (i.e., rational numbers that are not integers) values implying nontrivial SIs.
While $\Z_n$SRSIs cannot be directly assigned to topological bands, they can still serve as indicators of topological gaps that cannot be diagnosed using SIs.
Specifically, we investigate the situation where disconnected parts of a split EBR appear indistinguishable from atomic limits based on momentum-space symmetry data (SIs) alone.
Even in such cases, the real-space framework of TQC reveals an obstruction to both split parts being atomic limits~\cite{Cano17-2,BarryFragile,Bouhon18Fragile,AlexandradinataTheorem20,Schindler21Noncompact}.
Using SRSIs we analyze all 211 split EBRs in 51 SGs where TQC implies band topology beyond SIs~\cite{Sander2DTCI2019,Sun20Pfaffian,Slager20BeyondSIs,Kooi21RealSpace,Wezel21,Wan21Homotopy,Cano22Invisible,Chen22}.
Notably, we show that constraints arising from $\Z_n$SRSIs can be used to diagnose band topology in all split EBR cases except only 8 cases across 5 SGs, highlighting the power of SRSIs in capturing band topology beyond the scope of SIs.
This demonstrates the existence of many topological-only systems where at least some bands have to be topological (either stable or fragile), regardless of the model parameters.
Finally, we illustrate our results using simple tight-binding models, where we use the Wilson loop to confirm the band topology.
In addition, we demonstrate the material-level applicability of SRSIs by presenting a concrete example of the symmetry-based diagnosis of an obstructed atomic insulator in Y$_3$Al$_2$.

\section*{Results}
\subsection*{Real-space invariants}
\label{sec:rsi_review}
Before we introduce the SRSIs, let us review the local RSIs derived in Refs.~\cite{ZhidaFragileTwist1,ZhidaFragileTwist2}.
To define RSIs, let us consider topologically trivial, atomic bands, which transform under a representation of a space group induced from site-symmetry representations (or exponentially localized and symmetric Wannier orbitals) at WPs within a unit cell.
Recall from Refs.~\cite{Bradlyn17-2,jenbarryreview21,elcoro2017,Cano17} that for each WP $W$, the site-symmetry group $G_W$ consists of symmetries that leave $W$ invariant, and the site-symmetry irreducible representations (irreps) $\rho^i_W$ ($i=1, \dots, N^\rho_W$) are defined for $G_W$.

Local RSIs are defined locally at each WP $W$ in the unit cell, in terms of the multiplicities $m(\rho^i_W)$ of site-symmetry irreps at $W$ for a given set of bands.
For a WP $W$, we first consider all WPs connected to $W$, meaning those WPs with coordinates that can be continuously tuned to those of $W$ while preserving all crystal symmetries~\cite{elcoro2017,jenbarryreview21}.
The local RSIs remain invariant under adiabatic processes that deform occupied site-symmetry irreps between $W$ and the connected lower-symmetry WPs without breaking the symmetries of the SG $G$ and without closing a gap to other bands.

The full set of local RSIs in the 1651 magnetic SGs was computed in Ref.~\cite{xu2021threedimensional}, where it was found that there are both $\Z$- and $\Z_n$-valued ($n=2$) local RSIs.
Earlier works~\cite{van2018higher,Hwang:2019aa} identified related local RSIs for specific point-group settings, such as rotation and (roto)inversion symmetries.
These works provide early examples of local RSIs in restricted symmetry settings.
To compute the local RSIs at a WP $W$, we first analyze the induction relations between the site-symmetry group $G_W$ of $W$ and the site symmetry groups of all connected lower-symmetry WPs.
These relations determine how direct sums of irreps of $G_W$ decompose into irreps at lower-symmetry WPs, corresponding to possible adiabatic processes.
As detailed in Section 2 of the Supplementary Note (SN 2), local RSIs can be systematically derived using the Smith decomposition of the matrix enumerating these adiabatic processes.

Local RSIs fall into two distinct categories: those defined at maximal Wyckoff positions (WPs) and those defined at non-maximal WPs.
A WP $W$ is maximal if all connected WPs have lower site-symmetry groups than $G_W$; otherwise, $W$ is non-maximal.
The local RSIs at maximal WPs are topological invariants under all adiabatic processes, whereas the local RSIs at non-maximal WPs are invariant only under a subset of adiabatic processes.
That is, local RSIs associated to non-maximal WPs and are not strictly topological invariants when arbitrary adiabatic processes are considered.
In contrast, we will demonstrate that SRSIs are robust topological invariants under all adiabatic processes, addressing these subtleties comprehensively.
This motivates the need for SRSIs as a comprehensive framework for real-space topology.

\begin{figure}[t]
\centering
\includegraphics[width=0.48\textwidth]{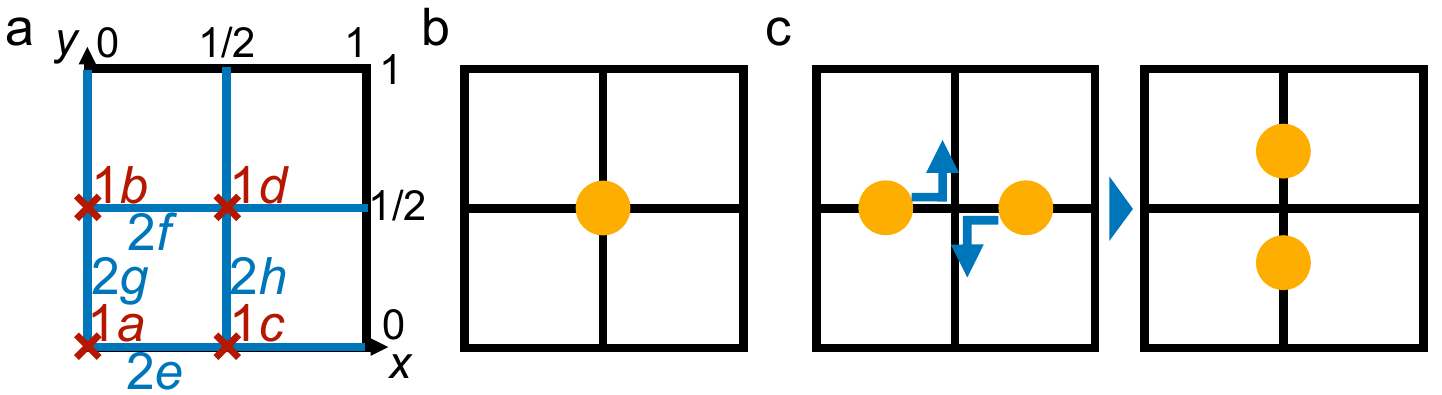}
\caption{Real-space invariants (RSIs) in the SG $Pmm21'$ (No. 25).
\textbf{a} Unit cell structure in 2D.
There are four maximal WPs ($1a$, $1b$, $1c$, $1d$), four non-maximal WPs ($2e, 2f, 2g, 2h$), and non-maximal, general WP $4i$ (not shown).
\textbf{b} A single Kramers pair of site-symmetry irreps forms the irrep $(\bar{E})_{1d}$ at the maximal WP $1d$~\cite{Aroyo2011183}, which cannot be moved away from $1d$.
This is encoded in a nonzero local RSI, $\delta_{1d} = m[(\bar{E})_{1d}] \mod 2 = 1$, which is also an SRSI.
\textbf{c} A pair of Kramers pairs forms the irrep ${}^1\bar{E} {}^2\bar{E}$ at the WP $2f$, which can be adiabatically moved to the WP $2h$.
This process changes the local RSIs at WPs $2f$ and $2h$, $\{m[({}^1\bar{E} {}^2\bar{E})_{2f}],m[({}^1\bar{E} {}^2\bar{E})_{2h}]\} \mod 2$, from (1,0) to (0,1).}
\label{fig:rsi}
\end{figure}

As an example, let us consider the SG $Pmm21'$ (No. 25) with spin-orbit coupling (SOC) in two dimensions (2D).
This SG is generated by $x$- and $y$-flipping mirror symmetries, $M_{x,y}=\{m_{x,y}|\bb 0\}$, time-reversal symmetry $T$, and two translations $\{E|1,0\}$ and $\{E|0,1\}$.
Here, $m_x$ ($m_y$) flips the $x$ ($y$) coordinate, and $E$ is the trivial point-group element.
(Note that while $Pmm21'$ is a 3D SG, it can be viewed as the trivial direct product of the 2D wallpaper group $p2mm1'$ with translations along the $z$-direction.
As such, in what follows we suppress the $z$-translations for simplicity.)
See Fig.~\ref{fig:rsi}\textbf{a} for the unit cell structure.
The WPs and site-symmetry irreps are summarized in Table~\ref{table:WPs}.
In this SG, the direct sum of two $(\bar{E})_{1d}$ irreps at maximal WP $1d$ can be adiabatically deformed to $(^1\bar{E} ^2\bar{E})_{2f}$ or $(^1\bar{E} ^2\bar{E})_{2h}$ at connected non-maximal WP $2f$ or $2h$, as shown in Figs.~\ref{fig:rsi}\textbf{b,c}.
Thus, we define the equivalence relation, $2(\bar{E})_{1d} \Leftrightarrow (^1\bar{E} ^2\bar{E})_{2f/2h}$.
For these adiabatic processes, $m[(\bar{E})_{1d}]$ mod 2 remains invariant and serve as a local RSI at WP $1d$.
Similarly, by considering the adiabatic deformation between $(^1\bar{E} ^2\bar{E})_{2f} \oplus (^1\bar{E} ^2\bar{E})_{2f}$ and $(\bar{A}\bar{A})_{4i}$, the local RSI at WP $2f$ is defined as $m[(^1\bar{E} ^2\bar{E})_{2f}]$ mod 2.
In the same way, the local RSI at WP $2h$ is defined as $m[(^1\bar{E} ^2\bar{E})_{2h}]$ mod 2.

\renewcommand{\arraystretch}{1.4}
\begin{table}[t!]
\centering
\begin{minipage}{0.48\textwidth}
\caption{
Wyckoff positions (WPs) and site-symmetry irreps in the SG $Pmm21'$ (No. 25) in 2D.
The first column indicates the type of WP and its representative position for each WP $W$.
When $W$ has a multiplicity larger than 1, other positions are generated by full symmetry group.
The second and third columns denote the site-symmetry group generators at the representative position and the site-symmetry irreps, respectively.
For the second column, nontrivial generators are provided.
The WPs $1a,1b,1c,1d$ are maximal.
For non-maximal WPs $2e,2f,2g,2h,4i$, $x$ and $y$ are free real parameters.}
\label{table:WPs}
\end{minipage}
\begin{tabular*}{0.4\textwidth}{@{\extracolsep{\fill}}c| c c}
\hline \hline
WP $W$ & $G_W$ & $\rho_W$ \\
\hline
$1a(0,0)$ & $\{m_x|\bb 0\},\{m_y|\bb 0\}$ & $(\bar{E})_{1a}$ \\
$1b(0,1/2)$ & $\{m_x|\bb 0\},\{m_y|0,1\}$ & $(\bar{E})_{1b}$ \\
$1c(1/2,0)$ & $\{m_x|1,0\},\{m_y|\bb 0\}$ & $(\bar{E})_{1c}$ \\
$1d(1/2,1/2)$ & $\{m_x|1,0\},\{m_y|0,1\}$ & $(\bar{E})_{1d}$ \\
$2e(x,0)$ & $\{m_y|\bb 0\}$ & $(^1\bar{E} ^2\bar{E})_{2e}$ \\
$2f(x,1/2)$ & $\{m_y|0,1\}$ & $(^1\bar{E} ^2\bar{E})_{2f}$ \\
$2g(0,y)$ & $\{m_x|\bb 0\}$ & $(^1\bar{E} ^2\bar{E})_{2g}$ \\
$2h(1/2,y)$ & $\{m_x|1,0\}$ & $(^1\bar{E} ^2\bar{E})_{2h}$ \\
$4i(x,y)$ & $\{E|0,0\}$ & $(\bar{A}\bar{A})_{4i}$ \\
\hline \hline
\end{tabular*}
\end{table}

There is a crucial difference between the local RSIs at maximal and non-maximal WPs.
Local RSIs at maximal WPs remain invariant under all adiabatic processes and are thus topological invariants.
In contrast, local RSIs at non-maximal WPs $W'$ may change when site-symmetry irreps are moved from $W'$ to another connected maximal WP $W$.
The irreps at WP $W$ may further be moved to another non-maximal WP $W''$ connected to $W$.
For example, $(^1\bar{E} ^2\bar{E})_{2f}$ at $2f(x,1/2)$ can be moved to $1d(1/2,1/2)$ by tuning the free real parameter $x$ without breaking any symmetry, as shown in Fig.~\ref{fig:rsi}\textbf{c}.
Then, this irrep can be deformed to $2(\bar{E})_{1d}$.
Moreover, $2(\bar{E})_{1d}$ can be further deformed to $(^1\bar{E} ^2\bar{E})_{2h}$, as the WP $2h$ is connected to WP $1a$ (see Fig.~\ref{fig:rsi}\textbf{c}).
During this adiabatic process, the local RSI at WP $2f$ ($2h$), $m[(^1\bar{E} ^2\bar{E})_{2f}]$ ($m[(^1\bar{E} ^2\bar{E})_{2h}]$) mod 2, changes from 1 (0) to 0 (1).

The above example demonstrates why local RSIs at non-maximal WPs cannot serve as topological invariants.
Below, we introduce the SRSIs that address this issue.
For a detailed review of local RSIs, see SN 2.

\subsection*{Stable Real-Space Invariants}
\label{sec:srsi}
To define true topological invariants, we introduce and calculate SRSIs that remain invariant under all adiabatic processes in a SG.
SRSIs fully contain the local RSIs at maximal WPs and also include additional topological invariants formed as linear combinations of site-symmetry-irrep multiplicities at multiple WPs, including both maximal and non-maximal positions.
Similar to local RSIs, both $\Z$- and $\Z_n$-valued SRSIs exist.
However, while $\Z_n$-valued local RSIs exist only for $n=2$ both with and without SOC, $\Z_n$-valued SRSIs exist for $n=2,4$.
Notably, $\Z_4$-valued SRSIs appear in only 4 SGs, all with SOC.
While $\Z_2$-valued local RSIs arise due to either time-reversal symmetry (protecting Kramers pairs if SOC exists) or the presence multiple adiabatic processes (that hinder the emergence of $\Z$-valued RSIs)~\cite{ZhidaFragileTwist2,xu2021threedimensional}, $\Z_4$-valued SRSIs require both (though their presence is not guaranteed even when both conditions are met).
For a detailed examination on $\Z_4$-valued SRSIs, see SN 3C.

To calculate the SRSIs, we first define symmetry-representation vector $p$ associated to a set of trivial (atomic limit) bands.
The vector $p$ lists all site-symmetry irrep multiplicities at all WPs:
\ba
p = \left[ m(\rho^1_{W_1}), m(\rho^2_{W_1}), \dots, m(\rho^1_{W_2}), \dots \right]^T,
\label{eq:pvectordef}
\ea
where the irrep multiplicity $m(\rho^j_{W})$ ($j=1, \dots, N_W^\rho$) is the multiplicity of the $j$-th irrep of the site-symmetry group at the WP $W$, and $m(\rho^j_{W})$ is a non-negative integer.
The total length of $p$ is $\sum_W N_W^\rho = N_{\rm UC}^\rho$, which is the total number of distinct site-symmetry irreps defined within a unit cell (UC).
Adiabatic processes can be characterized by changes in the irrep multiplicities by $\Delta m(\rho^j_W)$ associated with moving orbitals, consistent with the symmetries; collecting the $\Delta m(\rho^j_W)$, we can associate to each adiabatic process a column vector $q_m$ which represents the change to $p$ under the adiabatic process $m$, i.e $p\rightarrow p + q_m$.

For any SG (either with and without SOC), we can find a finite basis set that generates all adiabatic processes.
Such a basis is constructed as follows: site-symmetry irreps at non-maximal WPs $W'$ can always be deformed into direct sums of irreps at the WPs connected to $W'$ with larger site-symmetry groups.
For all site-symmetry irreps at all non-maximal WPs $W'$, we list the adiabatic processes that deform the irrep from $W'$ to all connected high-symmetry WPs.
Then, we define a matrix of adiabatic processes $q$, whose component $(q)_{im}$ represents the change to the $i$-th entry of the symmetry-representation vector $p$ by $m$-th basis of adiabatic process.
Here, $i=1, \dots, N_{\rm UC}^\rho$ runs over all site-symmetry irreps for every WPs in the unit cell (including the general WP), and $m=1, \dots, N_{\rm adia}$ runs over all adiabatic processes.
The entries of the columns of $q$ are integers that specify the change in irrep multiplicities under the adiabatic process: when an adiabatic process deforms an irrep at $W'$ into a linear combination of irreps at a connected high-symmetry WP $W$, the corresponding column of $q$ contains a negative integer reflecting the decrease of irrep multiplicity at $W'$ and positive integers indicating the increases of irrep multiplicities at $W$.
Note that only integer multiples of the columns of $q$ correspond to valid adiabatic processes.

To extract meaningful invariants that remain unchanged under adiabatic processes, we perform the Smith decomposition:
\ba
q = L \cdot \Lambda \cdot R,
\label{eq:qvecdef_global}
\ea
where $L$ and $R$ are $N^{\rho}_{\rm UC} \times N^{\rho}_{\rm UC}$ and $N_{\rm adia} \times N_{\rm adia}$ unimodular matrices, respectively.
$\Lambda$ is an $N^{\rho}_{\rm UC} \times N_{\rm adia}$ matrix with non-negative integers on the diagonal and zeros for all entries off the diagonal.
We will always choose $L$ and $R$ such that the distinct diagonal entries $\lambda, \lambda', \lambda'', \dots$ of $\Lambda$ are sorted in decreasing order, $\lambda > \lambda' > \lambda'' \dots \geq 0$.
The nonzero diagonal entries of $\Lambda$ are known as the elementary divisors of $q$.
Notably, the number of elementary divisors of $q$ is equal to ${\rm rank}(q)$.

A linear combination of adiabatic processes changes the symmetry-representation vector $p$ by $\Delta p = q \cdot z$ with $z \in \Z^{N_{\rm adia}}$ a vector of integers specifying the adiabatic process.
It follows that under any adiabatic process the quantities
\ba
\theta_i = 
\begin{cases}
(L^{-1} \cdot p)_i \mod \Lambda_{ii} & \trm{for } i \le {\rm rank}(q) \\
(L^{-1} \cdot p)_i & \trm{for } i > {\rm rank}(q),
\end{cases}
\label{eq:SRSI_definition}
\ea
are invariant under all adiabatic processes, i.e., $\Delta \theta_i = 0$.
This follows from the fact that
\ba
\Delta \theta_i = (L^{-1} \cdot q \cdot z)_i = \Lambda_{ii} (R \cdot z)_i.
\ea
Since $R$ is unimodular, $(R \cdot z)_i \in \Z$ and thus $\Delta \theta_i \in \Lambda_{ii} \Z$, which vanishes for all $i$ according to Eq.~\eqref{eq:SRSI_definition}.
Eq.~\eqref{eq:SRSI_definition} constitute our definition for the SRSIs.
The first and second cases in Eq.~\eqref{eq:SRSI_definition} correspond to $\Z_{\Lambda_{ii}}$SRSIs and $\Z$SRSIs, respectively.
We refer to $\Z_{\Lambda_{ii}}$SRSIs collectively as $\Z_n$SRSIs.
Practically, for $\Z_n$SRSIs, we only consider $i$ with $\Lambda_{ii} > 1$, as $\Z_1$SRSIs are trivial.
For further details on the algorithm obtaining SRSIs and the exhaustive listing of SRSIs in all SGs with and without SOC, see SN 3 and SN 7A, respectively.

In the example of SG $Pmm21'$, the algorithm described above yields one $\Z$SRSI and three $\Z_2$SRSIs.
(For a step-by-step procedure, see Constructing SRSIs from adiabatic processes in Methods and SN 3C.)
These SRSIs are given by
\ba
\theta_1 =& m[(\bar{E})_{1a}] + m[(\bar{E})_{1b}] + m[(\bar{E})_{1c}] + m[(\bar{E})_{1d}]
\nn
&+ 2m[({}^1\bar{E} {}^2\bar{E})_{2e}] + 2m[({}^1\bar{E} {}^2\bar{E})_{2f}] + 2m[({}^1\bar{E} {}^2\bar{E})_{2g}] \nn
&+ 2m[({}^1\bar{E} {}^2\bar{E})_{2h}] + 4m[(\bar{A}\bar{A})_{4i}],
\nn
\theta_2 =& m[(\bar{E})_{1b}] \mod 2,
\quad
\theta_3 = m[(\bar{E})_{1c}] \mod 2,
\nn
\theta_4 =& m[(\bar{E})_{1d}] \mod 2.
\label{eq:srsi_pmm2}
\ea
A unique $\Z$SRSI, $\theta_1$, corresponds to half the total number of electrons per unit cell, which is invariant under any adiabatic process and fixed by the number of bands.
In other SGs, multiple $\Z$SRSIs may exist, providing information beyond this physical invariant.
More generally, SRSIs are linear combinations of site-symmetry irrep multiplicities that can involve contributions from multiple WPs, including both maximal and non-maximal ones.

Finally, let us comment that local RSIs at maximal WPs are a subset of SRSIs.
In the case of SG $Pmm21'$, the $\Z_2$SRSIs $\theta_2, \theta_3$ and $\theta_4$ are identical to the local RSIs at the maximal WPs $1b$, $1c$, and $1d$, respectively.
The local RSI at the maximal WP $1a$ can be obtained from the linear combination,
\ba
\theta_1 + \theta_2 + \theta_3 + \theta_4 \mod 2 = m[(\bar{E})_{1a}] \mod 2.
\ea
While a local RSI at a non-maximal WP is not a topological invariant, the multiplicities of irreps at non-maximal WPs appear in the definition of SRSIs (such as $\theta_1$), reflecting the consideration of all adiabatic processes that involve multiple WPs.

\begin{figure}[t]
\centering
\includegraphics[width=0.48\textwidth]{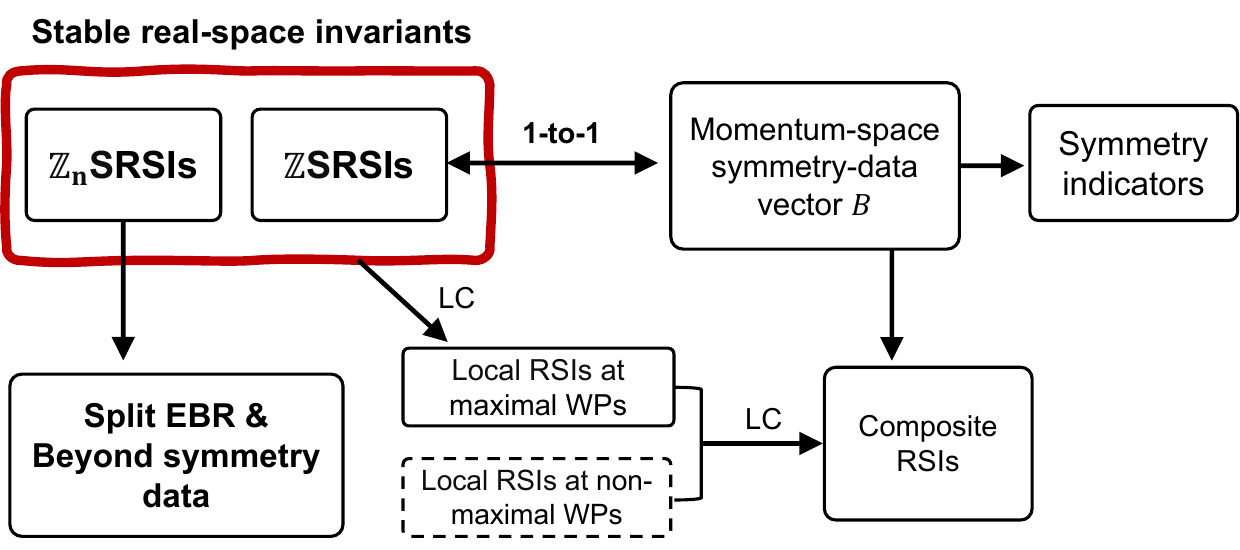}
\caption{Schematic relationship between SRSIs, local RSIs, composite RSIs, symmetry indicators (SIs), and symmetry-data vectors.
$\Z$SRSIs are in one-to-one correspondence with the symmetry-data vector and therefore fully determine all SIs.
$\Z$SRSIs also encode information that can be further processed into indicators for quantities such as fragile topology~\cite{Hwang:2019aa,Song20,ZhidaFragileTwist2} or filling anomaly~\cite{benalcazar2018quantization,velury2025global}.
$\Z_n$SRSIs may capture additional topological information that is invisible to momentum-space data, including topology associated with split EBRs and other non-symmetry-indicated phases.
Both local RSIs at maximal and non-maximal WPs, as well as composite RSIs formed from them, can be obtained as linear combinations (LCs) of SRSIs.}
\label{fig:diagram}
\end{figure}

\subsection*{Stable equivalence of band topology}
\label{sec:stable}
We will now argue that the SRSIs are intimately connected to the notion of stable equivalence of bands.
We will start by considering atomic insulators induced from localized orbitals that transform as direct sums of site-symmetry irreps.
Consider two site symmetry representations, $\rho_1$ and $\rho_2$, and their corresponding (E)BRs, $\rho_1 \uparrow G$ and $\rho_2 \uparrow G$.
These BRs are stably equivalent if they can be adiabatically deformed into each other in the presence of auxiliary trivial bands corresponding to the (E)BR $\rho_{aux} \uparrow G$.
Importantly, $\rho_1 \uparrow G$ and $\rho_2 \uparrow G$ can be stably equivalent even if they are inequivalent, i.e., even if they cannot be deformed into each other (and are therefore topologically distinct).
The stable equivalence between $\rho_1 \uparrow G$ and $\rho_2 \uparrow G$ can be denoted as $\rho_1 \uparrow G \oplus \rho_{aux} \uparrow G \equiv \rho_2 \uparrow G \oplus \rho_{aux} \uparrow G$.

One crucial property of SRSIs is their ability to indicate the stable equivalence between atomic insulators.
We prove in SN 3B that two atomic insulators are stably equivalent if and only if they have matching $\Z$SRSIs and $\Z_n$SRSIs.
This property justifies the name of SRSIs: if $\rho_1\uparrow G$ and $\rho_2\uparrow G$ are two atomic insulators that are not equivalent but have matching SRSIs, then they are (adiabatically and hence topologically) equivalent in the presence of additional trivial bands.
This is analogous to the case of fragile topology, which become equivalent to trivial bands in the presence of additional (trivial) bands.
Topologically trivial bands in the same stable equivalence class must have the same SRSIs.

To illustrate the idea of stable equivalence, we will consider two examples.
Let us first consider the simplest case where the adiabatic deformation between two sets of site-symmetry irreps with matching SRSIs does not require auxiliary bands.
In the SG $Pmm21'$ with SOC, let us focus two band representations induced from distinct site-symmetry irreps, $(\bar{E})_{1a}$ and $(\bar{E})_{1b}$.
Their $\Z$SRSI and $\Z_2$SRSIs, computable from Eq.~\eqref{eq:srsi_pmm2}, are given as
\bg
(\bar{E})_{1a} \uparrow G: \quad
(\theta_1,\theta_2,\theta_3,\theta_4) = (1,0,0,0),
\nn
(\bar{E})_{1b} \uparrow G: \quad
(\theta_1,\theta_2,\theta_3,\theta_4) = (1,1,0,0).
\eg
Since the SRSI vectors differ, we know that these irreps cannot be deformed into one another, even with the addition of auxiliary irreps $\rho_{aux}$, i.e., $(\bar{E})_{1a} \oplus \rho_{aux} \nLeftrightarrow (\bar{E})_{1b} \oplus \rho_{aux}$ for any $\rho_{aux}$.
Physically, a single $(\bar{E})_{1a}$ irrep corresponds to a Kramers pair of two orbitals with opposite mirror ($M_x$ or $M_y$) eigenvalues located at $(0,0)$.
At any position, time-reversal symmetry imposes that orbitals form Kramers pairs.
Thus, moving orbitals at $(0,0)$ to another position while preserving SG symmetries requires at least two Kramers pairs to move together, such as to WP $2g$ with positions $(0,y)$ and $(0,-y)$ (see also Fig.~\ref{fig:rsi}\textbf{a}).
Hence, a single Kramers pair at WP $1a$ cannot move to any other position.
The same argument applies to the irrep $(\bar{E})_{1b}$.
However, because $\theta_2$ is a $\Z_2$SRSIs, the SRSIs of $2(\bar{E}_{1a})$ and $2(\bar{E}_{1b})$ are same, meaning that they can be deformed to each other.
For example, $2(\bar{E}_{1a})$ at (0,0) can be deformed to $({}^1\bar{E} {}^2\bar{E})_{2h}$ at $(0,y)$ and $(0,-y)$, and further deformed to $2(\bar{E}_{1b})$ at $(0,1/2)$ by tuning $y$ from 0 to $1/2$.

\renewcommand{\arraystretch}{1.4}
\begin{table}[t!]
\centering
\begin{minipage}{0.48\textwidth}
\caption{
WPs $12d$ and $24i$ and site-symmetry irreps in the SG $I432$ (No. 211).
The second and third columns denote the generators of site-symmetry group and site-symmetry irreps at the representative position, respectively.
Note that $C_{2x} = \{2_{100}|0,1,0\}$, $C_{2,y+z} = \{2_{011}|1/2,1/2,-1/2\}$, $C_{2,y-z} = \{2_{01\bar{1}}|1/2,1/2,1/2\}$, which transform the coordinates $(x,y,z)$ to $(x,-y+1,-z)$, $(-x+1/2,z+1/2,y-1/2)$, and $(-x+1/2,-z+1/2,-y+1/2)$, respectively.
$z$ is a free real parameter.}
\label{table:sg211}
\end{minipage}
\begin{tabular*}{0.48\textwidth}{@{\extracolsep{\fill}}c| c c}
\hline \hline
WP $W$ & $G_W$ & $\rho_W$ \\
\hline
\multirow{2}{*}{$12d(1/4,1/2,0)$} & \multirow{2}{*}{$C_{2x}, C_{2,y+z}, C_{2,y-z}$} & $(A_1)_{12d},(B_1)_{12d},$ \\
& & $(B_2)_{12d},(B_3)_{12d}$ \\
$24i(1/4,z+1/2,z)$ & $C_{2,y+z}$ & $(A)_{24i},(B)_{24i}$ \\
\hline \hline
\end{tabular*}
\end{table}

Now, we consider a case with two distinct band representations that are inequivalent but stably equivalent, requiring auxiliary irrep(s) for deformation between them.
Such band representations can be induced from either same WP or different WPs.
For simplicity, we focus on the case where they are induced from the same WP, specifically in SG $I432$.
(For cases where stably equivalent band representations are induced from different WPs, see SN 4D.)

Specifically, consider the two band representations induced from the irreps $(B_2)_{12d}$ and $(B_3)_{12d}$ at the maximal WP $12d(1/4,1/2,0)$.
As distinct irreps at the same maximal WP, they cannot be deformed into each other, making them topologically inequivalent.
In SG $I432$, we can define seven $\Z$SRSIs ($\theta_\Z$) and three $\Z_2$SRSIs ($\theta_{\Z_2}$) (see also SN 3C).
Notably, the band representations induced from both $(B_2)_{12d}$ and $(B_3)_{12d}$ have matching SRSIs:
\bg
\theta_\Z = (1,0,1,1,1,-1,0)^T, \quad
\theta_{\Z_2} = (0,0,0)^T,
\label{eq:sg211_srsi_12d}
\eg
indicating their stable equivalence.

\begin{figure}[t]
\centering
\includegraphics[width=0.48\textwidth]{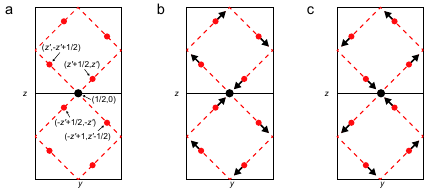}
\caption{
Stable equivalence in SG $I432$ (No. 211).
The site-symmetry irreps at WP $12d$ and $24i$ are shown on the $(y,z)$ plane with a fixed $x$-coordinate, $x=1/4$.
Unit cells are indicated by the solid black lines.
The schematic is depicted in reduced coordinates, where the length in both the $x$- and $y$-directions is 1.
\textbf{a} A representative position $\bb x_{12d}=(1/2,0)$ of WP $12d$ (marked as a black dot) is connected to four positions of WP $24i$ (marked by red dots with black arrows).
\textbf{b} A pair of $(A)$ irreps at $(y,z)=(z'+1/2,z')$ and $(-z'+1/2,-z')$ are deformed to a sum of $(A_1)$ and $(B_3)$ irreps at $\bb x_{12d}$, by tuning $z'$ to $0$.
\textbf{c} A pair of $(A)$ irreps at $(z',-z'+1/2)$ and $(-z'+1,z'-1/2)$ are deformed to a sum of $(A_1)$ and $(B_2)$ irreps at $\bb x_{12d}$, by tuning $z'$ to $1/2$.}
\label{fig:stable}
\end{figure}

The stable equivalence can be understood through the connectivity between the maximal WP $12d$ and the non-maximal WP $24i$.
(For the site-symmetry groups and irreps at these WPs, see Table~\ref{table:sg211}.)
For the WP $12d$, with representative position $\bb x_{12d}=(1/4,1/2,0)$, there are four points in the WP $24i$ that are connected to $\bb x_{12d}$, explicitly given as $\bb x_{24i,1,\dots,4} = (1/4,z+1/2,z),(1/4,-z+1/2,-z),(1/4,z,-z+1/2),(1/4,-z+1,z-1/2)$, where $z$ is a free real parameter.
(See Fig.~\ref{fig:stable}\textbf{a}.)
A pair of site-symmetry irreps $(A)$ at $\bb x_{24i,1}$ and $\bb x_{24i,2}$ can be deformed to the linear combination of $(A_1)$ and $(B_3)$ irreps at $\bb x_{12d}$, as shown in Fig.~\ref{fig:stable}\textbf{b}.
Alternatively, as shown in Fig.~\ref{fig:stable}\textbf{c}, a pair of $(A)$ irreps at $\bb x_{24i,3}$ and $\bb x_{24i,4}$ can be deformed to the $(A_1)$ and $(B_2)$ irreps at $\bb x_{12d}$.
Thus, the two different deformation processes considered involving the $(A)_{24i}$ irrep imply the following equivalence relation,
\bg
(B_2)_{12d} \oplus \rho_{aux} \Leftrightarrow (B_3)_{12d} \oplus \rho_{aux},
\eg
where $\rho_{aux}=(A)_{12d}$.
This establishes the stable equivalence between the EBRs $(B_2)_{12d} \uparrow G$ and $(B_3)_{12d} \uparrow G$ in the presence of bands $\rho_{aux} \uparrow G$ induced from the auxiliary site-symmetry irrep $\rho_{aux}$.
This example demonstrates how SRSIs capture subtle aspects of band topology, such as stable equivalence, that go beyond simple adiabatic deformation arguments.
We note that this $I432$ example was tabulated in Refs.~\cite{Cano22Invisible,zaktqcprecursor88} as a case where distinct site-symmetry irreps at the same WP induce band representations with matching momentum space little group representations, i.e., they are irrep equivalent in the language of Ref.~\cite{Cano22Invisible}.
Here we see that the physical origin of the irrep equivalence comes from the stable equivalence of the band representations.

\subsection*{$\Z$SRSIs are determined by symmetry-data vector}
\label{sec:mapping}
We now establish a crucial property of SRSIs: the $\Z$SRSIs are in one-to-one correspondence with the momentum-space symmetry data of bands in every non-magnetic space group.
We first explain this mapping for atomic insulators and then extend this concept to show how $\Z$SRSIs can also be assigned to topological phases.
To construct the mapping explicitly, let us define the symmetry-data vector $B$.
The symmetry-data vector $B$ is composed of the multiplicities of little-group irreps at all high-symmetry momenta in the BZ, which we write as
\ba
B = \left[
m(\rho^1_{K_1}), m(\rho^2_{K_1}), \dots, m(\rho^1_{K_2}), m(\rho^2_{K_2}), \dots
\right]^T,
\label{eq:symmetrydatavec_definition}
\ea
where the nonnegative integers $m(\rho^\alpha_K)$ denote the multiplicity of the irrep $\rho^\alpha_K$ ($\alpha=1, \dots, N_K^\rho$).
The length of $B$ is $N_{\rm BZ}^\rho = \sum_K N_K^\rho$.
The vector $B$ is determined by the symmetry-representation vector $p$ in Eq.~\eqref{eq:pvectordef} as $B = BR \cdot p$, where $BR$ is $N_{\rm BZ}^\rho \times N_{\rm UC}^\rho$ band-representation (BR) matrix.
The columns of $BR$ are indexed by each irrep $\rho_W^i$ of the site-symmetry group of every WP $W$, and the corresponding column vectors give the symmetry-data vector for the BR induced from $\rho_W^i$.
By introducing a pseudoinverse $BR^\ddagger$ of $BR$, we find
\bg
p = BR^\ddagger \cdot B + p_{{\rm ker},BR},
\label{eq:p_from_B}
\eg
where $p_{{\rm ker},BR}$ is a general vector in $\ker BR$, i.e., $BR \cdot p_{{\rm ker},BR}=\bb 0$.
By expressing the $\Z$SRSIs in the second line of Eq.~\eqref{eq:SRSI_definition} as $\theta_\Z = \Theta^{(0)} \cdot p$, collectively, we have
\ba
\theta_\Z = \Theta^{(0)} \cdot BR^\ddagger \cdot B + \Theta^{(0)} \cdot p_{{\rm ker},BR}.
\ea
In a similar way, we represent $B$ as
\bg
B = BR \cdot [\Theta^{(0)}]^\ddagger \cdot \theta_\Z + BR \cdot p_{{\rm ker},\Theta^{(0)}}.
\label{eq:B_from_p}
\eg
In all SGs with and without SOC, $\Theta^{(0)} \cdot p_{{\rm ker},BR}=\bb 0$ and $BR \cdot p_{{\rm ker},\Theta^{(0)}}=\bb 0$ (see SN 3B).
This establishes a one-to-one mapping between symmetry-data vector $B$ and $\Z$SRSIs for atomic insulators.
We give further details of the construction and explicit examples of the one-to-one mapping between $\Z$SRSIs and symmetry data vectors in SN 3B and SN 3C.

Let us comment on three key implications of this one-to-one mapping.
First, for any atomic insulators $AI_1$ and $AI_2$ with the same symmetry-data vector, there exists some integer $\mc{N}$ such that $\mc{N} AI_1$ and $\mc{N} AI_2$, are stably equivalent.
We can choose $\mc{N}$ by demanding that $\mc{N}\theta_{\Z_n}=0$ for all the $\Z_n$SRSIs in the space group.

Second, all composite RSIs are determined by $\Z$SRSIs.
Local RSIs, particularly at non-maximal WPs, are not mapped uniquely to symmetry-data vectors due to their failure as topological invariants.
To address this, Ref.~\cite{xu2021threedimensional} introduced composite RSIs, which are linear combinations of local RSIs across multiple WPs that can be determined from symmetry-data vector.
However, the definitions of composite RSIs rely on ad hoc methods leaving their organizing principles still to be found.
By contrast, $\Z$SRSIs incorporate composite RSIs as a subset, providing a systematic framework based on the one-to-one mapping.

Finally, while our discussion so far has focused on atomic insulators with symmetry-representation vectors composed of non-negative integers, the linear nature of the one-to-one mapping ensures that $\Z$SRSIs can be assigned to topological phases.
This includes symmetry-indicated fragile or stable topological phases, where Eq.~\eqref{eq:p_from_B} yields a symmetry-representation vector $p$ having negative integers or rational fractions for a given symmetry-data vector, respectively.
As $\Z$SRSIs are directly calculable from symmetry-data vectors, they remain well-defined even for symmetry-indicated topological phases.

\subsection*{SRSIs determine SIs}
\label{sec:si}
The SIs determine whether a given set of bands are necessarily (symmetry-indicated) stable topological bands.
They are expressed in terms of the symmetry-data vector $B$ as
\ba
\kappa = A \cdot B \mod w,
\label{eq:SImatrixAdef}
\ea
where $\kappa$ is the $N_\kappa$-dimensional vector of SIs, $A$ is a $N_\kappa \times N_{\rm BZ}^\rho$ matrix defining the SIs (as derived in Ref.~\cite{Song_2018}), and $w \in \N$ is a $N_\kappa$-dimensional vector.
A non-zero $\kappa$ indicates that the corresponding bands must be stably topologically nontrivial.
For symmetry-indicated stable topological phases, where the symmetry-representation vector $p$ takes fractional values, the $\Z$SRSIs calculated from $B$, take fractional values~\cite{Hwang:2019aa,ZhidaFragileTwist2}.
Since $\Z$SRSIs are in one-to-one correspondence with symmetry-data vectors, the fractional values of $\Z$SRSIs fully specify all nontrivial SIs.
In general, using Eq.~\eqref{eq:B_from_p} combined with $BR \cdot p_{{\rm ker},\Theta^{(0)}} = 0$, we can write
\bg
\kappa= A \cdot BR \cdot [\Theta^{(0)}]^\ddag \cdot \theta_\Z \mod w.
\label{eq:SI_Bvec_correspondence}
\eg
In SN 7B, we provide a comprehensive list of this quantitative relationship between $\Z$SRSIs and SIs in all 230 SGs with and without SOC.

\subsection*{Topological band splitting indicated by $\Z_n$SRSIs}
\label{sec:znsrsi}
In contrast to $\Z$SRSIs, there is no straightforward way to extend $\Z_n$SRSIs to topological bands since they are not determined by momentum space symmetry data
Despite this, we demonstrate that $\Z_n$SRSIs provide a sufficient criterion for band topology, even in cases where symmetry-data vectors or SIs fail diagnose nontrivial topology.
The key idea is as follows:
Suppose we assume that a given set of bands is topologically trivial, allowing the assignment of position space orbital multiplicities and $\Z_n$SRSIs.
If the assigned $\Z_n$SRSIs violate certain consistency conditions, this assumption of trivial topology is invalidated, indicating that the bands must exhibit nontrivial topology.

As a concrete illustration, let us consider a set of valence and conduction bands separated from a band gap.
Denote the Hilbert spaces spanned by the valence and conduction bands as $V$ and $C$, respectively.
We assume that the full Hilbert space $\mc{H}_{Lat} = V \oplus C$ is topologically trivial, and so induced from orbitals at some WPs.
However, the valence and conduction bands corresponding to $V$ and $C$ can individually be atomic, fragile topological, or stable topological.
(If either of $V$ or $C$ is stable topological, then both must be stable topological.)

Now, let us compare the symmetry-data vectors of the full, valence, and conduction bands.
They must satisfy
\ba
B[\mc{H}_{Lat}] = B[V] + B[C].
\label{eq:bvecdecomposition}
\ea
Similarly, let us denote the SRSIs of a set of bands $X$ by $\theta[X]$ ($X=C,V,\mc{H}_{Lat}$).
The $\Z$SRSIs $\theta_\Z[X]$ can be computed from the symmetry-data vector of $X$.
If at least one of $\theta_\Z[X]$ $(X=C,V)$ takes fractional value, the corresponding bands must be stable topological and have nontrivial SIs.
Otherwise, the band splitting is either trivial or non-symmetry-indicated topological (NSIT), implying that either $V$ or $C$ is topological, but its topology is not indicated by the symmetry-data vector.

We now show how $\Z_n$SRSIs can provide sufficient conditions to detect NSIT band splittings.
Suppose that $V$ and $C$ have integer-valued $\Z$SRSIs.
Let us temporarily assume that both $V$ and $C$ correspond to atomic insulators with well-defined $\Z_n$SRSIs.
Under this assumption, the full set of SRSIs is well-defined for $V$, $C$, and $\mc{H}_{Lat}$, and must satisfy
\ba
\theta_{\Z_n}[\mc{H}_{Lat}] = \theta_{\Z_n}[V] + \theta_{\Z_n}[C] \mod n.
\label{eq:rsidecomposition}
\ea
If Eq.~\eqref{eq:rsidecomposition} is not satisfied, it indicates that our initial assumption -- that both $V$ and $C$ are trivial -- is incorrect.
Instead, the gap between $V$ and $C$ must result in band topology not captured by SIs, implying that either $V$ or $C$ must be fragile or NSIT.
Below, we explicitly demonstrate how $\Z_n$SRSIs can identify a topological band gap using a tight-binding model in the SG $P41'$.

\subsection*{$\Z_n$RSIs and Split EBRs}
\label{sec:split_ebr}
A central result of Ref.~\cite{Bradlyn17} is that when any EBR is separated into multiple sets of bands by band gap(s), at least one of these sets must be topological.
This holds true even if the symmetry-data vector of the gapped sets of bands can be decomposed into those of other EBRs.
Previous works~\cite{Bradlyn17,Cano17,Bouhon18Fragile,BarryFragile} have studied cases where the bands constituting an EBR are split by a band gap, and the (stable or fragile) SIs of each resulting gapped sets of bands are nontrivial.

However, there are instances where an EBR (denoted as $EBR$) has a symmetry data vector that is identical to the symmetry data vector for a sum of other EBRs.
This is expressed as
\ba
B[EBR] = B[EBR_1] + B[EBR_2] + \dots.
\label{eq:ebr_split_symdata}
\ea
Nontrivial solutions to Eq.~\eqref{eq:ebr_split_symdata} exist in 51 SGs.
In such cases, the splitting of $EBR$ is NSIT since symmetry-data vectors fail to capture the nontrivial topology of gapped band sets of split EBRs.
In SN 7C, a detailed list of all 211 split EBRs with NSIT band splitting (with and without SOC) is provided, occurring in 51 of the 230 SGs.

Since $\Z_n$SRSIs provide a sufficient criteria even for NSIT band splitting, it is natural to ask how effectively $\Z_n$SRSIs can diagnose all split EBR cases.
For this, we generalize Eq.~\eqref{eq:rsidecomposition} for the putative EBR decomposition implied by Eq.~\eqref{eq:ebr_split_symdata} as
\ba\label{eq:zn_equiv_condition}
\theta_{\Z_n}[EBR] \stackrel{?}{\equiv} \theta_{\Z_n}[EBR_1] + \theta_{\Z_n}[EBR_2] \dots \mod n.
\ea
We find that violations of this condition are sufficient to diagnose NSIT in all but 8 solutions to Eq.~\eqref{eq:ebr_split_symdata} in 5 SGs (all in the presence of SOC).
That is, except for these 8 exceptions, violation of Eq.~\eqref{eq:zn_equiv_condition} implies that the splitting Eq.~\eqref{eq:ebr_split_symdata} cannot imply an (adiabatic) equivalence of band representations.
The eight exceptional cases arise in cases where $EBR$ is stably equivalent (but not equivalent) to a sum of EBRs.
In these cases the fact that the split EBR [$EBR$ in Eq.~\eqref{eq:ebr_split_symdata}] and its hypothetical EBR decompositions, $EBR_1 \oplus EBR_2 \oplus \dots$, have matching SRSIs, including $\Z_n$SRSIs, means that there exists some auxiliary band representation $BR_{\rm aux}$ such that $EBR\oplus BR_{\rm aux} \equiv EBR_1 \oplus EBR_2 \oplus \dots \oplus BR_{\rm aux}$ could be true (see SN 4D).
Nonetheless, the ability of $\Z_n$SRSIs to diagnose almost all split EBR cases -- except 8 exceptions -- underscores the power of SRSIs in capturing band topology beyond the limitations of symmetry-data vectors and SIs.
Note, however, that while SRSIs allow us to exclude the existence of an adiabatic equivalence of split EBRs (in all but 8 cases), they alone do not rule out the theoretical possibility of rewriting an EBR as a sum of other EBRs through a large gauge transformation (i.e., a non-trivial change of basis that cannot be carried out adiabatically).
In many cases, Wilson loop techniques such as those employed in Ref.~\cite{Cano22Invisible} can be used to rule out the existence of such large gauge transformations.
We currently know of no cases where a pair of EBRs is equivalent under such a large gauge transformation, and we conjecture that no such cases exist.
We defer a more thorough investigation of this point to future work.

\begin{figure*}[t]
\centering
\includegraphics[width=0.98\textwidth]{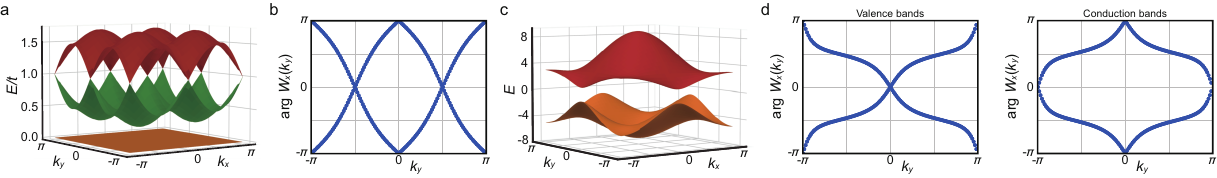}
\caption{Tight-binding models in SG $P41'$.
\textbf{a} Energy spectrum of $H(\bb k)$ in Eq.~\eqref{eq:bloch_hamiltonian}.
There are two degenerate zero-energy flat bands and two dispersive bands.
The dispersive bands form the EBR $({}^1\bar{E}_1{}^2\bar{E}_1)_{1a} \uparrow G$.
\textbf{b} Spectrum of the $k_x$-directed Wilson loop $W_x(k_y)$ for the two flat bands.
The double winding, protected by $C_{2z}T$ symmetry, implies fragile topology.
The Wilson loop of the two dispersive bands does not wind (not shown).
\textbf{c} Energy spectrum of the model in the $\Z_2$-odd phase discussed in SN 5A.
Each band is doubly-degenerate.
\textbf{d} $k_x$-directed Wilson loop spectra of the valence and the conduction bands.
The windings indicate the $\Z_2$-odd topology of both the valence and conduction bands. Wilson loops were computed using the algorithm of Ref.~\cite{devescovi2024tutorial}.}
\label{fig:tb_75_main}
\end{figure*}

\subsection*{Tight-binding model in SG $P41'$}
\label{sec:tbmodel}
We now discuss an example of a split EBR with NSIT and demonstrate how $\Z_n$SRSIs diagnose this case.
Specifically, we consider the SG $P41'$ (No. 75) with SOC and time-reversal symmetry (TRS) in 2D.
The spatial symmetries of this SG include twofold and fourfold rotations, $C_{2z}$ and $C_{4z}$, as well as two translations, $\{E|\hat{x}\}$ and $\{E|\hat{y}\}$.
Note that $C_{2z}$ and $C_{4z}$ transform $(x,y)$ to $(-x,-y)$ and $(-y,x)$, respectively.
We focus on the maximal WP $1a$ and the non-maximal WP $2c$.
The WP $1a$ is located at (0,0), and its site-symmetry group, generated by $C_{4z}$, has two irreps: $({}^1\bar{E}_1 {}^2\bar{E}_1)_{1a}$ and $({}^1\bar{E}_2 {}^2\bar{E}_2)_{1a}$.
The WP $2c$ consists of the positions $(0,1/2)$ and $(1/2,0)$.
At these positions, the site-symmetry group are generated by $\{E|\hat{y}\} C_{2z}$ and $\{E|\hat{x}\} C_{2z}$, respectively.
This group has a unique irrep $({}^1\bar{E} {}^2\bar{E})_{2c}$.

For a split EBR with NSIT band splitting, let us consider a putative EBR decomposition:
\ba
({}^1\bar{E} {}^2\bar{E})_{2c} \uparrow G \stackrel{?}{\equiv} ({}^1\bar{E}_1 {}^2\bar{E}_1)_{1a} \uparrow G \, \oplus \, ({}^1\bar{E}_2 {}^2\bar{E}_2)_{1a} \uparrow G.
\label{eq:p75_decomp_candidate}
\ea
This decomposition satisfies the symmetry-data vector condition in Eq.~\eqref{eq:bvecdecomposition}.
However, a unique $\Z_2$SRSI (see SN 7A), $\theta_5 = m[({}^1\bar{E} {}^2\bar{E})_{2c}] \mod 2$, reveals that the decomposition in Eq.~\eqref{eq:p75_decomp_candidate} is invalid.
Specifically, we find
\bg
\theta_5[({}^1\bar{E} {}^2\bar{E})_{2c}] = 1,
\quad
\theta_5[({}^1\bar{E}_1 {}^2\bar{E}_1)_{1a}] = 0,
\nn
\theta_5[({}^1\bar{E}_2 {}^2\bar{E}_2)_{1a}] = 0,
\label{eq:p4z2srsi}
\eg
contradicting Eq.~\eqref{eq:rsidecomposition}.
Consequently, a gap separating the two sets of bands identified as $({}^1\bar{E}_1 {}^2\bar{E}_1)_{1a} \uparrow G$ and $({}^1\bar{E}_2 {}^2\bar{E}_2)_{1a} \uparrow G$ requires at least one set of bands to be either fragile or non-symmetry-indicated stable topological.
Physically, the decomposition in Eq.~\eqref{eq:p75_decomp_candidate} is impossible because two $({}^1\bar{E} {}^2\bar{E})_{2c}$ irreps can be symmetrically moved away from the WP $2c$; however, a single $({}^1\bar{E} {}^2\bar{E})_{2c}$ irrep alone cannot.

As a concrete an example, consider a tight-binding Hamiltonian where the full Hilbert space $\mc{H}_{Lat}$ is spanned by the irrep $({}^1\bar{E} {}^2\bar{E})_{2c}$.
The basis orbitals for this irrep consist of two Kramers pairs of $s$-like orbitals possessing opposite spins, located at $\hat{x}/2=(1/2,0)$ and $\hat{y}/2=(0,1/2)$ of WP $2c$ in each unit cell.
Denoting the unit cells by Bravais lattice vectors $\bb R$, we label the basis orbitals in each unit cell as $\ket{\bs{R}, \alpha} = (\ket{\bs{R} + \hat{x}/2, \uparrow}, \ket{\bs{R} + \hat{x}/2, \downarrow},\ket{\bs{R} + \hat{y}/2, \uparrow},\ket{\bs{R} + \hat{y}/2, \downarrow})_\alpha$ ($\alpha = 1, \dots, 4$).
To achieve a band splitting described in Eq.~\eqref{eq:p75_decomp_candidate}, we construct trial Wannier states transforming under the EBR $({}^1\bar{E}_1 {}^2\bar{E}_1)_{1a} \uparrow G$ (see SN 5A).
These states can be used to construct a tight-binding Hamiltonian~\cite{Schindler21Noncompact,hwang2021general,graf2021designing}.
We choose the trial states $\ket{\Phi^\pm_{\bb R}}$ with the following (unnormalized) Fourier transforms:
\ba
u^{+}_{\bb k \alpha} &= \frac{1}{\sqrt{2}} \left(\cos \frac{k_x}{2}, -i \sin \frac{k_x}{2}, \cos \frac{k_y}{2}, -\sin \frac{k_y}{2} \right)_\alpha,
\\
u^{-}_{\bb k \alpha} &= \frac{1}{\sqrt{2}} \left(-i \sin \frac{k_x}{2}, -\cos \frac{k_x}{2}, \sin \frac{k_y}{2}, -\cos \frac{k_y}{2} \right)_\alpha,
\ea
where $\bb k=(k_x,k_y)$ lies in the BZ.
The tight-binding Hamiltonian is then constructed as
\ba
[H(\bb k)]_{\alpha \beta} = t(u^{+}_{\bb k \alpha} u^{+*}_{\bb k \beta} + u^{-}_{\bb k \alpha} u^{-*}_{\bb k \beta}),
\label{eq:bloch_hamiltonian}
\ea
with the energy scale $t$.

This model hosts two exact zero-energy flat bands and two dispersive bands, as shown in Fig.~\ref{fig:tb_75_main}\textbf{a}.
In position space, Eq.~\eqref{eq:bloch_hamiltonian} represents the hopping processes between trial states, i.e., $[H(\bb R)]_{\alpha\beta} = \sum_{\bb R} t (\ket{\Phi^+_{\bb R}} \bra{\Phi^+_{\bb R}} + \ket{\Phi^-_{\bb R}} \bra{\Phi^-_{\bb R}})$.
This means that the bands spanned by these trial states become dispersive.
Thus, the dispersive bands correspond to the EBR $({}^1\bar{E}_1 {}^2\bar{E}_1)_{1a} \uparrow G$, as determined by the representation of the trial states $\ket{\Phi^\pm_{\bb R}}$.
Since the full Hilbert space corresponds to $({}^1\bar{E} {}^2\bar{E})_{2c} \uparrow G$, the gapped flat bands, which are spanned by the complement of the trial states correspond to $({}^1\bar{E} {}^2\bar{E})_{2c} \uparrow G \ominus ({}^1\bar{E}_1 {}^2\bar{E}_1)_{1a} \uparrow G$, whose symmetry-data vector matches that of $({}^1\bar{E}_2 {}^2\bar{E}_2)_{1a} \uparrow G$.
As indicated by Eq.~\eqref{eq:p4z2srsi}, this band must be topological -- either with fragile or non-symmetry-indicated stable topology.

The Wilson loop spectrum of the two flat bands exhibits double winding, indicating fragile topology protected by $C_{2z}T$ symmetry~\cite{Benalcazar17,wieder2018axion,KoreanTBG,Ahn:2019aa,Ahn_2019_stiefels,Schindler2022Defects} (see Fig.~\ref{fig:tb_75_main}\textbf{b}).
In contrast, the Wilson loop of the dispersive bands is trivial, consistent with the fact that they correspond to an EBR.
Notably, a phase transition could subsequently induce stable $\Z_2$ quantum spin hall topology in each of the valence and conductance bands, as shown in Fig.~\ref{fig:tb_75_main}\textbf{d} (see SN 5A for more details).
However, unlike the standard $\Z_2$ models like the Kane-Mele model~\cite{Kane05a} and the BHZ model~\cite{Bernevig06}, where both sets of bands can be trivialized via band inversions, it is impossible to remove the windings of both the sets as this model still realizes a split EBR and violates Eq.~\eqref{eq:zn_equiv_condition}.
A hint of this can be seen in the relative $\pi$-phase shift between the spectra of the valence and the conduction band Wilson loops, as shown in Fig.~\ref{fig:tb_75_main}\textbf{d}.
A comprehensive study of this phenomenon is deferred to a future work.

\subsection*{SRSI-based diagnosis of a real material}
To illustrate how SRSIs enable the identification of symmetry-protected phases beyond SIs, we consider obstructed atomic insulators (OAIs).
Although OAIs are characterized by trivial SIs, they can exhibit nontrivial boundary phenomena, which have been linked to enhanced catalytic functionalities~\cite{xu2021threedimensional,li2022obstructed}.
OAIs therefore provide a well-defined and physically meaningful class of materials for which SRSI-based indicators can be constructed systematically.

In general, once the atomic WPs of a material are specified, one can construct $\Z$- or $\Z_m$-valued ($m \ge 2$) OAI indicators as linear combinations of SRSIs that necessarily vanish for any unobstructed phase in that material.
A nonzero value of such an indicator implies that at least one Wannier orbital must reside on a WP not occupied by the constituent atoms, thereby diagnosing an obstructed limit.
The general construction of these SRSI-based OAI indicators is described in SN 6.

As a concrete material example, we consider the compound Y$_3$Al$_2$ [Inorganic Crystal Structure Database (ICSD)~\cite{bergerhoff1983inorganic} collection code 609643], crystallizing in SG $P4_2nm$ (No. 102).
In a conventional unit cell, the crystal contains twelve Y atoms and eight Al atoms, occupying the $4b$ and $4c$ WPs of this SG.
Using the symmetry data for this material, we evaluate the relevant $\Z$SRSI for SG $P4_2nm$.
Combining this $\Z$SRSI with the information that the atoms occupy the $4b$ and $4c$ WPs, we obtain a $\Z_2$-valued OAI indicator, which is nonzero for Y$_3$Al$_2$.
This establishes that Y$_3$Al$_2$ realizes an OAI.
See SN 6 for the detailed analysis.
We note that Y$_3$Al$_2$ has been previously identified as a filling-enforced OAI~\cite{xu2024filling} by exploiting local RSIs along with a set of filling constraints.
While such approaches rely on specific filling constraints, the SRSI-based diagnosis presented here does not depend on filling-enforced conditions and applies more generally.
Specifically, a main advantage of the SRSI-based diagnosis is that it depends only on symmetry data and the atomic WP configuration of a material within any SG.
This highlights the broader applicability of SRSIs as a real-space framework for diagnosing topology beyond SIs.

\section*{Discussion}
\label{sec:discussion}
In this work, we introduced SRSIs as a novel framework that extends the capabilities of TQC beyond SIs.
While SIs have been instrumental in identifying topological materials using momentum-space symmetry data, SRSIs offer a complementary perspective by incorporating position-space information.
Both $\Z$SRSIs and $\Z_n$SRSIs provide more information than SIs;
$\Z$SRSIs are one-to-one mapped to symmetry-data vectors and determine the SIs while symmetry data cannot determine $\Z_n$SRSIs meaning that $\Z_n$SRSIs have information not encoded in the symmetry-data vector.
Moreover, $\Z_n$SRSIs enable the diagnosis of NSIT band splittings (with only 8 rare exceptions in 5 SGs), which cannot be diagnosed by SIs or symmetry-data vectors.

In addition, SRSIs enrich the TQC framework by highlighting the importance of stable equivalence of band representations and offering a more comprehensive classification of both topological and trivial phases.
We demonstrated how distinct EBR can be shown to be adiabatically connected in the presence of auxiliary trivial bands, implying their stable equivalence, based on matching SRSIs.
This stable equivalence is important in several situations.
For example, when studying the topology of bands, extending the set of bands with additional trivial bands (by considering energetically close bands) necessitates a classification scheme based on stable equivalence, ensuring that the classification remains independent of such extensions.
Also, in 3D photonic crystals, Maxwell's equations often give rise to singular band structures for transversely polarized modes~\cite{watanabe2018space,christensen2022location,wang2023non}.
To regularize this singularity, additional auxiliary bands can be introduced~\cite{christensen2022location,wang2023non,morales2025transversality,devescovi2024axion}.
The stable equivalence relation between bands thus naturally arises in the classification of photonic bands, ensuring the independence of the classification from the choice of auxiliary bands.
The application of SRSIs provides a framework connecting stable equivalence and topological classification, which is directly applicable to bands in 3D photonic crystals as well~\cite{hwang2026building}.
This underscores the practical utility of SRSIs in identifying and classifying topological phases in both electronic and metamaterial systems.
Beyond classification, we note that SRSIs are connected to known topological and geometric responses in crystalline systems.
Fractional SRSIs are tied to bulk topological response coefficients via nonzero SIs~\cite{fang2012,Song_2018}, while local RSIs contained within SRSIs control quantized defect-bound and geometric responses~\cite{herzog2022superfluid,herzog2024interacting}.

We also briefly comment on the relation between our approach and the classification scheme based on K theory~\cite{shiozaki2022atiyah,shiozaki2023atiyah}.
The number of $\Z$SRSIs and $\Z_n$SRSIs matches the group structure of the $E^2_{0,0}$ page in the real-space Atiyah–Hirzebruch spectral sequence introduced in Ref.~\cite{shiozaki2023atiyah}, where the equivalence classes are defined by adiabatic deformations of site-symmetry orbitals.
Thus, the SRSIs provide a systematic way to compute this $E^2$-page, and our formulas for the SRSIs in SN 7A give an explicit construction of the elements of the $E^2_{0,0}$ page that was not provided in earlier references.

A further natural follow-up to our work would be the generalization of SRSIs to magnetic space groups~\cite{Watanabe18Magnetic,Ono21Magnetic,Bouhon21Magnetic,Elcoro21MagneticTQC}.
As a concrete example, we analyzed the magnetic SG $F\bar{4}3c$ in SN 3C, demonstrating that our formalism can be directly extended to magnetic systems.
More broadly, recent works based on K theory~\cite{wada2024general} suggest that magnetic SGs may host new types of adiabatic equivalences beyond those in nonmagnetic cases.
Connecting these ideas with SRSIs may offer an interesting direction.
It would also be interesting to develop an efficient algorithm for calculating the SRSIs from density functional theory, enabling high-throughput material searches based on SRSIs and SRSI-derived OAI indicators, as well as the split EBR tables provided in SN 7.
Since SRSIs do not distinguish between topologically distinct but stably equivalent (E)BRs with auxiliary trivial bands, the derivation and characterization of fragile RSIs, which address these distinctions, represent an important avenue for future study.
Finally, additional research is required to definitively exclude the existence of large gauge transformations that could signify equivalences between split EBRs and a linear combination of EBRs.

\section*{Methods}
\subsection*{Constructing SRSIs from adiabatic processes}
We illustrate the construction of stable real-space invariants (SRSIs) using the space group (SG) $Pmm21'$ with spin-orbit coupling.
The Wyckoff positions (WPs) and site-symmetry irreps are summarized in Table~\ref{table:WPs}.
The distinct site-symmetry irreps defining the site-symmetry representation vector $p$ are
\bg
(\bar{E})_{1a}, (\bar{E})_{1b}, (\bar{E})_{1c}, (\bar{E})_{1d},
({}^1\bar{E} {}^2\bar{E})_{2e},
\nn
({}^1\bar{E} {}^2\bar{E})_{2f}, ({}^1\bar{E} {}^2\bar{E})_{2g}, ({}^1\bar{E} {}^2\bar{E})_{2h}, (\bar{A} \bar{A})_{4i},
\label{eq:irreps_method}
\eg
giving $N^\rho_{\rm UC}=9$.
To construct the adiabatic-process matrix $q$, we identify 12 adiabatic processes that span the space of all possible adiabatic deformations in this SG.
As explained in the main text, for example, two $(\bar{E})_{1d}$ irreps can be adiabatically deformed into either a $({}^1\bar{E} {}^2\bar{E})_{2f}$ or a $({}^1\bar{E} {}^2\bar{E})_{2h}$ irrep.
Similarly, two $(\bar{E})$ irreps at the maximal WPs $1a,1b,1c$ can be adiabatically deformed into a $({}^1\bar{E} {}^2\bar{E})$ irrep at the connected non-maximal WPs $(2e,2g)$, $(2f,2g)$, and $(2e,2h)$, respectively.
The remaining four adiabatic processes deform the $(\bar{A}\bar{A})_{4i}$ irrep at the general WP $4i$ into two copies of the irrep ${}^1\bar{E} {}^2\bar{E}$ at a WP $2y$ ($y=e,f,g,h$).
We therefore find $N_{\rm adia} = 12$.
For these 12 processes, the adiabatic-process matrix $q$ is
\ba
q =
\left( \begin{smallmatrix}
2 & 2 & 0 & 0 & 0 & 0 & 0 & 0 & 0 & 0 & 0 & 0 \\
0 & 0 & 2 & 2 & 0 & 0 & 0 & 0 & 0 & 0 & 0 & 0 \\
0 & 0 & 0 & 0 & 2 & 2 & 0 & 0 & 0 & 0 & 0 & 0 \\
0 & 0 & 0 & 0 & 0 & 0 & 2 & 2 & 0 & 0 & 0 & 0 \\
-1 & 0 & 0 & 0 & -1 & 0 & 0 & 0 & 2 & 0 & 0 & 0 \\
0 & 0 & -1 & 0 & 0 & 0 & -1 & 0 & 0 & 2 & 0 & 0 \\
0 & -1 & 0 & -1 & 0 & 0 & 0 & 0 & 0 & 0 & 2 & 0 \\
0 & 0 & 0 & 0 & 0 & -1 & 0 & -1 & 0 & 0 & 0 & 2 \\
0 & 0 & 0 & 0 & 0 & 0 & 0 & 0 & -1 & -1 & -1 & -1
\end{smallmatrix} \right),
\label{eq:qmatrix_method}
\ea
where the row basis is ordered as in Eq.~\eqref{eq:irreps_method}, and each column represents an adiabatic process described above.
For instance, the first column of $q$ represents the adiabatic process, $2(\bar{E})_{1a} \Leftrightarrow (^1 \bar{E} ^2 \bar{E})_{2e}$, which creates (removes) two $(\bar{E})_{1a}$ irreps while removing (creating) a $(^1 \bar{E} ^2 \bar{E})_{2e}$ irrep.

The Smith decomposition of the integer matrix $q$ is given by $L \cdot \Lambda \cdot R$, where $L$ and $R$ are integer unimodular matrices.
In the present example, the matrices $L$, $\Lambda$, and $R$ are given by
\bg
L = \left( \begin{smallmatrix}
2 & 2 & 0 & -4 & 2 & -1 & 0 & 2 & -1 \\
0 & 0 & 2 & 0 & -2 & 1 & -1 & 0 & 0 \\
0 & 0 & 0 & 4 & -2 & 2 & 0 & -3 & 0 \\
0 & 0 & 0 & 2 & 0 & 0 & 1 & -1 & 0 \\
-1 & 0 & 0 & 2 & 0 & 0 & 0 & 0 & -1 \\
0 & 0 & -1 & 0 & 0 & 0 & 0 & 0 & 0 \\
0 & -1 & 0 & 0 & 0 & 0 & 0 & 0 & 0 \\
0 & 0 & 0 & -1 & 1 & -1 & 0 & 1 & 0 \\
0 & 0 & 0 & -1 & 0 & 0 & 0 & 0 & 1
\end{smallmatrix} \right),
\nn
\Lambda = \left( \begin{smallmatrix}
1 & 0 & 0 & 0 & 0 & 0 & 0 & 0 & 0 & 0 & 0 & 0 \\
0 & 1 & 0 & 0 & 0 & 0 & 0 & 0 & 0 & 0 & 0 & 0 \\
0 & 0 & 1 & 0 & 0 & 0 & 0 & 0 & 0 & 0 & 0 & 0 \\
0 & 0 & 0 & 1 & 0 & 0 & 0 & 0 & 0 & 0 & 0 & 0 \\
0 & 0 & 0 & 0 & 1 & 0 & 0 & 0 & 0 & 0 & 0 & 0 \\
0 & 0 & 0 & 0 & 0 & 2 & 0 & 0 & 0 & 0 & 0 & 0 \\
0 & 0 & 0 & 0 & 0 & 0 & 2 & 0 & 0 & 0 & 0 & 0 \\
0 & 0 & 0 & 0 & 0 & 0 & 0 & 2 & 0 & 0 & 0 & 0 \\
0 & 0 & 0 & 0 & 0 & 0 & 0 & 0 & 0 & 0 & 0 & 0
\end{smallmatrix} \right),
\nn
R = \left( \begin{smallmatrix}
1 & 0 & 0 & 0 & 1 & 0 & 0 & 0 & 0 & 2 & 2 & 2 \\
0 & 1 & 0 & 1 & 0 & 0 & 0 & 0 & 0 & 0 & -2 & 0 \\
0 & 0 & 1 & 0 & 0 & 0 & 1 & 0 & 0 & -2 & 0 & 0 \\
0 & 0 & 0 & 0 & 0 & 0 & 0 & 0 & 1 & 1 & 1 & 1 \\
0 & 0 & 0 & -2 & 0 & 1 & 0 & -1 & 1 & -3 & 1 & -1 \\
0 & 0 & 0 & -1 & -1 & 1 & 0 & 1 & 1 & -1 & 1 & -3 \\
0 & 0 & 0 & 0 & -1 & 0 & 1 & 2 & 0 & 0 & 0 & -2 \\
0 & 0 & 0 & 0 & -1 & 0 & 0 & 1 & 1 & 1 & 1 & -1 \\
0 & 0 & 0 & 0 & -1 & 0 & 0 & 2 & 0 & 0 & 0 & -2 \\
0 & 0 & 0 & 0 & 0 & 0 & 0 & 0 & 0 & 1 & 0 & 0 \\
0 & 0 & 0 & 0 & 0 & 0 & 0 & 0 & 0 & 0 & 1 & 0 \\
0 & 0 & 0 & 0 & 0 & 0 & 0 & 0 & 0 & 0 & 0 & 1
\end{smallmatrix} \right).
\eg
The diagonal entries of $\Lambda$ are
\bg
{\rm diag} (\Lambda) = (1, 1, 1, 1, 1,2,2,2,0)^T.
\eg
From this, we find that ${\rm rank}(q)=8$.
According to the definition in Eq.~\eqref{eq:SRSI_definition}, this decomposition yields three $\Z_2$SRSIs and a single $\Z$SRSI, since $N^\rho_{\rm UC}-{\rm rank} (q) = 1$.
We discard the rows $i$ corresponding to $\Lambda_{ii}=1$, as these generate trivial $\Z_1$SRSIs and therefore contain no physical information.
The remaining three rows, corresponding to $\Lambda_{ii}=2$, imply the existence of three $\Z_2$SRSIs.
To construct the SRSI matrices explicitly, we compute the inverse of $L$:
\bg
L^{-1} = 
\left( \begin{smallmatrix}
1 & 1 & 1 & 1 & 1 & 2 & 2 & 2 & 2 \\
0 & 0 & 0 & 0 & 0 & 0 & -1 & 0 & 0 \\
0 & 0 & 0 & 0 & 0 & -1 & 0 & 0 & 0 \\
1 & 1 & 1 & 1 & 2 & 2 & 2 & 2 & 3 \\
1 & 0 & 1 & 0 & 2 & 0 & 2 & 1 & 3 \\
2 & 1 & 1 & 1 & 4 & 2 & 4 & 0 & 6 \\
0 & 0 & -1 & 1 & 0 & 0 & 0 & -2 & 0 \\
2 & 2 & 1 & 2 & 4 & 4 & 4 & 2 & 6 \\
1 & 1 & 1 & 1 & 2 & 2 & 2 & 2 & 4
\end{smallmatrix} \right).
\eg

Let us denote the three $\Z_2$SRSIs $(\theta_1,\theta_2,\theta_3)$ and the $\Z$SRSI $\theta_4$ as
\bg
(\theta_1, \theta_2, \theta_3)^T = \Theta^{(2)} \cdot p,
\quad
\theta_4 = \Theta^{(0)} \cdot p.
\eg
Then, following the definition of SRSIs in Eq.~\eqref{eq:SRSI_definition}, we obtain
\ba
\Theta^{(2)} &= (L^{-1})_{6,7,8} \text{ mod } 2 =
\bpm
0 & 1 & 1 & 1 & 0 & 0 & 0 & 0 & 0 \\
0 & 0 & 1 & 1 & 0 & 0 & 0 & 0 & 0 \\
0 & 0 & 1 & 0 & 0 & 0 & 0 & 0 & 0
\epm,
\nn
\Theta^{(0)} &= (L^{-1})_{9} =
\bpm
1 & 1 & 1 & 1 & 2 & 2 & 2 & 2 & 4
\epm.
\ea
Note that the matrices $\Theta^{(0)}$ and $\Theta^{(2)}$ are not unique, since they may always be redefined by unimodular transformations.
For example, by performing the Hermite decomposition~\cite{cohen2013course} of $\Theta^{(2)}$, we obtain the redefined matrix
\ba
\Theta^{(2)}_{\rm new} =&
\bpm 1 & -1 & 0 \\ 0 & 0 & 1 \\ 0 & 1 & -1 \epm \cdot \Theta^{(2)}
\nn
=& \bpm
0 & 1 & 0 & 0 & 0 & 0 & 0 & 0 & 0 \\
0 & 0 & 1 & 0 & 0 & 0 & 0 & 0 & 0 \\
0 & 0 & 0 & 1 & 0 & 0 & 0 & 0 & 0
\epm.
\ea
Using the $\Z$SRSI together with the $\Z_2$SRSIs defined by $\Theta^{(0)}$ and $\Theta^{(2)}_{\rm new}$, we recover the same expressions for the SRSIs in SG $Pmm21'$ as those presented in Eq.~\eqref{eq:srsi_pmm2} of the main text.
A more detailed discussion of the algorithm for obtaining SRSIs and their possible redefinitions is provided in SN 3A and additional examples for other SGs are presented in SN 3C.

\section*{Data availability}
The data supporting the findings of this study are available from the corresponding author upon request.

\section*{Code availability}
The code supporting the findings of this study are available from the corresponding author upon request.

\bibliography{refs}

\section*{Acknowledgements}
The authors thank Ken Shiozaki for fruitful discussions.

\section*{Funding}
The initial work of Y.H., V.G., and B.B. was supported by the Air Force Office of Scientific Research under award number FA9550-21-1-0131 and the National Science Foundation under grant No. DMR-1945058.
Y.H. received additional support from the US Office of Naval Research (ONR) Multidisciplinary University Research Initiative (MURI) grant N00014-20-1-2325 on Robust Photonic Materials with High-Order Topological Protection.
F.S. and Y.H. were supported by a UKRI Future Leaders Fellowship MR/Y017331/1.
Z.-D. S. were supported by National Natural Science Foundation of China (General Program No. 12274005), National Key Research and Development Program of China (No. 2021YFA1401900), and Innovation Program for Quantum Science and Technology (No. 2021ZD0302403).
L.E. was supported by the Government of the Basque Country (Project IT1458-22) and the Spanish Ministry of Science and Innovation (PID2019-106644GB-I00).
B.B. received additional support from the National Science Foundation under grant No. DMR-2510219.
B.A.B. was supported by NSF-MRSEC Grant No. DMR-2011750, Simons Investigator Grant Nos. 404513 and SFI-MPS-NFS-00006741-01, ONR Grant No. N00014-20-1-2303, the Schmidt Fund for Innovative Research, the BSF Israel US Foundation Grant No. 2018226, the Gordon and Betty Moore Foundation through Grant No. GBMF8685 towards the Princeton theory program and Grant No. GBMF11070 towards the EPiQS Initiative, and the Princeton Global Network Fund.
L.E. and B.A.B. acknowledge additional support through the European Research Council (ERC) under the European Union's Horizon 2020 research and innovation program (Grant Agreement No. 101020833).
\\

\section*{Author contributions}
F.S., Z.S., B.B., and B.A.B. conceived of the project. Y.H. and F.S. derived the properties of the SRSIs and analyzed specific examples with input from B.B., B.A.B., Z.S., and L.E.. The SRSIs were computed for all space groups by L.E.. The SRSI mismatch criteria for band topology were formulated by F.S. and Y.H., and concrete examples were analyzed by Y.H. and V.G.. The manuscript was prepared by F.S., Y.H., V.G., and L.E. with input from all authors. B.B. and B.A.B. oversaw the work and were responsible for the overall research direction.

\section*{Competing interests}
The authors declare no competing interest.

\end{document}


\title{Supplementary Information:\\ Stable Real-Space Invariants and Topology Beyond Symmetry Indicators}

\author{Yoonseok Hwang}
\affiliation{Department of Physics, University of Illinois Urbana-Champaign, Urbana IL 61801, USA}
\affiliation{Anthony J. Leggett Institute for Condensed Matter Theory, University of Illinois Urbana-Champaign, Urbana IL 61801, USA}
\affiliation{Blackett Laboratory, Imperial College London, London SW7 2AZ, United Kingdom}

\author{Vaibhav Gupta}
\affiliation{Department of Physics, University of Illinois Urbana-Champaign, Urbana IL 61801, USA}
\affiliation{Anthony J. Leggett Institute for Condensed Matter Theory, University of Illinois Urbana-Champaign, Urbana IL 61801, USA}

\author{Frank Schindler}
\affiliation{Blackett Laboratory, Imperial College London, London SW7 2AZ, United Kingdom}

\author{Luis Elcoro}
\affiliation{Department of Physics, University of the Basque Country UPV/EHU, Apartado 644, 48080 Bilbao, Spain}
\affiliation{EHU Quantum Center, University of the Basque Country UPV/EHU, Apartado 644, 48080 Bilbao, Spain}

\author{Zhida Song}
\affiliation{International Center for Quantum Materials, School of Physics, Peking University, Beijing 100871, China}

\author{B. Andrei Bernevig}
\affiliation{Department of Physics, Princeton University, Princeton, NJ 08544, USA}
\affiliation{Donostia International Physics Center, P. Manuel de Lardizabal 4, 20018 Donostia-San Sebastian, Spain}
\affiliation{IKERBASQUE, Basque Foundation for Science, Bilbao, Spain}

\author{Barry Bradlyn}
\affiliation{Department of Physics, University of Illinois Urbana-Champaign, Urbana IL 61801, USA}
\affiliation{Anthony J. Leggett Institute for Condensed Matter Theory, University of Illinois Urbana-Champaign, Urbana IL 61801, USA}

\maketitle
\begin{center}
{\large \bf Supplementary Notes}
\end{center}

\tableofcontents

\section{Review of basic notation}
We begin by recalling the central concepts of Topological Quantum Chemistry (TQC)~\cite{Bradlyn17} and symmetry indicators (SIs)~\cite{Po_2017}.

\subsection{Elementary band representations}
\label{sec:ebr_intro}
Consider the electronic band structure of a crystal with space group (SG) $G$, with or without spin-orbit coupling (SOC).
The site-symmetry group of a submanifold (point, line, or plane) in the unit cell of the crystalline lattice is comprised of all operations in $G$ which leave the submanifold invariant.
The collection of all points with conjugate site-symmetry groups is referred to as a Wyckoff position (WP): the smaller the site-symmetry group (i.e., the larger the index of the site-symmetry group in the point group of the space group), the larger the multiplicity of the WP.
The site-symmetry group of a maximal WP is, by definition, not contained as a proper subgroup in the site-symmetry groups of any neighboring WPs connected to that maximal WP.
An elementary band representation (EBR) is a representation of $G$ that is induced from an irreducible representation (irrep) of the site-symmetry group (which we will refer to as a site-symmetry irrep for convenience) at a maximal WP~\cite{ZakEBR1,ZakEBR2,Slager17,Po_2017,Bradlyn17}, if this irrep is not contained in the list of exceptions of Supplementary Reference (SRef.)~\cite{Bradlyn17}\footnote{We will give an explicit example of such an exceptional case in Supplementary Note~\ref{sec:rsi_algorithm}.}.
Physically, an EBR corresponds to the electronic bands that arise from placing localized orbitals transforming in a given site-symmetry irrep at a maximal WP.
We write
\bg
{\rm EBR} = \rho_W \uparrow G, \nonumber
\eg
which indicates the $G$ representation that is induced from the irrep $\rho$ of the site-symmetry group at maximal WP $W$.
Throughout this work, we will follow the notation of the Bilbao Crystallographic Server (BCS)~\cite{bilbao-server,Aroyo2011183} for denoting the SGs, WPs, site-symmetry irreps, and EBRs.

A band representation (BR) is a direct sum of EBRs,
\bg
{\rm BR} = {\rm EBR} \oplus {\rm EBR}' \oplus \dots, \nonumber
\eg
and describes an {\it atomic insulator}.
The occupied band subspace (valence bands) of an atomic insulator admits exponentially localized and symmetric Wannier states~\cite{Bradlyn17}, which transform as the respective EBR or BR under the action of $G$.
When the Wannier states are centered on the lattice orbitals (the WPs occupied by the atoms) and share their symmetry representation, the atomic insulator is called {\it unobstructed}.
On the other hand, the atomic insulator is called {\it obstructed} when its Wannier centers cannot be made to coincide with the location of the lattice orbitals or their symmetry transformation.

We define adiabatic transformations so that they neither close the band gap between occupied (valence) and unoccupied (conductance) bands of an insulator nor break any symmetries.
In the following, the only relevant symmetries will be the SG $G$ and, when indicated, time-reversal symmetry (TRS).
Adiabatic transformations do not change the exponential localizability and symmetry of the Wannier states of an atomic insulator.
However, they may move the Wannier centers around the unit cell in a symmetric fashion if this is compatible with their site-symmetry representations.
By applying adiabatic transformations, the Wannier centers of an atomic insulator can always be localized on maximal WPs~\cite{Bradlyn17}.
As such, unless otherwise stated we will focus our attention on Wannier states centered at maximal WPs.

A set of occupied bands that is not an atomic insulator, and so does not allow for any representation in terms of exponentially localized and symmetric Wannier states, is called {\it topological}.
A classic example is the Chern insulator in two spatial dimensions (2D), which is stable even in the absence of nontrivial SG symmetries.
Many more types of topological bands exist for different choices of $G$, with and without TRS~\cite{Po_2017,Bradlyn17}.

\subsection{Momentum-space symmetry data}
One of the central accomplishments of TQC is establishing a link between the position space description of band representations presented in Supplementary Note (SN)~\ref{sec:ebr_intro} and the symmetry properties of electronic bands in momentum space.
For a set of bands, the momentum-space symmetry data is encapsulated by the vector of irrep multiplicities
\ba
B = \left[
m(\rho^1_{K_1}), m(\rho^2_{K_1}), \dots, m(\rho^1_{K_2}), m(\rho^2_{K_2}), \dots
\right]^T,
\label{eq:symmetrydatavec_definition}
\ea
which we call the symmetry-data vector.
Here, the nonnegative integers $m(\rho)$ denote the multiplicity of the irrep $\rho$.
Also, $\rho^i_{K}$ for $i=1, \dots, N_K^\rho$ is the $i$-th irrep of the little group\footnote{The little group is comprised of all crystal symmetry operations that leave the given crystal momentum invariant up to a reciprocal lattice vector.} at the high-symmetry momentum (HSM) $K$ in the Brillouin Zone (BZ).
Similar to maximal WPs, we define HSMs as BZ submanifolds where each point has the same little group, and this little group is not a proper subgroup of the little groups of any connected momenta~\cite{Bradlyn17-2}.
We define the total number of distinct BZ irreps by $N_{\rm BZ}^\rho = \sum_K N_K^\rho$.
We will follow the notation of the BCS~\cite{bilbao-server} for denoting BZ irreps.

\subsubsection{Atomic insulators}
For an atomic insulator (unobstructed or obstructed), the symmetry-data vector $B$ follows directly from the orbital or site-symmetry irrep content of the unit cell.
The site-symmetry irrep content for WP $W$ is encoded in the site-symmetry representation vector $p_W$.
We further denote by $p$ the collection of $p_W$ for all WPs $W$:
\bg
p_W = \left[ m(\rho^1_W), m(\rho^2_W), \dots, m \left(\rho^{N^\rho_W}_W\right) \right]^T,
\quad
p = \bpm p_{W_1} \\ p_{W_2} \\ \vdots \epm,
\label{eq:p_W}
\eg
where $m(\rho^i_W)$ is the multiplicity of site-symmetry irrep $\rho^i_W$ at a WP $W$ ($i=1,\dots,N^\rho_W$).
In other words, $m(\rho^i_W)$ counts the occupied Wannier states of the atomic insulator that are located at the WP $W$ and transform in the irrep $\rho^i_W$.
(In the case of an unobstructed atomic insulator, Wannier states can be constructed such that their positions coincide with the positions of the lattice orbital basis states.)
The total number of distinct real-space irreps at all WPs of the unit cell is
\ba
N_{\rm UC}^\rho = \sum_{W \in {\rm WPs}} N_W^\rho.
\label{eq:nucmaxdef}
\ea

The symmetry-data vector $B$ for an atomic insulator is determined by the following linear relation,
\bg
B = \BR \cdot p.
\label{eq:B_from_p}
\eg
Here, the band representation matrix~\cite{Bradlyn17,Song20,ZhidaFragileTwist2} $\BR$ (sometimes colloquially referred to as the EBR matrix when entries of $p$ and the rows of $\BR$ are restricted to site-symmetry irreps that induce EBRs) is a $N_{\rm BZ}^\rho \times N_{\rm UC}^\rho$ matrix.
The columns are indexed by each irrep $\rho_W^i$ of the site-symmetry group of every WP $W$, and the corresponding column vectors give the symmetry-data vector for the BR induced from $\rho_W^i$.
Note that we considered all WPs to define the site-symmetry representation vector $p$ in Supplementary Equation (SEq.)~\eqref{eq:p_W}.
The rows of $p$ or the columns of $\BR$ corresponding to non-maximal WPs do not correspond to EBRs but instead to BRs that decompose linearly into EBRs.
We nevertheless include these linearly dependent columns in $\BR$ in order to study the adiabatic deformation processes between site-symmetry irreps at maximal and non-maximal WPs.
Thus, we refer to $\BR$ the BR matrix in this work.
Note that the linearly dependent columns in $\BR$ corresponding to non-maximal WPs were not included in SRefs.~\cite{Bradlyn17,Song20,ZhidaFragileTwist2} that originally defined the EBR matrix.
The BR matrix $\BR$ does {\it not} have full rank, because different orbital configurations in real space, when arranged in a periodic lattice, may induce the same representation of $G$.

\subsubsection{General band structures}
\label{subsubsec:genbandstruct_pvec}
The relationship $B = \BR \cdot p$ [SEq.~\eqref{eq:B_from_p}] can be generalized beyond atomic insulators to define a corresponding $p$ for non-atomic insulating bands.
However, because the BR matrix $\BR$ generically is not an invertible matrix, the site-symmetry representation vector $p$ is not uniquely determined from the symmetry-data vector $B$.
Thus, there are multiple different choices of $p$ for a given $B$ (their relationship will be made precise in SN~\ref{sec:realspacedata}).
Moreover, the entries of $p$ cannot always be chosen as non-negative integers:
\begin{itemize}
\item[(1)] When $p$ is integer-valued but at least some entries of $p$, given $B$, cannot be chosen positive or zero, the corresponding symmetry-data vector $B$ belongs to a symmetry-indicated fragile topological phase~\cite{AshvinFragile,Song20} that can be transformed into an atomic insulator only upon the addition of further atomic insulator bands to the occupied subspace.
\item[(2)] When some entries of $p$, given $B$, cannot be chosen as integers (so that they must take fractional values), the corresponding symmetry-data vector $B$ belongs to a symmetry-indicated stable topological phase.
Bands with stable topology cannot be expressed in terms of exponentially-localized and symmetric Wannier states even when atomic bands are added.
\end{itemize}

From the symmetry-data vector $B$, a set of symmetry indicators (SIs) can be computed that give sufficient conditions for a set of bands belong to case (1) or (2)~\cite{Po_2017,Bradlyn17} (see also SN~\ref{sec:stableindexdef}).
Crucially, such SIs do not fully capture band topology, in that they only provide sufficient but not necessary criteria for a given set of bands to be fragile or stable topological~\cite{Bradlyn17,jenbarryreview21}.
In SN~\ref{sec:realspacecrits}, we discuss types of band structures where all entries of $p$ can be chosen as non-negative integers, yet these bands must still be topological as guaranteed by real-space information.

\section{Review of local real-space invariants (RSIs)}
\label{sec:realspacedata}
We next revisit the definition of local real space invariants (RSIs)~\cite{ZhidaFragileTwist2}.
First, we introduce our notation for $\Z$- and $\Z_n$-valued {\it local} RSIs in SNs~\ref{sec:zindicators_intro} and \ref{sec:znindicators_intro}.
Then we explain how the local RSIs can be derived in SN~\ref{sec:rsi_algorithm}, and illustrate more details by studying some illustrative SGs in SN~\ref{sec:examples}.
We also show that both $\Z$- and $\Z_n$-valued local RSIs cannot be uniquely determined from the symmetry-data vector $B$ [SEq.~\eqref{eq:symmetrydatavec_definition}] in general by showing explicit counterexamples.
Relatedly, we show how local RSIs corresponding to non-maximal Wyckoff positions can fail to be adiabatic invariants of a set of bands.
These observations motivate us to develop the notion of {\it stable} RSIs in SN~\ref{sec:globalrsis}, which are invariant under all adiabatic deformations and whose $\Z$-valued subset have a one-to-one mapping to $B$.
We will also show that the $\Z_n$-valued stable RSIs capture extra information not contained in $B$.

\subsection{$\Z$-valued local RSIs}
\label{sec:zindicators_intro}
At each WP $W$ (either maximal or non-maximal), we can define a set of $\Z$-valued local RSIs ($\Z$RSIs) $(\delta_{\Z W})_\mu$ for $\mu=1, \dots, N^\delta_{\Z W}$, that are linear functions of the multiplicities $m(\rho^i_W)$ of site-symmetry irreps $\rho^i_W$ at WP $W$ ($i=1,\dots,N^\rho_W$).
We write
\ba
\delta_{\Z W} = \Delta_{\Z W} \cdot p_W \in \Z^{N^\delta_{\Z W}},
\quad
p_W = \left[ m(\rho^1_W), m(\rho^2_W), \dots \right]^T,
\label{eq:zrsi_introduction}
\ea
where the $W$-position $\Z$RSI matrix $\Delta_{\Z W}$ is an integer-valued $N^\delta_{\Z W} \times N^\rho_W$ matrix.
This matrix has the defining property that the $\delta_{\Z W}$ calculated via SEq.~\eqref{eq:zrsi_introduction} are fully invariant under bringing in orbitals to $W$ from lower-symmetric WPs connected to $W$ in a symmetric fashion.
(Here, a lower-symmetric WP connected to $W$ is defined by having a site-symmetry group that is a proper subgroup of the site-symmetry group of $W$.)
In other words, the $W$-position site-symmetry representation vectors $p_W$ that correspond to orbitals that can be symmetrically moved away from $W$ must lie in the kernel of $\Delta_{\Z W}$.
The derivation of the $W$-position $\Z$RSI matrix $\Delta_{\Z W}$ from this property was first given in SRef.~\cite{ZhidaFragileTwist2} and is reviewed in SN~\ref{sec:rsi_algorithm}.

We can collect the $\delta_{\Z W}$ for all WPs $W$ to introduce the set of all $\Z$RSIs defined in the unit cell:
\ba
\delta_\Z
= \bpm \delta_{\Z W_1} \\ \delta_{\Z W_2} \\ \vdots \epm
= \Delta_{\Z} \cdot p \in \Z^{N^\delta_\Z},
\label{eq:fullzrsidef}
\ea
where $N^\delta_\Z = \sum_{W \in {\rm WPs}} N^\delta_{\Z W}$ and $p$ is the site-symmetry representation vector defined in SEq.~\eqref{eq:p_W}.
According to SEq.~\eqref{eq:fullzrsidef}, the $\Z$RSI matrix
\ba
\Delta_{\Z} = \bigoplus_{W \in {\rm WPs}} \Delta_{\Z W}
\ea
is an $N^\delta_{\Z} \times N_{\rm UC}^\rho$ matrix that is block-diagonal with different blocks belonging to different WPs.
[Recall from SEq.~\eqref{eq:nucmaxdef} that $N_{\rm UC}^\rho = \sum_{W \in {\rm WPs}} N_W^\rho$.]

Because we defined the $\Z$RSIs in real space, one can ask if they can be determined from the symmetry-data vector $B$ [SEq.~\eqref{eq:symmetrydatavec_definition}], which can be directly obtained from a given band structure.
However, we will show that the $\Z$RSI cannot be uniquely determined from the symmetry-data vector $B$ for general SGs.
The map from a site-symmetry representation vector $p$ to a symmetry-data vector $B$ [SEq.~\eqref{eq:B_from_p}] implies that
\ba
p = \BR^\ddagger \cdot B + p_0,
\label{eq:pinverserelation}
\ea
where $\BR^\ddagger$ is the Moore-Penrose pseudoinverse of the BR matrix $\BR$, and $p_0$ is an arbitrary vector in the kernel of $\BR$ [see SEq.~\eqref{eq:pseudoinv} and the surrounding discussion for the definition of the pseudoinverse].
By combining SEqs.~\eqref{eq:fullzrsidef} and \eqref{eq:pinverserelation}, we obtain
\ba
\delta_\Z = \Delta_\Z \cdot \BR^\ddagger \cdot B + \Delta_\Z \cdot p_0, \quad \BR \cdot p_0 = 0.
\label{eq:zrsis_fromb}
\ea
Hence, in any SG where $\Delta_{\Z} \cdot p_0 = 0$ does not hold for all $p_0$ in the kernel of $\BR$, then the $\Z$RSIs in that SG do not follow uniquely from the symmetry-data vector $B$, and multiple inequivalent $\Z$RSI configurations $\delta_\Z$ may be compatible with the same symmetry-data vector $B$ SG.
We show in SN~\ref{sec:examples} through an explicit example that there exist SGs where the $\Z$RSIs are not uniquely determined by the symmetry-data vector.
For a list of such local RSIs in all wallpaper groups and more details, we refer the readers again to SRef.~\cite{ZhidaFragileTwist1}.
Moreover, SRef.~\cite{xu2021threedimensional} introduced linear combinations of the above local RSIs called composite RSIs which can also be determined from the symmetry-data vector.
We show in SN~\ref{sec:globalrsis} that those local RSIs which can be determined by the symmetry-data vector and all the composite RSIs can be obtained from stable RSIs defined below.
In this sense, the stable RSI framework already encompasses the information provided by the local RSIs.

\subsection{$\Z_n$-valued local RSIs}
\label{sec:znindicators_intro}
In addition to $\Z$RSIs, there also exist $\Z_n$-valued real space invariants ($\Z_n$RSIs) $(\delta_{\Z_n W})_\mu$ for $\mu=1, \dots, N^\delta_{\Z_n W}$, defined modulo $n$.
Note that SRef.~\cite{xu2021threedimensional} found that only $\Z$- and $\Z_2$-valued RSIs exist in any of the 1651 magnetic SG, meaning that only $n=2$ is allowed for $\Z_n$RSIs.
Nevertheless, it will be useful for the discussion here and in SN~\ref{sec:globalrsis} to formulate the theory for general $n$.
At each WP $W$, the $\Z_n$RSI $\delta_{\Z_n W}$ follows the relation 
\ba
\delta_{\Z_n W} = \Delta_{\Z_n W} \cdot p_W \mod n \in \{ 0,1, \dots, n-1 \}^{N^\delta_{\Z_n W}},
\label{eq:znrsi_introduction}
\ea
where the $W$-position $\Z_n$RSI matrix $\Delta_{\Z_n W}$ is an integer-valued $N^\delta_{\Z_n W} \times N^\rho_W$ matrix.
Like the $\Z$RSIs, the $\delta_{\Z_n W}$ are invariant under symmetrically bringing in orbitals to $W$ from lower-symmetric WPs connected to $W$.
The derivation of the $W$-position $\Z_n$RSI matrix $\Delta_{\Z_n W}$ is reviewed in SN~\ref{sec:rsi_algorithm}.
Using the site-symmetry representation vector $p$ defined in SEq.~\eqref{eq:B_from_p}, we may again summarize SEq.~\eqref{eq:znrsi_introduction} for all WPs ($W_1, W_2, \dots$) in the unit cell as
\ba
\delta_{\Z_n} =
\bpm
\delta_{\Z_n W_1} \\ \delta_{\Z_n W_2} \\ \vdots \epm
= \Delta_{\Z_n} \cdot p \mod n \in \{ 0,1, \dots, n-1 \}^{N^\delta_{\Z_n}},
\label{eq:zn_summary_calc}
\ea
where $N^\delta_{\Z_n} = \sum_{W \in {\rm WPs}} N^\delta_{\Z_n W}$, so that $\Delta_{\Z_n}$ is an $N^\delta_{\Z_n} \times N_{\rm UC}^\rho$ matrix that is block-diagonal with different blocks belonging to different WPs.

From SEq.~\eqref{eq:pinverserelation}, we have that
\ba
\delta_{\Z_n} = \Delta_{\Z_n} \cdot \BR^\ddagger \cdot B + \Delta_{\Z_n} \cdot p_0 \mod n.
\label{eq:znrsi_from_b}
\ea
Like the $\Z$RSIs $\delta_\Z$, the $\Z_n$RSIs $\delta_{\Z_n}$ cannot be uniquely determined from symmetry-data vector $B$ in general, as shown in SN~\ref{sec:examples}.
This follows from the fact that, in some SGs, $\Delta_{\Z_n} \cdot p_0 \mod n \neq 0$ for $p_0$ in the kernel of $\BR$.
Thus, multiple inequivalent $\Z_n$RSI configurations $\delta_{\Z_n}$ may be compatible with the same symmetry-data vector $B$.
This is analogous to the ambiguity that exists for $\Z$RSIs in SEq.~\eqref{eq:zrsis_fromb}; as in that case, we refer the reader to SRef.~\cite{xu2021threedimensional} and SN~\ref{sec:globalrsis} for more details.

\subsection{Derivation of the local RSIs}
\label{sec:rsi_algorithm}
For completeness, we briefly review the method for constructing the RSIs developed in SRef.~\cite{ZhidaFragileTwist2}.
The calculation of the local RSIs in this SN and the SRSIs in SN~\ref{sec:globalrsis} make use of several databases hosted in the BCS~\cite{bilbao-server,Aroyo2011183}.
In particular, the lists of WPs and their site-symmetry groups have been obtained from the \texttt{WYCKPOS} tool~\cite{wyckpos_tool}, the irreps of the site-symmetry groups from the \texttt{REPRESENTATIONS DPG} tool~\cite{rep_tool}, and the EBRs from the \texttt{BANDREP} tool~\cite{bandrep_tool}.
Our first step is to define, at each WP $W$ (which can be either maximal or non-maximal WP), a matrix $q_W$ representing a list of ``trivial'' configurations.
To define $q_W$, we consider adiabatic processes where orbitals at lower-symmetry WPs $(W'_1, W'_2, \dots)$ connected to $W$ are adiabatically deformed to orbitals at $W$.
Each configuration of low-symmetry orbitals at $W'_m$ induces an orbital configuration at $W$ through the adiabatic process of moving orbitals from $W_m'$ to $W$; we refer to these configurations of orbitals at $W$ as trivial.
This means that $N^{\rm triv}_W = N^\rho_{W'_1} + N^\rho_{W'_2} + \cdots$, the total number of distinct site-symmetry irreps defined at lower-symmetric WPs connected to $W$.
Then, matrix $q_W$ of trivial configurations is thus an $N^\rho_W \times N^{\rm triv}_W$ integer-valued matrix where each element $(q_W)_{im}$ gives the multiplicity of site-symmetry irrep $\rho^i_W$ induced by $m$-th adiabatic process ($i=1,\dots,N^\rho_W$ and $m=1, \dots, N^{\rm triv}_W$).
In other words, each column of $q_W$ denotes a trivial configuration of site-symmetry irreps at $W$ that can be adiabatically moved to lower-symmetric WPs connected to $W$.
The calculation of $(q_W)_{im}$ can be simplified by using the Frobenius reciprocity theorem: the integers $(q_W)_{im}$ are obtained in the opposite process.
Then, we calculate the multiplicity of the irrep $\rho_{W'}^m$ after the subduction (or reduction) of the irrep $\rho_W^i$ into the site-symmetry group of $W'$.

To see how this works in a simple example, let us consider the one-dimensional SG $Pm$ in one dimension without SOC.
This SG is generated by a mirror reflection $m_x$ and a translation $t_x$.
This group has two maximal WPs labeled $1a$ and $1b$, and one non-maximal WP labeled $2c$, as shown in Supplementary Figure (SFig.)~\ref{fig:pm}\textbf{a}.
The WP $1a$ is located at $x=0$ and it remains invariant under the mirror $m_x: x \to -x$.
Thus, there are two site-symmetry group irreps at WP $1a$: $(A')_{1a}$ and $(A'')_{1a}$ with mirror eigenvalues $+1$ and $-1$, respectively (see SFig.~\ref{fig:pm}\textbf{b}); as such we have $N^\rho_{1a} = 2$.
The WP $1a$ is connected to lower-symmetric WP $2c$, which corresponds to a pair of positions $\{x_0,-x_0\}$ with continuous parameter $x_0 \ne 0, 1/2$.
Since the site-symmetry group of the $2c$ WP is trivial, this WP has only a single site-symmetry irrep $(A)_{2c}$, yielding $N^{\rm triv}_{1a} = N^\rho_{2c} = 1$.
This irrep then induces the trivial configuration $(A')_{1a} + (A'')_{1a}$ at the WP $1a$.
The relevant adiabatic processes are illustrated in SFigs.~\ref{fig:pm}\textbf{c},\textbf{d}.
This defines $q_{1a} = (1,1)^T$ whose rows correspond to irrep multiplicities $m[(A')_{1a}]$ and $m[(A'')_{1a}]$ in this trivial configuration.
Similarly, the WP $1b$, located at $x = 1/2$, remains invariant under the mirror symmetry $\td{m}_x = t_x m_x: x \to -x + 1$.
Again, $N^\rho_{1b} = 2$ because only two site-symmetry irreps $(A')_{1b}$ and $(A'')_{1b}$ exist which have $\td{m}_x$ eigenvalues of $+1$ and $-1$ respectively.
This WP is also connected to the WP $2c$ which means that we can symmetrically move the trivial configuration $(A')_{1b} + (A'')_{1b}$ orbitals to two $(A)_{2c}$ orbitals, one at position $1/2 - x'_0$ and the other at $-1/2 + x'_0$ (see SFigs.~\ref{fig:pm}\textbf{e},\textbf{f}).
This means that $q_{1b} = (1, 1)^T$.
We will revisit the SG $Pm$ and give full details about the derivation of RSIs in SN~\ref{sec:examples}.

\begin{figure}[t]
\centering
\includegraphics[width=0.8\textwidth]{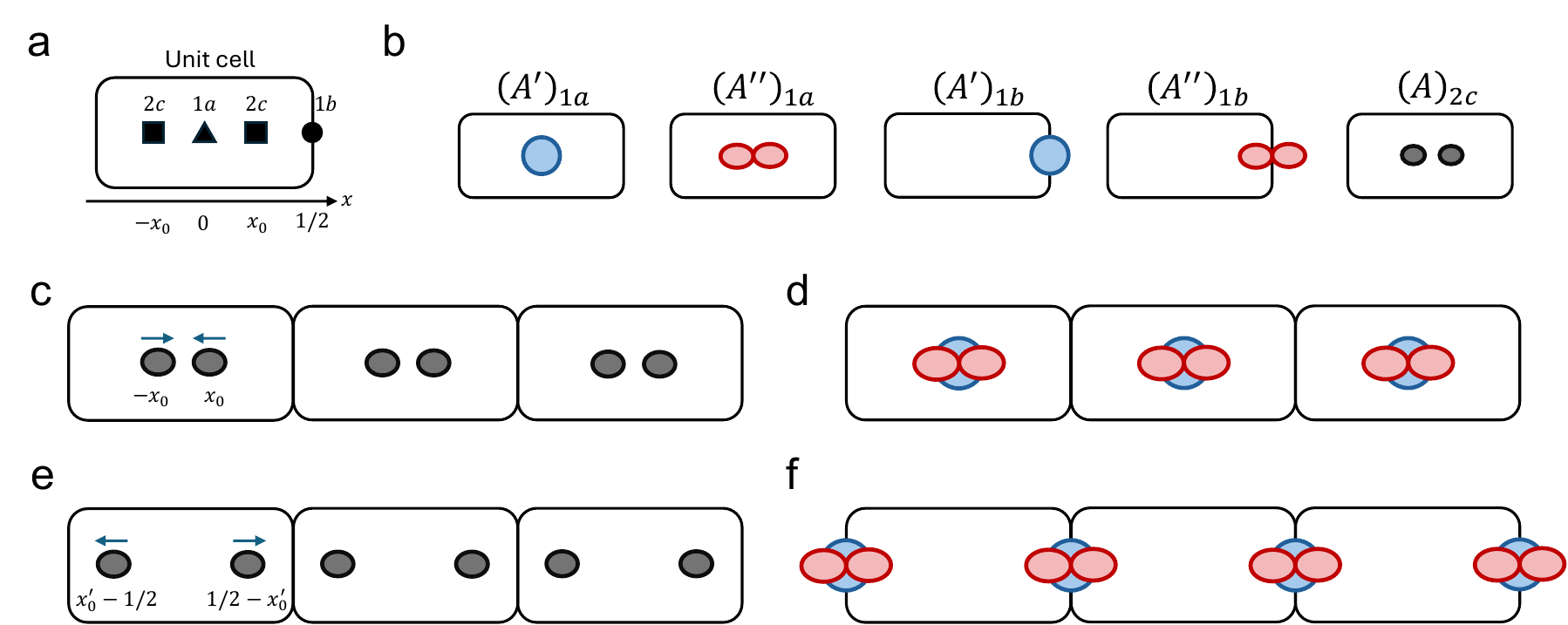}
\caption{Trivial configurations in SG $Pm$ in one dimension without SOC.
\textbf{a} Unit cell and Wyckoff positions (WPs).
The two maximal WPs, $1a$ (triangle) and $1b$ (circle), are located at $x=0$ and $1/2$, respectively.
The non-maximal, general WP $2c$ consists of two points, $x_0$ and $-x_0$, where $x_0 \ne 0,1/2$ is a continuous real parameter.
\textbf{b} Five distinct site-symmetry irreps are defined: $(A')_{1a}$, $(A'')_{1a}$, $(A')_{1b}$, $(A'')_{1b}$, and $(A)_{2c}$.
$A'$ orbitals are even under mirror symmetry ($s$-like), while $A''$ orbitals are odd under mirror symmetry ($p$-like).
\textbf{c} Configuration of orbitals induced from the site-symmetry irrep $(A)_{2c}$.
By tuning the free parameter $x_0$ to 0, the $(A)_{2c}$ irreps can be adiabatically moved to WP $1a$, with the mirror and translation symmetries of the SG preserved throughout the process.
\textbf{d} As a result of the adiabatic process in \textbf{c}, the $(A')_{1a}$ and $(A'')_{1a}$ irreps are obtained from even and odd linear combinations of the two $(A)_{2c}$ orbitals with $x_0=0$.
This constitutes a trivial configuration of site-symmetry irreps at WP $1a$.
\textbf{e} Similarly, the $(A)_{2c}$ irrep can be adiabatically moved to WP $1b$ by tuning a free parameter $x'_0$ to 0.
\textbf{f} As a result of the adiabatic process in \textbf{e}, the $(A')_{1b}$ and $(A'')_{1b}$ irreps are obtained by taking even and odd linear combinations of $(A)_{2c}$ orbitals with $x_0'=0$.
This gives a trivial configuration at WP $1b$.}
\label{fig:pm}
\end{figure}

Now, given a $W$-position trivial configuration matrix $q_W$, we want to extract information that remains invariant under the adiabatic addition or removal of Wannier orbitals at the WP $W$.
Since the columns of $q_W$ are the simplest adiabatic processes, the column space of $q_W$ (i.e., the vector space spanned by {\it integer} linear combinations of the columns) contains all such adiabatic additions and removals.
To find an independent basis for the column space of $q_W$, we perform a Smith decomposition
\ba
q_W = L_W \cdot \Lambda_W \cdot R_W
\quad \rightarrow \quad
L^{-1}_W \cdot q_W \cdot R^{-1}_W = \Lambda_W,
\label{eq:qvecdef}
\ea
where $L_W$ and $R_W$ are $N^\rho_W \times N^\rho_W$ and $N^{\rm triv}_W \times N^{\rm triv}_W$ unimodular matrices, respectively, and $\Lambda_W$ is an $N^\rho_W \times N^{\rm triv}_W$ diagonal matrix whose diagonal entries ${\rm diag}(\Lambda_W)$ are non-negative integers.
Note that first ${\rm rank}(q_W)$ elements of ${\rm diag}(\Lambda_W)$ are nonzero, and ${\rm rank}(q_W) \leq \min (N^\rho_W, N^{\rm triv}_W)$.
Since $R_W$ is unimodular, the lattice spanned by the columns of $q_W \cdot R^{-1}_W$ is the same as that spanned by the columns of $q_W$.
Each lattice point $(q_W \cdot R_W^{-1}) \cdot v$ for $v \in \Z^{N^{\rm triv}_W}$ corresponds to the addition or removal of orbitals at $W$ in a trivial configuration, and corresponds to a change of the $W$-position site-symmetry representation vector
\ba
p_W \rightarrow p_W'= p_W + (q_W \cdot R_W^{-1}) \cdot v.
\label{eq:adiabaticchangeofpw}
\ea
To define the RSIs $\delta_W$ at WP $W$, we construct linear combinations of the site-symmetry multiplicities that is invariant under SEq.~\eqref{eq:adiabaticchangeofpw}.
To do so, we make use of
\bg
L_W^{-1} \cdot (q_W \cdot R_W^{-1}) \cdot v = \Lambda_W \cdot v
\eg
which follows from SEq.~\eqref{eq:qvecdef}.
Since $v$ is a integer-valued vector and only first ${\rm rank}(q_W)$ diagonal elements of $\Lambda_W$ are nonzero, $(\Lambda_W \cdot v)_i = 0 \mod (\Lambda_W)_{ii}$ if $i \le {\rm rank}(q_W)$ and $(\Lambda_W \cdot v)_i = 0$ if $i > {\rm rank}(q_W)$.
If we multiply SEq.~\eqref{eq:adiabaticchangeofpw} on the left by $L_W^{-1}$, we see that under an adiabatic addition or removal of orbitals at $W$ in a trivial configuration,
\ba
L_W^{-1} \cdot p_W' = L_W^{-1} \cdot p_W + \Lambda_W \cdot v.
\ea
Since $v$ is an integer-valued matrix, this means that under adiabatic processes, $(L_W^{-1} \cdot p_W)_i$ can only change by integer multiples of $(\Lambda_W)_{ii}$.
Thus, we conclude that, given a $W$-position site-symmetry representation vector $p_W$,
\ba
(\delta_W)_i =
\begin{cases}
(L^{-1}_W \cdot p_W)_i \mod (\Lambda_W)_{ii} & \trm{for } i \le {\rm rank}(q_W)
\\
(L^{-1}_W \cdot p_W)_i & \trm{for } i > {\rm rank}(q_W)
\end{cases}
\label{eq:rsialg_finalresult}
\ea
is a good set of quantum numbers that remains invariant under the addition of trivial orbital configurations to $p_W$.
We then have several cases:
\begin{enumerate}
\item $i>{\rm rank}(q_W)$, in which case no adiabatic process can change $(\delta_W)_i$.
Thus $(\delta_W)_i$ is a $\Z$RSI.
This means that the number of $W$-position $\Z$RSIs is equal to the number of zero rows of $\Lambda_W$.
\item {$i< {\rm rank}(q_W)$ and } $(\Lambda_W)_{ii}$ is equal to $1$.
In this case, $(\delta_W)_i$ is defined modulo 1.
This means that $(\delta_W)_i =0$ trivially because $p_W$ is integer-valued and $(L_W)^{-1}$ is unimodular.
Thus, $(\delta_W)_i$ does not yield a nontrivial RSI.
\item $i< {\rm rank}(q_W)$ and $(\Lambda_W)_{ii} \ge 2$.
Then, $(\delta_W)_i$ is a $\Z_n$RSI.
\end{enumerate}
The first and third cases allow us to define the $\Z$RSIs $\delta_{\Z W}$ and $\Z_n$RSIs $\delta_{\Z_n W}$ respectively based on the value of $\Lambda_{ii}$.
Therefore, we obtain the $W$-position $\Z$RSI and $\Z_n$RSIs matrices, $\Delta_{\Z W}$ and $\Delta_{\Z_n W}$, introduced in SEqs.~\eqref{eq:zrsi_introduction} and \eqref{eq:znrsi_introduction}, from the rows of $(L_W)^{-1}$.
Finally, by considering all WPs, we relate the set of all $\Z$RSIs $\delta_\Z$ and site-symmetry representation vector $p$ according to SEq.~\eqref{eq:fullzrsidef}.
This relation defines the $\Z$RSI matrix $\Delta_\Z$ in terms of $\Delta_{\Z W}$.
The set of all $\Z_n$RSIs $\delta_{\Z_n}$ and the $\Z_n$RSI matrix $\Delta_{\Z_n}$ are defined in a similar way.
We provide detailed examples of obtaining these matrices in the next SN~\ref{sec:examples}.

We close by remarking on one practical application of RSIs.
As mentioned in SN~\ref{sec:ebr_intro}, an EBR is induced from a site-symmetry group irrep at a maximal WP, if this irrep is not an exception tabulated in SRef.~\cite{Bradlyn17}.
When an irrep corresponds to one of these tabulated exceptional cases, it induces a composite BR that can be adiabatically deformed, following a continuous path along non-maximal WPs, to a sum of EBRs at another maximal WP.
This situation can be understood through the following example: consider the SG $I4/m$ (No. 87) without SOC.
In this SG, a point $(0,1/2,z)$ with continuous parameter $z$ belongs to the non-maximal WP $8g$.
By varying $z$, this point is connected to $(0,1/2,0)$ and $(0,1/2,1/4)$, which belong to the maximal $4c$ and $4d$ WPs, respectively.
Because of the connectivity between WPs $4c,4d,8g$, the two-dimensional co-irrep $(B)_{8g}$ at the WP $8g$ induces either $(B_g)_{4c}+(B_u)_{4c}$ at the WP $4c$, or $({}^1E {}^2E)_{4d}$ at the WP $4d$.
This indicates that $({}^1E {}^2E)_{4d}$ can be adiabatically deformed to $(B)_{8g}$, and then to $(B_g)_{4c}+(B_u)_{4c}$.
Thus, the BR induced from $({}^1E {}^2E)_{4d}$, which is a site-symmetry irrep at maximal WP, can be decomposed into a linear combination of EBRs induced from $(B_g)_{4c}$ and $(B_u)_{4c}$ at another maximal WP.

In these exceptional cases, the associated Wannier states can be moved away from the maximal WP and do not contribute nonzero RSIs.
Hence, the set of exceptional EBRs can only be induced from irreps that have vanishing RSIs.
This provides a necessary (but not sufficient) condition to verify whether a given site-symmetry irrep can induce an exceptional band representation.
Indeed, in SG $I4/m$ without SOC, the unique RSI at $4d$ is defined as $\delta_{\Z,4d} = -m[(A)_{4d}] + m[(B)_{4d}]$, and it does not depend on the multiplicity of exceptional irrep $({}^1E {}^2E)_{4d}$.
However, it must be noted that not all irreps at maximal WPs with vanishing RSIs must induce composite BRs: as we will show in SN~\ref{sec:matchingSRSIs}, some EBRs can be written as differences (but not sums) of BRs induced from the connected non-maximal WPs.
They therefore have vanishing RSIs without being movable away from the respective maximal WP.
As we will discuss in detail in SN~\ref{sec:zsrsi_proof}, such pairs of band representations are {\it stably equivalent} in a sense that we will make precise.

\subsection{Examples: Local RSIs in SGs $Pm$ (No. 6) and $P4_332$ (No. 212)}
\label{sec:examples}
To demonstrate how the RSIs are derived, we consider two example SGs, $Pm$ (No. 6) and $P4_332$ (No. 212).
We also study the mapping from symmetry-data vectors to $\Z$RSIs and $\Z_n$RSIs in these SGs.
The SG $Pm$ with and without SOC serves as the simplest example.
We consider non-magnetic space groups, i.e., TRS is assumed to be present for all examples (unless stated otherwise).
In SG $Pm$, $\Z$RSIs are uniquely determined by the symmetry-data vector $B$ but $\Z_n$RSIs are not.
On the other hand, in SG $P4_332$ with SOC, neither the $\Z$RSIs nor the $\Z_n$RSIs are uniquely determined from the symmetry-data vector.
Based on this example, we conclude that local RSIs cannot be uniquely determined by symmetry-data vector in general SGs, consistent with the analysis of SRef.~\cite{xu2021threedimensional}.
Below, we follow the BCS~\cite{Aroyo2011183} for denoting the labeling of WPs, HSMs, and irreps, unless noted otherwise.

\subsubsection{SG $Pm$ (No. 6) without SOC}
The SG $Pm$ is generated from lattice translations and a single mirror symmetry $M$ that reverses one spatial direction; thus all the essential properties of the RSIs in this SG can be illustrated by considering a one-dimensional system.
Without SOC, mirror symmetry satisfies $M^2=E$ (where $E$ denotes the identity element), so that in any representation its eigenvalues must be either $\pm 1$.
At each of the two HSMs of the 1D BZ, $\Gamma$ $(k=0)$ and $X$ $(k=\pi)$, there are then two possible little group irreps, $\Gamma_1$ ($X_1$) and $\Gamma_2$ ($X_2$), with mirror eigenvalues $+1$ and $-1$, respectively.
Similarly, at each maximal WP $1a$ ($x=0$) and $1b$ ($x=1/2$), there are two possible site-symmetry irreps $A'$ and $A''$, again with mirror eigenvalues $+1$ and $-1$, respectively.
At the general WP $2c$, there is only one irrep $A$, which is the trivial representation.
The BR matrix for site-symmetry irreps defined at all WPs can be computed using the induction procedure of SRefs.~\cite{Bradlyn17,Cano17,jenbarryreview21}, or directly accessed using the \texttt{BANDREP} tool from the BCS~\cite{elcoro2017,bandrep_tool}.
It is given by
\ba
\BR =
\begin{pNiceArray}[first-row,code-for-first-row=\scriptstyle,last-col,code-for-last-col=\scriptstyle]{ccccc}
(A')_{1a} & (A'')_{1a} & (A')_{1b} & (A'')_{1b} & (A)_{2c} & \\
1 & 0 & 1 & 0 & 1 & \Gamma_1 \\ 
0 & 1 & 0 & 1 & 1 & \Gamma_2 \\
1 & 0 & 0 & 1 & 1 & X_1 \\
0 & 1 & 1 & 0 & 1 & X_2
\end{pNiceArray}
\label{eq:pmbrmatrix}
\ea
where the row basis is ordered as $\{ \Gamma_1, \Gamma_2, X_1, X_2 \}$ and the column basis as $\{ (A')_{1a}, (A'')_{1a}, (A')_{1b}, (A'')_{1b},(A)_{2c} \}$.

Next, let us construct the $q_W$ matrices [SEq.~\eqref{eq:qvecdef}] that contains the trivial configurations of site-symmetry irreps at each WP $W$.
As mentioned in the previous SN~\ref{sec:rsi_algorithm}, bringing in a pair of orbitals transforming in the unique irrep $A$ at the WP $2c$ to either the maximal WPs $1a$ or $1b$ induces one $A'$ and one $A''$ irrep at that maximal WP, corresponding to bonding and anti-bonding superpositions of orbitals.
Hence, at the $1a$ maximal WP, the $q_W$ matrix is given by
\ba
q_{1a} = \bpm 1 \\ 1 \epm
= \bpm 1 & 0 \\ 1 & 1 \epm
\bpm 1 \\ 0 \epm
\bpm 1 \epm
= L_{1a} \cdot \Lambda_{1a} \cdot R_{1a}, \quad
L_{1a}^{-1} = \bpm 1 & 0 \\ -1 & 1 \epm,
\label{eq:smith_rsi_pm}
\ea
where the row basis is ordered as $\{ (A')_{1a}, (A'')_{1a} \}$.
Note additionally that $q_{1b}=q_{1a}$, and so $q_{1b}$ has the same Smith decomposition as in SEq.~\eqref{eq:smith_rsi_pm}.
Similar to $q_{1a}$, the row basis of $q_{1b}$ is $\{ (A')_{1b}, (A'')_{1b} \}$.
Examining the divisor matrices $\Lambda_{1a}=\Lambda_{1b}=(1,0)^T$, we see from SEq.~\eqref{eq:rsialg_finalresult} that there is unique $\Z$RSI at each maximal WPs $1a$ and $1b$, derived from the second row of $L_{1a}^{-1}$ and $L_{1b}^{-1}$, respectively [since ${\rm rank}(\Lambda_{1a})=1,$ and $(\Lambda_{1a})_{11}=1$ does not give a nontrivial RSI].
From SEqs.~\eqref{eq:zrsi_introduction}, and~\eqref{eq:rsialg_finalresult}, we find the $1a$- and $1b$-position RSI matrices,
\ba
\Delta_{\Z, 1a} = (L_{1a}^{-1})_{2,*} = \bpm -1 & 1 \epm, \quad
\Delta_{\Z, 1b} = (L_{1b}^{-1})_{2,*} = \bpm -1 & 1
\epm.
\label{eq:pmzdeltas}
\ea
where we use $A_{i,*}$ to denote the $i$-th row of matrix $A$.
From $\Delta_{\Z, 1a}$ and $\Delta_{\Z, 1b}$ in SEq.~\eqref{eq:pmzdeltas}, we can read off the two $\Z$RSIs defined at the WP $1a$ and $1b$,
\ba
\delta_{\Z, 1a} &= -m[(A')_{1a}] + m[(A'')_{1a}] \nn
\delta_{\Z, 1b} &= -m[(A')_{1b}] + m[(A'')_{1b}].
\ea

For the general WP $2c$, we find no RSI since there is no lower-symmetric WP connected to the WP $2c$.
In summary, following SEq.~\eqref{eq:fullzrsidef}, we construct the $\Z$RSI matrix,
\ba
\Delta_{\Z} = \bpm
\Delta_{\Z, 1a} & 0 & 0 \\
0 & \Delta_{\Z, 1b} & 0 \epm
= \bpm -1 & 1 & 0 & 0 & 0
\\ 0 & 0 & -1 & 1 & 0 \epm,
\label{eq:ssh_nosocnotrs_rsis}
\ea
where the row basis is ordered as $\{ \delta_{\Z, 1a}, \delta_{\Z, 1b} \}$ and the column basis is ordered as $\{ (A')_{1a}, (A'')_{1a}, (A')_{1b}, (A'')_{1b},(A)_{2c} \}$.

Next, we study the mapping between $\Z$RSIs and symmetry-data vector $B$ [SEq.~\eqref{eq:zrsis_fromb}].
The kernel of $\BR$ in SEq.~\eqref{eq:pmbrmatrix} is spanned by two basis vectors, $p^{(1)}_0 = (-1,-1,1,1,0)^T$ and $p^{(2)}_0 = (-1,-1,0,0,1)^T$.
Since $\Delta_\Z \cdot p^{(1)}_0 = \Delta_\Z \cdot p^{(2)}_0 = 0$, we can deduce that $\Delta_\Z \cdot p_0 = 0$ holds for any $p_0$ in the kernel of $\BR$.
Thus from SEq.~\eqref{eq:zrsis_fromb} it follows that in the SG $Pm$ without SOC, the $\Z$RSIs are uniquely determined by the symmetry-data vector $B$.
Explicitly, we compute the pseudoinverse $BR^\ddag$ of $BR$ to obtain the mapping from $B$ to $\Z$-valued local RSIs, given by $\delta_\Z=\Delta_\Z \cdot p$.
For the integer-valued matrix $BR$, the pseudoinverse $B^\ddagger$ can be determined as follows.
Using the Smith decomposition, $BR = L_{BR} \cdot \Lambda_{BR} \cdot R_{BR}$, where $\Lambda_{BR}$ is an $N^\rho_{\rm BZ} \times N^\rho_{\rm UC}$ diagonal matrix, we define the $N^\rho_{\rm UC} \times N^\rho_{\rm BZ}$ pseudoinverse matrix $\Lambda_{BR}^\ddagger$ of $\Lambda_{BR}$ such that $(\Lambda_{BR}^\ddagger)_{ii}=1/(\Lambda_{BR})_{ii}$ for $i=1,\dots,{\rm rank}(BR)$ and $(\Lambda_{BR}^\ddagger)_{ij}=0$ otherwise.
The pseudoinverse of $BR$ is then given by
\bg
BR^\ddagger = (R_{BR})^{-1} \cdot \Lambda_{BR}^\ddagger \cdot (L_{BR})^{-1}.
\label{eq:pseudoinv}
\eg
Following this definition, we obtain $BR^\ddagger$:
\bg
BR
= L_{BR} \cdot \Lambda_{BR} \cdot R_{BR}
= \bpm 1 & 0 & 1 & 0 \\
0 & 1 & 0 & 0 \\
1 & 0 & 0 & 0 \\
0 & 1 & 1 & 1 \epm
\cdot \bpm 1 & 0 & 0 & 0 & 0 \\
0 & 1 & 0 & 0 & 0 \\
0 & 0 & 1 & 0 & 0 \\
0 & 0 & 0 & 0 & 0 \epm
\cdot \bpm 1 & 0 & 0 & 1 & 1 \\
0 & 1 & 0 & 1 & 1 \\
0 & 0 & 1 & -1 & 0 \\
0 & 0 & 0 & 1 & 0 \\
0 & 0 & 0 & 0 & 1 \epm,
\nn
BR^\ddagger
= (R_{BR})^{-1} \cdot \Lambda_{BR}^\ddagger \cdot (L_{BR})^{-1}
= \bpm 1 & 0 & 0 & -1 & -1 \\
0 & 1 & 0 & -1 & -1 \\
0 & 0 & 1 & 1 & 0 \\
0 & 0 & 0 & 1 & 0 \\
0 & 0 & 0 & 0 & 1 \epm
\cdot \bpm 1 & 0 & 0 & 0 \\
0 & 1 & 0 & 0 \\
0 & 0 & 1 & 0 \\
0 & 0 & 0 & 0 \\
0 & 0 & 0 & 0 \epm
\cdot \bpm 0 & 0 & 1 & 0 \\
0 & 1 & 0 & 0 \\
1 & 0 & -1 & 0 \\
-1 & -1 & 1 & 1 \epm
= \bpm 0 & 0 & 1 & 0 \\
0 & 1 & 0 & 0 \\
1 & 0 & -1 & 0 \\
0 & 0 & 0 & 0 \\
0 & 0 & 0 & 0 \epm.
\eg
The pseudoinverse satisfies
\ba
(BR) \cdot (BR)^\ddagger \cdot (BR) = BR 
\ea
by construction.
Using the pseudoinverse, we can rewrite the equation $B=(BR) \cdot p$ for the symmetry-data vector $B$ as
\ba
BR \cdot \left[ BR^\ddagger \cdot B - p \right] = 0,
\ea
from which we deduce that $p=BR^\ddagger \cdot B + p_0$ for some $p_0$ in the kernel of $BR$.
Since $\Delta_\Z \cdot p_0=0$ for space group $Pm$, we find that the $\Z$-valued local RSIs $\delta_\Z$ are determined as
\ba
\delta_{\Z}
= \Delta_{\Z} \cdot \BR^\ddagger \cdot B
= \bpm 0 & 1 & -1 & 0 \\
-1 & 0 & 1 & 0 \epm \cdot B.
\ea

\subsubsection{SG $Pm$ (No. 6) with SOC}
With SOC, mirror symmetry $M$ satisfies $M^2 = \bar{E}$, where $\bar{E}$ represents a $2\pi$ spin rotation that must be represented by minus the identity matrix in any representation.
This implies that in any representation $M$ will have eigenvalue either $\pm i$.
Moreover, TRS $T$ (which, recall, we always consider) is represented by an anti-unitary operator and enforces Kramers pairing via $T^2=\bar{E}$, and so all irreps must be paired into time-reversal preserving co-irreps with eigenvalue sets that are invariant under complex conjugation.
At each HSM $\Gamma$ $(k=0)$ and $X$ $(k=\pi)$, there is then only one possible little group co-irrep, $\bar{\Gamma}_3 \bar{\Gamma}_4$ ($\bar{X}_3 \bar{X}_4$), corresponding to a Kramers pair of orbitals with mirror eigenvalues $\pm i$.
Similarly, at each maximal WP $1a$ ($x=0$) and $1b$ ($x=1/2$), there is one possible site-symmetry co-irrep ${}^1\bar{E} {}^2\bar{E}$.
The only co-irrep at the general WP $2c$ is $\bar{A} \bar{A}$.
The BR matrix is given, found either via explicit induction or using the \texttt{BANDREP} tool~\cite{bandrep_tool}, by
\ba
\BR = \bpm 
1 & 1 & 2 \\ 
1 & 1 & 2
\epm,
\label{eq:pmsocbr}
\ea
where the row basis is ordered as $\{ \bar{\Gamma}_3 \bar{\Gamma}_4, \bar{X}_3 \bar{X}_4 \}$ and the column basis is ordered as $\{ ({}^1\bar{E} {}^2\bar{E})_{1a}, ({}^1\bar{E} {}^2\bar{E})_{1b},(\bar{A} \bar{A})_{2c} \}$.

Next, we consider adiabatic processes that bring orbitals at the general WP $2c$ to either maximal WPs $W=1a$ or $1b$ in order to construct the $q_W$ matrices encoding trivial configurations of site-symmetry irreps at each WP $W$.
A mirror-related pair of Kramers pairs of orbitals transforming in the site-symmetry co-irrep $(\bar{A} \bar{A})_{2c}$ can be deformed to the (reducible) site-symmetry corepresentations $2({}^1\bar{E} {}^2\bar{E})_{1a}$ or $2({}^1\bar{E} {}^2\bar{E})_{1b}$.
Hence, we define the $q_W$ matrices [SEq.~\eqref{eq:qvecdef}] at the WP $W=1a,1b$,
\ba
q_{1a} = q_{1b} = q_W = \bpm 2 \epm
= \bpm 1 \epm \bpm 2 \epm \bpm 1 \epm
= L_W \cdot \Lambda_W \cdot R_W, \quad
L_W^{-1} = \bpm 1 \epm,
\ea
where the row basis is $\{ ({}^1\bar{E} {}^2\bar{E})_W \}$.
Because ${\rm diag}(\Lambda_W) = 2$, there is one $\Z_2$RSI at each WPs $1a$ and $1b$ but no $\Z$RSIs.
From SEqs.~\eqref{eq:znrsi_introduction} and~\eqref{eq:rsialg_finalresult}, we find the $W$-position $\Z_2$RSI matrices
\ba
\Delta_{\Z_2, 1a} = \Delta_{\Z_2, 1b}
= (L_W^{-1})_{1,*} = \bpm 1 \epm.
\ea
Thus, the $\Z_n$RSIs are given by
\ba
\delta_{\Z_2, 1a} &= m[({}^1 \bar{E} {}^2 \bar{E})_{1a}] \mod 2 \nn
\delta_{\Z_2, 1b} &= m[({}^1 \bar{E} {}^2 \bar{E})_{1b}] \mod 2.
\ea
Lastly, there are no RSIs at the general WP $2c$ because it does not have any connected lower-symmetric WP.
In summary, we define the $\Z_n$RSI matrix $\Delta_{\Z_2}$ [SEq.~\eqref{eq:zn_summary_calc}],
\ba
\Delta_{\Z_2}
= \bpm \Delta_{\Z_2, 1a} & 0 & 0 \\
0 & \Delta_{\Z_2, 1b} & 0 \epm
= \bpm 1 & 0 & 0 \\
0 & 1 & 0 \epm,
\label{eq:ssh_soctrs_rsis}
\ea
where the row basis is ordered as $\{ \delta_{\Z_2, 1a}, \delta_{\Z_2, 1b} \}$ and the column basis is ordered as $\{ ({}^1\bar{E} {}^2\bar{E})_{1a}, ({}^1\bar{E} {}^2\bar{E})_{1b}, (\bar{A} \bar{A})_{2c} \}$.

We now examine to what extend these $\Z_2$RSIs can be determined from the symmetry-data vector.
The kernel of $\BR$ in SEq.~\eqref{eq:pmsocbr} is spanned by $p^{(1)}_0 = (-1,1,0)^T$ and $p^{(2)}_0 = (-2,0,1)^T$.
Furthermore, we see from SEq.~\eqref{eq:ssh_soctrs_rsis} that although $\Delta_{\Z_2} \cdot p^{(2)}_0 
\mod 2=\bb 0$, we have
\ba
\Delta_{\Z_2} \cdot p^{(1)}_0 
\mod 2
&= \bpm 1 \\ 1 \epm \neq \bb 0.
\ea
Using SEq.~\eqref{eq:znrsi_from_b}, this implies that $p^{(1)}_0$ corresponds to an ambiguity in determining the $\Z_2$RSIs from the symmetry-data vector $B$.
Intuitively, this ambiguity results from the fact that at both the HSM, there is only one little group co-irrep forcing the columns of the $BR$ matrix to be linearly dependent.
Thus, in SG $Pm$ with SOC and TRS, the $\Z_2$RSIs do not follow uniquely from the symmetry-data vector, and instead represent additional information.
The additional information provided by the $\Z_2$RSIs is the number of Kramers pairs $m[({}^1\bar{E} {}^2\bar{E})_W]$ mod 2 at the WP $W=1a,1b$.
Note that because the EBRs induced from to $({}^1\bar{E} {}^2\bar{E})_{1a}$ and $({}^1\bar{E} {}^2\bar{E})_{1b}$ have the same symmetry-data vector, the individual multiplicities $m[({}^1\bar{E} {}^2\bar{E})_W]$ mod 2 cannot be inferred from the symmetry-data vector.

\subsubsection{SG $P4_332$ (No. 212) with SOC}
\label{sec:lrsi_212}
{We next examine the local RSIs in space group $P4_332$ with time-reversal symmetry.
Although this is a cubic nonsymmorphic group, we will see that it has only four nontrivial local RSIs and illustrates many of the important properties discussed in SN~\ref{sec:rsi_algorithm}.}
SG $P4_332$ with SOC is generated from lattice translations $\{E|\bb v \in \Z^3 \}$, threefold rotation $\{3^+_{111}|\bb 0\}$, and fourfold screw rotation $\{4^+_{001}|3/4,1/4,3/4\}$.
In this SG, there are 5 types of WPs: two maximal WPs $4a$ and $4b$, two non-maximal WPs $8c$ and $12d$, and the general WP $24e$.
The site-symmetry co-irreps defined at all WPs are
\bg
\{ ({}^1 \bar{E} {}^2 \bar{E})_{4a}, (\bar{E}_1)_{4a}, ({}^1 \bar{E} {}^2 \bar{E})_{4b}, (\bar{E}_1)_{4b}, (\bar{E} \bar{E})_{8c}, ({}^1 \bar{E} {}^2 \bar{E})_{8c}, ({}^1 \bar{E} {}^2 \bar{E})_{12d}, (\bar{A} \bar{A})_{24e} \}.
\label{eq:sg212_irrep}
\eg
Supplementary Table~\ref{tab:sg212_ssg_coirreps} gives the full details of these co-irreps including their representative basis orbitals.
To give a physical example, let us focus on the $4a$ WP.
It has representative position $(1/8,1/8,1/8)$ which is invariant under the site-symmetry group generated by the threefold rotation $\{3^+_{111}|\bb 0\}$ and the twofold rotation $\{2_{1\bar{1}0}|1/4,1/4,1/4\}$.
For convenience, we adopt a rotated coordinate system such that these operations correspond to a threefold rotation $C_{3\bar{z}}$ along the $\bar{z}$ direction and a twofold rotation $C_{2\bar{y}}$ along the $\bar{y}$ direction.
In this frame, consider the spinful $s$ orbitals $\ket{s,\uparrow / \downarrow} = \ket{s} \otimes \ket{\uparrow / \downarrow}$, where $\uparrow$ and $\downarrow$ denotes the up and down spin aligned along the $\bar{z}$ direction.
These orbitals transform under $C_{3\bar{z}}$ with eigenvalues $e^{-i\pi/3}$ and $e^{i\pi/3}$, respectively.
Under $C_{2\bar{y}}$, these states are exchanged as $C_{2\bar{y}} \ket{s, \uparrow} = \ket{s, \downarrow}$ and $C_{2\bar{y}} \ket{s, \downarrow} = - \ket{s, \uparrow}$.
Thus, these $\ket{s,\uparrow}$ and $\ket{s,\downarrow}$ form a two-dimensional irrep $\bar{E}_1$ with character 1 for $C_{3\bar{z}}$ and 0 under $C_{2\bar{y}}$.
Alternatively, we may construct a co-irrep using spinful $p$ orbitals aligned in the $(\bar{x},\bar{y})$ plane (in the rotated coordinate system).
These orbitals form the basis of the co-irrep ${}^1 \bar{E} {}^2 \bar{E}$.
In the absence of time-reversal symmetry, one can distinguish two separate irreps: ${}^1\bar{E}$, with eigenvalues $-1$ and $i$ under $C_{3\bar{z}}$ and $C_{2\bar{y}}$, respectively, and ${}^2\bar{E}$, with eigenvalues $-1$ and $-i$.
However, time-reversal symmetry combines these two irreps into a single {\it co}-irrep ${}^1 \bar{E} {}^2 \bar{E}$, which has character $-2$ for $C_{3\bar{z}}$ and 0 under $C_{2\bar{y}}$.

\let\oldaddcontentsline\addcontentsline
\renewcommand{\addcontentsline}[3]{}
\begin{table}[]
\renewcommand{\arraystretch}{1.4}
\centering
\begin{tabular}{c|c|c|c|c|c|c|c|c|c|c|c}
\hline \hline
\multirow{2}{*}{Wyckoff position} & \multirow{2}{*}{Site-symmetry group} & \multirow{2}{*}{Co-irrep} & \multicolumn{8}{c|}{Conjugacy class} & \multirow{2}{*}{Basis orbitals}
\\ \cline{4-11}
& & & 1 & $3^+_{111}$ & $3^-_{111}$ & $2_{1\bar{1}0}$ & ${}^d 1$ & ${}^d 3^+_{111}$ & ${}^d 3^-_{111}$ & ${}^d 2_{1\bar{1}0}$ \\
\hline
\multirow{2}{*}{$4a \ (\frac{1}{8}, \frac{1}{8}, \frac{1}{8})$} & \multirow{2}{*}{$321'$} & ${}^1 \bar{E} {}^2 \bar{E}$ & 2 & $-2$ & $-2$ & 0 & $-2$ & 2 & $2$ & 0 & $\ket{p_{\bar x} + i p_{\bar y}, \uparrow_{\bar z}}, \ket{p_{\bar x} - i p_{\bar y}, \downarrow_{\bar z}}$
\\
& & $\bar{E}_1$ & 2 & 1 & 1 & 0 & $-2$ & $-1$ & $-1$ & 0 & $\ket{s, \uparrow_{\bar z}}, \ket{s, \downarrow_{\bar z}}$
\\ \hline
\multirow{2}{*}{$4b \ (\frac{5}{8}, \frac{5}{8}, \frac{5}{8})$} & \multirow{2}{*}{$321'$} & ${}^1 \bar{E} {}^2 \bar{E}$ & 2 & $-2$ & $-2$ & 0 & $-2$ & 2 & $2$ & 0 & $\ket{p_{\bar x} + i p_{\bar y}, \uparrow_{\bar z}}, \ket{p_{\bar x} - i p_{\bar y}, \downarrow_{\bar z}}$
\\
& & $\bar{E}_1$ & 2 & 1 & 1 & 0 & $-2$ & $-1$ & $-1$ & 0 & $\ket{s, \uparrow_{\bar z}}, \ket{s, \downarrow_{\bar z}}$
\\ \hline
\multirow{2}{*}{$8c \ (x,x,x)$} & \multirow{2}{*}{$31'$} & $\bar{E} \bar{E}$ & 2 & $-2$ & $-2$ & - & $-2$ & 2 & 2 & - & $\ket{p_{\bar x} + i p_{\bar y}, \uparrow_{\bar z}}, \ket{p_{\bar x} - i p_{\bar y}, \downarrow_{\bar z}}$
\\
& & ${}^1\bar{E} {}^2\bar{E}$ & 2 & 1 & 1 & - & $-2$ & $-1$ & $-1$ & - & $\ket{s, \uparrow_{\bar z}}, \ket{s, \downarrow_{\bar z}}$
\\ \hline
$12d \ (\frac{1}{8}, y, -y+\frac{1}{4})$ & $21'$ & ${}^1 \bar{E} {}^2 \bar{E}$ & 2 & - & - & 0 &$ -2$ & - & - & 0 & $\ket{s, \uparrow_{\bar z}}, \ket{s, \downarrow_{\bar z}}$
\\ \hline
$24e \ (x,y,z)$ & $11'$ & $\bar{A} \bar{A}$ & 2 & - & - & - & $-2$ & - & - & - & $\ket{s, \uparrow_{\bar z}}, \ket{s, \downarrow_{\bar z}}$ \\
\hline \hline
\end{tabular}
\caption{The character tables for the site-symmetry groups of all the WPs of SG $P4_332$ (No. 212).
The first column lists the WP label and a representative point in that WP, where $(x,y,z)$ are real free parameters.
The second column shows the (double, non-magnetic) point group isomorphic to the site-symmetry group of the WP.
(The symbol $1'$ denotes the presence of time-reversal symmetry.)
The sub-columns in the fourth column list representative symmetry elements of the conjugacy classes of the point group.
The hyphen ($-$) in an entry indicates that the representative symmetry element (and hence the conjugacy class) is absent in the point group.
${}^d1$ denotes a spin rotation by $2\pi$, and for any $g$, ${}^dg \equiv {}^d1 \circ g$, where $\circ$ indicates group multiplication.
Note that $3^+_{111}$ (${}^d3^+_{111}$) and $3^-_{111}$ (${}^d3^-_{111}$) are conjugate to each other in the point group $321'$ but not in the point group $31'$.
The last column shows the basis orbitals that span the co-irreps~\cite{altmann1994point,BigBook}.
For convenience, we introduce a rotated coordinate system $(\bar{x}, \bar{y}, \bar{z})$ such that the threefold rotation $3^+_{111}$ is aligned along $\hat{\bar{z}}=\frac{1}{\sqrt{3}}(1,1,1)$, the unit vector in the $\bar{z}$ direction.
The remaining two orthonormal unit vectors, $\hat{\bar{x}}$ and $\hat{\bar{y}}$, are perpendicular to $\hat{\bar{z}}$.
The basis $p$ orbitals, $p_{\bar x, \bar y}$ are defined in the rotated coordinate system $(\bar{x}, \bar{y}, \bar{z})$.
The spin basis states $\uparrow_{\bar z}$ and $\downarrow_{\bar z}$ are up and down spin aligned along $\hat{\bar{z}}$.}
\label{tab:sg212_ssg_coirreps}
\end{table}
\let\addcontentsline\oldaddcontentsline

The BR matrix $\BR$ for these irreps is given by
\bg
\BR = \bpm
0 & 1 & 0 & 1 & 0 & 2 & 2 & 4 \\
0 & 1 & 0 & 1 & 0 & 2 & 2 & 4 \\
2 & 1 & 2 & 1 & 4 & 2 & 4 & 8 \\
1 & 0 & 1 & 0 & 2 & 0 & 1 & 2 \\
0 & 1 & 0 & 1 & 0 & 2 & 2 & 4 \\
1 & 1 & 1 & 1 & 2 & 2 & 3 & 6 \\
2 & 2 & 2 & 2 & 4 & 4 & 6 & 12 \\
1 & 1 & 1 & 1 & 2 & 2 & 3 & 6 \\
1 & 1 & 1 & 1 & 2 & 2 & 3 & 6 \\
2 & 2 & 2 & 2 & 4 & 4 & 6 & 12
\epm,
\label{eq:sg212_BR}
\eg
where the row basis is ordered in terms of the little group representations at HSM as
\bg
\{ \bar{\Gamma}_6, \bar{\Gamma}_7, \bar{\Gamma}_8, \bar{R}_4 \bar{R}_5, \bar{R}_6, \bar{R}_7 \bar{R}_8, \bar{M}_6 \bar{M}_7, \bar{X}_3 \bar{X}_6, \bar{X}_4 \bar{X}_5, \bar{X}_7 \},
\label{eq:sg212_row}
\eg
and the column basis is ordered according to SEq.~\eqref{eq:sg212_irrep}.

Now, let us study adiabatic processes between site-symmetry irreps at WPs and connected lower-symmetric WPs.
The non-maximal WP $8c$ [with representative point $(x,x,x)$] is connected to the maximal WPs $4a$ and $4b$ [with representative points $(1/8,1/8,1/8)$ and $(5/8,5/8,5/8)$, respectively].
The site-symmetry irrep $(\bar{E} \bar{E})_{8c}$ can be deformed to either $2({}^1 \bar{E} {}^2 \bar{E})_{4a}$ or $2({}^1 \bar{E} {}^2 \bar{E})_{4b}$.
From now on, we denote such adiabatic processes as
\bg
(\bar{E} \bar{E})_{8c} \Leftrightarrow 2({}^1 \bar{E} {}^2 \bar{E})_{4a},
\quad
(\bar{E} \bar{E})_{8c} \Leftrightarrow 2({}^1 \bar{E} {}^2 \bar{E})_{4b}.
\label{eq:sg212_adia1_8c}
\eg
The remaining adiabatic processes can be determined using the same procedure, and we find they are given by
\bg
({}^1 \bar{E} {}^2 \bar{E})_{8c} \Leftrightarrow 2 (\bar{E}_1)_{4a},
\quad
({}^1 \bar{E} {}^2 \bar{E})_{8c} \Leftrightarrow 2 (\bar{E}_1)_{4b},
\label{eq:sg212_adia2_8c}
\\
({}^1 \bar{E} {}^2 \bar{E})_{12d} \Leftrightarrow ({}^1 \bar{E} {}^2 \bar{E})_{4a} + 2 (\bar{E}_1)_{4a},
\quad
({}^1 \bar{E} {}^2 \bar{E})_{12d} \Leftrightarrow ({}^1 \bar{E} {}^2 \bar{E})_{4b} + 2 (\bar{E}_1)_{4b},
\label{eq:sg212_adia_12d}
\\
(\bar{A} \bar{A})_{24e} \Leftrightarrow (\bar{E} \bar{E})_{8c} + 2 ({}^1 \bar{E} {}^2 \bar{E})_{8c},
\quad
(\bar{A} \bar{A})_{24e} \Leftrightarrow 2 ({}^1 \bar{E} {}^2 \bar{E})_{12d}.
\label{eq:sg212_adia_24e}
\eg

Based on the adiabatic processes in SEqs.~\eqref{eq:sg212_adia1_8c}-\eqref{eq:sg212_adia_24e}, we construct the trivial-configuration matrices $q_W$ at each WP $W$.
At each of the maximal WPs $4a$ and $4b$, SEqs.~\eqref{eq:sg212_adia1_8c}-\eqref{eq:sg212_adia_12d} determine
\bg
q_{4a}=q_{4b}= \bpm 2 & 0 & 1 \\
0 & 2 & 2 \epm
= \bpm 1 & 0 \\ 0 & 1 \epm
\bpm 1 & 0 & 0 \\ 0 & 2 & 0 \epm
\bpm 2 & 0 & 1 \\ 0 & 1 & 1 \\ -1 & 0 & 0 \epm
= L_W \cdot \Lambda_W \cdot R_W, \quad
L_W^{-1}
= \bpm 1 & 0 \\ 0 & 1 \epm,
\label{eq:sg212_4ab_q}
\eg
where the row basis is ordered as $\{ ({}^1 \bar{E} {}^2 \bar{E})_W, (\bar{E}_1)_W \}$ for $W=4a,4b$.
Applying SEq.~\eqref{eq:rsialg_finalresult} to SEq.~\eqref{eq:sg212_4ab_q}, we conclude that there is one $\Z_2$RSI at each of the maximal WPs $4a$ and $4b$, derived from the second row of $L_W^{-1}$, and no $\Z$RSIs {since ${\rm rank}(\Lambda_W)=2$, which implies there are no zero rows of $\Lambda_W$}.
From SEqs.~\eqref{eq:znrsi_introduction} and~\eqref{eq:rsialg_finalresult}, we find the $W$-position $\Z_2$RSI matrices,
\bg
\Delta_{\Z_2, W} = (L_W)^{-1}_{2,*} = (0, 1),
\eg
and the corresponding $\Z_2$RSI is given by
\ba
\delta_{\Z_2, W}
= (L_W)^{-1}_{2,*} \cdot p_W
= m[(\bar{E}_{1})_W] \mod 2
\ea
at the WPs $W=4a,4b$.

Similarly, we can construct the trivial-configuration matrices $q_W$ at the WPs $8c$ and $12d$ based on SEq.~\eqref{eq:sg212_adia_24e},
\bg
q_{8c}
= \bpm 1 \\ 2 \epm
= \bpm 1 & 0 \\ 2 & 1 \epm
\bpm 1 \\ 0 \epm
\bpm 1 \epm
= L_{8c} \cdot \Lambda_{8c} \cdot R_{8c}, \quad
L_{8c}^{-1}
= \bpm 1 & 0 \\ -2 & 1 \epm, \\
q_{12d}
= \bpm 2 \epm =
\bpm 1 \epm \bpm 2 \epm \bpm 1 \epm 
= L_{12d} \cdot \Lambda_{12d} \cdot R_{12d}, \quad
L_{12d}^{-1}
= \bpm 1 \epm,
\eg
where the row bases are ordered as $\{ (\bar{E} \bar{E})_{8c}, ({}^1 \bar{E} {}^2 \bar{E})_{8c} \}$ for $q_{8c}$ and $\{ ({}^1 \bar{E} {}^2 \bar{E})_{12s} \}$ for $q_{12d}$.
Applying SEq.~\eqref{eq:rsialg_finalresult} we see that there is one $\Z$RSI at the WP $8c$ because $\Lambda_{8c}$ has one zero row.
There is also one $\Z_2$RSI at the WP $12d$ since $\Lambda_{12d}=(2)$.
From SEqs.~\eqref{eq:znrsi_introduction} and~\eqref{eq:rsialg_finalresult}, we find
\bg
\Delta_{\Z,8c} = (L_{8c})^{-1}_{2,*} = (-2, 1),
\quad
\Delta_{\Z_2,12d} = (L_{12d})^{-1}_{1,*} = (1).
\eg
Thus, we define 
\bg
\delta_{\Z, 8c} = -2m[(\bar{E} \bar{E})_{8c}] + m[({}^1 \bar{E} {}^2 \bar{E})_{8c}],
\quad
\delta_{\Z_2, 12d} = m[({}^1 \bar{E} {}^2 \bar{E})_{12d}].
\label{eq:212_lrsis_non-maximal}
\eg
In summary, we determine the $\Z$RSI and $\Z_2$RSI matrices [SEqs.~\eqref{eq:fullzrsidef} and \eqref{eq:zn_summary_calc}]
\bg
\Delta_{\Z} = \bpm 0 & 0 & 0 & 0 & \Delta_{\Z, 8c} & 0 & 0 \epm
= \bpm 0 & 0 & 0 & 0 & -2 & 1 & 0 & 0 \epm \\
\Delta_{\Z_2} = \bpm \Delta_{\Z_2, 4a} & 0 & 0 & 0 & 0 & 0 \\
0 & \Delta_{\Z_2, 4b} & 0 & 0 & 0 & 0 \\
0 & 0 & 0 & 0 & \Delta_{\Z_2,12d} & 0 \epm
= \bpm 0 & 1 & 0 & 0 & 0 & 0 & 0 & 0 \\
0 & 0 & 0 & 1 & 0 & 0 & 0 & 0 \\
0 & 0 & 0 & 0 & 0 & 0 & 1 & 0 \epm,
\eg
where the row basis is ordered as in SEq.~\eqref{eq:sg212_irrep}.

Now let us study the mapping between the RSIs and symmetry-data vectors in SG $P4_332$.
The kernel of $\BR$ defined in SEq.~\eqref{eq:sg212_BR} is spanned by six basis vectors, 
\ba
p^{(1)}_0 &= (1,0,0,0,0,1,1,-1)^T, \\
p^{(2)}_0 &= (0,1,0,1,0,-1,0,0)^T, \\
p^{(3)}_0 &= (0,0,1,0,0,1,1,-1)^T, \\
p^{(4)}_0 &= (0,0,0,2,0,-1,0,0)^T, \\
p^{(5)}_0 &= (0,0,0,0,1,2,0,-1)^T, \\
p^{(6)}_0 &= (0,0,0,0,0,0,2,-1)^T.
\ea
To see if the $\Z$RSIs or $\Z_n$RSIs are determined uniquely from a general symmetry-data vector $B$, we compute
\bg
\Delta_{\Z} \cdot [ p^{(1)}_0, p^{(2)}_0, p^{(3)}_0, p^{(4)}_0, p^{(5)}_0, p^{(6)}_0]
= \bpm 1 & -1 & 1 & -1 & 0 & 0 \epm
\neq \bb 0,
\\
\Delta_{\Z_2} \cdot [ p^{(1)}_0, p^{(2)}_0, p^{(3)}_0, p^{(4)}_0, p^{(5)}_0, p^{(6)}_0] \mod 2
= \bpm 0 & 1 & 0 & 0 & 0 & 0 \\
0 & 1 & 0 & 0 & 0 & 0 \\
1 & 0 & 1 & 0 & 0 & 0 \epm
\mod 2 \neq \bb 0.
\eg
Thus, $\Delta_\Z \cdot p_0 \ne 0$ and $\Delta_{\Z_2} \cdot p_0 \ne 0$ for general $p_0$ in the kernel of $\BR$.
This implies from SEqs.~\eqref{eq:zrsis_fromb} and \eqref{eq:znrsi_from_b} that neither the $\Z$RSIs nor $\Z_2$RSIs can be uniquely determined from the symmetry-data vector in this SG.

\section{Stable real space invariants (SRSIs)}
\label{sec:globalrsis}
The local RSIs derived in SN~\ref{sec:realspacedata}, which are defined {\it locally} at a WP $W$, do not change under adiabatic processes that exchange site-symmetry irreps between $W$ and the {\it lower}-symmetric WPs connected to $W$.
This motivates and justifies their name, local RSIs, because they are directly defined by the multiplicities of site-symmetry irreps at $W$, or the $W$-position site-symmetry representation vector $p_W$.

However, in order to use the local RSIs as topological invariants, they must be good quantum numbers that are left invariant under all possible adiabatic processes, including adiabatic processes that involve WPs connected to $W$ that have {\it higher} symmetry than $W$.
The local RSIs at maximal WPs are topological invariants by definition since maximal WPs have the connected WPs only with lower site-symmetry groups.
In contrast, the local RSIs defined at non-maximal WPs are not topological invariants in general.
This is because a site-symmetry irrep at a non-maximal WP $W'$ can be adiabatically moved to maximal WPs connected to $W'$.
However, the local RSIs at $W'$ are defined such that they are left invariant only under the subset of adiabatic processes where a site-symmetry irrep at $W'$ is moved to connected lower-symmetric WPs.
They are not necessarily invariant under adiabatic processes where a site-symmetry irrep at $W'$ is moved to connected WPs with higher-symmetry.

We can see a concrete example of this in our analysis in SN~\ref{sec:lrsi_212} of the local RSIs in SG $P4_332$.
We saw in SEq.~\eqref{eq:212_lrsis_non-maximal} that there is a $\Z$RSI at the non-maximal WP $8c$ defined by $\delta_{\Z,8c} = -2 m[(\bar{E} \bar{E})_{8c}] + m[({}^1 \bar{E} {}^2 \bar{E})_{8c}]$.
By the construction, $\delta_{\Z,8c}$ is invariant under the adiabatic process $(\bar{E} \bar{E})_{8c} + 2 ({}^1 \bar{E} {}^2 \bar{E})_{8c} \Leftrightarrow (\bar{A} \bar{A})_{24e}$ where a linear combination of site-symmetry irreps at $8c$ is deformed to a site-symmetry irrep at the lower-symmetry $24e$ WP connected to $8c$.
However, for adiabatic processes where a site-symmetry irrep at $8c$ is deformed to a linear combination of site-symmetry irreps at the maximal $4a$ or $4b$ WP, $\delta_{\Z,8c}$ changes in general.
For example, $\delta_{\Z,8c}$ decreases by 1 when $({}^1 \bar{E} {}^2 \bar{E})_{8c}$ is moved to $2(\bar{E}_1)_{4a}$ via the adiabatic process in SEq.~\eqref{eq:sg212_adia2_8c}.

To obtain topological invariants capturing the full real-space data, we must find {\it stable} RSIs (SRSIs) that remain invariant under {\it all} possible adiabatic processes.
Then, as shown in SN~\ref{sec:rsi_relations}, the SRSIs will fully specify the local RSIs at maximal WPs\footnote{This means that the local RSIs at maximal WPs are a subset of SRSIs.}, but will also contain additional information expressed in terms of adiabatically invariant combinations of site-symmetry irrep multiplicities at both maximal and non-maximal WPs.
Similar to local RSIs, we will find both $\Z$SRSIs and $\Z_n$SRSIs.
We will show that the $\Z$SRSIs can be fully expressed in terms of momentum-space symmetry data, and that they capture the $\Z$-valued local RSIs at maximal WPs as well as the composite RSIs~\cite{xu2021threedimensional}.
Similarly, the $\Z_n$-valued local RSIs at maximal WPs can be expressed as combinations of $\Z$SRSIs and $\Z_n$SRSIs.
Furthermore, we will show that the $\Z_n$SRSIs cannot be expressed in terms of momentum-space data, and that they capture adiabatic invariants generally beyond the local and composite RSIs.
Note that unlike the local RSIs, where only $n=2$ $\Z_n$-valued RSIs occur, $\Z_n$SRSIs exist for $n=2$ for SGs without SOC and $n=2,4$ for SGs with SOC.

\subsection{Derivation of the SRSIs}
\label{sec:srsi_derivation}
We now describe the algorithm to calculate the SRSIs which is similar to the one used to compute the local RSIs in SN~\ref{sec:rsi_algorithm}.
The main difference is that we now must consider all possible adiabatic processes and track changes in the site-symmetry irrep multiplicities at all the WPs in every process.
In this SN, the derivation proceeds in two steps.
First, we define the adiabatic-process matrix $q$, which encodes all symmetry-preserving adiabatic processes within the unit cell.
Second, using $q$, we determine the linear combinations of site-symmetry irrep multiplicities that remain invariant under arbitrary integer linear combinations of these processes.
This is achieved by performing a Smith decomposition of $q$, which yields the SRSI matrix defining a linear map from the site-symmetry representation vector $p$ [SEq.~\eqref{eq:p_W}] to the SRSIs.
We explain each step in detail below.

\subsubsection{Construction of the adiabatic-process matrix $q$}
The adiabatic-process matrix $q$ encodes {\it all} symmetry-preserving adiabatic processes within the unit cell and can be defined as follows.
Using a notation similar to that used in SN~\ref{sec:rsi_algorithm} to define $q_W$, we denote the total number of distinct site-symmetry irreps within the unit cell by $N^\rho_{\rm UC}$ and the total number of distinct adiabatic processes by $N_{\rm adia}$.
Then, $q$ is a $N^\rho_{\rm UC} \times N_{\rm adia}$ integer-valued matrix.
Below, we detail how to construct the matrix $q$ for a general SG.
In particular, the columns of $q$ are generated by adiabatic processes that move site-symmetry irreps between connected Wyckoff positions while preserving all symmetries.

The calculation of the $q$ matrix involves all pairs of connected WPs.
We now explain the general algorithm to construct the $q$ matrix and, to illustrate the procedure, we apply it to the SG $Pmm21'$ with SOC, whose resulting SRSIs are explicitly given in the main text.
Let $W$ be a WP in a given SG and $W'$ another WP such that $W$ lies within $W'$ (for instance, $W'$ may represent a line and $W$ a point belonging to that line, or $W'$ a plane and $W$ a point or line belonging to that plane).
Let us also assume that the site-symmetry group $G_{W'}$ of $W'$ is a proper subgroup of the site-symmetry group of $W$, i.e., $G_{W'} < G_W$.
Every irrep $\rho_{W'}$ of $G_{W'}$ induces a representation in $G_W$ which is, in general, reducible into irreps $\rho_W^j$ of $G_W$ with multiplicities $m(\rho_W^j)$,
\bg
\rho_{W'} \uparrow G_W
= \sum_j \, m(\rho_W^j) \rho_W^j.
\eg
In the opposite direction, every irrep $\rho_W$ of $G_W$ subduces into a representation of $G_{W'}$ that is also, in general, reducible into irreps of $G_{W'}$,
\bg
\rho_W \downarrow G_{W'}
= \sum_i \, m(\rho_{W'}^i) \rho_{W'}^i.
\eg

For instance, let us take the WP $W = 1a$ with representative position $(0,0,z)$ and site-symmetry group $mm21'$, and the WP $W' = 2e$ with representative position $(x,0,z)$ and site-symmetry group $m1'$.
The site-symmetry group $m1'$ of the WP $2e$ has one double-valued irrep ${}^1 \bar{E} {}^2 \bar{E}$ of dimension 2, and the site-symmetry group $mm21'$ of the WP $1a$ also has only one double-valued irrep $\bar{E}$ of dimension 2.
The corresponding induction and subduction relations are
\bg
{}^1 \bar{E} {}^2 \bar{E} \uparrow G_{1a} = 2 \bar{E},
\label{eq:induction}
\\
\bar{E} \downarrow G_{2e} = {}^1 \bar{E} {}^2 \bar{E},
\label{eq:subduction}
\eg
respectively.
These relations indicate that the irrep ${}^1 \bar{E} {}^2 \bar{E}$ of $G_{2e}$, which is denoted as $({}^1 \bar{E} {}^2 \bar{E})_{2e}$) in our notation, is self-conjugated by the operations in $G_{1a}$ that are not contained in $G_{2e}$.
In the context of electronic band structures, this means that if there is a set of orbitals centered at $(x,0,z)$ belonging to the WP $2e$ and transforming as the irrep $({}^1 \bar{E} {}^2 \bar{E})_{2e}$, there must be another set of orbitals centered at $(-x,0,z)$, which corresponds to the other branch of the same WP $2e$.
This second set also transforms as the same irrep $({}^1 \bar{E} {}^2 \bar{E})_{2e}$.
The induction relation in SEq.~\eqref{eq:induction} shows that the two sets of orbitals located on the two branches of the WP $2e$ can be continuously moved towards the WP $1a$, where they merge.
The resulting set of orbitals centered at the WP $1a$ transforms as the induced representation ${}^1 \bar{E} {}^2 \bar{E} \uparrow G_{1a}$, which decomposes into two copies of the irrep $\bar{E}$ of $G_{1a}$, i.e., $2 (\bar{E})_{1a}$.

As stressed before, to construct the full set of adiabatic processes of this type, we must consider all pairs of WPs whose site-symmetry groups are related by a group-subgroup relation.
For each such pair, every irrep of the site-symmetry group of lower symmetry can be associated with an adiabatic process, which is represented by a column in the $q$ matrix.
However, not all of these processes are independent; if $W$, $W'$, and $W''$ are three WPs whose site-symmetry groups satisfy the group-subgroup relations $G_{W''} < G_{W'} < G_W$, then the adiabatic process connecting $W''$ to $W$ can be realized in two steps: first by moving the orbitals from $W''$ to $W'$, and then from $W'$ to $W$.
Accordingly, in the $q$ matrix, the columns associated with the adiabatic processes $W'' \to W$ are integer linear combinations of the columns associated with the processes $W'' \to W'$ and $W' \to W$.
Including such redundant columns does not change the integer column span of $q$, and therefore does not give rise to additional independent SRSIs (as the SRSIs are obtained from $q$ via a Smith decomposition, as will be explained below).
We therefore restrict the construction to pairs of WPs for which the site-symmetry group of lower symmetry $G_{W'}$ is a maximal subgroup of $G_W$.
This choice yields a minimal generating set of adiabatic processes and simplifies the subsequent Smith decomposition without affecting the final SRSIs.

In SG $Pmm21'$, there are four maximal WPs that represent lines, $1a \, (0,0,z)$, $1b \, (0,1/2,z)$, $1c \, (1/2,0,z)$, and $1d \, (1/2,1/2,z)$; four nonmaxial WPs that represent planes, $2e: \{(x,0,z), (-x,0,z)\}$, $2f: \{(x,1/2,z), (-x,1/2,z)\}$, $2g: \{(0,y,z), (0,-y,z)\}$, and $2h: \{(1/2,y,z), (1/2,-y,z)\}$; and the general position $4i: \, \{(x,y,z)$, $(-x,-y,z)$, $(x,-y,z)$, $(-x,y,z)\}$.
The site-symmetry groups of the lines, planes, and general position are isomorphic to the point groups $mm21'$, $m1'$, and $11'$, respectively.
These point groups each admit a single double-valued irrep: $\bar{E}$, ${}^1 \bar{E} {}^2 \bar{E}$, and $\bar{A} \bar{A}$, respectively, all of dimension 2.
Therefore, the total number of distinct site-symmetry irreps within the unit cell is $N^\rho_{\rm UC}=9$ for this SG.

Every maximal WP is connected to two plane WPs ($1a$ to $2e$ and $2g$, $1b$ to $2f$ and $2g$, $1c$ to $2e$ and $2h$, and $1d$ to $2f$ and $2h$).
Accordingly, we consider eight adiabatic processes (eight columns in the $q$ matrix) corresponding to symmetry-preserving interchanges of orbitals between line and layer WPs.
In addition, we must include four adiabatic processes corresponding to the connections between the four plane WPs and the general position $4i$.
As stressed before, it is not necessary to consider direct connections between the maximal WPs and the general position, since they would only introduce redundant columns that do not affect the final SRSIs.
Therefore, the total number of adiabatic processes considered is $N_{\rm adia}=12$.
The resulting $q$ matrix for the SG $Pmm21'$ with SOC is
\bg
q=
\begin{pNiceArray}[last-col,code-for-last-col=\scriptstyle]{cccccccccccc}
2 & 2 & 0 & 0 & 0 & 0 & 0 & 0 & 0 & 0 & 0 & 0 & (\bar{E})_{1a} \\
0 & 0 & 2 & 2 & 0 & 0 & 0 & 0 & 0 & 0 & 0 & 0 & (\bar{E})_{1b} \\
0 & 0 & 0 & 0 & 2 & 2 & 0 & 0 & 0 & 0 & 0 & 0 & (\bar{E})_{1c} \\
0 & 0 & 0 & 0 & 0 & 0 & 2 & 2 & 0 & 0 & 0 & 0 & (\bar{E})_{1d} \\
-1 & 0 & 0 & 0 & -1 & 0 & 0 & 0 & 2 & 0 & 0 & 0 & ({}^1 \bar{E} {}^2 \bar{E})_{2e} \\
0 & 0 & -1 & 0 & 0 & 0 & -1 & 0 & 0 & 2 & 0 & 0 & ({}^1 \bar{E} {}^2 \bar{E})_{2f} \\
0 & -1 & 0 & -1 & 0 & 0 & 0 & 0 & 0 & 0 & 2 & 0 & ({}^1 \bar{E} {}^2 \bar{E})_{2g} \\
0 & 0 & 0 & 0 & 0 & -1 & 0 & -1 & 0 & 0 & 0 & 2 & ({}^1 \bar{E} {}^2 \bar{E})_{2h} \\
0 & 0 & 0 & 0 & 0 & 0 & 0 & 0 & -1 & -1 & -1 & -1 & (\bar{A} \bar{A})_{4i}
\end{pNiceArray}.
\label{eq:adiamatrix_pmm2}
\eg
As indicated by the labels shown to the right of the matrix, the $N^\rho_{\rm UC}=9$ site-symmetry irreps are ordered in a fixed manner.
Each column of $q$ represents a specific adiabatic process.
The positions of the entries $-1$ and $2$ within a given column identify the irreps involved in the corresponding induction process.
For example, in the first column, the entries $-1$ and $2$ indicate that one set of orbitals transforming as
$({}^1 \bar{E} {}^2 \bar{E})_{2e}$ is removed from the WP $2e$, and a set of orbitals transforming as two copies of $(\bar{E})_{1a}$ is added at the WP $1a$.

The example discussed above illustrates how the columns of the adiabatic-process matrix $q$ can be constructed from induction relations between site-symmetry representations.
For an arbitrary SG, the same procedure can be carried out algorithmically by considering all pairs of Wyckoff positions related by group-subgroup relations.
All possible induction relations required for this construction [as exemplified by the relation in SEq.~\eqref{eq:induction}] can be found, for instance, in the Supplementary Materials of SRef.~\cite{xu2021threedimensional}, for both non-magnetic and magnetic point groups.

From a linear-algebraic perspective, the adiabatic-process matrix $q$ collects all symmetry-preserving adiabatic processes as integer vectors acting on the site-symmetry representation vector $p$ [SEq.~\eqref{eq:p_W}].
In particular, the $m$-th column $q_{*,m}=\Delta p^{(m)}$ represents an adiabatic process where the site-symmetry representation vector $p$ changes by $\Delta p^{(m)}$ as orbitals are adiabatically moved from one WP to a connected WP.
Each element $q_{im} = (\Delta p^{(m)})_i \in \Z$ captures the change of multiplicity of $i$-th site-symmetry irrep defined in the unit cell as a result of the $m$-th adiabatic process.
Note that row and column indices run over $i = 1, \dots, N^\rho_{\rm UC}$ and $m = 1, \dots, N_{\rm adia}$.
This interpretation of $q$ will be the starting point for constructing the SRSI matrix in the following.

Before constructing the SRSIs, let us first comment on two properties of adiabatic-process matrix $q$.
First, recall that the $m$-th adiabatic process changes a site-symmetry representation vector $p$ to $p'=p + \Delta p^{(m)}$.
By definition, any adiabatic process must leave the symmetry-data vector $B$ invariant.
This implies $\Delta B = \BR \cdot (p'-p) = \BR \cdot \Delta p^{(m)}=0$.
This relation holds for all $m$ and thus matrix $q$ satisfies
\bg
\BR \cdot q = 0.
\label{eq:BR_q_relation}
\eg
Second, we note that there exists a close relationship between the adiabatic-process matrix $q$ and the $W$-position trivial configuration matrices $q_W$, which introduced in SN~\ref{sec:rsi_algorithm} to derive the local RSIs.
In particular, $q_W$ defined for a WP $W$ can be obtained as the submatrix of $q$ where we keep the rows correspond to the site-symmetry irreps at $W$ and those columns with non-negative entries.
However, $q$ contains additional information beyond that contained in the collection of $q_W$ due to the existence of adiabatic processes that decrease the number of site-symmetry irreps at connected WPs.
As such, $q$ is not simply a direct sum of the $q_W$.
We illustrate a concrete example in SN~\ref{sec:srsi_212} where we compute the SRSIs for SG $P4_332$.

\subsubsection{Construction of the SRSI matrix}
Now, given a site-symmetry representation vector $p$, we aim to extract the information that remains invariant under any adiabatic processes.
Importantly, every adiabatic process must be representable as a linear combination of the columns of $q$ with integer coefficients.
To construct a basis for the space of adiabatic processes, we can compute the Smith decomposition of $q$ as
\ba
q = L \cdot \Lambda \cdot R \quad
\rightarrow \quad L^{-1} \cdot q \cdot R^{-1}
= \Lambda,
\label{eq:qvecdef_global}
\ea
where $L$ and $R$ are $N^\rho_{\rm UC} \times N^\rho_{\rm UC}$ and $N_{\rm adia} \times N_{\rm adia}$ unimodular matrices, respectively, and $\Lambda$ is an $N^\rho_{\rm UC} \times N_{\rm adia}$ diagonal matrix whose diagonal entries are non-negative integers known as the elementary divisors of $q$.
As in SN~\ref{sec:rsi_algorithm}, we order the basis of elementary divisors such that the diagonal part of $\Lambda$ is represented as
\bg
{\rm diag}(\Lambda)
= (\lambda_1, \dots, \lambda_1, \lambda_2, \dots, \lambda_2, \dots, \lambda_M, \dots, \lambda_M, 0, \dots, 0),
\label{eq:lambda_form}
\eg
where $\lambda_1 \le \lambda_2 \le \dots \le \lambda_M$ such that $\lambda_a$ divides $\lambda_b$ for $1 \le a < b \le M$.
Of these, only ${\rm rank}(q)$ elements are non-zero.
Since $R$ is unimodular, the lattice spanned by the columns of $q \cdot R^{-1}$ is the same as the lattice spanned by the columns of $q$, so that the columns of $\tilde{q} \equiv q \cdot R^{-1}$ form an equivalent basis for the adiabatic processes.
We can then follow similar logic as for the local RSIs in SN~\ref{sec:rsi_algorithm} to identify linear combinations of site-symmetry irrep multiplicities that cannot change under adiabatic processes.
A linear combination of adiabatic processes changes the site-symmetry representation vector $p$ by $\Delta p = \tilde{q} \cdot z$ with $z \in \Z^{N_{\rm adia}}$ (where we have used the fact that $R$ is unimodular).
Multiplying by $L^{-1}$, we then have that $L^{-1} \cdot p$ changes under an adiabatic process by
\ba
(L^{-1} \cdot \Delta p)_i = (\Lambda \cdot z)_i =
\begin{cases}
0 \mod \Lambda_{ii} & \trm{for } i \le {\rm rank}(q)
\\
0 & \trm{for } i > {\rm rank}(q).
\end{cases}
\label{eq:srsi_change_in_p}
\ea
SEq.~\eqref{eq:srsi_change_in_p} allows us to define a set of invariants
\bg
\tilde{\theta}_i =
\begin{cases}
(L^{-1} \cdot p)_i \mod \Lambda_{ii} & \trm{for } i \le {\rm rank}(q) \\
(L^{-1} \cdot p)_i & \trm{for } i > {\rm rank}(q)
\end{cases}
\label{eq:nonsimplifiedthetas}
\eg
that give a good set of quantum numbers that remain invariant under all adiabatic processes.
We see that for $i\leq {\rm rank}(q)$, $\tilde{\theta}_i$ is an $\Z_{\Lambda_{ii}}$SRSI, while the $\tilde{\theta}_i$ for $i>{\rm rank}(q)$ are $\Z$SRSIs.
We emphasize that while this procedure parallels our construction of the local RSIs in SN~\ref{sec:rsi_algorithm}, the differences between the local adiabatic-process matrices $q_W$ and the matrix $q$ imply that the SRSIs are not simply linear combinations of the local RSIs in each space group.

To go further, we can use SEq.~\eqref{eq:nonsimplifiedthetas} to define a simplified set of SRSI matrices.
To start, we denote the distinct elementary divisors of $q$ and their multiplicities as $\lambda_1, \lambda_2, \dots, \lambda_M, 0$ and $n_{\lambda_1}, n_{\lambda_2}, \dots, n_{\lambda_M}, n_0$, respectively.
[Note that $n_0=N^\rho_{\rm UC}-{\rm rank}(q)$ and $n_{\lambda_1} + \dots + n_{\lambda_M} = {\rm rank}(q)$.]
According to SEq.~\eqref{eq:nonsimplifiedthetas}, we define
\ba
& \tilde{\Theta}^{(\lambda_1)}
= (L^{-1})_{1, \dots, n_{\lambda_1}, *} \mod \lambda_1, \nn
& \tilde{\Theta}^{(\lambda_2)}
= (L^{-1})_{n_{\lambda_1}+1, \dots, n_{\lambda_1}+n_{\lambda_2}, *} \mod \lambda_2, \quad \dots \quad \nn
& \tilde{\Theta}^{(\lambda_M)}
= (L^{-1})_{n_{\lambda_1}+\dots+n_{\lambda_{M-1}}+1, \dots, N^\rho_{\rm UC}-n_0, *} \mod \lambda_M, \nn
& \tilde{\Theta}^{(0)}
= (L^{-1})_{N^\rho_{\rm UC}-n_0+1, \dots, N^\rho_{\rm UC}, *},
\label{eq:thetatilde_definition}
\ea
which correspond to $\Z_{\lambda_1}$SRSIs, \dots, $\Z_{\lambda_M}$SRSIs, and $\Z$SRSIs, respectively.
In SEq.~\eqref{eq:thetatilde_definition}, $(M)_{a, \dots, b,*}$ denotes a submatrix of $M$ defined by taking from $a$-th to $b$-th rows.
Then, we define the $\Z$SRSIs $\tilde{\theta}_\Z = \tilde{\Theta}^{(0)} \cdot p$.
For the $\Z_n$SRSIs, because $\Z_1$SRSIs are trivial, we only take $\tilde{\Theta}^{(n)}$ with $n \ne 1$ and define $\Z_n$SRSIs $\tilde{\theta}_{\Z_n} = \tilde{\Theta}^{(n)} \cdot p$ mod $n$.

Although the matrices in SEq.~\eqref{eq:thetatilde_definition} give valid SRSI matrices, they need not give the simplest expressions for each adiabatic invariant.
That is, we look for a change of basis for the SRSIs such that they are functions of as few site-symmetry irrep multiplicities as possible and the coefficients of these multiplicities should be as small in magnitude as possible.
One choice of simplified SRSI matrices can be obtained using the {\it Hermite decomposition}~\cite{cohen2013course} to define 
\ba
\Theta^{(0)}
&= U^{(0)} \cdot \tilde{\Theta}^{(0)}, \nn
\Theta^{(\mu)}
&= U^{(\mu)} \cdot \tilde{\Theta}^{(\mu)} \mod \mu, \quad
\mu = \lambda_1, \lambda_2, \dots, \lambda_M,
\label{eq:hermite_normal_form}
\ea
where $U^{(0)}$ and $U^{(\mu)}$ are unimodular matrices and $\Theta^{(0)}$ and $\Theta^{(\mu)}$ are upper-triangular matrices.
$\Theta^{(0)}$ and $\Theta^{(\mu)}$ are known as the {\it Hermite normal forms} of $\tilde{\Theta}^{(0)}$ and $\tilde{\Theta}^{(\mu)}$, respectively.
(In SN~\ref{sec:SRSI_example} we will see an explicit example of how the Hermite decomposition simplifies the form of the SRSIs.)
We next define the SRSI matrix $\Theta$,
\ba
\Theta = 
\bpm
\Theta^{(0)} \\
\Theta^{(\lambda)} \\
\vdots \\
\Theta^{(\lambda')}
\epm,
\label{eq:simplifiedSRSI_final}
\ea
by taking $\Theta^{(0)}$ and $\Theta^{(\mu=\lambda, \cdots, \lambda')}$ in SEq.~\eqref{eq:hermite_normal_form} with $\mu \ne 1$.
[Recall that the $\Z_1$RSIs are trivial quantities and thus we exclude $\Theta^{(1)}$ in SEq.~\eqref{eq:simplifiedSRSI_final}.]
Finally, the simplified $\Z$SRSIs and $\Z_n$SRSIs are defined as
\bg
\theta_\Z = \Theta^{(0)} \cdot p, \quad
\theta_{\Z_n} = \Theta^{(n)} \cdot p \mod n, \quad
\theta = \bpm \theta_\Z \\ \theta_{\Z_\lambda} \\ \vdots \\ \theta_{\Z_{\lambda'}} \epm,
\label{eq:SRSI_definition}
\eg
for $n=\lambda, \lambda', \dots$.

We next highlight an important structural property of the simplified SRSI matrices $\Theta^{(m)}$ with $m=0$ or $n$: the pseudoinverse of $\Theta^{(m)}$ is integer-valued.
To prove this, let us denote the Smith decomposition of $\Theta^{(m)}$ as $\Theta^{(m)} = L_\Theta \cdot \Lambda_\Theta \cdot R_\Theta$.
Specifically, we will show that the Smith normal form $\Lambda_\Theta$ and its pseudoinverse $[\Lambda_\Theta]^\ddagger$ have only 1's for their elementary divisors, meaning that all the nonzero diagonal elements of $\Lambda_\Theta$ and $[\Lambda_\Theta]^\ddagger$ are 1.
To see this, recall that the non-simplified SRSI matrix is defined as $\tilde{\Theta}^{(m)} = [U^{(m)}]^{-1} \cdot \Theta^{(m)}$ [SEq.~\eqref{eq:hermite_normal_form}], and consists of rows of the unimodular matrix $L^{-1}$ [SEq.~\eqref{eq:thetatilde_definition}].
Let $\{r_1, r_2, \dots, r_{N_\Theta}\}$ be the indices of the rows of $L^{-1}$ forming $\tilde{\Theta}^{(m)}$.
Then, we can write
\bg
\tilde{\Theta}^{(m)} = \bpm \bb e_{r_1} \\ \bb e_{r_2} \\ \vdots \\ \bb e_{r_{N_\Theta}} \epm \cdot L^{-1} \equiv \mc{R} \cdot L^{-1},\label{eq:thetamrows}
\eg
where $\bb e_i$ is the standard-basis row vector of dimension $N^\rho_{\rm UC}$ satisfying $(\bb e_i)_j=\delta_{ij}$ for $i,j=1,\dots,N^\rho_{\rm UC}$, and we have defined the row-selector matrix $\mc{R}$.
Now, we observe that the row-selector matrix $\mc{R}$ satisfies
\bg
\mc{R} = \bpm \mathbb{1}_{N_\Theta} | \bb 0_{N_\Theta \times (N^\rho_{\rm UC}-N_\Theta)} \epm \cdot P,
\quad
P = \bpm \mc{R} \\ \bb e_{\bar{r}_1} \\ \bb e_{\bar{r}_2} \\ \vdots \\ \bb e_{\bar{r}_{N^\rho_{\rm UC}-N_\Theta}} \epm,\label{eq:rowselectordef}
\eg
where $\{\bar{r}_1, \bar{r}_2, \dots, \bar{r}_{N^\rho_{\rm UC}-N_\Theta}\}$ are the complementary row indices not included in $\tilde{\Theta}^{(m)}$, and $\mathbb{1}_{N_1}$ and $\bb 0_{N_1 \times N_2}$ are the $N_1 \times N_1$ identity matrix and the $N_1 \times N_2$ zero matrix, respectively.
Importantly, $P$ is unimodular with $|{\rm Det} P|=1$ since it reduces to the identity matrix $\mathbb{1}_{N^\rho_{\rm UC}}$ by row permutations.
Combining SEqs.~\eqref{eq:thetamrows} and \eqref{eq:rowselectordef} with SEq.~\eqref{eq:hermite_normal_form}, we find
\bg
\Theta^{(m)} = U^{(m)} \cdot \bpm \mathbb{1}_{N_\Theta} | \bb 0_{N_\Theta \times (N^\rho_{\rm UC}-N_\Theta)} \epm \cdot P \cdot L^{-1},\label{eq:thetam_smith}
\eg
Since both $U^{(m)}$ and $P \cdot L^{-1}$ are unimodular, SEq.~\eqref{eq:thetam_smith} gives the Smith decomposition of $\Theta^{(m)}$, We thus identify Smith normal form $\Lambda_\Theta$ of $\Theta^{(m)}$ and its pseudoinverse $[\Lambda_\Theta]^\ddagger$ as
\bg
\Lambda_\Theta = \bpm \mathbb{1}_{N_\Theta} | \bb 0_{N_\Theta \times (N^\rho_{\rm UC}-N_\Theta)} \epm,
\quad
[\Lambda_\Theta]^\ddagger = [\Lambda_\Theta]^T.
\eg
Therefore, the pseudoinverse of $\Theta^{(m)}$, $[\Theta^{(m)}]^\ddagger = [R_\Theta]^{-1} \cdot [\Lambda_\Theta]^\ddagger \cdot [L_\Theta]^{-1}$, is integer-valued, as all three matrices, $L_\Theta^{-1}$, $R_\Theta^{-1}$, and $[\Lambda_\Theta]^\ddagger$ are integer matrices:
\bg
[\Theta^{(m)}]^\ddagger \in \Z^{N_\Theta \times N^\rho_{\rm UC}}.
\label{eq:srsi_inverse_integer}
\eg
This integer-valuedness is important for two main reasons.
First, it ensures that the mapping from $\Z$SRSIs to symmetry-data vector, which we will discuss in SN~\ref{sec:onetoone}, maps integers to integers, rather than integers to rational numbers.
Second, it becomes essential when extending $\Z$SRSIs to topological bands and establishing their correspondence with symmetry indicators, as discussed in SN~\ref{sec:stableindexdef}.

Lastly, we comment on the nonuniqueness of the SRSI matrices.
In general, the definition of SRSIs is still nonunique due to the freedom we have in permuting the rows or columns of $q$.
However, all choices of defining the SRSIs are equivalent up to the following redefinitions:
Recall that the SRSI matrices were redefined using the Hermite decomposition, as shown in SEq.~\eqref{eq:hermite_normal_form}.
Beyond this, a broader class of redefinition is possible, provided that the original $\Z$- and $\Z_n$-valuedness of SRSIs remain unchanged.
In particular, the $\Z$SRSI matrix and the $\Z_n$SRSI matrices can each be redefined by multiplication with a unimodular matrix.
Furthermore, note also that the $\Z_n$SRSIs can be redefined by adding any linear combination of $\Z$SRSIs and $\Z_{m>n}$SRSIs, as the redefined $\Z_n$SRSIs remain well-defined modulo $n$, and no information is lost.
To see this, recall that for $m>n$, $n$ divides any $m$, i.e., $n|m$, because elementary divisors in the Smith normal form of $q$ [SEq.~\eqref{eq:lambda_form}] always satisfy $\Lambda_{i,i}|\Lambda_{i+1,i+1}$ for $1 \le i < {\rm rank}(q)$.
Further details on this type of redefinition and its use can be found in SN~\ref{sec:I432_stable}.

In SN~\ref{sec:SRSI_example}, we will illustrate how to derive the SRSIs in some example SGs.
In SN~\ref{sec:srsitables} we provide exhaustive tables for the SRSIs in all 230 SGs with and without SOC.
From these exhaustive tables, we found that in SGs without SOC only $n=2$ $\Z_n$SRSIs occur; in the double SGs (those with SOC), both $n=2$ and $n=4$ $\Z_n$SRSIs can occur.
In particular, $\Z_4$SRSIs appear in only four SGs, all with SOC: $F222$ (No. 22), $I\bar{4}c2$ (No. 120), $F23$ (No. 196), and $F\bar{4}3c$ (No. 219).
Although this work primarily focuses on nonmagnetic SGs where time-reversal symmetry exists, the algorithm for obtaining SRSIs can be directly extended to magnetic SGs as well.
As a concrete example, in SN~\ref{sec:examples}, we study a magnetic space group where time-reversal symmetry does not exist, illustrating how SRSIs can be computed in such cases.
We now examine the physical interpretation of the SRSIs and their relationship to the symmetry-data vector.

\subsection{Key properties of SRSIs}
\label{sec:zsrsi_proof}
In this SN, we prove that the $\Z$SRSIs for a set of bands can be uniquely determined from the symmetry-data vector.
We will use a combination of analytic and numerical methods.
First, in SN~\ref{sec:zrsiimpliesb} we show analytically that for any SG both with and without SOC, if two sets of bands have the same $\Z$SRSIs, then the sets of bands must have the same symmetry-data vector.
Thus, the symmetry-data vector is uniquely determined by the $\Z$SRSIs.
Second, in SN~\ref{sec:bimplieszsrsi} we rely on explicit and exhaustive computation for proving that band representations with matching symmetry-data vectors have matching $\Z$SRSIs, such that the $\Z$SRSIs are uniquely determined from the the symmetry-data vector in all 230 SGs with and without SOC.
We combine these results to establish a one-to-one mapping between the space of symmetry-data vectors and $\Z$SRSIs in every SG in SN~\ref{sec:onetoone}.
However, unlike $\Z$SRSIs, we demonstrate that $\Z_n$SRSIs are not uniquely determined by symmetry-data vector, nor does the symmetry-data vector uniquely determine the $\Z_n$SRSIs in all SGs.
We demonstrate this with an explicit example.
Third, in SN~\ref{sec:stableeq} we define stable equivalence between BRs and show that BRs are stably equivalent if and only if all their SRSIs (both $\Z$SRSIs and $\Z_n$SRSIs) match.
Finally, in SN~\ref{sec:rsi_relations} we examine the relationship between SRSIs and other RSIs, with a particular focus on how local RSIs at maximal WPs and composite RSIs can be expressed in terms of SRSIs.

\subsubsection{Symmetry-data vector is uniquely determined by $\Z$SRSIs}
\label{sec:zrsiimpliesb}
We will now show that symmetry-data vector is uniquely determined by $\Z$SRSIs.
In the process, we will also demonstrate that the $\Z_n$SRSIs do not determine the symmetry-data vector.
In particular, we prove that two atomic insulators with the same values of all $\Z$SRSIs ($\Z_n$SRSIs) have (do not necessarily have) the same symmetry-data vector.
First, to prove these statements, consider two atomic insulators (which can be obstructed atomic limits) $AI_1$ and $AI_2$.
Each atomic insulator is described by a site-symmetry representation vector $p_{AI_a}$, the corresponding symmetry-data vector $B_{AI_a} = \BR \cdot p_{AI_a}$, and $\Z$SRSIs $\theta_{\Z, AI_a}$ ($a=1,2$).
We assume that $AI_1$ and $AI_2$ have the same $\Z$SRSIs such that $\theta_{\Z, AI_1} = \theta_{\Z, AI_2}$.
To prove that $AI_1$ and $AI_2$ have the same symmetry-data vector, we first prove an important result that lends a physical interpretation to the $\Z$SRSIs: if the $\Z$SRSIs of $AI_1$ and $AI_2$ match, there exists an integer $\mc{N}{\geq 1}$ and a symmetry-representation vector $p_{aux}$ (representing auxiliary Wannier orbitals or trivial bands) such that two atomic insulators with site-symmetry representation vectors $\mc{N} p_{AI_1} + p_{aux}$ and $\mc{N} p_{AI_2} + p_{aux}$ are adiabatically deformable to each other (i.e., such that $\mc{N} p_{AI_1} + p_{aux}$ and $\mc{N} p_{AI_2} + p_{aux}$ induce equivalent band representations).
This implies that $\mc{N} p_{AI_1} + p_{aux}$ and $\mc{N} p_{AI_2} + p_{aux}$ yield the same symmetry-data vector, and thus $p_{AI_1}$ and $p_{AI_2}$ must also yield the same symmetry-data vector, as the symmetry-data vector is linearly determined by the symmetry-representation vector [see SEq.~\eqref{eq:B_from_p}].
Also note that the adiabatic deformability of two atomic insulators in the presence of auxiliary orbitals or trivial bands will be formally defined with the notion of stable equivalence between BRs in SN~\ref{sec:stableeq}.
We prove this statement by showing that the Diophantine equation
\bg
(\mc{N} p_{AI_1} + p_{aux}) - (\mc{N} p_{AI_2} + p_{aux}) = q \cdot \bb n
\label{eq:diophantine1}
\eg
has integer-valued solutions $\bb n \in \Z^{N_{\rm adia}}$.
Note that $q$ is the $N^\rho_{\rm UC} \times N_{\rm adia}$ adiabatic-process matrix defined in SEq.~\eqref{eq:qvecdef_global}.
From the Smith decomposition $q= L \cdot \Lambda \cdot R$, SEq.~\eqref{eq:diophantine1} can be written as
\bg
\mc{N} L^{-1} \cdot \Delta p
= \Lambda \cdot (R \cdot \bb n)
= (\Lambda_{11} \, v_1, \Lambda_{22} \, v_2, \dots, \Lambda_{rr} \, v_r, 0, \dots, 0)^T,
\label{eq:diophantine2}
\eg
where $\Delta p = p_{AI_1} - p_{AI_2}$, $v_{i=1,\dots,r} = (R \cdot \bb n)_i$, and $r$ is the rank of $q$.
Since $R$ is unimodular, the existence integer-valued solutions for $\bb n$ is equivalent to the existence of integer $v_i \in \Z$.
Similarly, $L^{-1} \cdot \Delta p \in \Z^{N^\rho_{\rm UC}}$ because the assumption that $AI_1$ and $AI_2$ are atomic insulators means $p_{AI_1}$ and $p_{AI_2}$ are integer-valued vectors.

The first $r$ rows in SEq.~\eqref{eq:diophantine2} are
\bg
\mc{N} (L^{-1} \cdot \Delta p)_i = \Lambda_{ii} \, v_i \in \Lambda_{ii} \Z
\label{eq:diophantine_zn}
\eg
for $i=1, \dots, r$.
There always exists an integer $\mc{N}$ such that $(v_1, \dots, v_r) \in \Z^r$ and a solution exists.
In this sense, we say that SEq.~\eqref{eq:diophantine_zn} is solvable.
For example, one can take $\mc{N}$ as the least common multiple of $(\Lambda_{11}, \dots, \Lambda_{rr})$.
Note that $(L^{-1})_{i,:}$ with $i=1,\dots,r$ defines the $\Z_n$SRSI matrices $\tilde{\Theta}^{(n)}$ in SEq.~\eqref{eq:thetatilde_definition} as explained in SN~\ref{sec:srsi_derivation}.
Thus, the existence of integer solutions to SEq.~\eqref{eq:diophantine_zn} implies that the two atomic insulators with site-symmetry representation vectors $(\mc{N} p_{AI_1} + p_{aux})$ and $(\mc{N} p_{AI_2} + p_{aux})$ have the same $\Z_n$SRSIs.

The remaining $(N^\rho_{\rm UC} - r)$ rows in SEq.~\eqref{eq:diophantine2} can be written as
\bg
0 = \mc{N} (\tilde{\Theta}^{(0)} \cdot \Delta p) \quad \trm{and thus} \quad
\tilde{\Theta}^{(0)} \cdot \Delta p = 0,
\label{eq:diophantine3}
\eg
where $\tilde{\Theta}^{(0)}$ is the $\Z$SRSI matrix defined in SEq.~\eqref{eq:thetatilde_definition}, and we have used $\mc{N}\geq 1$.
This also implies $\Theta^{(0)} \cdot \Delta p =0$ where $\Theta^{(0)} = U^{(0)} \cdot \tilde{\Theta}^{(0)}$ [SEq.~\eqref{eq:hermite_normal_form}] is the simplified $\Z$SRSI matrix obtained from the Hermite decomposition of $\tilde{\Theta}^{(0)}$ with a unimodular matrix $U^{(0)}$.
The assumption that $\theta_{\Z, AI_1} = \theta_{\Z, AI_2}$, implies that
\bg
\mc{N} (\tilde{\Theta}^{(0)} \cdot \Delta p)
= \mc{N} [U^{(0)}]^{-1} \cdot \Theta^{(0)} \cdot (p_{AI_1} - p_{AI_2})
= \mc{N} [U^{(0)}]^{-1} \cdot (\theta_{\Z, AI_1} - \theta_{\Z, AI_2}) = 0,
\eg
and so SEq.~\eqref{eq:diophantine3} holds.
Thus, SEq.~\eqref{eq:diophantine1} is solvable for any two atomic insulators $AI_1$ and $AI_2$ with matching $\Z$SRSIs, completing the proof.

Finally, note that the solvability of SEq.~\eqref{eq:diophantine1} implies that the symmetry-data vectors $B_{AI_1}$ and $B_{AI_2}$ for the two sets of bands must be equal.
To see this, note that
\bg
\mc{N} (B_{AI_1} - B_{AI_2})
= \mc{N} \BR \cdot (p_{AI_1} - p_{AI_2})
= \BR \cdot q \cdot \bb n =0.
\label{eq:difference_in_sym_data_is_0}
\eg
Here, we used the fact that $\BR \cdot q = 0$ [SEq.~\eqref{eq:BR_q_relation}] which follows from the fact that an adiabatic process cannot change the symmetry-data vector by definition.
Now suppose that $\theta_{\Z, AI_1} = \theta_{\Z, AI_2}$.
This implies that SEq.~\eqref{eq:diophantine3} has integer solutions, and so by choosing $\mc{N}>1$ large enough, we can find integer solutions to SEq.~\eqref{eq:diophantine2} and hence SEq.~\eqref{eq:diophantine1} is solvable.
By SEq.~\eqref{eq:difference_in_sym_data_is_0}, $AI_1$ and $AI_2$ have the same symmetry-data vector.
This proves that if the $\Z$SRSIs of two atomic insulators match, their symmetry-data vectors are equal.
In other words, symmetry-data vector is uniquely determined from $\Z$SRSIs.

An important result which follows from the arguments above is that $\BR \cdot p_{\theta_\Z = 0} = 0$ for any vector $p_{\theta_\Z =0}$ in the kernel of the $\Z$SRSI matrix $\Theta^{(0)}$, i.e., for any vector satisfying $\Theta^{(0)} \cdot p_{\theta_\Z=0} = 0$.
To understand this, let us consider the Diophantine equation, $\theta_\Z = \Theta^{(0)} \cdot p$.
For a given $\Z$SRSIs $\theta_\Z$, we can solve for the symmetry-representation vector $p$ by introducing a pseudoinverse of the $\Z$SRSI matrix $\Theta^{(0)}$:
\bg
p = [\Theta^{(0)}]^\ddagger \cdot \theta_\Z + p_{\theta_\Z=0}.
\label{eq:p_from_ZSRSI}
\eg
From the relation between $p$ and the symmetry-data vector $B$, i.e., $B = \BR \cdot p $ [SEq.~\eqref{eq:B_from_p}], we have
\bg
B = \BR \cdot [\Theta^{(0)}]^\ddagger \cdot \theta_\Z + \BR \cdot p_{\theta_\Z = 0}.
\label{eq:B_from_ZSRSI}
\eg
Since we showed that $B$ is uniquely determined, we conclude that
\bg
\BR \cdot p_{\theta_\Z = 0} = 0
\label{eq:ker_rel_BR}
\eg
in all SGs with and without SOC.

As mentioned previously, the symmetry-data vector for an atomic insulator is not uniquely determined from $\Z_n$SRSIs in general (when all SGs are considered).
Suppose $AI_1$ and $AI_2$ do have matching $\Z_n$SRSIs.
As we have proved above, the symmetry-data vectors $B_{AI_1}$ and $B_{AI_2}$ of $AI_1$ and $AI_2$ can match if their $\Z$SRSI match.
This implies that there can exist SGs where $AI_1$ and $AI_2$ with matching $\Z_n$SRSIs have different $\Z$SRSIs and thus different symmetry-data vectors.
In SN~\ref{sec:SRSI_example}, we provide explicit examples of such cases, which demonstrate that the symmetry-data vector is not uniquely determined by $\Z_n$SRSIs.

\subsubsection{$\Z$SRSIs are uniquely determined by symmetry-data vector}
\label{sec:bimplieszsrsi}
We now examine the converse: given two identical symmetry-data vectors corresponding to atomic insulators, we show that they must have identical $\Z$SRSIs.
While we do not have an analytical proof of this statement, we can reduce it to a concrete criterion that can be verified in all 230 SGs with and without SOC.
To do this, for a given SG, we first find an independent set of basis vectors spanning ${\rm ker} \, \BR$, the kernel of the BR matrix $\BR$.
A linear combination of basis vectors forms a general vector $p_0$ in ${\rm ker} \, \BR$.
Then, we solve the Diophantine equation, $B = \BR \cdot p$ [SEq.~\eqref{eq:B_from_p}] by introducing a pseudoinverse of the BR matrix $\BR^\ddagger$.
That is, for given symmetry-data vector $B$, the site-symmetry representation vector $p$ is determined as [SEq.~\eqref{eq:pinverserelation}]
\bg
p = \BR^\ddagger \cdot B + p_0.
\eg
From the relation between $p$ and the $\Z$SRSIs $\theta_\Z$, i.e., $\theta_\Z = \Theta^{(0)} \cdot p$ [SEq.~\eqref{eq:SRSI_definition}], we have
\bg
\theta_\Z = \Theta^{(0)} \cdot \BR^\ddagger \cdot B + \Theta^{(0)} \cdot p_0.
\label{eq:ZSRSI_from_B}
\eg
{Thus, if we are able to show that
\bg
\Theta^{(0)} \cdot p_0 =0
\label{eq:ker_rel_theta}
\eg
(i.e., if we are able to show that the kernel of $\BR$ is contained in the kernel of $\Theta^{(0)}$),} then we will have proved that atomic insulators with identical symmetry-data vectors must have identical $\Z$SRSIs.
We numerically checked that $\Theta^{(0)} \cdot p_0 = 0$ holds for all SG with and without SOC.
We leave a proof of this result analytically as a future research.

Contrary to the situation with $\Z$SRSIs, we show that $\Z_n$SRSIs are not uniquely determined by symmetry-data vector.
This can be proven by explicitly showing that there is at least one SG where two (E)BRs with the same symmetry-data vector have different $\Z_n$SRSIs.
We provide this example in SN~\ref{sec:SRSI_example} with SG $Pmm2$ (No. 25) with SOC where all EBRs have the same symmetry-data vector.

\subsubsection{One-to-one mapping between $\Z$SRSIs and symmetry-data vector}
\label{sec:onetoone}
Our results in SNs~\ref{sec:zrsiimpliesb} and \ref{sec:bimplieszsrsi} imply that there exists a one-to-one mapping between symmetry-data vectors and $\Z$SRSIs for atomic insulators.
To construct this mapping explicitly, we recall that the general solutions for a site-symmetry representation vector $p$ in terms of the symmetry-data vector $B$ and the $\Z$SRSIs $\theta_\Z$ [SEqs.~
\eqref{eq:pinverserelation} and \eqref{eq:p_from_ZSRSI}] for an atomic insulator is given as
\ba
p &= \BR^\ddagger \cdot B + p_0,
\label{eq:p_from_B} \\
p &= [\Theta^{(0)}]^\ddagger \cdot \theta_\Z + p_{\theta_\Z =0}
\label{eq:p_from_thetaZ},
\ea
where $p_0 \in {\rm ker} \, \BR$ and $p_{\theta_\Z=0} \in {\rm ker} \, \Theta^{(0)}$ are general vectors in the kernels of the BR matrix $\BR$ and the $\Z$SRSI matrix $\Theta^{(0)}$, respectively.
In SNs~\ref{sec:zrsiimpliesb} and \ref{sec:bimplieszsrsi}, we showed in SEqs.~\eqref{eq:ker_rel_BR} and \eqref{eq:ker_rel_theta} that
\bg
\BR \cdot p_{\theta_\Z=0} = 0,
\quad
\Theta^{(0)} \cdot p_0 = 0,
\label{eq:kernels_vanish}
\eg
in all SGs with and without SOC.
This also implies that the number of $\Z$SRSIs, $N^\rho_{\rm UC}-{\rm rank}(q)$, is equal to the rank of BR matrix, $r_{BR} = {\rm rank}(BR)$, i.e.,
\bg
r_{BR} = N^\rho_{\rm UC}-{\rm rank}(q), \quad {\rm rank}(q) = N^\rho_{\rm UC}-r_{BR}.
\eg
Hence, we can multiply SEq.~\eqref{eq:p_from_B} by $\Theta^{(0)}$ on the left to obtain
\ba
\theta_\Z ={\Theta^{(0)}\cdot p=} \Theta^{(0)} \cdot \BR^\ddagger \cdot B, \label{eq:maps_B_SRSI_1}
\ea
which determines $\Z$SRSIs from the symmetry-data vector for an atomic insulator.
Similarly, we can multiply SEq.~\eqref{eq:p_from_thetaZ} on the left by $\BR$ to obtain
\ba
B ={\BR\cdot p = } \BR \cdot [\Theta^{(0)}]^\ddagger \cdot \theta_\Z,
\label{eq:maps_B_SRSI_2}
\ea
which determines the symmetry-data vector for an atomic insulator from its $\Z$SRSIs.
SEqs.~\eqref{eq:maps_B_SRSI_1} and \eqref{eq:maps_B_SRSI_2} define a one-to-one mapping between symmetry-data vectors and $\Z$SRSIs for atomic insulators.
The mapping matrix $\BR \cdot [\Theta^{(0)}]^\ddagger$ is an integer matrix, since both $\BR$ and $[\Theta^{(0)}]^\ddagger$ are integer-valued [see SEq.~\eqref{eq:srsi_inverse_integer}].
Thus, any $\theta_\Z \in \Z$ is mapped to integer-valued vector $B$ via SEq.~\eqref{eq:maps_B_SRSI_2}.

It is important to note that the mapping matrix $\Theta^{(0)} \cdot BR^\ddagger$ in SEq.~\eqref{eq:maps_B_SRSI_1} is not unique due to an ambiguity associated with compatibility relations in momentum space.
Compatibility relations impose a set of linear constraints that any symmetry-data vector must satisfy.
One intuitive example of such a condition is the filling constraint.
The filling, or the number of bands, must remain the same at each HSM.
Since the filling at each HSM is determined by the total dimension of all little group irreps that appear, the filling constraint gives a quantitative relations among the multiplicities of little group irreps at different HSM, constraining the symmetry-data vector.
For any SG, the compatibility relations can be systematically obtained as follows:
Using the Smith decomposition of the BR matrix, $BR = L_{BR} \cdot \Lambda_{BR} \cdot R_{BR}$, we can write $(L_{BR})^{-1}$ as
\bg
(L_{BR})^{-1} = \bpm SI \\ C \epm,
\label{eq:si_and_comp}
\eg
where $SI$ and $C$ are $r_{BR} \times N^\rho_{\rm BZ}$ and $(N^\rho_{\rm BZ} - r_{BR}) \times N^\rho_{\rm BZ}$ matrices, respectively.
The matrix $SI$ determines the formulae for symmetry-indicators, whose group structure is determined by the elementary divisors of $\Lambda_{BR}$ as $\Z_{(\Lambda_{BR})_{1,1}} \times \Z_{(\Lambda_{BR})_{2,2}} \times \Z_{(\Lambda_{BR})_{r_{BR},r_{BR}}}$, while $C$ defines the compatibility relations~\cite{Po_2017,elcoro2020application}.
Since $C$ is in the block of $L_{\BR}^{-1}$ that spans the kernel of $\Lambda_{\BR}$, we see that symmetry-data vector must satisfy $C \cdot B = 0$.
Then, SEq.~\eqref{eq:maps_B_SRSI_1} can be redefined via
\bg
\theta_\Z = \Theta^{(0)} \cdot \BR^\ddagger \cdot B = \left( \Theta^{(0)} \cdot \BR^\ddagger + M_C \cdot C \right) \cdot B,
\label{eq:maps_B_SRSI_redef}
\eg
where $M_C$ is an arbitrary $r_{BR} \times (N^\rho_{\rm BZ} - r_{BR})$ integer-valued matrix.
Since the symmetry-data vector is already constrained to lie in the kernel of $C$, this redefinition has no effect.
Rather, our ability to redefine the mapping SEq.~\eqref{eq:maps_B_SRSI_2} is a reflection of the fact that the dimension of the space spanned by allowed symmetry-data vectors $B$ is generally smaller than the length $N^\rho_{\rm BZ}$ of the symmetry data vector itself; in particular the mapping SEq.~\eqref{eq:maps_B_SRSI_2} is still one-to-one for any choice of $M_C$.

Finally, we remark that because the symmetry indicators (SIs) are determined by a symmetry-data vector, SIs are determined by $\Z$SRSIs.
So far, we considered atomic insulators with integer-valued site-symmetry representation vectors $p_{AI_{1,2}}$ and $\Z$RSIs $\theta_{\Z, AI_{1,2}}$.
However, because the mappings between symmetry-data vector $B$, $\Z$SRSIs $\theta_\Z$, and the SIs (which can be defined for non-atomic topological bands) are all linear, $\Z$SRSIs can be uniquely defined from symmetry-data vector even for topological phases in which some entries of the site-symmetry representation vector $p$ can take negative integers or rational numbers.
We will discuss the relation between the SIs and the $\Z$SRSIs in SN~\ref{sec:stableindexdef} in detail.

\subsubsection{Stable equivalence between band representations}
\label{sec:stableeq}
We now define the concept of {\it stable equivalence} between atomic insulators (or their corresponding BRs), and show SRSIs indicate this equivalence.
Consider two atomic insulators with BRs $AI_1=\rho_1 \uparrow G$ and $AI_2=\rho_2 \uparrow G$, induced from site-symmetry representations $\rho_1$ and $\rho_2$, respectively.
If $AI_1$ and $AI_2$ can be adiabatically deformed into each other, they are (topologically) equivalent, which we write as $\rho_1 \uparrow G = \rho_2 \uparrow G$.
However, even when $AI_1$ and $AI_2$ are not equivalent, they can still be {\it stably equivalent} if they become equivalent in the presence of additional trivial bands $AI_{aux}=\rho_{aux} \uparrow G$.
That is, we say $AI_1$ and $AI_2$ are stably equivalent if
\bg
AI_1 \oplus AI_{aux} \ = AI_2 \oplus AI_{aux}.
\label{eq:stable_equiv}
\eg
Note, importantly, that it is the {\it same} set of trivial bands on both sides of this equivalence.

Next, we establish that two atomic insulators are stably equivalent if and only if they have matching $\Z$SRSIs and $\Z_n$SRSIs.
To show this, we consider two atomic insulators $AI_1$ and $AI_2$ with corresponding BRs $\rho_1 \uparrow G$ and $\rho_2 \uparrow G$.
First, we show that if $AI_1$ and $AI_2$ have matching $\Z$SRSIs and $\Z_n$SRSIs, then they are stably equivalent.
To this end, we introduce symmetry-representation vectors, $p_{AI_1}$ and $p_{AI_2}$ for each insulator, along with the symmetry-representation vector $p_{aux}$ for the auxiliary trivial bands.
To prove the stable equivalence between $AI_1$ and $AI_2$, we need to show that there exists a choice of $p_{aux}$ such that there is an adiabatic process connecting $p_{AI_1}+p_{aux}$ to $p_{AI_2}+p_{aux}$.
Using our definition of the adiabatic-process matrix $q$ from SN~\ref{sec:rsi_algorithm}, we can find such an adiabatic process by looking for integer-valued solutions $n \in \Z^{N_{\rm adia}}$ to the equation
\ba
p_{AI_1} + p_{aux}
&= p_{AI_2} + p_{aux} + q \cdot \bb n,
\\
\implies q \cdot \bb n
&= (p_{AI_1} + p_{aux}) - (p_{AI_2} + p_{aux}) .
\label{eq:sym_vector_compare}
\ea
Note that $p_{aux}$ is required to keep all site-symmetry irrep multiplicities as nonnegative integers during the adiabatic process.
Also note that SEq.~\eqref{eq:sym_vector_compare} corresponds to SEq.~\eqref{eq:diophantine1} with $\mc{N}=1$.

We already showed in SN~\ref{sec:zrsiimpliesb} that SEq.~\eqref{eq:sym_vector_compare} reduces to [see SEqs.~\eqref{eq:diophantine_zn} and \eqref{eq:diophantine3} with $\mc{N}=1$]:
\bg
[L^{-1} \cdot (p_{AI_1}-p_{AI_2})]_i
= \Lambda_{ii} (R \cdot \bb n)_i \trm{ for } i=1,\dots,r,
\label{eq:diophantine4} \\
\tilde{\Theta}^{(0)} \cdot (p_{AI_1}-p_{AI_2}) = 0.
\label{eq:diophantine5}
\eg
(Recall the Smith decomposition of adiabatic-process matrix $q=L \cdot \Lambda \cdot R$ and $r={\rm rank}(q)$.)
Indeed, there exist integer-valued solutions $\bb n$ for SEq.~\eqref{eq:diophantine4} when $AI_1$ and $AI_2$ have matching $\Z_n$SRSIs, and SEq.~\eqref{eq:diophantine5} implies that $AI_1$ and $AI_2$ must have matching $\Z$SRSIs.
To see this explicit, note that if $AI_1$ and $AI_2$ have matching $\Z_n$SRSIs then from the definition of the $\Z_n$SRSIs [SEq.~\eqref{eq:nonsimplifiedthetas}], $p_{AI_1}$ and $p_{AI_2}$ must satisfy $[L^{-1} \cdot (p_{AI_1}-p_{AI_2})]_i \pmod \Lambda_{ii} = 0$, i.e.,
\bg
[L^{-1} \cdot (p_{AI_1}-p_{AI_2})]_i = \Lambda_{ii} m_i
\eg
for some integers $m_i \in \Z$ with $i=1,\dots,r$.
Since $R$ is unimodular, we can find an integer vector $\bb n$ such that $(R \cdot \bb n)_i = m_i$ for $i=1,\dots,r$:
\bg
\bb n = R^{-1} \cdot (m_1, m_2, \dots, m_r, z_1, z_2, \dots, z_{N_{\rm adia}-r}).
\label{eq:diophantine_sol}
\eg
In the above expression, we introduced arbitrary integers $z_{1, \dots, N_{\rm adia}-r}$, which allow multiple solutions for $\bb n$ parametrized by these integers.
Also, from the fact that $\Z$SRSIs match for $AI_1$ and $AI_2$, and the definition of $\Z$SRSIs [SEq.~\eqref{eq:nonsimplifiedthetas}], SEq.~\eqref{eq:diophantine5} is satisfied.
Since we found an explicit solution $\bb n$ [SEq.~\eqref{eq:diophantine_sol}] for SEq.~\eqref{eq:sym_vector_compare}, we have thus shown that two atomic insulators with matching SRSIs are stably equivalent.

We further show that two stably equivalent atomic insulators must have matching SRSIs, following the definition of SRSIs.
Assuming that $AI_1$ and $AI_2$ are stably equivalent, we know from SEq.~\eqref{eq:sym_vector_compare} that we must have $p_{AI_1} - p_{AI_2} = q \cdot \bb n$ for $\bb n \in \Z^{N_{\rm adia}}$.
Since SRSIs are topological invariants that remain unchanged under adiabatic process described by $q \cdot \bb n$, it follows that $p_{AI_1}$ and $p_{AI_2}$ yield matching $\Z$SRSIs and $\Z_n$SRSIs.

Finally, we remark on an important implication of the one-to-one relationship between the $\Z$SRSIs and symmetry-data vectors, which follows from the discussion of stable equivalence above: any atomic insulators with the same symmetry-data vector are stably equivalent if their $\Z_n$SRSIs match.
This follows directly from our one-to-one mapping between $\Z$SRSIs and symmetry-data vectors, and the ability of SRSIs to indicate stable equivalence.
In particular, two atomic insulators with the same symmetry-data vector also have matching $\Z$SRSIs due to the one-to-one mapping.
Consequently, if their $\Z_n$SRSIs also match, they have matching SRSIs and are therefore stably equivalent.

As a corollary, if two atomic insulators, $AI_1$ and $AI_2$, have the same symmetry-data vector but have different $\Z_n$SRSIs, we can consider $\mc{N}$ copies of $AI_1$ and $AI_2$.
Then, $\mc{N} AI_1$ and $\mc{N} AI_2$ becomes stably equivalent.
Here, $\mc{N}$ is any integer that ensures the $\Z_n$SRSIs of $\mc{N} AI_1$ and $\mc{N} AI_2$ match.

As an explicit example, we investigate the stable equivalence of two EBRs in SG $I432$ without SOC in SN~\ref{sec:I432_stable}.
Further, in SN~\ref{sec:ebrsplit}, we show how mismatching SRSIs provide a sufficient criterion to indicate the nontrivial topology of almost all disconnected EBRs.

\subsubsection{SRSIs determine local RSIs at maximal WPs and composite RSIs}
\label{sec:rsi_relations}
Here, we examine the relationship between SRSIs and the local (and composite) RSIs.
Specifically, we show that (i) any local RSI at a maximal WP can be expressed in terms of SRSIs and (ii) composite RSIs can be determined from $\Z$SRSIs.

First, we demonstrate that SRSIs capture the local RSIs at maximal WPs by explicitly identifying the linear combination of SRSIs that corresponds to a given local RSI at a maximal WP.
Recall that local RSIs at maximal WPs remain unchanged under all possible adiabatic processes, as discussed in SN~\ref{sec:globalrsis}, and can take values either in $\Z$ or $\Z_2$.

\paragraph{When a local RSI at a maximal WP is $\Z$-valued:} Let us first consider a $\Z$-valued local RSI $\delta$ at a maximal WP, defined as $\delta = \Delta_\delta \cdot p$ with a row vector $\Delta_\delta$ of length $N^\rho_{\rm UC}$ [similar to SEq.~\eqref{eq:fullzrsidef}], where $p$ represents the symmetry-representation vector for all distinct site-symmetry irreps within a unit cell.
Recall that the Smith decomposition of the adiabatic-process matrix, $q=L \cdot \Lambda \cdot R$, defines the SRSI matrices $\tilde{\Theta}^{(\lambda=0,1,2,4)}$ through the rows of $L^{-1}$, as given in SEq.~\eqref{eq:thetatilde_definition}, prior to performing the Hermite decomposition via SEq.~\eqref{eq:hermite_normal_form}.
Since $L$ is unimodular with a nonzero determinant, i.e., $|{\rm Det} L| = 1$, the rows of $L^{-1}$ are linearly independent with integer coefficients.
Thus, any integer-valued vector $\bb v \in \Z^\rho_{\rm UC}$ is in the span of the rows of $L^{-1}$: for a given $\bb v$, we can write $\bb v = \sum_i \, c_{\bb v,i} (L^{-1})_{i,*}$ where $(L^{-1})_{i,*}$ denotes the $i$-th row of $L^{-1}$ and the integers $c_{\bb v,i}=(\bb v \cdot L)_{i}$ $(i=1,\dots,N^\rho_{\rm UC})$ are uniquely determined.

Applying this to the integer-valued vector $\Delta_\delta$, we have
\bg
\Delta_\delta = \sum_{i=1}^{N^\rho_{\rm UC}} \, c_i \, (L^{-1})_{i,*}.
\eg
We will explicitly determine the coefficients $c_i$ for $i=1,\dots,N^\rho_{\rm UC}$ in this linear combination and demonstrate that $\Z$-valued local RSIs can be derived from $\Z$SRSIs.

Since $\delta$ is a $\Z$-valued local RSI, it must remain unchanged under any arbitrary adiabatic process.
This implies that $\Delta_\delta \cdot q \cdot \bb n = 0$ for an integer-valued vector $\bb n \in \Z^{N_{\rm adia}}$.
This equation can be rewritten using the Smith decomposition of $q$ [see SEq.~\eqref{eq:qvecdef_global}] as
\bg
(\Delta_\delta \cdot L) \cdot \Lambda \cdot (R \cdot \bb n)
= (\Delta_\delta \cdot L) \cdot
(\Lambda_{11} \td{n}_1, \Lambda_{22} \td{n}_2, \dots, \Lambda_{rr} \td{n}_r, 0, 0, \dots, 0)^T
=0,
\eg
where $r={\rm rank}(q)$ and $\td{n}_j = (R \cdot \bb n)_j \in \Z$ for $j=1,\dots,r$.
Since this equation must hold for arbitrary $\td{n}_j$, it follows that
\bg
(\Delta_\delta \cdot L) = (0, 0, \dots, 0, z_1, z_2, \dots, z_{N^\rho_{\rm UC}-r}).
\label{eq:delta_expansion_coeffs}
\eg
Here, $z_l$ ($l=1,\dots,N^\rho_{\rm UC}-r$) are integers since $\Delta_\delta$ and $L$ are integer matrices.
The $z_l$'s are our desired expansion coefficients $c_l$.
Multiplying SEq.~\eqref{eq:delta_expansion_coeffs} on the right by $L^{-1}$, we have that
\bg
\Delta_\delta
= \sum_{l=1}^{N^\rho_{\rm UC}-r} \, z_l \, (L^{-1})_{l,*}
= \sum_{l=1}^{N^\rho_{\rm UC}-r} \, z_l \, \left( \tilde{\Theta}^{(0)} \right)_{l,*}
= \bb z \cdot \tilde{\Theta}^{(0)},
\eg
where we define $\bb z = (z_1, \dots, z_{N^\rho_{\rm UC}-r}) \in \Z^{N^\rho_{\rm UC}-r}$.
In the second equality, we used the definition of the $\Z$SRSI matrix from SEq.~\eqref{eq:thetatilde_definition}.
Further using the relation between $\tilde{\Theta}^{(0)}$ and the simplified $\Z$SRSI matrix $\Theta^{(0)}$ [SEq.~\eqref{eq:hermite_normal_form}], we obtain
\bg
\Delta_\delta
= \bb z \cdot \tilde{\Theta}^{(0)}
= \bb z \cdot [U^{(0)}]^{-1} \cdot \Theta^{(0)},
\label{eq:stable_to_local_z}
\eg
implying that the given $\Z$-valued local RSI can be expressed as the following integer linear combination of $\Z$SRSIs: $\delta = (\bb z \cdot [U^{(0)}]^{-1}) \cdot \theta_\Z$.

\paragraph{When a local RSI at a maximal WP is $\Z_2$-valued:} For a $\Z_2$-valued local RSI at a maximal WP, we can follow a similar analysis as in the case of a $\Z$-valued local RSI.
Let us suppose that a $\Z_2$-valued local RSI $\delta$ is defined as $\delta = \Delta_\delta \cdot p$, where $\Delta_\delta \in \Z_2^{N^\rho_{\rm UC}}$, meaning that each element of the row vector $\Delta_\delta$ takes value in $\{0,1\}$.
Since $\delta$ is a $\Z_2$-valued local RSI, it remains invariant under adiabatic processes up to even integers, i.e., $\Delta_\delta \cdot q \cdot \bb n = 0 \mod 2$ for $\bb n \in \Z^{N_{\rm adia}}$.
Thus, in this case, $\Delta_\delta$ must satisfy
\bg
(\Delta_\delta \cdot L) \cdot (\Lambda_{11} \td{n}_1, \dots, \Lambda_{rr} \td{n}_r, 0, \dots, 0)^T =0 \pmod 2.
\label{eq:z2local_constraint1}
\eg
Since $\Z_n$SRSIs exist only for $n=2,4$, the elementary divisors $\Lambda_{ii}$ ($i=1,\dots,r$) take values only in $\{1,2,4\}$.
Let us denote the multiplicities of elementary divisors equal to 1, 2, and 4 as $N_1$, $N_2$, and $N_4$, respectively.
(Note that $N_1+N_2+N_4=r$.)

Then, SEq.~\eqref{eq:z2local_constraint1} can be rewritten as
\bg
\sum_{i=1}^{N_1} \, (\Delta_\delta \cdot L)_i \, \td{n}_i
+ \sum_{i=N_1+1}^{N_1+N_2} \, 2 (\Delta_\delta \cdot L)_i \, \td{n}_i
+ \sum_{i=N_1+N_2+1}^r \, 4(\Delta_\delta \cdot L)_i \, \td{n}_i \in 2\Z.
\label{eq:z2local_constraint2}
\eg
Thus, the first $N_1$ elements of $(\Delta_\delta \cdot L)$ must be even integers, while the remaining elements can take arbitrary integers values since SEq.~\eqref{eq:z2local_constraint2} must hold for arbitrary integers $\td{n}_{1,\dots,r}$.
Therefore, the most general form of $\Delta_\delta \cdot L$ is given by
\bg
\Delta_\delta \cdot L = (2 x_1, 2x_2, \dots, 2x_{N_1}, y_1, y_2, \dots, y_{N_2}, z_1, z_2, \dots, z_{N_4}, w_1, w_2, \dots, w_{N^\rho_{\rm UC}-r}) := (2 \bb x, \bb y, \bb z, \bb w),
\label{eq:z2local_constraint3}
\eg
where $x_{1,\dots,N_1}$, $y_{1,\dots,N_2}$, $z_{1,\dots,N_4}$, and $w_{1,\dots,N^\rho_{\rm UC}-r}$ are integers that can be computed from a given $\Delta_\delta$.
Multiplying both sides of SEq.~\eqref{eq:z2local_constraint3} by $L^{-1}$ on the right, we express $\delta$ in terms of SRSIs.
Recall that $L^{-1}$ is given by [SEq.~\eqref{eq:nonsimplifiedthetas}]:
\bg
L^{-1} = \bpm \tilde{\Theta}^{(1)} \\ \tilde{\Theta}^{(2)} \\ \tilde{\Theta}^{(4)} \\ \tilde{\Theta}^{(0)} \\ \epm,
\eg
where $\tilde{\Theta}^{(\lambda=1,2,4)}$ are $N_\lambda \times N^\rho_{\rm UC}$ integer-valued matrices and $\tilde{\Theta}^{(0)}$ is an $(N^\rho_{\rm UC}-r) \times N^\rho_{\rm UC}$ integer-valued matrix.
Consequently, $\Delta_\delta \in \Z_2^{N^\rho_{\rm UC}}$ can be expressed as
\ba
\Delta_\delta =& 2 \bb x \cdot \td{\Theta}^{(1)} + \bb y \cdot \td{\Theta}^{(2)} + \bb z \cdot \td{\Theta}^{(4)} + \bb w \cdot \td{\Theta}^{(0)} \pmod 2
\nn
=& \bb y \cdot [U^{(2)}]^{-1} \cdot \Theta^{(2)} + \bb z \cdot [U^{(4)}]^{-1} \cdot \Theta^{(4)} + \bb w \cdot [U^{(0)}]^{-1} \cdot \Theta^{(0)} \pmod 2,
\label{eq:stable_to_local_z2}
\ea
where we introduced $U^{(0,2,4)}$, which transform $\tilde{\Theta}^{(0,2,4)}$ into the simplified SRSI matrices $\Theta^{(0,2,4)}$ [SEq.~\eqref{eq:hermite_normal_form}].
Thus, we conclude that a $\Z_2$-valued local RSI at a maximal WP can be extracted from a linear combination of both $\Z$SRSIs and $\Z_n$SRSIs.

In SN~\ref{sec:examples}, we apply SEqs.~\eqref{eq:stable_to_local_z} and \eqref{eq:stable_to_local_z2} to provide a practical way to relate SRSIs and local RSIs at maximal WPs.
Additionally, the logic we followed here can be used to quantitatively compare different equivalent redefinitions of SRSIs, as discussed in SN~\ref{sec:srsi_derivation}.

We now turn to the connection between $\Z$SRSIs and composite RSIs.
In SRef.~\cite{xu2021threedimensional}, the composite RSIs are introduced as specific linear combinations of local RSIs, taking values in $\Z$, $\Z_2$, $\Z_4$.
Since composite RSIs are determined by the symmetry-data vector, which is one-to-one mapped to $\Z$SRSIs, they can be determined by $\Z$SRSIs.
Meanwhile, $\Z_n$SRSIs encode additional information about band topology that neither symmetry-data vector nor composite RSIs capture.

\subsection{Examples: SRSIs in SGs $Pmm2$ (No. 25), $I432$ (No. 211), $P4_332$ (No. 212), and $F\bar{4}3c$ (No. 219)}
\label{sec:SRSI_example}
In this SN, we study the SRSIs in some illustrative SGs.
We focus on SG $Pmm2$ (No. 25) with SOC, SG $P4_332$ (No. 212) with SOC, and SG $I432$ (No. 211) without SOC.
In all cases, we also have TRS (i.e., we consider non-magnetic space groups) unless stated otherwise.
In these three examples, we demonstrate the derivation of SRSIs and construct the mapping between $\Z$SRSIs and symmetry-data vector.
We also compare the SRSIs and the composite RSIs proposed in SRef.~\cite{xu2021threedimensional}.
In all examples, we illustrate how $\Z_n$SRSIs provide additional information on band topology of a given system beyond what is encoded in the symmetry-data vector and hence beyond the symmetry indicators.
In the third example of SG $I432$, we focus on a specific case with two distinct EBRs that are stably equivalent.
The relevant EBRs are induced from a different site-symmetry irreps at the same maximal WP, but they can be adiabatically deformed into each other in the presence of auxiliary trivial bands because of an intriguing connectivity between WPs.
This example highlights the implication of stable equivalence between EBRs and the usefulness of SRSIs for characterizing the band topology with real-space information.
Finally, in the last example in SG $F\bar{4}3c$, we explicitly demonstrate how a $\Z_4$SRSI arises in this group and analyze the role of time-reversal symmetry by comparing the SRSIs in this group with those in the same SG but without time-reversal symmetry.
This example also illustrates that our algorithm for obtaining SRSIs applies equally well to magnetic SGs, demonstrating its broader applicability beyond nonmagnetic SGs.

\subsubsection{SG $Pmm2$ (No. 25) with SOC}
\label{sec:srsi_25}
\begin{figure}[t]
\centering
\includegraphics[width=0.3\textwidth]{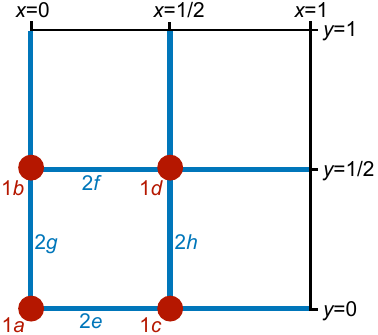}
\caption{Unit cell and Wyckoff positions (WPs) in SG $Pmm2$ (No. 25).
The four maximal WPs ($1a, 1b, 1c, 1d$) have multiplicity $m=1$ (red).
The four non-maximal WPs ($2e, 2f, 2g, 2h$), with exception of the general WP, have multiplicity $m=2$ (blue).
The general WP $4i$ has multiplicity $m=4$ (white).
The $z$-axis extends perpendicular to the plane (not shown).}
\label{fig:uc_25}
\end{figure}

We now calculate the SRSIs for SG $Pmm2$ with SOC, and compare them with the local RSIs, as well as with the composite RSIs of SRef.~\cite{xu2021threedimensional}.
The unit cell and WPs of SG $Pmm2$ are shown in SFig.~\ref{fig:uc_25}.
Following the conventions of the BCS~\cite{Aroyo2011183}, the distinct site-symmetry irreps (which provide the basis for the site-symmetry representation vector $p$) in this SG are:
\ba
(\bar{E})_{1a}, (\bar{E})_{1b}, (\bar{E})_{1c}, (\bar{E})_{1d}, ({}^1\bar{E} {}^2\bar{E})_{2e}, ({}^1\bar{E} {}^2\bar{E})_{2f}, ({}^1\bar{E} {}^2\bar{E})_{2g}, ({}^1\bar{E} {}^2\bar{E})_{2h}, (\bar{A} \bar{A})_{4i}.
\label{eq:basisforqmatsg25}
\ea
Note that each site-symmetry irrep is two-dimensional (recalling that we include time-reversal symmetry in the site-symmetry group).
Taking into account the multiplicity of each Wyckoff position, the band representations induced from these site-symmetry irreps contain $2,2,2,2,4,4,4,4$ or $8$ electrons per unit cell, respectively.
We therefore have $N^\rho_{\rm UC} = 9$.
To construct the adiabatic-process matrix $q$, we note that there are 12 adiabatic processes that span the space of all possible adiabatic processes in this SG: eight of them induce two irreps $\bar{E}$ at a WP $1x$ ($x=a,b,c,d$) by moving in an irrep ${}^1\bar{E} {}^2\bar{E}$ from a non-maximal WP $2y$ ($y=e,f,g,h$) connected to WP $1x$.
For example, two $(\bar{E})_{1a}$ irreps can be adiabatically deformed into either a $({}^1\bar{E} {}^2\bar{E})_{2e}$ or a $({}^1\bar{E} {}^2\bar{E})_{2g}$ irrep.
The remaining four adiabatic processes induce two copies of the irrep ${}^1\bar{E} {}^2\bar{E}$ at a WP $2y$ from the $(\bar{A} \bar{A})_{4i}$ irrep at the general WP $4i$.
Note that further adiabatic processes, which correspond to inducing four irreps $\bar{E}$ at a WP $1x$ from the general WP $4i$, are not linearly independent from these.
We therefore find $N_{\rm adia} = 12$.
Now, the adiabatic-process matrix $q$ reads
\ba
q =
\bpm
2 & 2 & 0 & 0 & 0 & 0 & 0 & 0 & 0 & 0 & 0 & 0 \\
0 & 0 & 2 & 2 & 0 & 0 & 0 & 0 & 0 & 0 & 0 & 0 \\
0 & 0 & 0 & 0 & 2 & 2 & 0 & 0 & 0 & 0 & 0 & 0 \\
0 & 0 & 0 & 0 & 0 & 0 & 2 & 2 & 0 & 0 & 0 & 0 \\
-1 & 0 & 0 & 0 & -1 & 0 & 0 & 0 & 2 & 0 & 0 & 0 \\
0 & 0 & -1 & 0 & 0 & 0 & -1 & 0 & 0 & 2 & 0 & 0 \\
0 & -1 & 0 & -1 & 0 & 0 & 0 & 0 & 0 & 0 & 2 & 0 \\
0 & 0 & 0 & 0 & 0 & -1 & 0 & -1 & 0 & 0 & 0 & 2 \\
0 & 0 & 0 & 0 & 0 & 0 & 0 & 0 & -1 & -1 & -1 & -1
\label{eq:pmm2_q}
\epm,
\ea
where the row basis is ordered as in SEq.~\eqref{eq:basisforqmatsg25}.
Each column of $q$ represents an adiabatic process.
For example, the first column of $q$ represents the adiabatic process, $2(\bar{E})_{1a} \Leftrightarrow (^1 \bar{E} ^2 \bar{E})_{2e}$.
This process creates (removes) $2(\bar{E})_{1a}$ and removes (creates) $(^1 \bar{E} ^2 \bar{E})_{2e}$.
Since each column of $q$ encodes the adiabatic processes, the matrix $q$ contains a $W$-position trivial configuration matrix $q_W$ (introduced to compute local RSIs in SN~\ref{sec:rsi_algorithm}) as a submatrix.
For example, to extract the $q_W$ at the WP $1a$, we can restrict the rows of $q$ to the those representing the site-symmetry irreps at WP $1a$.
As only the first row is relevant to the WP $1a$, such block matrix becomes $(2, 2,0,0,0,\dots,0)$.
By removing repeated columns and trivial columns (whose elements are zero), we obtain $q_{W=1a}=(2)$.
Crucially, while $q_W$ tracks how the site-symmetry irrep multiplicities change at the WP $W$ under adiabatic processes, it does not contain information about any other WP.
In that sense, $q$ contains strictly more information than the collection of $q_W$ for all WPs, as illustrated here.

From the Smith decomposition in SEq.~\eqref{eq:qvecdef_global}, we find
\bg
{\rm diag} (\Lambda) = (1, 1, 1, 1, 1,2,2,2,0)^T,
\label{eq:sg_25_lambda}
\eg
and
\bg
L = \bpm
2 & 2 & 0 & -4 & 2 & -1 & 0 & 2 & -1 \\
0 & 0 & 2 & 0 & -2 & 1 & -1 & 0 & 0 \\
0 & 0 & 0 & 4 & -2 & 2 & 0 & -3 & 0 \\
0 & 0 & 0 & 2 & 0 & 0 & 1 & -1 & 0 \\
-1 & 0 & 0 & 2 & 0 & 0 & 0 & 0 & -1 \\
0 & 0 & -1 & 0 & 0 & 0 & 0 & 0 & 0 \\
0 & -1 & 0 & 0 & 0 & 0 & 0 & 0 & 0 \\
0 & 0 & 0 & -1 & 1 & -1 & 0 & 1 & 0 \\
0 & 0 & 0 & -1 & 0 & 0 & 0 & 0 & 1
\epm,
\quad
R = \bpm
1 & 0 & 0 & 0 & 1 & 0 & 0 & 0 & 0 & 2 & 2 & 2 \\
0 & 1 & 0 & 1 & 0 & 0 & 0 & 0 & 0 & 0 & -2 & 0 \\
0 & 0 & 1 & 0 & 0 & 0 & 1 & 0 & 0 & -2 & 0 & 0 \\
0 & 0 & 0 & 0 & 0 & 0 & 0 & 0 & 1 & 1 & 1 & 1 \\
0 & 0 & 0 & -2 & 0 & 1 & 0 & -1 & 1 & -3 & 1 & -1 \\
0 & 0 & 0 & -1 & -1 & 1 & 0 & 1 & 1 & -1 & 1 & -3 \\
0 & 0 & 0 & 0 & -1 & 0 & 1 & 2 & 0 & 0 & 0 & -2 \\
0 & 0 & 0 & 0 & -1 & 0 & 0 & 1 & 1 & 1 & 1 & -1 \\
0 & 0 & 0 & 0 & -1 & 0 & 0 & 2 & 0 & 0 & 0 & -2 \\
0 & 0 & 0 & 0 & 0 & 0 & 0 & 0 & 0 & 1 & 0 & 0 \\
0 & 0 & 0 & 0 & 0 & 0 & 0 & 0 & 0 & 0 & 1 & 0 \\
0 & 0 & 0 & 0 & 0 & 0 & 0 & 0 & 0 & 0 & 0 & 1
\epm.
\eg
From the form of the $\Lambda$ matrix in SEq.~\eqref{eq:sg_25_lambda}, we conclude that the rank of $q$ is 8.
Since $N^\rho_{\rm UC} = 9$, this means that we have $N^\rho_{\rm UC} - {\rm rank} (q) = 1$ $\Z$SRSI.
We discard the rows $i$ corresponding to $\Lambda_{ii} = 1$, as these contribute to trivial $\Z_1$SRSIs that provide no information.
The remaining three rows which correspond to $\Lambda_{ii} = 2$ imply that we have three $\Z_2$SRSIs.
To construct the SRSI matrices explicitly, we compute the inverse of $L$:
\bg
L^{-1} = 
\bpm
1 & 1 & 1 & 1 & 1 & 2 & 2 & 2 & 2 \\
0 & 0 & 0 & 0 & 0 & 0 & -1 & 0 & 0 \\
0 & 0 & 0 & 0 & 0 & -1 & 0 & 0 & 0 \\
1 & 1 & 1 & 1 & 2 & 2 & 2 & 2 & 3 \\
1 & 0 & 1 & 0 & 2 & 0 & 2 & 1 & 3 \\
2 & 1 & 1 & 1 & 4 & 2 & 4 & 0 & 6 \\
0 & 0 & -1 & 1 & 0 & 0 & 0 & -2 & 0 \\
2 & 2 & 1 & 2 & 4 & 4 & 4 & 2 & 6 \\
1 & 1 & 1 & 1 & 2 & 2 & 2 & 2 & 4
\epm.
\eg
From SEq.~\eqref{eq:thetatilde_definition}, we find
\ba
\tilde{\Theta}^{(0)} &= (L^{-1})_{9} =
\bpm
1 & 1 & 1 & 1 & 2 & 2 & 2 & 2 & 4
\epm, \\
\tilde{\Theta}^{(2)} &= (L^{-1})_{6,7,8} \mod 2 =
\bpm
0 & 1 & 1 & 1 & 0 & 0 & 0 & 0 & 0 \\
0 & 0 & 1 & 1 & 0 & 0 & 0 & 0 & 0 \\
0 & 0 & 1 & 0 & 0 & 0 & 0 & 0 & 0
\epm.
\ea

We can further simplify the definition of SRSI matrices $\tilde{\Theta}^{(0,2)}$ by using the Hermite decomposition in SEq.~\eqref{eq:hermite_normal_form}.
The Hermite decomposition yields the simplified SRSI matrices $\Theta^{(0,2)}$ according to SEq.~\eqref{eq:simplifiedSRSI_final}:
\ba
\Theta = \bpm \Theta^{(0)} \\ \Theta^{(2)} \epm
= \bpm
1 & 1 & 1 & 1 & 2 & 2 & 2 & 2 & 4 \\
0 & 1 & 0 & 0 & 0 & 0 & 0 & 0 & 0 \\
0 & 0 & 1 & 0 & 0 & 0 & 0 & 0 & 0 \\
0 & 0 & 0 & 1 & 0 & 0 & 0 & 0 & 0
\epm.
\ea
From this and SEq.~\eqref{eq:SRSI_definition}, we define four nontrivial SRSIs,
\ba
\theta_1 =& m[(\bar{E})_{1a}] + m[(\bar{E})_{1b}] + m[(\bar{E})_{1c}] + m[(\bar{E})_{1d}] \\
&+ 2m[({}^1\bar{E} {}^2\bar{E})_{2e}] + 2m[({}^1\bar{E} {}^2\bar{E})_{2f}] + 2m[({}^1\bar{E} {}^2\bar{E})_{2g}] + 2m[({}^1\bar{E} {}^2\bar{E})_{2h}] + 4m[(\bar{A} \bar{A})_{4i}],
\label{eq:pmm2theta_1} \\
\theta_2 =& m[(\bar{E})_{1b}] \mod 2, \\ 
\theta_3 =& m[(\bar{E})_{1c}] \mod 2, \\ 
\theta_4 =& m[(\bar{E})_{1d}] \mod 2,
\label{eq:pmm2theta5}
\ea
where $m[(\rho)_W]$ denotes the multiplicity of site-symmetry irrep $(\rho)_W$ defined at WP $W$.
The definitions of SRSIs in SEqs.~\eqref{eq:pmm2theta_1}-\eqref{eq:pmm2theta5} match the SRSIs given in our exhaustive tables in SN~\ref{sec:srsitables}.
Note that $\theta_1$ can be equivalently expressed as half the total number of electrons per unit cell, where we recall that due to time-reversal symmetry, the number of electrons per unit cell is always even.
This ensures that $\theta_1 \in \Z$.

Now, we establish the mapping between $\Z$SRSIs and symmetry-data vectors according to SEqs.~\eqref{eq:maps_B_SRSI_1} and \eqref{eq:maps_B_SRSI_2}.
The BR matrix $\BR$ and its pseudoinverse $\BR^\ddagger$ in the SG $Pmm2$ are given by
\bg
\BR = \bpm
1 & 1 & 1 & 1 & 2 & 2 & 2 & 2 & 4 \\
1 & 1 & 1 & 1 & 2 & 2 & 2 & 2 & 4 \\
1 & 1 & 1 & 1 & 2 & 2 & 2 & 2 & 4 \\
1 & 1 & 1 & 1 & 2 & 2 & 2 & 2 & 4 \\
1 & 1 & 1 & 1 & 2 & 2 & 2 & 2 & 4 \\
1 & 1 & 1 & 1 & 2 & 2 & 2 & 2 & 4 \\
1 & 1 & 1 & 1 & 2 & 2 & 2 & 2 & 4 \\
1 & 1 & 1 & 1 & 2 & 2 & 2 & 2 & 4
\epm,
\quad
\BR^\ddagger= \bpm
1 & 0 & 0 & 0 & 0 & 0 & 0 & 0 \\
0 & 0 & 0 & 0 & 0 & 0 & 0 & 0 \\
0 & 0 & 0 & 0 & 0 & 0 & 0 & 0 \\
0 & 0 & 0 & 0 & 0 & 0 & 0 & 0 \\
0 & 0 & 0 & 0 & 0 & 0 & 0 & 0 \\
0 & 0 & 0 & 0 & 0 & 0 & 0 & 0 \\
0 & 0 & 0 & 0 & 0 & 0 & 0 & 0 \\
0 & 0 & 0 & 0 & 0 & 0 & 0 & 0 \\
0 & 0 & 0 & 0 & 0 & 0 & 0 & 0
\epm.
\label{eq:sg25_BR}
\eg
Here, for the BR matrix $\BR$, we take the row basis as $\{ \bar{\Gamma}_5, \bar{R}_5, \bar{S}_5, \bar{T}_5, \bar{U}_5, \bar{X}_5, \bar{Y}_5, \bar{Z}_5 \}$ and the column basis same to SEq.~\eqref{eq:basisforqmatsg25}.
As there is a unique little group irrep at each HSM, the BR matrix in SEq.~\eqref{eq:sg25_BR} is particularly simple.
By computing $\Theta^{(0)} \cdot \BR^\ddagger$,
\bg
\Theta^{(0)} \cdot \BR^\ddagger
= \bpm 1 & 0 & 0 & 0 & 0 & 0 & 0 & 0 \epm,
\label{eq:pmm2_theta0_from_B}
\eg
we obtain $\theta_1 = m(\bar{\Gamma}_5)$.
Note that there are other equivalent expressions for $\theta_1$.
For example, $\theta_1 = m(\bar{R}_5)$.
It is because there is only a single 2d irrep at each HSM and the multiplicity of such irreps are same, i.e., $m(\bar{\Gamma}_5)=m(\bar{R}_5)=m(\bar{S}_5)= \cdots = m(\bar{Z}_5)$.
Otherwise, the number of bands are not same at all HSM, which is not the case of insulator.
As such, we see that counting the multiplicity $m(\Gamma_5)$ immediately yields half the number of electrons per unit cell, which is $\theta_1$, and simultaneously determines the symmetry-data vector for an isolated set of bands.
Also, for an inverse map, we compute
\bg
\BR \cdot [\Theta^{(0)}]^\ddagger = (1,1,1,1,1,1,1,1)^T,
\eg
which implies the symmetry-data vector $B$ is given by $B = (\theta_1, \theta_1, \theta_1, \theta_1, \theta_1, \theta_1, \theta_1, \theta_1)$.

In this SG, we can immediately see that neither $\Z_n$SRSIs are uniquely determined by the symmetry-data vector, nor is the symmetry-data vector uniquely determined by $\Z_n$SRSIs.
To illustrate this, consider two EBRs induced from $(\bar{E})_{1a}$ and $(\bar{E})_{1b}$.
As shown in the BR matrix in SEq.~\eqref{eq:sg25_BR}, both have the same symmetry-data vector, $(1,1,1,1,1,1,1,1)^T$.
However, their $\Z_2$SRSIs differ: $(\bar{E})_{1a}$ has $\theta_{\Z_2}=(\theta_2,\theta_3,\theta_4)=(0,0,0)$, while $(\bar{E})_{1b}$ has $\theta_{\Z_2}=(1,0,0)$.
Conversely, we compare a BR induced from $(\bar{E})_{1a} \oplus (\bar{E})_{1b}$ with an EBR induced from $(\bar{E})_{1b}$.
These two representations share the same $\Z_2$SRSIs with $\theta_{\Z_2}=(1,0,0)$, yet they have different symmetry-data vectors: the former has $(2,2,2,2,2,2,2,2)^T$, while the latter has $(1,1,1,1,1,1,1,1)^T$.

Comparing with the eight local RSIs in SG $Pmm2$ that can be obtained from the BCS~\cite{Aroyo2011183}, we find that the $\Z_2$RSIs $\theta_2, \theta_3$ and $\theta_4$ are identical to the local RSIs at the maximal WPs $1b$, $1c$, and $1d$, respectively.
The local RSI at the maximal WP $1a$ can be obtained from the linear combination 
\ba
\theta_1 + \theta_2 + \theta_3 + \theta_4 \mod 2 = m[(\bar{E})_{1a}] \mod 2.
\ea
Hence, all local RSIs at maximal WPs can be determined from the SRSIs.
However, as expected, this is not the case for the local RSIs at non-maximal WPs.
Instead, the $\Z$SRSI $\theta_1$ is an adiabatically invariant determined from irrep multiplicities at both maximal and nonmaximal WPs.
Moreover, we note that the $\Z_2$SRSIs $\theta_2,\theta_3$ and $\theta_4$ derived here contain strictly more information than the two composite RSIs~\cite{xu2021threedimensional} of SG $Pmm2$ accessible from the BCS.
This is because $\Z_n$SRSIs cannot be determined from symmetry-data vector while the composite RSIs can.
Explicitly, both composite RSIs listed in the BCS are given by $\eta_1 = m(\bar{\Gamma}_5)$ mod 2 and $\zeta_1 = m(\bar{\Gamma}_5)$ mod 4.
By comparing SEq.~\eqref{eq:pmm2_theta0_from_B} which gives $\Z$RSI $\theta_1$ in terms of little group irreps, we see that the two composite RSIs are obtained from $\theta_1 \mod 2$ and $\theta_1 \mod 4$, respectively.
Thus, two composite RSIs provide partial information about $\theta_1$.
On the other hand, $\theta_2,\theta_3$ and $\theta_4$, which cannot be determined from the symmetry-data vector, provide additional information about the system beyond what is contained in the composite RSIs.

\subsubsection{SG $P4_332$ (No. 212) with SOC}
\label{sec:srsi_212}
We now turn to SG $P4_332$ (No. 212) to see an example of the SRSIs in a 3D SG.
In SN~\ref{sec:lrsi_212}, we have studied the adiabatic processes between site-symmetry irreps in the SG $P4_332$.
By collecting the adiabatic processes in SEqs.~\eqref{eq:sg212_adia1_8c}-\eqref{eq:sg212_adia_24e}, we construct the adiabatic-process matrix $q$,
\bg
q=
\bpm
2 & 0 & 0 & 0 & 1 & 0 & 0 & 0 \\
0 & 2 & 0 & 0 & 2 & 0 & 0 & 0 \\
0 & 0 & 2 & 0 & 0 & 1 & 0 & 0 \\
0 & 0 & 0 & 2 & 0 & 2 & 0 & 0 \\
-1 & 0 & -1 & 0 & 0 & 0 & 1 & 0 \\
0 & -1 & 0 & -1 & 0 & 0 & 2 & 0 \\
0 & 0 & 0 & 0 & -1 & -1 & 0 & 2 \\
0 & 0 & 0 & 0 & 0 & 0 & -1 & -1
\epm,
\label{eq:sg212_q}
\eg
where the row basis is ordered as
\bg
\{ ({}^1 \bar{E} {}^2 \bar{E})_{4a}, (\bar{E}_1)_{4a}, ({}^1 \bar{E} {}^2 \bar{E})_{4b}, (\bar{E}_1)_{4b}, (\bar{E} \bar{E})_{8c}, ({}^1 \bar{E} {}^2 \bar{E})_{8c}, ({}^1 \bar{E} {}^2 \bar{E})_{12d}, (\bar{A} \bar{A})_{24e} \}.
\label{eq:columnbasis_sg212}
\eg

The Smith decomposition of $q= L \cdot \Lambda \cdot R$ [SEq.~\eqref{eq:qvecdef_global}] yields
\bg
{\rm diag} (\Lambda) = (1, 1, 1, 1, 1, 2, 0, 0)^T,
\eg
and
\bg
L = \bpm
2 & 0 & -1 & 1 & 1 & -1 & 0 & -1 \\
0 & 2 & 0 & 0 & 2 & -1 & -1 & 0 \\
0 & 0 & 1 & 0 & 0 & 0 & 0 & 0 \\
0 & 0 & 0 & 2 & 0 & -1 & 0 & 0 \\
-1 & 0 & 0 & 1 & 0 & 0 & 0 & 0 \\
0 & -1 & 0 & 2 & 0 & 0 & 1 & -1 \\
0 & 0 & 0 & -1 & -1 & 1 & 0 & 0 \\
0 & 0 & 0 & -1 & 0 & 0 & 0 & 1
\epm,
\quad
R = \bpm
1 & 0 & 1 & 0 & 0 & 0 & 0 & 1 \\
0 & 1 & 0 & 1 & 0 & 0 & 0 & 2 \\
0 & 0 & 2 & 0 & 0 & 1 & 0 & 0 \\
0 & 0 & 0 & 0 & 0 & 0 & 1 & 1 \\
0 & 0 & 0 & -2 & 1 & -1 & 1 & -1 \\
0 & 0 & 0 & -1 & 0 & -1 & 1 & 1 \\
0 & 0 & -1 & -1 & 0 & -1 & 0 & 0 \\
0 & 0 & 0 & -1 & 0 & -1 & 0 & 1
\epm.
\label{eq:sg212_lambda_matrix}
\eg
From the explicit form of $L$ given above, we obtain its inverse:
\ba
L^{-1} = 
\bpm
1 & 0 & 1 & 0 & 1 & 0 & 1 & 1 \\
-1 & 1 & -1 & 1 & -2 & 1 & 1 & 0 \\
0 & 0 & 1 & 0 & 0 & 0 & 0 & 0 \\
1 & 0 & 1 & 0 & 2 & 0 & 1 & 1 \\
1 & 0 & 1 & -1 & 2 & 0 & 0 & 1 \\
2 & 0 & 2 & -1 & 4 & 0 & 2 & 2 \\
-2 & 1 & -2 & 1 & -4 & 2 & 0 & 0 \\
1 & 0 & 1 & 0 & 2 & 0 & 1 & 2
\epm.
\label{eq:sg212_Linv}
\ea
SEq.~\eqref{eq:sg212_lambda_matrix} shows that ${\rm rank} (q) = 6$ and that we have 2 $\Z$SRSIs and one $\Z_2$SRSI.
For the rows $\{i\}$ of SEq.~\eqref{eq:sg212_Linv} corresponding to $\Lambda_{ii} \ne 1$, we find the SRSI matrices $\tilde{\Theta}^{(0,2)}$, 
\ba
\tilde{\Theta}^{(0)}
&= (L^{-1})_{7,8}
= \bpm
-2 & 1 & -2 & 1 & -4 & 2 & 0 & 0 \\
1 & 0 & 1 & 0 & 2 & 0 & 1 & 2
\epm, \\
\tilde{\Theta}^{(2)}
&= (L^{-1})_{6} \mod 2
= \bpm
0 & 0 & 0 & 1 & 0 & 0 & 0 & 0
\epm.
\ea
The Hermite decomposition yields the simplified SRSI matrix $\Theta$ according to SEq.~\eqref{eq:simplifiedSRSI_final}:
\ba
\Theta = \bpm \Theta^{(0)} \\ \Theta^{(2)} \epm
=
\bpm
1 & 0 & 1 & 0 & 2 & 0 & 1 & 2 \\
0 & 1 & 0 & 1 & 0 & 2 & 2 & 4 \\
0 & 0 & 0 & 1 & 0 & 0 & 0 & 0
\epm.
\ea
Hence, we define two $\Z$SRSIs $\theta_{1}$ and $\theta_2$ and one $\Z_2$SRSI $\theta_3$,
\ba
\theta_1 =& m[ ({}^1 \bar{E} {}^2 \bar{E})_{4a} ] + m[ ({}^1 \bar{E} {}^2 \bar{E})_{4b} ] + 2 m[ (\bar{E} \bar{E})_{8c} ] + m[ ({}^1 \bar{E} {}^2 \bar{E})_{12d} ] + 2 m[ (\bar{A} \bar{A})_{24e} ],
\\
\theta_2 =& m[ (\bar{E}_1)_{4a} ] + m[ (\bar{E}_1)_{4b} ] + 2 m[ ({}^1 \bar{E} {}^2 \bar{E})_{8c} ] + 2 m[ ({}^1 \bar{E} {}^2 \bar{E})_{12d} ] + 4 m[ (\bar{A} \bar{A})_{24e} ],
\\ 
\theta_3 =& m[ (\bar{E}_1)_{4b} ] \mod 2.
\ea

Now, let us establish the mapping between the two $\Z$SRSIs $\theta_1$ and $\theta_2$ and symmetry-data vectors according to SEqs.~\eqref{eq:maps_B_SRSI_1} and \eqref{eq:maps_B_SRSI_2}.
The BR matrix $\BR$ in the SG $P4_332$ is given in SEq.~\eqref{eq:sg212_BR}.
We can compute its pseudoinverse $\BR^\ddagger$ to find
\bg
\BR^\ddagger =
\bpm
0 & 0 & 0 & 1 & 0 & 0 & 0 & 0 & 0 & 0 \\
0 & 1 & 0 & 0 & 0 & 0 & 0 & 0 & 0 & 0 \\
0 & 0 & 0 & 0 & 0 & 0 & 0 & 0 & 0 & 0 \\
0 & 0 & 0 & 0 & 0 & 0 & 0 & 0 & 0 & 0 \\
0 & 0 & 0 & 0 & 0 & 0 & 0 & 0 & 0 & 0 \\
0 & 0 & 0 & 0 & 0 & 0 & 0 & 0 & 0 & 0 \\
0 & 0 & 0 & 0 & 0 & 0 & 0 & 0 & 0 & 0 \\
0 & 0 & 0 & 0 & 0 & 0 & 0 & 0 & 0 & 0
\epm.
\label{eq:sg212_BRinv}
\eg
Here, for the BR matrix $\BR$, we take the row basis as $\{ \bar{\Gamma}_6, \bar{\Gamma}_7, \bar{\Gamma}_8, \bar{R}_4 \bar{R}_5, \bar{R}_6, \bar{R}_7 \bar{R}_8, \bar{M}_6 \bar{M}_7, \bar{X}_3 \bar{X}_6, \bar{X}_4 \bar{X}_5, \bar{X}_7 \}$ and the column basis same to SEq.~\eqref{eq:columnbasis_sg212}.
By computing $\Theta^{(0)} \cdot \BR^\ddagger$, we find
\bg
\Theta^{(0)} \cdot \BR^\ddagger=
\bpm
0 & 0 & 0 & 1 & 0 & 0 & 0 & 0 & 0 & 0 \\
0 & 1 & 0 & 0 & 0 & 0 & 0 & 0 & 0 & 0
\epm.
\label{eq:sg212_br_inv}
\eg
Hence we deduce that $\theta_1 = m(\bar{R}_4 \bar{R}_5)$ and $\theta_2 = m(\bar{\Gamma}_7)$.
We can understand both these $\Z$SRSIs from the BR matrix in SEq.~\eqref{eq:sg212_BR} and its pseudoinverse SEq.~\eqref{eq:sg212_br_inv}.
Since the only non-zero columns of SEq.~\eqref{eq:sg212_br_inv} are the second and the fourth, this means that $m(\bar{\Gamma}_7)$ and $m(\bar{R}_4 \bar{R}_5)$ are the only two independent little group multiplicities.
In other words, the rank of $BR$ is two, and the symmetry data-vector for band representations in this space group can be uniquely determined from just $m(\bar{\Gamma}_7)$ and $m(\bar{R}_4 \bar{R}_5)$.
To see this concretely, we compute an inverse map, $B = \BR \cdot [\Theta^{(0)}]^\ddagger \cdot \theta_\Z$, with
\bg
\BR \cdot [\Theta^{(0)}]^\ddagger =
\bpm
0 & 0 & 2 & 1 & 0 & 1 & 2 & 1 & 1 & 2 \\
1 & 1 & 1 & 0 & 1 & 1 & 2 & 1 & 1 & 2
\epm^T.
\eg

Let us compare the local RSIs defined in SN~\ref{sec:lrsi_212} with the SRSIs defined here.
There are three $\Z_2$-valued local RSIs ($\delta_{\Z_2,4a} ,\delta_{\Z_2,4b}, \delta_{\Z_2,12d}$), defined at the WPs $4a$, $4b$, $12d$ respectively.
Also, there is a unique $\Z$-valued local RSIs $\delta_{\Z,8c}$ is defined at the WP $8c$.
As expected, only local RSIs at the maximal WPs $4a$ and $4b$ match to the SRSIs:
\ba
\delta_{\Z_2,4a} &= m[(\bar{E}_1)_{4a}] \mod 2
= \theta_2 + \theta_3 \mod 2, \\
\delta_{\Z_2,4b} &= m[(\bar{E}_1)_{4b}] \mod 2
= \theta_3.
\ea

We also compare the SRSIs with the composite RSIs.
In SG $P4_332$ with SOC, three composite RSIs are listed in the BCS~\cite{Aroyo2011183}:
\ba
\delta_1 &= 2m(\bar{\Gamma}_6) - m(\bar{\Gamma}_8), \\
\eta_1 &= m(\bar{\Gamma}_6) + m(\bar{X}_4 \bar{X}_5) \mod 2, \\
\eta_2 &= m(\bar{\Gamma}_8) \mod 2.
\label{eq:sg212_comp}
\ea
We find that all three composite RSIs are determined by the SRSIs.
For this, we note the compatibility relations, which are the relations between little group irreps (irreps at HSM) imposed by symmetries,
\ba
m(\bar{\Gamma}_8) &= m(\bar{\Gamma}_7) + 2 m(\bar{R}_4 \bar{R}_5),
\\
m(\bar{\Gamma}_6) &= m(\bar{\Gamma}_7),
\\
m(\bar{X}_4 \bar{X}_5) &= m(\bar{\Gamma}_7) + m(\bar{R}_4 \bar{R}_5).
\ea
Note that the compatibility relations can be obtained by SEq.~\eqref{eq:si_and_comp}.
Then, we can express the composite RSIs in different forms, $\delta_1 = -2 m(\bar{R}_4 \bar{R}_5) + m(\bar{\Gamma}_7)$, 
$\eta_1 = m(\bar{R}_4 \bar{R}_5)$ mod 2, $\eta_2 = m(\bar{\Gamma}_7)$ mod 2.
Thus, we find
\ba
\delta_1 &= -2 \theta_1 + \theta_2,
\label{eq:d1fromthetas} \\
\eta_1 &= \theta_1 \mod 2, \\
\eta_2 &= \theta_2 \mod 2.
\ea
We can gain further insight into the meaning of the two $\Z$SRSIs $\theta_1$ and $\theta_2$ by noting that in this SG, the total electron filling is always a multiple of $8$, i.e., the total number of bands is given by $N=8\nu$ for (non-negative) integer $\nu$.
In terms of the $\Z$SRSIs, we have 
\ba
\nu = \theta_1+\theta_2.
\ea
We can combine this with SEq.~\eqref{eq:d1fromthetas}, we see that the two $\Z$SRSIs $\theta_1$ and $\theta_2$ determine the composite RSI $\delta_1$ and the filling $\nu$ via
\ba
\bpm
\delta_1 \\ \nu \epm
= M_\theta \cdot \bpm \theta_1 \\ \theta_2 \epm
= \bpm -2 & 1 \\ 1 & 1 \epm \bpm \theta_1 \\ \theta_2 \epm.
\ea
While the matrix $M_\theta$ is invertible, it is {\it not} invertible over the integers.
This means that while there exist trivial band structures in which $\theta_1$ and $\theta_2$ can take any non-negative integer value, the values of $\delta_1$ and $\nu$ are constrained.
This shows that the SRSIs, rather than the local or composite RSIs, provide a linearly-independent characterization of the adiabatic invariants of atomic limit band structures.

\subsubsection{SG $I432$ (No. 211) without SOC}
\label{sec:I432_stable}
As another example, we will examine the SRSIs in SG $I432$ (No. 211) without SOC.
We focus on this SG because it features two EBRs induced from the $(B_2)_{12d}$ and $(B_3)_{12d}$ site-symmetry irreps at WP $12d$ that have the same symmetry-data vector but are not equivalent.
We will show using the theory of SRSIs that these two EBRS are {\it stably} equivalent.
To begin, recall that SG $I432$ without SOC is generated from lattice translations, $\{E|-1/2,1/2,1/2\}$, $\{E|1/2,-1/2,1/2\}$, $\{E|1/2,1/2,-1/2\}$, the fourfold rotation $\{4^+_{001}|\bb 0\}$, and the threefold rotation $\{3^+_{111}|\bb 0\}$.
There are four maximal WPs, $2a, 6b, 8c, 12d$, and six non-maximal WPs, $12e, 16f, 24g, 24h, 24i, 48j$, where $48j$ is the general WP.
At these WPs, the site-symmetry irreps are
\bg
\{
(A_1)_{2a}, (A_2)_{2a}, (E)_{2a}, (T_1)_{2a}, (T_2)_{2a}, (A_1)_{6b}, (A_2)_{6b}, (B_1)_{6b}, (B_2)_{6b}, (E)_{6b}, (A_1)_{8c}, (A_2)_{8c}, (E)_{8c}, \nn
(A_1)_{12d}, (B_1)_{12d}, (B_2)_{12d}, (B_3)_{12d}, (A)_{12e}, (B)_{12e}, ({}^1 E {}^2 E)_{12e}, (A_1)_{16f}, ({}^1 E {}^2 E)_{16f}, (A)_{24g}, (B)_{24g}, \nn
(A)_{24h}, (B)_{24h}, (A)_{24i}, (B)_{24i}, (A)_{48j}
\}.
\label{eq:sg211_irrep}
\eg
Thus, $N^\rho_{\rm UC}=29$.

Now, we construct the adiabatic-process matrix $q$ by collecting the full set of generators for adiabatic processes between site-symmetry irreps in SEq.~\eqref{eq:sg211_irrep}:
\bg
q =
\left( \begin{smallmatrix}
1 & 0 & 0 & 1 & 0 & 1 & 0 & 0 & 0 & 0 & 0 & 0 & 0 & 0 & 0 & 0 & 0 & 0 & 0 & 0 & 0 & 0 & 0 & 0 & 0 & 0 & 0 & 0 & 0 \\
0 & 1 & 0 & 1 & 0 & 0 & 1 & 0 & 0 & 0 & 0 & 0 & 0 & 0 & 0 & 0 & 0 & 0 & 0 & 0 & 0 & 0 & 0 & 0 & 0 & 0 & 0 & 0 & 0 \\
1 & 1 & 0 & 0 & 2 & 1 & 1 & 0 & 0 & 0 & 0 & 0 & 0 & 0 & 0 & 0 & 0 & 0 & 0 & 0 & 0 & 0 & 0 & 0 & 0 & 0 & 0 & 0 & 0 \\
1 & 0 & 2 & 1 & 2 & 1 & 2 & 0 & 0 & 0 & 0 & 0 & 0 & 0 & 0 & 0 & 0 & 0 & 0 & 0 & 0 & 0 & 0 & 0 & 0 & 0 & 0 & 0 & 0 \\
0 & 1 & 2 & 1 & 2 & 2 & 1 & 0 & 0 & 0 & 0 & 0 & 0 & 0 & 0 & 0 & 0 & 0 & 0 & 0 & 0 & 0 & 0 & 0 & 0 & 0 & 0 & 0 & 0 \\
0 & 0 & 0 & 0 & 0 & 0 & 0 & 1 & 0 & 0 & 1 & 0 & 1 & 0 & 0 & 0 & 0 & 0 & 0 & 0 & 0 & 0 & 0 & 0 & 0 & 0 & 0 & 0 & 0 \\
0 & 0 & 0 & 0 & 0 & 0 & 0 & 1 & 0 & 0 & 0 & 1 & 0 & 1 & 0 & 0 & 0 & 0 & 0 & 0 & 0 & 0 & 0 & 0 & 0 & 0 & 0 & 0 & 0 \\
0 & 0 & 0 & 0 & 0 & 0 & 0 & 0 & 1 & 0 & 0 & 1 & 1 & 0 & 0 & 0 & 0 & 0 & 0 & 0 & 0 & 0 & 0 & 0 & 0 & 0 & 0 & 0 & 0 \\
0 & 0 & 0 & 0 & 0 & 0 & 0 & 0 & 1 & 0 & 1 & 0 & 0 & 1 & 0 & 0 & 0 & 0 & 0 & 0 & 0 & 0 & 0 & 0 & 0 & 0 & 0 & 0 & 0 \\
0 & 0 & 0 & 0 & 0 & 0 & 0 & 0 & 0 & 2 & 1 & 1 & 1 & 1 & 0 & 0 & 0 & 0 & 0 & 0 & 0 & 0 & 0 & 0 & 0 & 0 & 0 & 0 & 0 \\
0 & 0 & 0 & 0 & 0 & 0 & 0 & 0 & 0 & 0 & 0 & 0 & 0 & 0 & 1 & 0 & 1 & 0 & 0 & 0 & 0 & 0 & 0 & 0 & 0 & 0 & 0 & 0 & 0 \\
0 & 0 & 0 & 0 & 0 & 0 & 0 & 0 & 0 & 0 & 0 & 0 & 0 & 0 & 1 & 0 & 0 & 1 & 0 & 0 & 0 & 0 & 0 & 0 & 0 & 0 & 0 & 0 & 0 \\
0 & 0 & 0 & 0 & 0 & 0 & 0 & 0 & 0 & 0 & 0 & 0 & 0 & 0 & 0 & 2 & 1 & 1 & 0 & 0 & 0 & 0 & 0 & 0 & 0 & 0 & 0 & 0 & 0 \\
0 & 0 & 0 & 0 & 0 & 0 & 0 & 0 & 0 & 0 & 0 & 0 & 0 & 0 & 0 & 0 & 0 & 0 & 1 & 0 & 1 & 0 & 1 & 0 & 0 & 0 & 0 & 0 & 0 \\
0 & 0 & 0 & 0 & 0 & 0 & 0 & 0 & 0 & 0 & 0 & 0 & 0 & 0 & 0 & 0 & 0 & 0 & 0 & 1 & 0 & 1 & 1 & 0 & 0 & 0 & 0 & 0 & 0 \\
0 & 0 & 0 & 0 & 0 & 0 & 0 & 0 & 0 & 0 & 0 & 0 & 0 & 0 & 0 & 0 & 0 & 0 & 0 & 1 & 1 & 0 & 0 & 1 & 0 & 0 & 0 & 0 & 0 \\
0 & 0 & 0 & 0 & 0 & 0 & 0 & 0 & 0 & 0 & 0 & 0 & 0 & 0 & 0 & 0 & 0 & 0 & 1 & 0 & 0 & 1 & 0 & 1 & 0 & 0 & 0 & 0 & 0 \\
-1 & 0 & 0 & 0 & 0 & 0 & 0 & -1 & 0 & 0 & 0 & 0 & 0 & 0 & 0 & 0 & 0 & 0 & 0 & 0 & 0 & 0 & 0 & 0 & 1 & 0 & 0 & 0 & 0 \\
0 & -1 & 0 & 0 & 0 & 0 & 0 & 0 & -1 & 0 & 0 & 0 & 0 & 0 & 0 & 0 & 0 & 0 & 0 & 0 & 0 & 0 & 0 & 0 & 1 & 0 & 0 & 0 & 0 \\
0 & 0 & -1 & 0 & 0 & 0 & 0 & 0 & 0 & -1 & 0 & 0 & 0 & 0 & 0 & 0 & 0 & 0 & 0 & 0 & 0 & 0 & 0 & 0 & 1 & 0 & 0 & 0 & 0 \\
0 & 0 & 0 & -1 & 0 & 0 & 0 & 0 & 0 & 0 & 0 & 0 & 0 & 0 & -1 & 0 & 0 & 0 & 0 & 0 & 0 & 0 & 0 & 0 & 0 & 1 & 0 & 0 & 0 \\
0 & 0 & 0 & 0 & -1 & 0 & 0 & 0 & 0 & 0 & 0 & 0 & 0 & 0 & 0 & -1 & 0 & 0 & 0 & 0 & 0 & 0 & 0 & 0 & 0 & 1 & 0 & 0 & 0 \\
0 & 0 & 0 & 0 & 0 & 0 & 0 & 0 & 0 & 0 & 0 & 0 & -1 & 0 & 0 & 0 & 0 & 0 & 0 & 0 & 0 & 0 & -1 & 0 & 0 & 0 & 1 & 0 & 0 \\
0 & 0 & 0 & 0 & 0 & 0 & 0 & 0 & 0 & 0 & 0 & 0 & 0 & -1 & 0 & 0 & 0 & 0 & 0 & 0 & 0 & 0 & 0 & -1 & 0 & 0 & 1 & 0 & 0 \\
0 & 0 & 0 & 0 & 0 & -1 & 0 & 0 & 0 & 0 & -1 & 0 & 0 & 0 & 0 & 0 & 0 & 0 & 0 & 0 & 0 & 0 & 0 & 0 & 0 & 0 & 0 & 1 & 0 \\
0 & 0 & 0 & 0 & 0 & 0 & -1 & 0 & 0 & 0 & 0 & -1 & 0 & 0 & 0 & 0 & 0 & 0 & 0 & 0 & 0 & 0 & 0 & 0 & 0 & 0 & 0 & 1 & 0 \\
0 & 0 & 0 & 0 & 0 & 0 & 0 & 0 & 0 & 0 & 0 & 0 & 0 & 0 & 0 & 0 & -1 & 0 & -1 & 0 & -1 & 0 & 0 & 0 & 0 & 0 & 0 & 0 & 1 \\
0 & 0 & 0 & 0 & 0 & 0 & 0 & 0 & 0 & 0 & 0 & 0 & 0 & 0 & 0 & 0 & 0 & -1 & 0 & -1 & 0 & -1 & 0 & 0 & 0 & 0 & 0 & 0 & 1 \\
0 & 0 & 0 & 0 & 0 & 0 & 0 & 0 & 0 & 0 & 0 & 0 & 0 & 0 & 0 & 0 & 0 & 0 & 0 & 0 & 0 & 0 & 0 & 0 & -1 & -1 & -1 & -1 & -1
\end{smallmatrix} \right)
\label{eq:sg211_q}
\eg
where the row basis is ordered as in SEq.~\eqref{eq:sg211_irrep}.

The Smith decomposition of $q= L \cdot \Lambda \cdot R$ [SEq.~\eqref{eq:qvecdef_global}] yields
\bg
{\rm diag} (\Lambda) = (1, 1, 1, 1, 1, 1, 1, 1, 1, 1, 1, 1, 1, 1, 1, 1, 1, 1, 1, 2, 2, 2, 0, 0, 0, 0, 0, 0, 0)^T,
\eg
which shows that ${\rm rank} (q) = 22$, and that there are 7 $\Z$SRSIs and 3 $\Z_2$SRSIs.
For completeness, we also provide the explicit forms of the matrices $L$ and $R$:
\bg
L = \left( \begin{smallmatrix}
1 & 0 & 0 & 1 & 0 & 1 & 0 & 0 & 0 & -2 & 0 & 0 & 0 & 0 & 0 & 0 & 0 & 0 & 0 & 0 & -1 & 1 & 0 & 0 & 0 & 0 & 0 & 0 & 0 \\
0 & 1 & 0 & 1 & 0 & 0 & 1 & 0 & 0 & -2 & 0 & 0 & 0 & 0 & 0 & 0 & 0 & 0 & 0 & 0 & -1 & 1 & 0 & 0 & 0 & 0 & 0 & 0 & 0 \\
1 & 1 & 0 & 0 & 2 & 1 & 1 & 0 & 0 & -4 & 0 & 0 & 0 & 0 & 0 & 0 & 0 & 0 & 0 & 0 & -1 & 2 & 0 & 0 & 0 & 0 & 0 & 0 & 0 \\
1 & 0 & 2 & 1 & 2 & 1 & 2 & 0 & 0 & -6 & 0 & 0 & 0 & 0 & 2 & 0 & 0 & 0 & 0 & -1 & -1 & 3 & -1 & 0 & 0 & 0 & 0 & 0 & 0 \\
0 & 1 & 2 & 1 & 2 & 2 & 1 & 0 & 0 & -8 & 0 & 0 & 0 & 0 & 2 & 0 & 0 & 0 & 0 & -1 & -2 & 4 & -1 & 0 & 0 & 0 & 0 & 0 & 0 \\
0 & 0 & 0 & 0 & 0 & 0 & 0 & 1 & 0 & 0 & 1 & 1 & 0 & 0 & -4 & 0 & 0 & 0 & 0 & 2 & -1 & -1 & 0 & 0 & 0 & 0 & 0 & 0 & 0 \\
0 & 0 & 0 & 0 & 0 & 0 & 0 & 1 & 0 & 1 & 0 & 0 & 1 & 0 & -2 & 0 & 0 & 0 & 0 & 1 & 0 & -1 & 0 & 0 & 1 & 0 & 0 & 0 & 0 \\
0 & 0 & 0 & 0 & 0 & 0 & 0 & 0 & 1 & -1 & 0 & 1 & 0 & 0 & -2 & 0 & 0 & 0 & 0 & 1 & -1 & 0 & 0 & 0 & 0 & 0 & 0 & 0 & 0 \\
0 & 0 & 0 & 0 & 0 & 0 & 0 & 0 & 1 & 0 & 1 & 0 & 1 & 0 & -4 & 0 & 0 & 0 & 0 & 2 & -1 & -1 & 0 & 0 & 0 & 0 & 0 & 0 & 0 \\
0 & 0 & 0 & 0 & 0 & 0 & 0 & 0 & 0 & 3 & 1 & 1 & 1 & 0 & -8 & 0 & 0 & 0 & 0 & 4 & -1 & -3 & 0 & 0 & 0 & 0 & 0 & 0 & 0 \\
0 & 0 & 0 & 0 & 0 & 0 & 0 & 0 & 0 & 0 & 0 & 0 & 0 & 1 & -3 & 1 & 0 & 0 & 0 & 1 & -1 & -1 & 0 & 0 & 0 & 1 & 0 & 0 & 0 \\
0 & 0 & 0 & 0 & 0 & 0 & 0 & 0 & 0 & 0 & 0 & 0 & 0 & 1 & -3 & 0 & 1 & 0 & 0 & 1 & -1 & -1 & 0 & 0 & 0 & 0 & 0 & 0 & 0 \\
0 & 0 & 0 & 0 & 0 & 0 & 0 & 0 & 0 & 2 & 0 & 0 & 0 & 0 & -2 & 1 & 1 & 0 & 0 & 0 & -1 & -2 & 0 & 0 & 0 & 0 & 0 & 0 & 0 \\
0 & 0 & 0 & 0 & 0 & 0 & 0 & 0 & 0 & 2 & 0 & 0 & 0 & 0 & -2 & 0 & 0 & 1 & 0 & 1 & 0 & -2 & 0 & 0 & 0 & 0 & 0 & 0 & 0 \\
0 & 0 & 0 & 0 & 0 & 0 & 0 & 0 & 0 & 1 & 0 & 0 & 0 & 0 & -1 & 0 & 0 & 0 & 1 & 0 & 0 & -1 & 0 & 0 & 0 & 0 & 0 & 0 & 0 \\
0 & 0 & 0 & 0 & 0 & 0 & 0 & 0 & 0 & 0 & 0 & 0 & 0 & 0 & -2 & 0 & 0 & 0 & 1 & 1 & -1 & -1 & 0 & 0 & 0 & 0 & 0 & 0 & 0 \\
0 & 0 & 0 & 0 & 0 & 0 & 0 & 0 & 0 & 1 & 0 & 0 & 0 & 0 & -1 & 0 & 0 & 1 & 0 & 0 & 0 & -1 & 0 & 0 & 0 & 0 & 1 & 0 & 0 \\
-1 & 0 & 0 & 0 & 0 & 0 & 0 & -1 & 0 & 1 & 0 & 0 & 0 & 0 & 0 & 0 & 0 & 0 & 0 & 0 & 0 & 0 & 0 & 1 & 0 & 0 & 0 & 0 & 0 \\
0 & -1 & 0 & 0 & 0 & 0 & 0 & 0 & -1 & 3 & 0 & 0 & 0 & 0 & 0 & 0 & 0 & 0 & 0 & 0 & 1 & -1 & 0 & 0 & 0 & 0 & 0 & 0 & 0 \\
0 & 0 & -1 & 0 & 0 & 0 & 0 & 0 & 0 & 1 & 0 & 0 & 0 & 0 & 0 & 0 & 0 & 0 & 0 & 0 & 0 & 0 & 1 & 0 & 0 & 0 & 0 & 0 & 0 \\
0 & 0 & 0 & -1 & 0 & 0 & 0 & 0 & 0 & 1 & 0 & 0 & 0 & -1 & 2 & 0 & 0 & 0 & 0 & -1 & 1 & 0 & 0 & 0 & 0 & 0 & 0 & 0 & 0 \\
0 & 0 & 0 & 0 & -1 & 0 & 0 & 0 & 0 & 0 & 0 & 0 & 0 & 0 & 0 & 0 & 0 & 0 & 0 & 0 & 0 & 0 & 0 & 0 & 0 & 0 & 0 & 0 & 0 \\
0 & 0 & 0 & 0 & 0 & 0 & 0 & 0 & 0 & -1 & 0 & -1 & 0 & 0 & 2 & 0 & 0 & 0 & 0 & -1 & 0 & 1 & 0 & 0 & 0 & 0 & 0 & 0 & 0 \\
0 & 0 & 0 & 0 & 0 & 0 & 0 & 0 & 0 & -1 & 0 & 0 & -1 & 0 & 2 & 0 & 0 & 0 & 0 & -1 & 0 & 1 & 0 & 0 & 0 & 0 & 0 & 0 & 0 \\
0 & 0 & 0 & 0 & 0 & -1 & 0 & 0 & 0 & 1 & -1 & 0 & 0 & 0 & 2 & 0 & 0 & 0 & 0 & -1 & 1 & 0 & 0 & 0 & 0 & 0 & 0 & 0 & 0 \\
0 & 0 & 0 & 0 & 0 & 0 & -1 & 0 & 0 & 0 & 0 & 0 & 0 & 0 & 0 & 0 & 0 & 0 & 0 & 0 & 0 & 0 & 0 & 0 & 0 & 0 & 0 & 0 & 0 \\
0 & 0 & 0 & 0 & 0 & 0 & 0 & 0 & 0 & -1 & 0 & 0 & 0 & 0 & 3 & -1 & 0 & -1 & 0 & -1 & 1 & 2 & 0 & 0 & 0 & 0 & 0 & 0 & 0 \\
0 & 0 & 0 & 0 & 0 & 0 & 0 & 0 & 0 & 0 & 0 & 0 & 0 & 0 & 2 & 0 & -1 & 0 & -1 & 0 & 1 & 1 & 0 & 0 & 0 & 0 & 0 & 1 & 0 \\
0 & 0 & 0 & 0 & 0 & 0 & 0 & 0 & 0 & -1 & 0 & 0 & 0 & 0 & 0 & 0 & 0 & 0 & 0 & 0 & 0 & 0 & 0 & 0 & 0 & 0 & 0 & 0 & 1
\end{smallmatrix} \right)
\eg
and
\bg
R = \left( \begin{smallmatrix}
1 & 0 & 0 & 0 & 0 & 0 & 0 & 0 & 0 & 0 & 0 & -1 & 0 & 0 & 0 & 0 & 0 & 0 & 0 & 0 & 0 & 0 & 0 & 1 & 0 & 1 & 0 & 1 & 1 \\
0 & 1 & 0 & 0 & 0 & 0 & 0 & 0 & 0 & 0 & 0 & -1 & 0 & 0 & 0 & 0 & 0 & 0 & 0 & 0 & 0 & 0 & 1 & 0 & 0 & 1 & 0 & 1 & 1 \\
0 & 0 & 1 & 0 & 0 & 0 & 0 & 0 & 0 & 1 & 0 & 0 & 0 & 0 & 0 & 0 & 0 & 0 & 0 & 0 & 0 & 0 & 0 & 0 & 0 & 1 & 1 & 1 & 1 \\
0 & 0 & 0 & 1 & 0 & 0 & 0 & 0 & 0 & 0 & 0 & 0 & 0 & 0 & 0 & 0 & 0 & 0 & 0 & 0 & 0 & 0 & 1 & -2 & 0 & -1 & 0 & 0 & -1 \\
0 & 0 & 0 & 0 & 1 & 0 & 0 & 0 & 0 & 0 & 0 & 0 & 0 & 0 & 0 & 1 & 0 & 0 & 0 & 0 & 0 & 0 & 0 & 0 & 0 & -1 & 0 & 0 & 0 \\
0 & 0 & 0 & 0 & 0 & 1 & 0 & 0 & 0 & 0 & 0 & 1 & 0 & 0 & 0 & 0 & 0 & 0 & 0 & 0 & 0 & 0 & 1 & -1 & 0 & 0 & 0 & -1 & 0 \\
0 & 0 & 0 & 0 & 0 & 0 & 1 & 0 & 0 & 0 & 0 & 1 & 0 & 0 & 0 & 0 & 0 & 0 & 0 & 0 & 0 & 0 & 0 & 0 & 0 & 0 & 0 & -1 & 0 \\
0 & 0 & 0 & 0 & 0 & 0 & 0 & 1 & 0 & 0 & 0 & 1 & 0 & 0 & 0 & 0 & 0 & 0 & 0 & 0 & 0 & 0 & 0 & -1 & 0 & 0 & 1 & 0 & 0 \\
0 & 0 & 0 & 0 & 0 & 0 & 0 & 0 & 1 & 0 & 0 & 1 & 0 & 0 & 0 & 0 & 0 & 0 & 0 & 0 & 0 & 0 & 1 & -2 & 0 & 0 & 1 & 0 & 0 \\
0 & 0 & 0 & 0 & 0 & 0 & 0 & 0 & 0 & 0 & 0 & 0 & 0 & 0 & 0 & 0 & 0 & 0 & 0 & 0 & 0 & 0 & 0 & 0 & 1 & 1 & 1 & 1 & 1 \\
0 & 0 & 0 & 0 & 0 & 0 & 0 & 0 & 0 & -2 & 1 & -3 & 0 & 0 & 0 & 0 & 0 & 0 & 0 & 0 & 0 & 0 & 1 & 1 & 1 & 1 & -1 & 1 & 1 \\
0 & 0 & 0 & 0 & 0 & 0 & 0 & 0 & 0 & -2 & 0 & -2 & 1 & 0 & 0 & 0 & 0 & 0 & 0 & 0 & 0 & 0 & 1 & 2 & 1 & 1 & -2 & 1 & 1 \\
0 & 0 & 0 & 0 & 0 & 0 & 0 & 0 & 0 & -2 & 0 & -2 & 0 & 1 & 0 & 0 & 0 & 0 & 0 & 0 & 0 & 0 & 0 & 3 & 1 & 1 & -2 & 1 & 1 \\
0 & 0 & 0 & 0 & 0 & 0 & 0 & 0 & 0 & -2 & 0 & -2 & 0 & 0 & 1 & 0 & 0 & 0 & 0 & 0 & 0 & 0 & 1 & 2 & 1 & 1 & -1 & 1 & 2 \\
0 & 0 & 0 & 0 & 0 & 0 & 0 & 0 & 0 & -2 & 0 & -2 & 0 & 0 & 0 & 0 & 0 & 0 & 0 & 0 & 1 & -1 & 1 & 1 & 1 & 1 & -1 & 1 & 1 \\
0 & 0 & 0 & 0 & 0 & 0 & 0 & 0 & 0 & -2 & 0 & -2 & 0 & 0 & 0 & -2 & 1 & 0 & 0 & 0 & 1 & -1 & 2 & -3 & 2 & 2 & 0 & 2 & -1 \\
0 & 0 & 0 & 0 & 0 & 0 & 0 & 0 & 0 & -2 & 0 & -2 & 0 & 0 & 0 & -2 & 0 & 1 & 0 & 0 & 1 & -1 & 2 & -3 & 2 & 2 & 0 & 2 & -1 \\
0 & 0 & 0 & 0 & 0 & 0 & 0 & 0 & 0 & -2 & 0 & -2 & 0 & 0 & 0 & -2 & 0 & 0 & 1 & 0 & 1 & 0 & 1 & 0 & 2 & 2 & 0 & 2 & 0 \\
0 & 0 & 0 & 0 & 0 & 0 & 0 & 0 & 0 & -2 & 0 & -2 & 0 & 0 & 0 & -2 & 0 & 0 & 0 & 1 & 1 & 0 & 2 & -1 & 2 & 2 & 0 & 2 & 0 \\
0 & 0 & 0 & 0 & 0 & 0 & 0 & 0 & 0 & -1 & 0 & -1 & 0 & 0 & 0 & -1 & 0 & 0 & 0 & 0 & 1 & -1 & 1 & -1 & 1 & 1 & 0 & 1 & 0 \\
0 & 0 & 0 & 0 & 0 & 0 & 0 & 0 & 0 & 0 & 0 & 0 & 0 & 0 & 0 & -1 & 0 & 0 & 0 & 0 & 0 & 0 & 1 & -2 & 0 & 0 & 0 & 0 & -1 \\
0 & 0 & 0 & 0 & 0 & 0 & 0 & 0 & 0 & 0 & 0 & 0 & 0 & 0 & 0 & -1 & 0 & 0 & 0 & 0 & 0 & 0 & 0 & -1 & 1 & 1 & 1 & 1 & 0 \\
0 & 0 & 0 & 0 & 0 & 0 & 0 & 0 & 0 & 0 & 0 & 1 & 0 & 0 & 0 & -1 & 0 & 0 & 0 & 0 & 0 & 0 & 0 & -1 & 0 & 0 & 0 & 0 & -1 \\
0 & 0 & 0 & 0 & 0 & 0 & 0 & 0 & 0 & 0 & 0 & 0 & 0 & 0 & 0 & -1 & 0 & 0 & 0 & 0 & 0 & 0 & 0 & 0 & 0 & 0 & 0 & 0 & -1 \\
0 & 0 & 0 & 0 & 0 & 0 & 0 & 0 & 0 & 0 & 0 & 0 & 0 & 0 & 0 & -1 & 0 & 0 & 0 & 0 & 0 & 1 & 0 & -1 & 0 & 0 & 0 & 0 & -1 \\
0 & 0 & 0 & 0 & 0 & 0 & 0 & 0 & 0 & 0 & 0 & 0 & 0 & 0 & 0 & -1 & 0 & 0 & 0 & 0 & 0 & 0 & 0 & -1 & 0 & 1 & 0 & 0 & -1 \\
0 & 0 & 0 & 0 & 0 & 0 & 0 & 0 & 0 & 0 & 0 & 0 & 0 & 0 & 0 & -1 & 0 & 0 & 0 & 0 & 0 & 0 & 0 & -1 & 0 & 0 & 1 & 0 & -1 \\
0 & 0 & 0 & 0 & 0 & 0 & 0 & 0 & 0 & 0 & 0 & 0 & 0 & 0 & 0 & -1 & 0 & 0 & 0 & 0 & 0 & 0 & 0 & -1 & 0 & 0 & 0 & 1 & -1 \\
0 & 0 & 0 & 0 & 0 & 0 & 0 & 0 & 0 & 0 & 0 & 0 & 0 & 0 & 0 & -1 & 0 & 0 & 0 & 0 & 0 & 0 & 0 & -1 & 0 & 0 & 0 & 0 & 0
\end{smallmatrix} \right).
\eg
Additionally, we find that
\ba
L^{-1} = 
\left( \begin{smallmatrix}
-1 & 0 & 1 & 1 & -1 & 0 & 0 & 1 & -1 & 0 & 0 & -1 & 1 & 1 & 0 & 0 & 0 & 0 & 0 & 0 & -1 & 2 & 1 & -1 & -1 & 2 & 1 & 0 & 0 \\
-1 & 0 & 1 & 0 & 0 & 0 & 0 & 0 & 0 & 0 & 0 & -1 & 1 & 1 & 0 & 0 & 0 & 0 & 0 & 0 & -1 & 2 & 0 & 0 & 0 & 1 & 1 & 0 & 0 \\
-1 & -1 & 0 & 1 & 0 & 0 & 0 & 0 & -1 & 1 & 0 & -1 & 1 & 1 & 0 & 0 & 0 & 0 & -1 & 1 & -1 & 2 & 1 & 0 & 0 & 1 & 1 & 0 & 0 \\
2 & 0 & -1 & -1 & 1 & 0 & 0 & -2 & 2 & 0 & 0 & 1 & -1 & -1 & 0 & 0 & 0 & 0 & 0 & 0 & 1 & -2 & -2 & 2 & 2 & -2 & -1 & 0 & 0 \\
0 & 0 & 0 & 0 & 0 & 0 & 0 & 0 & 0 & 0 & 0 & 0 & 0 & 0 & 0 & 0 & 0 & 0 & 0 & 0 & 0 & -1 & 0 & 0 & 0 & 0 & 0 & 0 & 0 \\
1 & -1 & 0 & -1 & 1 & 0 & 0 & -1 & 1 & 0 & 0 & 0 & 0 & 0 & 0 & 0 & 0 & 0 & 0 & 0 & 0 & 0 & -1 & 1 & 1 & -2 & 0 & 0 & 0 \\
0 & 0 & 0 & 0 & 0 & 0 & 0 & 0 & 0 & 0 & 0 & 0 & 0 & 0 & 0 & 0 & 0 & 0 & 0 & 0 & 0 & 0 & 0 & 0 & 0 & -1 & 0 & 0 & 0 \\
1 & -1 & 0 & -1 & 1 & 1 & 0 & -1 & 1 & 0 & 0 & 0 & 0 & 0 & 0 & 0 & 0 & 0 & 0 & 0 & 0 & 0 & 0 & 1 & 2 & -2 & 0 & 0 & 0 \\
1 & -1 & 0 & -1 & 1 & 0 & 0 & -1 & 2 & 0 & 0 & 0 & 0 & 0 & 0 & 0 & 0 & 0 & 0 & 0 & 0 & 0 & -1 & 2 & 2 & -2 & 0 & 0 & 0 \\
-1 & 0 & 1 & 0 & 0 & 0 & 0 & 1 & 0 & 0 & 0 & -1 & 1 & 1 & 0 & 0 & 0 & 0 & 1 & 0 & -1 & 2 & 1 & 0 & 0 & 1 & 1 & 0 & 0 \\
-2 & 1 & 1 & 1 & -1 & 0 & 0 & 1 & 0 & -1 & 0 & -1 & 1 & 1 & 0 & 0 & 0 & 0 & 1 & 0 & -1 & 2 & 0 & -1 & -2 & 3 & 1 & 0 & 0 \\
-2 & 1 & 1 & 1 & -1 & 0 & 0 & 2 & -1 & -1 & 0 & -1 & 1 & 1 & 0 & 0 & 0 & 0 & 1 & 0 & -1 & 2 & 0 & -2 & -2 & 3 & 1 & 0 & 0 \\
-2 & 1 & 1 & 1 & -1 & 0 & 0 & 2 & -1 & -1 & 0 & -1 & 1 & 1 & 0 & 0 & 0 & 0 & 1 & 0 & -1 & 2 & 1 & -3 & -2 & 3 & 1 & 0 & 0 \\
-3 & 0 & 2 & 1 & -1 & 0 & 0 & 2 & -1 & -1 & 0 & -2 & 2 & 2 & 0 & 0 & 0 & 0 & 1 & 0 & -3 & 4 & 1 & -2 & -2 & 3 & 2 & 0 & 0 \\
-1 & 0 & 1 & 0 & 0 & 0 & 0 & 0 & 1 & -1 & 0 & -1 & 1 & 1 & -1 & 1 & 0 & 0 & 1 & 0 & -1 & 2 & -1 & 0 & 0 & 1 & 1 & 0 & 0 \\
1 & -1 & 1 & -2 & 2 & 0 & 0 & -3 & 5 & -1 & 0 & 0 & 0 & -1 & -1 & 1 & 0 & 0 & 2 & 0 & 0 & 2 & -4 & 4 & 4 & -2 & -1 & 0 & 0 \\
2 & 0 & 0 & -2 & 2 & 0 & 0 & -3 & 5 & -1 & 0 & 2 & -1 & -1 & -1 & 1 & 0 & 0 & 2 & 0 & 2 & 0 & -4 & 4 & 4 & -2 & -1 & 0 & 0 \\
-1 & 1 & 1 & 0 & 0 & 0 & 0 & 1 & 1 & -1 & 0 & 0 & 0 & 1 & 0 & 0 & 0 & 0 & 2 & 0 & 0 & 2 & 0 & 0 & 0 & 2 & 0 & 0 & 0 \\
0 & 0 & 1 & -1 & 1 & 0 & 0 & -1 & 3 & -1 & 0 & 0 & 0 & 0 & 0 & 1 & 0 & 0 & 2 & 0 & 0 & 2 & -2 & 2 & 2 & 0 & 0 & 0 & 0 \\
1 & -1 & 1 & -2 & 2 & 0 & 0 & -3 & 5 & -1 & 0 & 0 & 0 & 0 & -2 & 2 & 0 & 0 & 2 & 0 & 0 & 2 & -4 & 4 & 4 & -2 & 0 & 0 & 0 \\
3 & -1 & -1 & -2 & 2 & 0 & 0 & -4 & 4 & 0 & 0 & 2 & -2 & -2 & 0 & 0 & 0 & 0 & 0 & 0 & 2 & -2 & -4 & 4 & 4 & -4 & -2 & 0 & 0 \\
0 & 0 & 1 & -1 & 1 & 0 & 0 & 0 & 2 & 0 & 0 & 0 & 0 & 0 & 0 & 0 & 0 & 0 & 2 & 0 & 0 & 2 & 0 & 2 & 2 & 0 & 0 & 0 & 0 \\
0 & -1 & -1 & 1 & 0 & 0 & 0 & -1 & -1 & 1 & 0 & 0 & 0 & 0 & 0 & 0 & 0 & 0 & -2 & 2 & 0 & 0 & 0 & 0 & 0 & 0 & 0 & 0 & 0 \\
1 & -1 & 0 & 0 & 0 & 1 & 0 & -1 & 0 & 0 & 0 & 0 & 0 & 0 & 0 & 0 & 0 & 1 & -1 & 0 & 0 & 0 & 0 & 0 & 1 & -1 & 0 & 0 & 0 \\
-1 & 1 & 0 & 1 & -1 & -1 & 1 & 1 & -1 & 0 & 0 & 0 & 0 & 0 & 0 & 0 & 0 & 0 & 0 & 0 & 0 & 0 & 0 & 0 & -2 & 2 & 0 & 0 & 0 \\
1 & 1 & -1 & 0 & 0 & 0 & 0 & 0 & 0 & 0 & 1 & 1 & -1 & 0 & 0 & 0 & 0 & 0 & 0 & 0 & 2 & -2 & 0 & 0 & 0 & 0 & 0 & 0 & 0 \\
1 & -1 & 0 & -1 & 1 & 0 & 0 & -2 & 2 & 0 & 0 & 0 & 0 & -1 & -1 & 1 & 1 & 0 & 0 & 0 & 0 & 0 & -2 & 2 & 2 & -2 & 0 & 0 & 0 \\
1 & 1 & -1 & 0 & 0 & 0 & 0 & 0 & 0 & 0 & 0 & 2 & -1 & -1 & 1 & 0 & 0 & 0 & 0 & 0 & 2 & -2 & 0 & 0 & 0 & 0 & -1 & 1 & 0 \\
-1 & 0 & 1 & 0 & 0 & 0 & 0 & 1 & 0 & 0 & 0 & -1 & 1 & 1 & 0 & 0 & 0 & 0 & 1 & 0 & -1 & 2 & 1 & 0 & 0 & 1 & 1 & 0 & 1
\end{smallmatrix} \right).
\ea
Keeping the rows $(L^{-1})_i$ of $L^{-1}$ for $\{i\}$ corresponding to $\Lambda_{ii} \ne 1$, we find the SRSI matrices $\tilde{\Theta}^{(0)} = (L^{-1})_{23,\dots,29}$ for the 7 $\Z$SRSIs and $\tilde{\Theta}^{(2)} = (L^{-1})_{20,21,22}$ mod 2 for the 3 $\Z_2$SRSIs.
By performing the Hermite decomposition, we define the simplified SRSI matrix according to SEq.~\eqref{eq:simplifiedSRSI_final} as
\ba
\Theta = \bpm \Theta^{(0)} \\ \Theta^{(2)} \epm
\ea
with
\bg
\Theta^{(0)}
=
\bpm
1 & 0 & 0 & 0 & 0 & 0 & 1 & -1 & 1 & 0 & 0 & 1 & 0 & -1 & 0 & 1 & 1 & 1 & 0 & 0 & 1 & 0 & -1 & 2 & 1 & 0 & 0 & 1 & 1 \\
0 & 1 & 0 & 0 & 0 & 0 & 0 & 1 & 0 & 0 & 0 & 1 & 0 & 0 & 1 & 0 & 0 & 0 & 1 & 0 & 1 & 0 & 1 & 0 & 0 & 1 & 0 & 1 & 1 \\
0 & 0 & 1 & 0 & 0 & 0 & 1 & 0 & 1 & 0 & 0 & 0 & 1 & 0 & 0 & 1 & 1 & 1 & 1 & 0 & 0 & 2 & 0 & 2 & 1 & 1 & 1 & 1 & 2 \\
0 & 0 & 0 & 1 & 0 & 0 & 1 & 0 & 0 & 1 & 0 & 1 & 1 & 0 & 1 & 1 & 1 & 1 & 0 & 2 & 1 & 2 & 1 & 2 & 1 & 2 & 1 & 2 & 3 \\
0 & 0 & 0 & 0 & 1 & 0 & 0 & 0 & 1 & 1 & 0 & 1 & 1 & 0 & 1 & 1 & 1 & 0 & 1 & 2 & 1 & 2 & 1 & 2 & 2 & 1 & 1 & 2 & 3 \\
0 & 0 & 0 & 0 & 0 & 1 & -1 & 1 & -1 & 0 & 0 & 0 & 0 & 1 & 1 & -1 & -1 & 0 & 0 & 0 & 0 & 0 & 2 & -2 & 0 & 0 & 0 & 0 & 0 \\
0 & 0 & 0 & 0 & 0 & 0 & 0 & 0 & 0 & 0 & 1 & -1 & 0 & 1 & -1 & 0 & 0 & 0 & 0 & 0 & 0 & 0 & 0 & 0 & 0 & 0 & 1 & -1 & 0
\epm, \\
\Theta^{(2)} =
\bpm
1 & 1 & 0 & 1 & 1 & 0 & 0 & 0 & 0 & 0 & 0 & 0 & 0 & 0 & 0 & 0 & 0 & 0 & 0 & 0 & 0 & 0 & 0 & 0 & 0 & 0 & 0 & 0 & 0 \\
0 & 0 & 1 & 1 & 1 & 0 & 0 & 0 & 0 & 0 & 0 & 0 & 0 & 0 & 0 & 0 & 0 & 0 & 0 & 0 & 0 & 0 & 0 & 0 & 0 & 0 & 0 & 0 & 0 \\
0 & 0 & 0 & 0 & 0 & 0 & 0 & 1 & 1 & 1 & 0 & 0 & 0 & 0 & 0 & 0 & 0 & 0 & 0 & 0 & 0 & 0 & 0 & 0 & 0 & 0 & 0 & 0 & 0
\epm.
\eg
Hence, we define seven $\Z$SRSIs with $\theta_\Z=(\theta_1,\theta_2,\dots,\theta_7)^T = \Theta^{(0)} \cdot p$ and three $\Z_2$SRSIs with $\theta_{\Z_2} = (\theta_8,\theta_9,\theta_{10})^T = \Theta^{(2)} \cdot p$ mod 2.
The 7 $\Z$SRSIs $\theta_\Z$ are given by
\ba
\theta_1 =& m[ (A_1)_{2a} ] + m[ (A_2)_{6b} ] - m[(B_1)_{6b}] + m[(B_2)_{6b}] + m[(A_2)_{8c}] - m[(A_1)_{12d}] + m[(B_2)_{12d}] + m[(B_3)_{12d}]
\nn
&+ m[(A)_{12e}] + m[(A_1)_{16f}] - m[(A)_{24g}] + 2m[(B)_{24h}] + m[(A)_{24h}] + m[(B)_{24i}] + m[(A)_{48j}],
\nn
\theta_2 =& m[(A_2)_{2a}] + m[(B_1)_{6b}] + m[(A_2)_{8c}] + m[(B_1)_{12d}] + m[(B)_{12e}] + m[(A_1)_{16f}]
\nn
&+ m[(A)_{24g}] + m[(B)_{24h}] + m[(B)_{24i}] + m[(A)_{48j}],
\nn
\theta_3 =& m[(E)_{2a}] + m[(A_2)_{6b}] + m[(B_2)_{6b}] + m[(E)_{8c}] + m[(B_2)_{12d}] + m[(B_3)_{12d}] + m[(A)_{12e}] + m[(B)_{12e}]
\nn
&+ 2m[({}^1E {}^2E)_{16f}] + 2m[(B)_{24g}] + m[(A)_{24h}] + m[(B)_{24h}] + m[(A)_{24i}] + m[(B)_{24i}] +2 m[(A)_{48j}],
\nn
\theta_4 =& m[(T_1)_{2a}] + m[(A_2)_{6b}] + m[(E)_{6b}] + m[(A_2)_{8c}] + m[(E)_{8c}] + m[(B_1)_{12d}] + m[(B_2)_{12d}] + m[(B_3)_{12d}]
\nn
&+ m[(A)_{12e}] + 2m[{{}^1E {}^2E}_{12e}] + m[(A_1)_{16f}] + 2m[({}^1E {}^2E)_{16f}] + m[(A)_{24g}]
\nn
&+ 2m[(B)_{24g}]+ m[(A)_{24h}]+ 2m[(B)_{24h}]+ m[(A)_{24i}]+ 2m[(B)_{24i}]+ 3m[(A)_{48j}],
\nn
\theta_5 =& m[(T_2)_{2a}] + m[(B_2)_{6b}] + m[(E)_{6b}] + m[(A_2)_{8c}] + m[(E)_{8c}] + m[(B_1)_{12d}] + m[(B_2)_{12d}] + m[(B_3)_{12d}]
\nn
&+ m[(B)_{12e}] + 2m[({}^1E {}^2E)_{12e}] + m[(A_1)_{16f}] + 2m[({}^1E {}^2E)_{16f}] + m[(A)_{24g}] + 2m[(B)_{24g}]
\nn
&+ 2m[(A)_{24h}] + m[(B)_{24h}] + m[(A)_{24i}] + 2m[(B)_{24i}] + 3m[(A)_{48j}],
\nn
\theta_6 =& m[(A_1)_{6b}] - m[(A_2)_{6b}] + m[(B_1)_{6b}] - m[(B_2)_{6b}] + m[(A_1)_{12d}]+ m[(B_1)_{12d}]- m[(B_2)_{12d}]- m[(B_3)_{12d}]
\nn
& + 2m[(A)_{24g}] - 2m[(B)_{24g}],
\nn
\theta_7 =& m[(A_1)_{8c}] - m[(A_2)_{8c}] + m[(A_1)_{12d}] - m[(B_1)_{12d}] + m[(A)_{24i}] -m[(B)_{24i}],
\ea
which coincide with the definitions given in Supplementary Table~\ref{tab: singlevaluedSRSIs}.
The 3 $\Z_2$SRSIs are given by
\ba
\theta_8 =& m[(A_1)_{2a}] + m[(A_2)_{2a}] + m[(T_1)_{2a}] + m[(T_2)_{2a}] \mod 2,
\nn
\theta_9 =& m[(E)_{2a}] + m[(T_1)_{2a}] + m[(T_2)_{2a}] \mod 2,
\nn
\theta_{10} =& m[(B_1)_{6b}] + m[(B_2)_{6b}] + m[(E)_{6b}] \mod 2.
\ea
This definition does not match with Supplementary Table~\ref{tab: singlevaluedSRSIs}.
However, as we discussed in SN~\ref{sec:srsi_derivation}, the $\Z_n$SRSIs can be redefined by adding to them any linear combination of $\Z$SRSIs.
Exploiting this, we can redefine $\theta_{\Z_2}$ to $\theta_{\Z_2}' =(\theta_8', \theta_9', \theta_{10}')^T$ with
\ba
\theta_8' =& \theta_1 +\theta_2 +\theta_4 +\theta_5 +\theta_6 + \theta_9 \mod 2
\nn
=& m[(A_1)_{2a}] + m[(A_2)_{2a}] + m[(T_1)_{2a}] + m[(T_2)_{2a}] + m[(A_1)_{6b}] + m[(A_2)_{6b}] + m[(E)_{6b}] \mod 2,
\nn
\theta_9' =& \theta_1 +\theta_2 +\theta_4 +\theta_5 +\theta_6 + \theta_8 + \theta_9 + \theta_{10} \mod 2
\nn
=& m[(E)_{2a}] + m[(T_1)_{2a}] + m[(T_2)_{2a}] + m[(A_1)_{6b}] + m[(A_2)_{6b}] + m[(E)_{6b}] \mod 2,
\nn
\theta_{10}' =& \theta_1 +\theta_2 +\theta_4 +\theta_5 + \theta_8 \mod 2
\nn
=& m[(A_1)_{12d}] + m[(B_1)_{12d}] + m[(B_2)_{12d}] + m[(B_3)_{12d}] \mod 2.
\label{eq:sg211_srsi_redef}
\ea
Then, $\theta_{8,9,10}'$ match with $\theta_{8,9,10}$ listed in Supplementary Table~\ref{tab: singlevaluedSRSIs}.

We now discuss the mapping between $\Z$SRSIs and symmetry-data vector.
To do this, we define the BR matrix $BR$ induced from the site-symmetry irreps in SEq.~\eqref{eq:sg211_irrep} ($BR$ is also available on the BCS~\cite{bilbao-server})
\bg
\BR = \left( \begin{smallmatrix}
1 & 0 & 0 & 0 & 0 & 1 & 0 & 0 & 0 & 0 & 1 & 0 & 0 & 1 & 0 & 0 & 0 & 1 & 0 & 0 & 1 & 0 & 1 & 0 & 1 & 0 & 1 & 0 & 1 \\
0 & 1 & 0 & 0 & 0 & 0 & 0 & 1 & 0 & 0 & 0 & 1 & 0 & 0 & 1 & 0 & 0 & 0 & 1 & 0 & 1 & 0 & 1 & 0 & 0 & 1 & 0 & 1 & 1 \\
0 & 0 & 1 & 0 & 0 & 1 & 0 & 1 & 0 & 0 & 0 & 0 & 1 & 1 & 1 & 0 & 0 & 1 & 1 & 0 & 0 & 2 & 2 & 0 & 1 & 1 & 1 & 1 & 2 \\
0 & 0 & 0 & 1 & 0 & 0 & 1 & 0 & 0 & 1 & 0 & 1 & 1 & 0 & 1 & 1 & 1 & 1 & 0 & 2 & 1 & 2 & 1 & 2 & 1 & 2 & 1 & 2 & 3 \\
0 & 0 & 0 & 0 & 1 & 0 & 0 & 0 & 1 & 1 & 1 & 0 & 1 & 1 & 0 & 1 & 1 & 0 & 1 & 2 & 1 & 2 & 1 & 2 & 2 & 1 & 2 & 1 & 3 \\
1 & 0 & 0 & 0 & 0 & 1 & 0 & 0 & 0 & 0 & 0 & 1 & 0 & 0 & 1 & 0 & 0 & 1 & 0 & 0 & 1 & 0 & 1 & 0 & 1 & 0 & 0 & 1 & 1 \\
0 & 1 & 0 & 0 & 0 & 0 & 0 & 1 & 0 & 0 & 1 & 0 & 0 & 1 & 0 & 0 & 0 & 0 & 1 & 0 & 1 & 0 & 1 & 0 & 0 & 1 & 1 & 0 & 1 \\
0 & 0 & 1 & 0 & 0 & 1 & 0 & 1 & 0 & 0 & 0 & 0 & 1 & 1 & 1 & 0 & 0 & 1 & 1 & 0 & 0 & 2 & 2 & 0 & 1 & 1 & 1 & 1 & 2 \\
0 & 0 & 0 & 1 & 0 & 0 & 1 & 0 & 0 & 1 & 1 & 0 & 1 & 1 & 0 & 1 & 1 & 1 & 0 & 2 & 1 & 2 & 1 & 2 & 1 & 2 & 2 & 1 & 3 \\
0 & 0 & 0 & 0 & 1 & 0 & 0 & 0 & 1 & 1 & 0 & 1 & 1 & 0 & 1 & 1 & 1 & 0 & 1 & 2 & 1 & 2 & 1 & 2 & 2 & 1 & 1 & 2 & 3 \\
1 & 1 & 0 & 0 & 0 & 0 & 1 & 0 & 1 & 0 & 1 & 1 & 0 & 0 & 0 & 1 & 1 & 1 & 1 & 0 & 2 & 0 & 0 & 2 & 1 & 1 & 1 & 1 & 2 \\
0 & 0 & 1 & 0 & 0 & 0 & 1 & 0 & 1 & 0 & 0 & 0 & 1 & 0 & 0 & 1 & 1 & 1 & 1 & 0 & 0 & 2 & 0 & 2 & 1 & 1 & 1 & 1 & 2 \\
0 & 0 & 1 & 0 & 0 & 0 & 1 & 0 & 1 & 0 & 0 & 0 & 1 & 0 & 0 & 1 & 1 & 1 & 1 & 0 & 0 & 2 & 0 & 2 & 1 & 1 & 1 & 1 & 2 \\
0 & 0 & 0 & 1 & 1 & 1 & 0 & 1 & 0 & 2 & 1 & 1 & 2 & 2 & 2 & 1 & 1 & 1 & 1 & 4 & 2 & 4 & 4 & 2 & 3 & 3 & 3 & 3 & 6 \\
1 & 0 & 1 & 0 & 1 & 1 & 1 & 0 & 2 & 1 & 1 & 1 & 2 & 1 & 1 & 2 & 2 & 2 & 2 & 2 & 2 & 4 & 2 & 4 & 4 & 2 & 3 & 3 & 6 \\
0 & 1 & 1 & 1 & 0 & 0 & 2 & 1 & 1 & 1 & 1 & 1 & 2 & 1 & 1 & 2 & 2 & 2 & 2 & 2 & 2 & 4 & 2 & 4 & 2 & 4 & 3 & 3 & 6 \\
0 & 0 & 0 & 1 & 1 & 1 & 0 & 1 & 0 & 2 & 2 & 0 & 2 & 3 & 1 & 1 & 1 & 1 & 1 & 4 & 2 & 4 & 4 & 2 & 3 & 3 & 4 & 2 & 6 \\
0 & 0 & 0 & 1 & 1 & 1 & 0 & 1 & 0 & 2 & 0 & 2 & 2 & 1 & 3 & 1 & 1 & 1 & 1 & 4 & 2 & 4 & 4 & 2 & 3 & 3 & 2 & 4 & 6
\end{smallmatrix} \right),
\label{eq:sg211_BR}
\eg
where the row basis (corresponding to multiplicity of each little group irrep) is ordered as
\bg
\{ \Gamma_1, \Gamma_2, \Gamma_3, \Gamma_4, \Gamma_5, H_1, H_2, H_3, H_4, H_5, P_1, P_2, P_3, P_4, N_1, N_2, N_3, N_4 \}
\label{eq:sg211_kirrep}
\eg
and the column basis follows the ordering in SEq.~\eqref{eq:sg211_irrep}.
Using this, we compute
\bg
\Theta^{(0)} \cdot \BR^\ddagger =
\bpm
0 & 0 & 0 & 0 & 0 & 0 & -1 & 0 & 0 & 0 & 1 & 0 & 0 & 0 & 0 & 0 & 0 & 0 \\
0 & 1 & 0 & 0 & 0 & 0 & 0 & 0 & 0 & 0 & 0 & 0 & 0 & 0 & 0 & 0 & 0 & 0 \\
-1 & -1 & 1 & 0 & 0 & 0 & 0 & 0 & 0 & 0 & 1 & 0 & 0 & 0 & 0 & 0 & 0 & 0 \\
0 & 0 & 0 & 1 & 0 & 0 & 0 & 0 & 0 & 0 & 0 & 0 & 0 & 0 & 0 & 0 & 0 & 0 \\
-1 & 1 & 0 & -1 & 0 & 0 & -2 & 0 & 0 & 0 & 1 & 0 & 0 & 0 & 0 & 0 & 1 & 0 \\
1 & 1 & 0 & 0 & 0 & 0 & 0 & 0 & 0 & 0 & -1 & 0 & 0 & 0 & 0 & 0 & 0 & 0 \\
0 & -1 & 0 & 0 & 0 & 0 & 1 & 0 & 0 & 0 & 0 & 0 & 0 & 0 & 0 & 0 & 0 & 0
\epm,
\label{eq:sg211_map1}
\eg
to determine the $\Z$SRSIs from symmetry-data vector $B$ according to $\theta_\Z = \Theta^{(0)} \cdot \BR^\ddagger \cdot B$.
Conversely, the symmetry-data vector $B$ can be obtained from $\theta_\Z$ via $B = BR \cdot [\Theta^{(0)}]^\ddagger \cdot \theta_\Z$ by computing
\bg
BR \cdot [\Theta^{(0)}]^\ddagger =
\bpm
1 & 0 & 0 & 0 & 0 & 1 & 0 & 0 & 0 & 0 & 1 & 0 & 0 & 0 & 1 & 0 & 0 & 0 \\
0 & 1 & 0 & 0 & 0 & 0 & 1 & 0 & 0 & 0 & 1 & 0 & 0 & 0 & 0 & 1 & 0 & 0 \\
0 & 0 & 1 & 0 & 0 & 0 & 0 & 1 & 0 & 0 & 0 & 1 & 1 & 0 & 1 & 1 & 0 & 0 \\
0 & 0 & 0 & 1 & 0 & 0 & 0 & 0 & 1 & 0 & 0 & 0 & 0 & 1 & 0 & 1 & 1 & 1 \\
0 & 0 & 0 & 0 & 1 & 0 & 0 & 0 & 0 & 1 & 0 & 0 & 0 & 1 & 1 & 0 & 1 & 1 \\
1 & 0 & 1 & 0 & 0 & 1 & 0 & 1 & 0 & 0 & 0 & 0 & 0 & 1 & 1 & 0 & 1 & 1 \\
1 & 0 & 0 & 0 & 1 & 0 & 1 & 0 & 1 & 0 & 1 & 0 & 0 & 1 & 1 & 1 & 2 & 0
\epm^T.
\label{eq:sg211_map2}
\eg

Let us compare the SRSIs with the local RSIs and the composite RSIs accessible in the BCS~\cite{bilbao-server}.
Similar to the previous examples SGs $Pmm2$ and $P4_332$ we considered in SNs~\ref{sec:srsi_25} and \ref{sec:srsi_212}, all 7 local RSIs defined at maximal WPs $2a, 6b, 8c, 12d$ are determined by appropriate linear combinations of $\Z$SRSIs and $\Z_2$SRSIs.
Also, all 17 composite RSIs are determined by $\Z$SRSIs.
Let us demonstrate the comparison between RSIs with a few examples.
At the WP $2a$, three local $\Z_2$ RSIs are defined.
One of them is expressed as $\delta_{\Z_2,2a}^{(1)} = -m[(A_1)_{2a}] + m[(T_1)_{2a}]$ mod 2.
This can be expressed as a linear combination of $\Z$SRSIs,
\bg
{\delta_{\Z_2,2a}^{(1)}}
= \theta_1 + \theta_3 + \theta_5 + \theta_6 + \theta_9' \mod 2.
\eg
This linear combination can be found using the method outlined in SEq.~\eqref{eq:stable_to_local_z2}.
A unique local RSI $\delta_{\Z_2,12d}$ at the WP $12d$ is defined as $m[(A)_{12d}] + m[(B_1)_{12d}] - m[(B_2)_{12d}] - m[(B_3)_{12d}]$ mod 2, which is exactly equal to $\theta_{10}'$.
Another example is the composite RSI $\delta_3$, whose relevant WPs are $2a$, $8c$, and $16f$.
$\delta_3$ is determined by the symmetry-data vector as $\delta_3 = m(\Gamma_3) - m(H_1) - m(H_2)$.
Using the compatibility relations, which can be derived according to SEq.~\eqref{eq:si_and_comp}, we find that
\bg
m(\Gamma_1) + m(\Gamma_2) = m(H_1) + m(H_2),
\eg
which lets us rewrite $\delta_3$ as $\delta_3 = -m(\Gamma_1) - m(\Gamma_2) + m(\Gamma_3)$.
From SEq.~\eqref{eq:sg211_map1}, we find that $\delta_3 = -\theta_1 - \theta_2 + \theta_3 - \theta_7$.
In general, instead of using the compatibility relations to determine the linear combination of SRSIs corresponding to a given composite RSI, we can alternatively take advantage of the direct mapping from $\theta_\Z$ to $B$ given in SEq.~\eqref{eq:sg211_map2}.
For $\mc{M}_{\delta_3}=(0,0,1,0,0,-1,-1,0,0,0,0,0,0,0,0,0,0,0)$, such that $\delta_3 = \mc{M}_{\delta_3} \cdot B$, we find:
\bg
\delta_3 = \mc{M}_{\delta_3} \cdot BR \cdot [\Theta^{(0)}]^\ddagger \cdot \theta_\Z
= (-1,-1,1,0,0,0,-1) \cdot \theta_\Z,
\eg
confirming that $\delta_3 = -\theta_1 - \theta_2 + \theta_3 - \theta_7$, as already derived using the compatibility relations.
For other local and composite RSIs, the relation between them and SRSIs can be established in a similar way.

Finally, we remark an interesting case where a pair of inequivalent EBRs have matching $\Z$SRSIs (and hence matching symmetry-data vector $B$) and $\Z_n$SRSIs.
For this, consider the two distinct EBRs induced by the site-symmetry irreps $(B_2)_{12d}$ and $(B_3)_{12d}$ at the $12d$ WP.
Both EBRs have the same SRSIs with
\bg
\theta_\Z = (1,0,1,1,1,-1,0)^T, \quad
\theta_{\Z_2} = (0,0,0)^T.
\label{eq:sg211_srsi_12d}
\eg
As we argued in SN~\ref{sec:srsi_derivation}, this implies that $(B_2)_{12d}$ and $(B_3)_{12d}$ are stably equivalent.
That is, in the presence of appropriate auxiliary trivial orbitals, $(B_2)_{12d}$ and $(B_3)_{12d}$ can be adiabatically deformed into each other.
To understand this phenomena, let us focus on adiabatic processes,
\bg
(A_1)_{12d} + (B_3)_{12d} \Leftrightarrow (A)_{24i}, \quad
(A_1)_{12d} + (B_2)_{12d} \Leftrightarrow (A)_{24i},
\label{eq:211adiabatic24i}
\eg
which correspond to 19th and 21th columns of adiabatic matrix $q$ in SEq.~\eqref{eq:sg211_q}.
To understand these adiabatic processes, let us focus on one representative point of the WP $12d$: $\bb x_{12d,1} = (1/4,1/2,0)$.
The site-symmetry group $G_{\bb x_{12d,1}}$ at $\bb x_{12d,1}$ is isomorphic to the point group 222 with three twofold rotations, $\{2_{100}|0,1,0\}$, $\{2_{011}|1/2,1/2,-1/2\}$, and $\{2_{01\bar{1}}|1/2,1/2,1/2\}$.
Points on these axes of rotation can be parameterized by $(x,1/2,0)$, $(1/4,z+1/2,z)$, and $(1/4,-z+1/2,z)$ respectively, with free real parameters $(x,z)$.
Interestingly, both $\bb x_{24i,1}(z)=(1/4,z+1/2,z)$ and $\bb x_{24i,2}(z)=(1/4,-z+1/2,z)$ belong to the WP $24i$.
The site-symmetry irrep $(A)$ at $\bb x_{24i,1}(z) = (1/4,z+1/2,z)$ can be deformed to the $(A_1)$ and $(B_3)$ site-symmetry irreps at $\bb x_{12d,1}$.
Equivalently, the site-symmetry irrep $(A)$ at $\bb x_{24i,2}(z) = (1/4,-z+1/2,z)$ can be deformed to the $(A_1)$ and $(B_2)$ site-symmetry irreps at $\bb x_{12d,1}$.
Crucially, note that the $(A)$ irrep at either of these points induces the same band representation.
In particular, the band representation induced by the $A$ irrep at $\bb x_{24i,1}$ contains an $A$ irrep at $\bb x_{24i,2}$, and vice-versa.

\begin{figure}[t]
\centering
\includegraphics[width=0.85\textwidth]{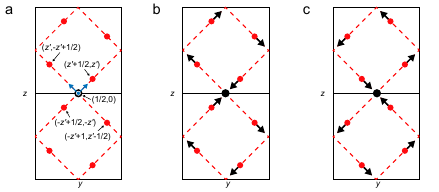}
\caption{Adiabatic processes in SG $I432$ (No. 211).
The site-symmetry irreps at WP $12d$ and $24i$ are shown on the $(y,z)$ plane with a fixed $x$-coordinate, $x=1/4$.
Unit cells are indicated by solid black lines.
The schematic is shown in reduced coordinates such that the length in both the $x$- and $y$-directions is 1.
\textbf{a} The $(A)$ site-symmetry irrep is located at $(y,z)=(z'+1/2,z')$ (marked by a red dot).
Other $(A)$ irreps must exist at $(y,z)=(-z'+1/2,-z')$, $(z',-z'+1/2)$, and $(-z'+1,z'-1/2)$ by the twofold and fourfold rotations, $\{2_{100}|0,1,0\}$ and $\{4^+_{100}|0,0,0\}$.
Note that $\{2_{100}|0,1,0\}: (z'+1/2,z') \to (-z'+1/2,-z')$ and $\{4^+_{100}|0,0,0\}: (-z'+1/2,-z') \to (z',-z'+1/2)$.
All four positions are connected to a representative point of the WP $12d$, $\bb x_{12d,1}=(1/4,1/2,0)$ (shown as a black dot), which lies at the intersection of three two-fold rotation axes.
The pair of diagonal blue arrows indicate the two-fold rotation axes $\{2_{011}|1/2,1/2,-1/2\}$ and $\{2_{01\bar{1}}|1/2,1/2,1/2\}$, while the out-of-plane axis corresponds to $\{2_{100}|0,1,0\}.$
\textbf{b} The pair of $(A)$ irreps at $(y,z)=(z'+1/2,z')$ and $(-z'+1/2,-z')$ are deformed to a sum of $(A_1)$ and $(B_3)$ irreps at $\bb x_{12d,1}$, by tuning $z'$ to $0$.
\textbf{c} The pair of $(A)$ irreps at $(y,z)=(z',-z'+1/2)$ and $(-z'+1,z'-1/2)$ are deformed to a sum of $(A_1)$ and $(B_2)$ irreps at $\bb x_{12d,1}$, by tuning $z'$ to $1/2$.}
\label{fig:sg211}
\end{figure}

To see this intuitively, note that placing an $s$ orbital (transforming in the $A$ representation) at $\bb x_{24i,1}(z') = (1/4,z'+1/2,z')$ implies that there is also an $s$ orbital (transforming in the $A$ representation) at $\bb x_{24i,1}(-z') = (1/4,-z'+1/2,-z')$ by symmetry $\{2_{100}|0,1,0\}$.
This is illustrated in SFig.~\ref{fig:sg211}\textbf{a}.
Both positions $\bb x_{24i,1}(\pm z')$ are on the same rotation axis of $\{2_{011}|1/2,1/2,-1/2\}$ and connected to $\bb x_{12d,1} = \bb x_{24i,1}(0)$.
Thus, the pair of $(A)$ irreps at $\bb x_{24i,1}(\pm z')$ can be moved to $\bb x_{12d,1}$ adiabatically by tuning $z'$ to 0, as shown in SFig.~\ref{fig:sg211}\textbf{b}.
The pair can then be deformed to the $(A_1\oplus B_3)$ site-symmetry irreps at $\bb x_{12d,1}$.
Simultaneously, by fourfold rotation $\{4^+_{100}|\bb 0\}$, the existence of an orbital transforming in the $(A)$ irrep at $\bb x_{24i,1}(-z')$ implies the existence of an orbital transforming in the $(A)$ irrep at $\bb x_{24i,2}(-z'+1/2) = (1/4,z',-z'+1/2)$ (see SFig.~\ref{fig:sg211}\textbf{a}).
Again, by $\{2_{100}|0,1,0\}$, the $(A)$ irrep at $\bb x_{24i,2}(-z'+1/2)$ has a symmetric partner irrep, the $(A)$ irrep at $\bb x_{24i,2}(z'-1/2) = (1/4,-z'+1,z'-1/2)$.
Both positions $\bb x_{24i,2}(z'-1/2)$ and $\bb x_{24i,2}(-z'+1/2)$ are connected to $\bb x_{12d,1} = \bb x_{24i,2}(0)$.
Thus, the pair of $(A)$ irreps at these positions can be moved to $\bb x_{12d,1}$ by tuning $z'$ to $1/2$, and induce the $(A_1\oplus B_2)$ site-symmetry irreps at $\bb x_{12d,1}$.
This adiabatic process is illustrated in SFig.~\ref{fig:sg211}\textbf{c}.
Thus, chaining together the two adiabatic processes in SEq.~\eqref{eq:211adiabatic24i} gives rise to a stable equivalence between the $(B_2)_{12d} \uparrow G$ and $(B_3)_{12d} \uparrow G$ EBRs.
As we computed, this stable equivalence is indicated by the matching SRSIs in SEq.~\eqref{eq:sg211_srsi_12d}.

As this example demonstrates, SRSIs can yield insight into the analysis of inequivalent EBRs (${\rm EBR}_1$ and ${\rm EBR}_2$) that have the same symmetry-data vector, as studied in SRef.~\cite{Cano22Invisible}.
According to the discussion in SN~\ref{sec:zsrsi_proof}, ${\rm EBR}_1$ and ${\rm EBR}_2$ are stably equivalent if their $\Z_n$SRSIs match.
If their $\Z_n$SRSIs do not match, one can consider $\mc{N}$ copies of them such that $\Z_n$SRSIs of $\mc{N}$ copies of ${\rm EBR}_1$ match to those of $\mc{N}$ copies of ${\rm EBR}_2$.
Then, $\mc{N}$ copies of ${\rm EBR}_1$ can be deformed to $\mc{N}$ copies of ${\rm EBR}_2$ in the presence of appropriate trivial bands, and vice versa, where $\mc{N}$ can always be chosen to be the least common multiple of values of $n$ defined for $\Z_n$SRSIs.
In some cases, a smaller multiple (due to the modulo operations) can be chosen provided that the $\Z_n$SRSIs of $\mc{N}$ copies of ${\rm EBR}_1$ and ${\rm EBR}_2$ match.

\subsubsection{SG $F\bar{4}3c$ (No. 219) with SOC}
\label{sec:sg219_stable}
As a final example, we examine the SRSIs in SG $F\bar{4}3c$ (No. 219) with SOC, which is one of the four SGs that host $\Z_4$SRSIs.
We focus on this SG because the other three SGs with $\Z_4$SRSIs are its subgroups.
This example illustrates that $\Z_4$SRSIs require both time-reversal symmetry (which protects Kramers pairs in the presence of SOC) and multiple adiabatic processes connecting a pair of WPs.
Together, these conditions can (though not necessarily) lead to a specific structure of the adiabatic matrix $q$, where an elementary divisor $4$ appears in its Smith normal form.

To explore this, we first derive the SRSIs for $F\bar{4}3c$ with both time-reversal and SOC, which corresponds to the magnetic SG $F\bar{4}3c1'$ (No. 219.86) in the Belov-Neronova-Smirnova (BNS) notation~\cite{belov1957Shubnikov}, where a $\Z_4$SRSI exists.
We then compare this to $F\bar{4}3c$ with SOC but without time-reversal symmetry, which corresponds to the magnetic SG $F\bar{4}3c$ (No. 219.85) in the BNS notation and does not host any $\Z_4$SRSI.
This comparison highlights how the absence of time-reversal symmetry alters the structure of adiabatic processes and, consequently, the SRSIs.

\paragraph{SRSIs in $F\bar{4}3c$ with SOC and time-reversal symmetry:} To begin, let us first explain the symmetries of SG $F\bar{4}3c$ with time-reversal symmetry.
This SG is generated by lattice translations, $\{E|0,1/2,1/2\}$, $\{E|1/2,0,1/2\}$, $\{E|1/2,1/2,0\}$, the fourfold roto-inversion $\{-4^+_{001}|1/2,1/2,1/2\}$, and the threefold rotation $\{3^+_{111}|\bb 0\}$.
There are four maximal WPs, $8a, 8b, 24c, 24d$, and four non-maximal WPs, $32e, 48f, 48g, 96h$, where $96h$ is the general WP.
The site-symmetry (co)irreps at these WPs are as follows:
\bg
\{
(\bar{E})_{8a}, ({}^1\bar{F} {}^2\bar{F})_{8a}, (\bar{E})_{8b}, ({}^1\bar{F} {}^2\bar{F})_{8b}, ({}^1\bar{E}_2 {}^2\bar{E}_2)_{24c}, ({}^1\bar{E}_1 {}^2\bar{E}_1)_{24c}, ({}^1\bar{E}_2 {}^2\bar{E}_2)_{24d},
\nn
({}^1\bar{E}_1 {}^2\bar{E}_1)_{24d}, (\bar{E}\bar{E})_{32e}, ({}^1\bar{E} {}^2\bar{E})_{32e}, ({}^1\bar{E} {}^2\bar{E})_{48f}, ({}^1\bar{E} {}^2\bar{E})_{48g}, (\bar{A}\bar{A})_{96h}
\}.
\label{eq:sg219_irrep}
\eg
Thus, $N^\rho_{\rm UC}=13$.

Now, we construct the adiabatic-process matrix $q$ by collecting the full set of generators for adiabatic processes between site-symmetry irreps in SEq.~\eqref{eq:sg219_irrep}:
\bg
q = \bpm
0 & 2 & 2 & 0 & 0 & 0 & 0 & 0 & 0 & 0 & 0 \\
2 & 1 & 2 & 0 & 0 & 0 & 0 & 0 & 0 & 0 & 0 \\
0 & 0 & 0 & 0 & 2 & 2 & 0 & 0 & 0 & 0 & 0 \\
0 & 0 & 0 & 2 & 1 & 2 & 0 & 0 & 0 & 0 & 0 \\
0 & 0 & 0 & 0 & 0 & 0 & 1 & 0 & 0 & 0 & 0 \\
0 & 0 & 0 & 0 & 0 & 0 & 1 & 0 & 0 & 0 & 0 \\
0 & 0 & 0 & 0 & 0 & 0 & 0 & 1 & 0 & 0 & 0 \\
0 & 0 & 0 & 0 & 0 & 0 & 0 & 1 & 0 & 0 & 0 \\
-1 & 0 & 0 & -1 & 0 & 0 & 0 & 0 & 1 & 0 & 0 \\
0 & -1 & 0 & 0 & -1 & 0 & 0 & 0 & 2 & 0 & 0 \\
0 & 0 & -1 & 0 & 0 & 0 & 0 & -1 & 0 & 2 & 0 \\
0 & 0 & 0 & 0 & 0 & -1 & -1 & 0 & 0 & 0 & 2 \\
0 & 0 & 0 & 0 & 0 & 0 & 0 & 0 & -1 & -1 & -1
\epm,
\label{eq:sg219_q}
\eg
where the row basis is ordered as in SEq.~\eqref{eq:sg219_irrep}.

The Smith decomposition of $q= L \cdot \Lambda \cdot R$ [SEq.~\eqref{eq:qvecdef_global}] yields
\bg
{\rm diag} (\Lambda) = (1,1,1,1,1,1,1,1,4,0,0)^T,
\nn
L = \left( \begin{smallmatrix}
0 & 2 & 2 & -2 & -4 & 0 & 0 & 0 & 1 & 0 & 0 & 0 & -1 \\
2 & 1 & 2 & -1 & 0 & 0 & 0 & 0 & 0 & 0 & 0 & 0 & -2 \\
0 & 0 & 0 & 2 & 6 & 2 & 0 & 0 & -3 & 0 & 0 & 0 & 0 \\
0 & 0 & 0 & 1 & 2 & 2 & 0 & 0 & -2 & 1 & 0 & 0 & 0 \\
0 & 0 & 0 & 0 & 1 & 0 & 1 & 0 & -1 & 0 & 0 & 0 & 0 \\
0 & 0 & 0 & 0 & 1 & 0 & 1 & 0 & -1 & 0 & 0 & 1 & 0 \\
0 & 0 & 0 & 0 & 1 & 0 & 0 & 1 & -1 & 0 & 1 & 0 & 0 \\
0 & 0 & 0 & 0 & 1 & 0 & 0 & 1 & -1 & 0 & 0 & 0 & 0 \\
-1 & 0 & 0 & 0 & 1 & 0 & 0 & 0 & 0 & 0 & 0 & 0 & 0 \\
0 & -1 & 0 & 0 & 2 & 0 & 0 & 0 & 0 & 0 & 0 & 0 & -1 \\
0 & 0 & -1 & 0 & -1 & 0 & 0 & -1 & 1 & 0 & 0 & 0 & 0 \\
0 & 0 & 0 & 0 & -2 & -1 & -1 & 0 & 2 & 0 & 0 & 0 & 0 \\
0 & 0 & 0 & 0 & -1 & 0 & 0 & 0 & 0 & 0 & 0 & 0 & 1
\end{smallmatrix} \right),
\quad
R = \left( \begin{smallmatrix}
1 & 0 & 0 & 1 & 0 & 0 & 0 & 0 & 0 & 1 & 1 \\
0 & 1 & 0 & 0 & 1 & 0 & 0 & 0 & 0 & 2 & 2 \\
0 & 0 & 1 & 0 & 0 & 0 & 0 & 0 & 0 & -2 & 0 \\
0 & 0 & 0 & 2 & 1 & 0 & 0 & 0 & 0 & 0 & 4 \\
0 & 0 & 0 & 0 & 0 & 0 & 0 & 0 & 1 & 1 & 1 \\
0 & 0 & 0 & 4 & 0 & 1 & 0 & 0 & 3 & 3 & 5 \\
0 & 0 & 0 & 4 & 0 & 0 & 1 & 0 & 3 & 3 & 7 \\
0 & 0 & 0 & 4 & 0 & 0 & 0 & 1 & 3 & 3 & 7 \\
0 & 0 & 0 & 1 & 0 & 0 & 0 & 0 & 1 & 1 & 2 \\
0 & 0 & 0 & 1 & 0 & 0 & 0 & 0 & 0 & 1 & 1 \\
0 & 0 & 0 & 1 & 0 & 0 & 0 & 0 & 0 & 0 & 2
\end{smallmatrix} \right).
\eg
This implies that the rank of $q$ is ${\rm rank}(q)=9$ and that four $\Z$SRSIs and a single $\Z_4$SRSI can be defined in this SG.
Also, recall that the number of $\Z$SRSIs is equal to $N^\rho_{\rm UC} - {\rm rank}(q)$ ($13-9=4$).
To define the SRSIs, we compute the inverse of $L$:
\ba
L^{-1} = 
\bpm
1 & 0 & 1 & 0 & 2 & 0 & 0 & 2 & -1 & 2 & 2 & 2 & 3 \\
1 & 0 & 1 & 0 & 2 & 0 & 0 & 2 & 0 & 1 & 2 & 2 & 2 \\
0 & 0 & 0 & 0 & 0 & 0 & 0 & -1 & 0 & 0 & -1 & 0 & 0 \\
1 & -1 & 1 & 0 & 2 & 0 & 0 & 0 & -2 & 1 & 0 & 2 & 0 \\
1 & 0 & 1 & 0 & 2 & 0 & 0 & 2 & 0 & 2 & 2 & 2 & 3 \\
5 & -2 & 4 & 0 & 7 & 0 & 0 & 6 & -4 & 8 & 6 & 7 & 9 \\
5 & -2 & 4 & 0 & 9 & 0 & 0 & 6 & -4 & 8 & 6 & 8 & 9 \\
5 & -2 & 4 & 0 & 8 & 0 & 0 & 7 & -4 & 8 & 6 & 8 & 9 \\
6 & -2 & 5 & 0 & 10 & 0 & 0 & 8 & -4 & 10 & 8 & 10 & 12 \\
-1 & 1 & -1 & 1 & 0 & 0 & 0 & 0 & 2 & -1 & 0 & 0 & 0 \\
0 & 0 & 0 & 0 & 0 & 0 & 1 & -1 & 0 & 0 & 0 & 0 & 0 \\
0 & 0 & 0 & 0 & -1 & 1 & 0 & 0 & 0 & 0 & 0 & 0 & 0 \\
1 & 0 & 1 & 0 & 2 & 0 & 0 & 2 & 0 & 2 & 2 & 2 & 4
\epm
\label{eq:sg219_inverse_l}
\ea
Collecting the rows $(L^{-1})_i$ of $L^{-1}$ for indices $\{i\}$ corresponding to $\Lambda_{ii} \ne 1$, we obtain the SRSI matrices $\tilde{\Theta}^{(0)} = (L^{-1})_{10,\dots,13}$ for the 4 $\Z$SRSIs and $\tilde{\Theta}^{(4)} = (L^{-1})_{9}$ mod 4 for the unique $\Z_4$SRSI:
\ba
\tilde{\Theta}^{(0)} =& \bpm
-1 & 1 & -1 & 1 & 0 & 0 & 0 & 0 & 2 & -1 & 0 & 0 & 0 \\
0 & 0 & 0 & 0 & 0 & 0 & 1 & -1 & 0 & 0 & 0 & 0 & 0 \\
0 & 0 & 0 & 0 & -1 & 1 & 0 & 0 & 0 & 0 & 0 & 0 & 0 \\
1 & 0 & 1 & 0 & 2 & 0 & 0 & 2 & 0 & 2 & 2 & 2 & 4
\epm,
\nn
\tilde{\Theta}^{(4)} =& (6, -2, 5, 0, 10, 0, 0, 8, -4, 10, 8, 10, 12) \mod 4
\nn
=& (2, 2, 1, 0, 2, 0, 0, 0, 0, 2, 0, 2, 0) \mod 4.
\ea

Applying the Hermite decomposition, we then define the simplified SRSI matrices according to SEq.~\eqref{eq:simplifiedSRSI_final}:
\ba
\Theta^{(0)}
=& U^{(0)} \cdot \tilde{\Theta}^{(0)}
= \bpm 
0 & 0 & 2 & 1 \\ 1 & 0 & 2 & 1 \\
0 & 0 & -1 & 0 \\ 0 & 1 & 0 & 0
\epm \cdot \tilde{\Theta}^{(0)}
= \bpm
1 & 0 & 1 & 0 & 0 & 2 & 0 & 2 & 0 & 2 & 2 & 2 & 4 \\
0 & 1 & 0 & 1 & 0 & 2 & 0 & 2 & 2 & 1 & 2 & 2 & 4 \\
0 & 0 & 0 & 0 & 1 & -1 & 0 & 0 & 0 & 0 & 0 & 0 & 0 \\
0 & 0 & 0 & 0 & 0 & 0 & 1 & -1 & 0 & 0 & 0 & 0 & 0 \epm,
\nn
\Theta^{(4)}
=& U^{(4)} \cdot \tilde{\Theta}^{(4)}
= \bpm 1 \epm \cdot \tilde{\Theta}^{(4)}
= (2, 2, 1, 0, 2, 0, 0, 0, 0, 2, 0, 2, 0) \mod 4.
\ea
Consequently, we define 4 $\Z$SRSIs with $\theta_\Z=(\theta_1,\theta_2,\theta_3,\theta_4)^T = \Theta^{(0)} \cdot p$ and the unique $\Z_4$SRSI with $\theta_{\Z_4} = \theta_5 = \Theta^{(4)} \cdot p$ mod 4.
The 4 $\Z$SRSIs $\theta_\Z$ are given by
\ba
\theta_1 =& m[ (\bar{E})_{8a} ] + m[ (\bar{E})_{8b} ] + 2m[({}^1\bar{E}_1 {}^2\bar{E}_1)_{24c}] + 2m[({}^1\bar{E}_1 {}^2\bar{E}_1)_{24d}] + 2m[({}^1\bar{E} {}^2\bar{E})_{32e}]
\nn
&+ 2m[({}^1\bar{E} {}^2\bar{E})_{48f}] + 2m[({}^1\bar{E} {}^2\bar{E})_{48g}] + 2m[(\bar{A} \bar{A})_{96h}],
\nn
\theta_2 =& m[({}^1\bar{F} {}^2\bar{F})_{8a}] + m[({}^1\bar{F} {}^2\bar{F})_{8b}] + 2m[({}^1\bar{E}_1 {}^2\bar{E}_1)_{24c}] + 2m[({}^1\bar{E}_1 {}^2\bar{E}_1)_{24d}] + 2m[(\bar{E} \bar{E})_{32e}]
\nn
&+ m[({}^1\bar{E} {}^2\bar{E})_{32e}] + 2m[({}^1\bar{E} {}^2\bar{E})_{48f}] + 2m[({}^1\bar{E} {}^2\bar{E})_{48g}] + 2m[(\bar{A} \bar{A})_{96h}],
\nn
\theta_3 =& -m[({}^1\bar{E}_1 {}^2\bar{E}_1)_{24c}] + m[({}^1\bar{E}_2 {}^2\bar{E}_2)_{24c}],
\nn
\theta_4 =& -m[({}^1\bar{E}_1 {}^2\bar{E}_1)_{24d}] + m[({}^1\bar{E}_2 {}^2\bar{E}_2)_{24d}].
\label{eq:sg219_zsrsi}
\ea
For the $\Z_4$SRSI, we obtain
\bg
\theta_5 = 2m[(\bar{E})_{8a}] + 2m[({}^1\bar{F} {}^2\bar{F})_{8a}] + m[(\bar{E})_{8b}] + 2m[({}^1\bar{E}_2 {}^2\bar{E}_2)_{24c}] + 2m[({}^1\bar{E} {}^2\bar{E})_{32e}] + 2m[({}^1\bar{E} {}^2\bar{E})_{48g}] \mod 4.
\label{eq:sg219_znsrsi}
\eg
Note that the definitions of $\Z$SRSIs and $\Z_4$SRSI coincide with those presented in Supplementary Table~\ref{tab: singlevaluedSRSIs}.

Now, let us turn to the discussion on the mapping between $\Z$SRSIs and symmetry-data vectors.
To proceed, we define the BR matrix $\BR$ induced from site-symmetry irreps in SEq.~\eqref{eq:sg219_irrep}, which is given by
\bg
\BR = \bpm
1 & 0 & 1 & 0 & 0 & 2 & 0 & 2 & 0 & 2 & 2 & 2 & 4 \\
1 & 0 & 1 & 0 & 2 & 0 & 2 & 0 & 0 & 2 & 2 & 2 & 4 \\
0 & 2 & 0 & 2 & 2 & 2 & 2 & 2 & 4 & 2 & 4 & 4 & 8 \\
0 & 1 & 0 & 1 & 1 & 1 & 1 & 1 & 2 & 1 & 2 & 2 & 4 \\
0 & 1 & 0 & 1 & 1 & 1 & 1 & 1 & 2 & 1 & 2 & 2 & 4 \\
1 & 1 & 1 & 1 & 2 & 2 & 2 & 2 & 2 & 3 & 4 & 4 & 8 \\
1 & 2 & 1 & 2 & 3 & 3 & 3 & 3 & 4 & 4 & 6 & 6 & 12 \\
1 & 2 & 1 & 2 & 3 & 3 & 3 & 3 & 4 & 4 & 6 & 6 & 12 \\
1 & 2 & 1 & 2 & 2 & 4 & 4 & 2 & 4 & 4 & 6 & 6 & 12 \\
1 & 2 & 1 & 2 & 4 & 2 & 2 & 4 & 4 & 4 & 6 & 6 & 12
\epm.
\label{eq:sg219_BR}
\eg
Each column of $BR$ represents the multiplicities of little group irreps at HSM, structured with a row basis ordered as
\bg
\{ \bar{\Gamma}_6, \bar{\Gamma}_7, \bar{\Gamma}_8, \bar{L}_4 \bar{L}_4, \bar{L}_5 \bar{L}_5, \bar{L}_6 \bar{L}_6, \bar{W}_5 \bar{W}_6, \bar{W}_7 \bar{W}_8, \bar{X}_6, \bar{X}_7 \}.
\label{eq:sg219_kirrep}
\eg
Thus, $N^\rho_{\rm BZ}=10$.
The column basis of $BR$ follows the ordering given in SEq.~\eqref{eq:sg219_irrep}.
The mapping between $\Z$SRSIs and symmetry-data vectors can be established by computing $\Theta^{(0)} \cdot \BR^\ddagger$ and $\BR \cdot [\Theta^{(0)}]^\ddagger$, as shown in SEqs.~\eqref{eq:maps_B_SRSI_1} and \eqref{eq:maps_B_SRSI_2}:
\ba
\Theta^{(0)} \cdot \BR^\ddagger
=& \bpm
1 & 0 & 0 & 0 & 0 & 0 & 0 & 0 & 0 & 0 \\
1 & 0 & 0 & 2 & 0 & -1 & 0 & 0 & 0 & 0 \\
-\frac{1}{2} & 0 & 0 & 0 & 0 & 1 & 0 & 0 & -\frac{1}{2} & 0 \\
-\frac{1}{2} & 0 & 0 & -1 & 0 & 0 & 0 & 0 & \frac{1}{2} & 0
\epm,
\quad
\BR \cdot [\Theta^{(0)}]^\ddagger
= \bpm
1 & 1 & 0 & 0 & 0 & 1 & 1 & 1 & 1 & 1 \\
0 & 0 & 2 & 1 & 1 & 1 & 2 & 2 & 2 & 2 \\
0 & 2 & 2 & 1 & 1 & 2 & 3 & 3 & 2 & 4 \\
0 & 2 & 2 & 1 & 1 & 2 & 3 & 3 & 4 & 2
\epm^T.
\label{eq:sg219_map}
\ea
From this, the $\Z$SRSIs can be determined from the symmetry-data vector $B$, and vice versa, i.e., $\theta_\Z = \Theta^{(0)} \cdot \BR^\ddagger \cdot B$ and $B = BR \cdot [\Theta^{(0)}]^\ddagger \cdot \theta_\Z$.

\paragraph{SRSIs in $F\bar{4}3c$ with SOC and {\it without} time-reversal symmetry:} To examine how time-reversal symmetry affects the structure of SRSIs, we now analyze the SRSIs in SG $F\bar{4}3c$ (No. 219) with SOC but without time-reversal symmetry.
This corresponds to the magnetic SG $F\bar{4}3c$ (No. 219.85) in the BNS notation~\cite{belov1957Shubnikov}, which does not host any $\Z_4$SRSI.

First, we repeat the analysis by constructing the adiabatic-process matrix, which now differs from SEq.~\eqref{eq:sg219_q} due to the absence of time-reversal symmetry.
Since the absence of time-reversal symmetry also alters the allowed site-symmetry irreps, we first list all irreps allowed in this magnetic SG:
\bg
\{
(\bar{E})_{8a}, ({}^2\bar{F})_{8a}, ({}^1\bar{F})_{8a}, (\bar{E})_{8b}, ({}^2\bar{F})_{8b}, ({}^2\bar{F})_{8b}, ({}^2\bar{E}_2)_{24c}, ({}^2\bar{E}_1)_{24c}, ({}^1\bar{E}_2)_{24c}, ({}^1\bar{E}_1)_{24c}, ({}^2\bar{E}_2)_{24d}, ({}^2\bar{E}_1)_{24d},
\nn
({}^1\bar{E}_2)_{24d}, ({}^1\bar{E}_1)_{24d}, (\bar{E})_{32e}, ({}^1\bar{E})_{32e}, ({}^2\bar{E})_{32e}, ({}^2\bar{E})_{48f}, ({}^1\bar{E})_{48f}, ({}^2\bar{E})_{48g}, ({}^1\bar{E})_{48g}, (\bar{A})_{96h}
\},
\label{eq:sg219mag_irrep}
\eg
corresponding to $N^\rho_{\rm UC}=22$.
The role of time-reversal symmetry is evident when comparing site-symmetry irreps with and without time-reversal symmetry [SEqs.~\eqref{eq:sg219_irrep} and \eqref{eq:sg219mag_irrep}].
For example, in the nonmagnetic case (where time-reversal symmetry exists), the irreps $(\bar{E}\bar{E})_{32e}$ and $(\bar{A}\bar{A})_{96h}$ in SEq.~\eqref{eq:sg219_irrep} are composed of Kramers pairs.
Without time-reversal symmetry, these Kramers pairs are not present and thus $(\bar{E})_{32e}$ and $(\bar{A})_{96h}$ are allowed in SEq.~\eqref{eq:sg219mag_irrep}.
Other irreps can be understood similarly.
For instance, the $({}^1\bar{F} {}^2\bar{F})$ irrep at WPs $8a$ and $8b$ (when time-reversal symmetry is present) is a co-irrep formed by the irreps ${}^1\bar{F}$ and ${}^2\bar{F}$.
These two irreps, which have conjugate eigenvalues $-(1+i\sqrt{3})/2$ and $-(1-i\sqrt{3})/2$ under threefold rotation $\{3^+_{111}|\bb 0\}$, respectively, are mapped into each other by time-reversal symmetry.

By considering adiabatic processes between irreps in SEq.~\eqref{eq:sg219mag_irrep}, we construct the adiabatic-process matrix $q$, given by:
\bg
q = \left( \begin{smallmatrix}
0 & 0 & 1 & 1 & 0 & 0 & 1 & 1 & 0 & 0 & 1 & 0 & 1 & 0 & 0 & 0 & 0 & 0 & 0 & 0 & 0 \\
1 & 0 & 1 & 0 & 0 & 0 & 1 & 0 & 0 & 0 & 1 & 0 & 1 & 0 & 0 & 0 & 0 & 0 & 0 & 0 & 0 \\
1 & 0 & 0 & 1 & 0 & 0 & 0 & 1 & 0 & 0 & 1 & 0 & 1 & 0 & 0 & 0 & 0 & 0 & 0 & 0 & 0 \\
0 & 0 & 0 & 0 & 1 & 1 & 0 & 0 & 1 & 1 & 0 & 0 & 0 & 0 & 1 & 0 & 1 & 0 & 0 & 0 & 0 \\
0 & 1 & 0 & 0 & 1 & 0 & 0 & 0 & 1 & 0 & 0 & 0 & 0 & 0 & 1 & 0 & 1 & 0 & 0 & 0 & 0 \\
0 & 1 & 0 & 0 & 0 & 1 & 0 & 0 & 0 & 1 & 0 & 0 & 0 & 0 & 1 & 0 & 1 & 0 & 0 & 0 & 0 \\
0 & 0 & 0 & 0 & 0 & 0 & 0 & 0 & 0 & 0 & 0 & 0 & 0 & 0 & 0 & 1 & 0 & 0 & 0 & 0 & 0 \\
0 & 0 & 0 & 0 & 0 & 0 & 0 & 0 & 0 & 0 & 0 & 0 & 0 & 0 & 0 & 1 & 0 & 0 & 0 & 0 & 0 \\
0 & 0 & 0 & 0 & 0 & 0 & 0 & 0 & 0 & 0 & 0 & 0 & 0 & 0 & 0 & 0 & 0 & 1 & 0 & 0 & 0 \\
0 & 0 & 0 & 0 & 0 & 0 & 0 & 0 & 0 & 0 & 0 & 0 & 0 & 0 & 0 & 0 & 0 & 1 & 0 & 0 & 0 \\
0 & 0 & 0 & 0 & 0 & 0 & 0 & 0 & 0 & 0 & 0 & 1 & 0 & 0 & 0 & 0 & 0 & 0 & 0 & 0 & 0 \\
0 & 0 & 0 & 0 & 0 & 0 & 0 & 0 & 0 & 0 & 0 & 1 & 0 & 0 & 0 & 0 & 0 & 0 & 0 & 0 & 0 \\
0 & 0 & 0 & 0 & 0 & 0 & 0 & 0 & 0 & 0 & 0 & 0 & 0 & 1 & 0 & 0 & 0 & 0 & 0 & 0 & 0 \\
0 & 0 & 0 & 0 & 0 & 0 & 0 & 0 & 0 & 0 & 0 & 0 & 0 & 1 & 0 & 0 & 0 & 0 & 0 & 0 & 0 \\
-1 & -1 & 0 & 0 & 0 & 0 & 0 & 0 & 0 & 0 & 0 & 0 & 0 & 0 & 0 & 0 & 0 & 0 & 1 & 0 & 0 \\
0 & 0 & -1 & -1 & -1 & -1 & 0 & 0 & 0 & 0 & 0 & 0 & 0 & 0 & 0 & 0 & 0 & 0 & 1 & 0 & 0 \\
0 & 0 & 0 & 0 & 0 & 0 & -1 & -1 & -1 & -1 & 0 & 0 & 0 & 0 & 0 & 0 & 0 & 0 & 1 & 0 & 0 \\
0 & 0 & 0 & 0 & 0 & 0 & 0 & 0 & 0 & 0 & -1 & -1 & 0 & 0 & 0 & 0 & 0 & 0 & 0 & 1 & 0 \\
0 & 0 & 0 & 0 & 0 & 0 & 0 & 0 & 0 & 0 & 0 & 0 & -1 & -1 & 0 & 0 & 0 & 0 & 0 & 1 & 0 \\
0 & 0 & 0 & 0 & 0 & 0 & 0 & 0 & 0 & 0 & 0 & 0 & 0 & 0 & -1 & -1 & 0 & 0 & 0 & 0 & 1 \\
0 & 0 & 0 & 0 & 0 & 0 & 0 & 0 & 0 & 0 & 0 & 0 & 0 & 0 & 0 & 0 & -1 & -1 & 0 & 0 & 1 \\
0 & 0 & 0 & 0 & 0 & 0 & 0 & 0 & 0 & 0 & 0 & 0 & 0 & 0 & 0 & 0 & 0 & 0 & -1 & -1 & -1
\end{smallmatrix} \right).
\label{eq:sg219mag_q}
\eg

The Smith decomposition of $q= L \cdot \Lambda \cdot R$ [SEq.~\eqref{eq:qvecdef_global}] yields
\bg
{\rm diag} (\Lambda) = (1, 1, 1, 1, 1, 1, 1, 1, 1, 1, 1, 1, 1, 1, 1, 2, 0, 0, 0, 0, 0)^T,
\eg
implying that ${\rm rank}(q)=16$.
The explicit forms of the matrices $L$ and $R$ are given by
\bg
\left( \begin{smallmatrix}
0 & 0 & 1 & 1 & 0 & 1 & 1 & 1 & 0 & 1 & 0 & 0 & 0 & 0 & 0 & -2 & 1 & 0 & 0 & 0 & 0 & -2 \\
1 & 0 & 1 & 0 & 0 & 1 & 1 & 1 & 0 & 1 & 0 & 0 & 0 & 0 & 0 & -2 & 0 & 0 & 0 & 0 & 0 & -2 \\
1 & 0 & 0 & 1 & 0 & 0 & 0 & 1 & 0 & 1 & 0 & 0 & 0 & 0 & 0 & -1 & 0 & 0 & 0 & 0 & 0 & -1 \\
0 & 0 & 0 & 0 & 1 & 2 & 0 & 0 & 0 & 0 & 0 & 1 & 0 & 1 & 0 & -2 & 0 & 0 & 0 & 0 & 0 & 0 \\
0 & 1 & 0 & 0 & 1 & 0 & 0 & 0 & 0 & 0 & 0 & 1 & 0 & 1 & 0 & -1 & 0 & 0 & 0 & 0 & 0 & 0 \\
0 & 1 & 0 & 0 & 0 & 2 & 0 & 0 & 0 & 0 & 0 & 1 & 0 & 1 & 0 & -2 & 0 & 0 & 0 & 0 & 0 & 0 \\
0 & 0 & 0 & 0 & 0 & 1 & 0 & 0 & 0 & 0 & 0 & 0 & 1 & 0 & 0 & -1 & 0 & 0 & 0 & 0 & 0 & 0 \\
0 & 0 & 0 & 0 & 0 & 1 & 0 & 0 & 0 & 0 & 0 & 0 & 1 & 0 & 0 & -1 & 0 & 1 & 0 & 0 & 0 & 0 \\
0 & 0 & 0 & 0 & 0 & 1 & 0 & 0 & 0 & 0 & 0 & 0 & 0 & 0 & 1 & -1 & 0 & 0 & 0 & 0 & 0 & 0 \\
0 & 0 & 0 & 0 & 0 & 1 & 0 & 0 & 0 & 0 & 0 & 0 & 0 & 0 & 1 & -1 & 0 & 0 & 1 & 0 & 0 & 0 \\
0 & 0 & 0 & 0 & 0 & 1 & 0 & 0 & 1 & 0 & 0 & 0 & 0 & 0 & 0 & -1 & 0 & 0 & 0 & 0 & 0 & 0 \\
0 & 0 & 0 & 0 & 0 & 1 & 0 & 0 & 1 & 0 & 0 & 0 & 0 & 0 & 0 & -1 & 0 & 0 & 0 & 1 & 0 & 0 \\
0 & 0 & 0 & 0 & 0 & 1 & 0 & 0 & 0 & 0 & 1 & 0 & 0 & 0 & 0 & -1 & 0 & 0 & 0 & 0 & 0 & 0 \\
0 & 0 & 0 & 0 & 0 & 1 & 0 & 0 & 0 & 0 & 1 & 0 & 0 & 0 & 0 & -1 & 0 & 0 & 0 & 0 & 1 & 0 \\
-1 & -1 & 0 & 0 & 0 & 3 & 0 & 0 & 0 & 0 & 0 & 0 & 0 & 0 & 0 & -1 & 0 & 0 & 0 & 0 & 0 & 0 \\
0 & 0 & -1 & -1 & -1 & 3 & 0 & 0 & 0 & 0 & 0 & 0 & 0 & 0 & 0 & -1 & 0 & 0 & 0 & 0 & 0 & 0 \\
0 & 0 & 0 & 0 & 0 & 0 & -1 & 0 & 0 & 0 & 0 & 0 & 0 & 0 & 0 & 1 & 0 & 0 & 0 & 0 & 0 & 0 \\
0 & 0 & 0 & 0 & 0 & -2 & 0 & -1 & -1 & 0 & 0 & 0 & 0 & 0 & 0 & 2 & 0 & 0 & 0 & 0 & 0 & 0 \\
0 & 0 & 0 & 0 & 0 & -2 & 0 & 0 & 0 & -1 & -1 & 0 & 0 & 0 & 0 & 2 & 0 & 0 & 0 & 0 & 0 & 0 \\
0 & 0 & 0 & 0 & 0 & -2 & 0 & 0 & 0 & 0 & 0 & -1 & -1 & 0 & 0 & 2 & 0 & 0 & 0 & 0 & 0 & 0 \\
0 & 0 & 0 & 0 & 0 & -2 & 0 & 0 & 0 & 0 & 0 & 0 & 0 & -1 & -1 & 2 & 0 & 0 & 0 & 0 & 0 & 0 \\
0 & 0 & 0 & 0 & 0 & -1 & 0 & 0 & 0 & 0 & 0 & 0 & 0 & 0 & 0 & 0 & 0 & 0 & 0 & 0 & 0 & 1
\end{smallmatrix} \right),
\, \,
\left( \begin{smallmatrix}
1 & 0 & 0 & 0 & 0 & 1 & 0 & 0 & 0 & 1 & 0 & 0 & 0 & 0 & 0 & 0 & 0 & 0 & 0 & 1 & 1 \\
0 & 1 & 0 & 0 & 0 & 1 & 0 & 0 & 0 & 1 & 0 & 0 & 0 & 0 & 0 & 0 & 0 & 0 & 0 & 0 & 2 \\
0 & 0 & 1 & 0 & 0 & 1 & 0 & -1 & -1 & 0 & 0 & 0 & 0 & 0 & 0 & 0 & 0 & 0 & 0 & 0 & 0 \\
0 & 0 & 0 & 1 & 0 & 1 & 0 & 1 & 0 & 1 & 0 & 0 & 0 & 0 & 0 & 0 & 0 & 0 & 0 & 1 & 1 \\
0 & 0 & 0 & 0 & 1 & 1 & 0 & 0 & 1 & 1 & 0 & 0 & 0 & 0 & 0 & 0 & 0 & 0 & 0 & 0 & 2 \\
0 & 0 & 0 & 0 & 0 & 0 & 0 & 0 & 0 & 0 & 0 & 0 & 0 & 0 & 0 & 0 & 0 & 0 & 1 & 1 & 1 \\
0 & 0 & 0 & 0 & 0 & -2 & 1 & 1 & 1 & -1 & 0 & 0 & 0 & 0 & 0 & 0 & 0 & 0 & 1 & 2 & 0 \\
0 & 0 & 0 & 0 & 0 & -2 & 0 & 0 & 0 & -2 & 1 & 0 & 0 & 0 & 0 & 0 & 0 & 0 & 1 & 0 & -1 \\
0 & 0 & 0 & 0 & 0 & -2 & 0 & 0 & 0 & -2 & 0 & 1 & 0 & 0 & 0 & 0 & 0 & 0 & 1 & 1 & -1 \\
0 & 0 & 0 & 0 & 0 & -2 & 0 & 0 & 0 & -2 & 0 & 0 & 1 & 0 & 0 & 0 & 0 & 0 & 1 & 0 & -1 \\
0 & 0 & 0 & 0 & 0 & -2 & 0 & 0 & 0 & -2 & 0 & 0 & 0 & 1 & 0 & 0 & 0 & 0 & 1 & 1 & -1 \\
0 & 0 & 0 & 0 & 0 & -2 & 0 & 0 & 0 & -2 & 0 & 0 & 0 & 0 & 1 & 0 & 0 & 0 & 1 & 1 & -2 \\
0 & 0 & 0 & 0 & 0 & -2 & 0 & 0 & 0 & -2 & 0 & 0 & 0 & 0 & 0 & 1 & 0 & 0 & 1 & 1 & -1 \\
0 & 0 & 0 & 0 & 0 & -2 & 0 & 0 & 0 & -2 & 0 & 0 & 0 & 0 & 0 & 0 & 1 & 0 & 1 & 1 & -2 \\
0 & 0 & 0 & 0 & 0 & -2 & 0 & 0 & 0 & -2 & 0 & 0 & 0 & 0 & 0 & 0 & 0 & 1 & 1 & 1 & -1 \\
0 & 0 & 0 & 0 & 0 & -1 & 0 & 0 & 0 & -1 & 0 & 0 & 0 & 0 & 0 & 0 & 0 & 0 & 1 & 1 & 0 \\
0 & 0 & 0 & 0 & 0 & -1 & 0 & 1 & 0 & -1 & 0 & 0 & 0 & 0 & 0 & 0 & 0 & 0 & 0 & 0 & -1 \\
0 & 0 & 0 & 0 & 0 & -1 & 0 & 0 & 1 & -1 & 0 & 0 & 0 & 0 & 0 & 0 & 0 & 0 & 0 & 0 & -1 \\
0 & 0 & 0 & 0 & 0 & -1 & 0 & 0 & 0 & 0 & 0 & 0 & 0 & 0 & 0 & 0 & 0 & 0 & 0 & 0 & -1 \\
0 & 0 & 0 & 0 & 0 & -1 & 0 & 0 & 0 & -1 & 0 & 0 & 0 & 0 & 0 & 0 & 0 & 0 & 0 & 1 & -1 \\
0 & 0 & 0 & 0 & 0 & -1 & 0 & 0 & 0 & -1 & 0 & 0 & 0 & 0 & 0 & 0 & 0 & 0 & 0 & 0 & 0
\end{smallmatrix} \right),
\eg
respectively.
From the structure of ${\rm diag} (\Lambda)$, we observe that the elementary divisors no longer contain 4, which confirms the absence of $\Z_4$SRSIs.
Instead, we find a single $\Z_2$SRSI and 6 $\Z$SRSIs.

For completeness, let us derive the SRSI matrices $\Theta^{(0,2)}$.
To achieve this, we first compute the inverse of $L$:
\ba
L^{-1} = 
\left( \begin{smallmatrix}
0 & 1 & 1 & 1 & 0 & 1 & 2 & 0 & 2 & 0 & 2 & 0 & 2 & 0 & 1 & 1 & 1 & 2 & 2 & 2 & 2 & 3 \\
0 & 0 & 0 & 0 & 0 & 1 & 1 & 0 & 1 & 0 & 0 & 0 & 0 & 0 & 0 & 0 & 0 & 0 & 0 & 1 & 1 & 0 \\
0 & 1 & 0 & 0 & 0 & 1 & 1 & 0 & 1 & 0 & 1 & 0 & 1 & 0 & 1 & 0 & 1 & 1 & 1 & 1 & 1 & 2 \\
0 & 0 & 1 & 0 & 0 & 1 & 1 & 0 & 1 & 0 & 1 & 0 & 1 & 0 & 1 & 0 & 0 & 1 & 1 & 1 & 1 & 1 \\
0 & 0 & 0 & 1 & 0 & 0 & 1 & 0 & 1 & 0 & 0 & 0 & 0 & 0 & 0 & 0 & 0 & 0 & 0 & 1 & 1 & 0 \\
0 & 1 & 1 & 0 & 1 & 1 & 2 & 0 & 2 & 0 & 2 & 0 & 2 & 0 & 2 & 1 & 1 & 2 & 2 & 2 & 2 & 3 \\
0 & 2 & 2 & -1 & 3 & 1 & 3 & 0 & 3 & 0 & 4 & 0 & 4 & 0 & 4 & 2 & 1 & 4 & 4 & 3 & 3 & 6 \\
0 & 1 & 1 & -1 & 2 & 0 & 1 & 0 & 1 & 0 & 1 & 0 & 2 & 0 & 2 & 1 & 1 & 1 & 2 & 1 & 1 & 3 \\
0 & 1 & 1 & -1 & 2 & 0 & 1 & 0 & 1 & 0 & 3 & 0 & 2 & 0 & 2 & 1 & 1 & 2 & 2 & 1 & 1 & 3 \\
0 & 1 & 1 & -1 & 2 & 0 & 1 & 0 & 1 & 0 & 2 & 0 & 1 & 0 & 2 & 1 & 1 & 2 & 1 & 1 & 1 & 3 \\
0 & 1 & 1 & -1 & 2 & 0 & 1 & 0 & 1 & 0 & 2 & 0 & 3 & 0 & 2 & 1 & 1 & 2 & 2 & 1 & 1 & 3 \\
0 & 1 & 1 & -1 & 2 & 0 & 0 & 0 & 1 & 0 & 2 & 0 & 2 & 0 & 2 & 1 & 1 & 2 & 2 & 0 & 1 & 3 \\
0 & 1 & 1 & -1 & 2 & 0 & 2 & 0 & 1 & 0 & 2 & 0 & 2 & 0 & 2 & 1 & 1 & 2 & 2 & 1 & 1 & 3 \\
0 & 1 & 1 & -1 & 2 & 0 & 1 & 0 & 0 & 0 & 2 & 0 & 2 & 0 & 2 & 1 & 1 & 2 & 2 & 1 & 0 & 3 \\
0 & 1 & 1 & -1 & 2 & 0 & 1 & 0 & 2 & 0 & 2 & 0 & 2 & 0 & 2 & 1 & 1 & 2 & 2 & 1 & 1 & 3 \\
0 & 2 & 2 & -1 & 3 & 1 & 3 & 0 & 3 & 0 & 4 & 0 & 4 & 0 & 4 & 2 & 2 & 4 & 4 & 3 & 3 & 6 \\
1 & 0 & 0 & 1 & 0 & 0 & 1 & 0 & 1 & 0 & 1 & 0 & 1 & 0 & 0 & 1 & 1 & 1 & 1 & 1 & 1 & 2 \\
0 & 0 & 0 & 0 & 0 & 0 & -1 & 1 & 0 & 0 & 0 & 0 & 0 & 0 & 0 & 0 & 0 & 0 & 0 & 0 & 0 & 0 \\
0 & 0 & 0 & 0 & 0 & 0 & 0 & 0 & -1 & 1 & 0 & 0 & 0 & 0 & 0 & 0 & 0 & 0 & 0 & 0 & 0 & 0 \\
0 & 0 & 0 & 0 & 0 & 0 & 0 & 0 & 0 & 0 & -1 & 1 & 0 & 0 & 0 & 0 & 0 & 0 & 0 & 0 & 0 & 0 \\
0 & 0 & 0 & 0 & 0 & 0 & 0 & 0 & 0 & 0 & 0 & 0 & -1 & 1 & 0 & 0 & 0 & 0 & 0 & 0 & 0 & 0 \\
0 & 1 & 1 & 0 & 1 & 1 & 2 & 0 & 2 & 0 & 2 & 0 & 2 & 0 & 2 & 1 & 1 & 2 & 2 & 2 & 2 & 4
\end{smallmatrix} \right)
\label{eq:sg219mag_inverse_l}
\ea
Collecting the rows $(L^{-1})_i$ of $L^{-1}$ for indices $\{i\}$ corresponding to $\Lambda_{ii} \ne 1$, we obtain the SRSI matrices $\tilde{\Theta}^{(0)} = (L^{-1})_{17,\dots,22}$ for the 6 $\Z$SRSIs and $\tilde{\Theta}^{(2)} = (L^{-1})_{16}$ mod 2 for the unique $\Z_2$SRSI.
Applying the Hermite decomposition, we then obtain the simplified SRSI matrices [SEq.~\eqref{eq:simplifiedSRSI_final}]:
\ba
\Theta^{(0)}
=& \bpm 
1 & 0 & 0 & 1 & 0 & 0 & 0 & 1 & 0 & 1 & 0 & 1 & 0 & 1 & 0 & 1 & 1 & 1 & 1 & 1 & 1 & 2 \\
0 & 1 & 1 & 0 & 1 & 1 & 0 & 2 & 0 & 2 & 0 & 2 & 0 & 2 & 2 & 1 & 1 & 2 & 2 & 2 & 2 & 4 \\
0 & 0 & 0 & 0 & 0 & 0 & 1 & -1 & 0 & 0 & 0 & 0 & 0 & 0 & 0 & 0 & 0 & 0 & 0 & 0 & 0 & 0 \\
0 & 0 & 0 & 0 & 0 & 0 & 0 & 0 & 1 & -1 & 0 & 0 & 0 & 0 & 0 & 0 & 0 & 0 & 0 & 0 & 0 & 0 \\
0 & 0 & 0 & 0 & 0 & 0 & 0 & 0 & 0 & 0 & 1 & -1 & 0 & 0 & 0 & 0 & 0 & 0 & 0 & 0 & 0 & 0 \\
0 & 0 & 0 & 0 & 0 & 0 & 0 & 0 & 0 & 0 & 0 & 0 & 1 & -1 & 0 & 0 & 0 & 0 & 0 & 0 & 0 & 0
\epm
\nn
\Theta^{(2)}
=& (0, 0, 0, 1, 1, 1, 1, 0, 1, 0, 0, 0, 0, 0, 0, 0, 0, 0, 0, 1, 1, 0).
\ea
Consequently, we define 6 $\Z$SRSIs with $\theta_\Z=(\theta_1,\theta_2,\dots,\theta_6)^T = \Theta^{(0)} \cdot p$ and the unique $\Z_2$SRSI with $\theta_{\Z_2} = \theta_7 = \Theta^{(2)} \cdot p$ mod 2.

Now, let us examine an interesting case of an adiabatic process between site-symmetry irreps at the same WP, revealing that seemingly distinct irreps can, in fact, be continuously deformed into one another.
While this resembles the example in SG $I432$ discussed in SN~\ref{sec:I432_stable}, where auxiliary orbitals were necessary, here the equivalence emerges without requiring auxiliary orbitals.
A particularly interesting case in the SG at hand is the adiabatic equivalence between $({}^1\bar{E})_{32e}$ and $({}^2\bar{E})_{32e}$.
At first glance, this equivalence appears counterintuitive from the perspective of site-symmetry groups: consider a representative position of WP $32e$, $\bb x_{32e,1}=(x_0,x_0,x_0)$, where $x_0$ is a continuous real-valued parameter.
Under the threefold rotation $\{3^+_{111}|\bb 0\}$, which maps $(x,y,z)$ to $(z,x,y)$, the irreps $({}^1\bar{E})_{32e}$ $({}^2\bar{E})_{32e}$ have eigenvalues $(1-i\sqrt{3})/2$ and $(1+i\sqrt{3})/2$, respectively, indicating that the two irreps are inequivalent.
To understand the equivalence between these irreps, let us define a state $\ket{\phi_1(\bb x_{32e,1})}$ representing a $({}^1\bar{E})$ irrep at $\bb x_{32e,1}$.
By definition of the irrep, we have
\bg
\{3^+_{111}|\bb 0\} \, \ket{\phi_1(\bb x_{32e,1})}
= \frac{1-i\sqrt{3}}{2} \, \ket{\phi_1(\bb x_{32e,1})}.
\eg
Meanwhile, in this SG, there is a glide mirror $\{m_{01\bar{1}}|1/2,1/2,1/2\}$, which maps $(x,y,z)$ to $(x+1/2,z+1/2,y+1/2)$.
The glide mirror transforms $\bb x_{32e,1}$ to another representative point in WP $32e$, $\bb x_{32e,2}=(x_0+1/2,x_0+1/2,x_0+1/2)$.
This glide mirror satisfies the commutation relation:
\bg
\{3^+_{111}|\bb 0\} \, \{m_{01\bar{1}}|1/2,1/2,1/2\}
= \{m_{01\bar{1}}|1/2,1/2,1/2\} \, \{3^+_{111}|\bb 0\}^{-1}.
\eg
Thus, defining the state at $\bb x_{32e,2}$ as
\bg
\ket{\phi_2(\bb x_{32e,2})}
:= \{m_{01\bar{1}}|1/2,1/2,1/2\} \, \ket{\phi_1(\bb x_{32e,1})},
\eg
we can evaluate its transformation under threefold rotation:
\ba
\{3^+_{111}|\bb 0\} \, \ket{\phi_2(\bb x_{32e,2})} 
=& \{3^+_{111}|\bb 0\} \, \{m_{01\bar{1}}|1/2,1/2,1/2\} \, \ket{\phi_1(\bb x_{32e,1})}
= \{m_{01\bar{1}}|1/2,1/2,1/2\} \, \{3^+_{111}|\bb 0\}^{-1} \, \ket{\phi_1(\bb x_{32e,1})}
\nn
=& \{m_{01\bar{1}}|1/2,1/2,1/2\} \, \left( \frac{1-i\sqrt{3}}{2} \right)^{-1} \, \ket{\phi_1(\bb x_{32e,1})}
= \frac{1+i\sqrt{3}}{2} \, \ket{\phi_2(\bb x_{32e,2})}.
\ea
Thus, $\ket{\phi_2(\bb x_{32e,2})}$ corresponds to the ${}^2\bar{E}$ irrep at $\bb x_{32e,2}$.
This result implies that the presence of ${}^1\bar{E}$ irrep at $\bb x_{32e,1}=(x_0,x_0,x_0)$ necessarily leads to the presence of ${}^2\bar{E}$ irrep at $\bb x_{32e,2} = (x_0+1/2,x_0+1/2,x_0+1/2)$.
By tuning the continuous parameter $x_0$, we can continuously deform one irrep into the other.
For instance, if $-1/2<x_0<0$, increasing $x_0$ causes ${}^1\bar{E}$ at $\bb x_{32e,1}$ to pass through $\bb x_{8a,1}=(0,0,0)$, a representative position of WP $8a$, and then move to $\bb x_{32e,2}$.
(For other values of $x_0$, the irrep passes through different representative positions of WPs $8a$ and $8b$.)
This process allows the two irreps $({}^1\bar{E})_{32e}$ and $({}^2\bar{E})_{32e}$ to be {\it exchanged} via an adiabatic deformation.
The equivalence of these irreps is also reflected in their SRSI values.
Evaluating the SRSIs for both irreps yields identical values:
\bg
(\theta_1,\dots,\theta_7) = (1,1,0,0,0,0,0),
\eg
demonstrating that their stable equivalence is fully captured by the SRSIs.
This example provides insight into adiabatic deformations involving the {\it exchange} of distinct irreps at the same WP $W'$ through a connected higher-symmetry WP $W$.
The previous study on the composite RSIs~\cite{xu2021threedimensional} classified such exchange deformation processes based on the relation between a magnetic point group and its subgroups, where the former and latter represent the site-symmetry groups of $W$ and $W'$, respectively.
However, rather than a group-subgroup relation, our example shows that such exchanges can also arise due to the interplay between symmetry operations connecting different representative positions of the same WP and the connectivity between them via a higher-symmetry WP, offering a broader perspective on the mechanisms that enable adiabatic equivalences between site-symmetry irreps.

Through a comparative analysis of nonmagnetic and magnetic SGs that share the same spatial symmetry but differ in the presence of time-reversal symmetry, we have demonstrated how the existence of a $\Z_4$SRSI can depend on time-reversal symmetry.
However, fully understanding the origin of the $\Z_4$SRSI remains challenging.
Determining when the number 4 appears as an elementary divisor in the Smith decomposition of q requires accounting for multiple factors: the presence or absence of SOC and time-reversal symmetry, the allowed site-symmetry groups at each WP, and the way in which WP connectivity influences possible adiabatic processes.
This intricate dependence makes the emergence of $\Z_4$SRSIs somewhat unintuitive.
Nonetheless, among these factors, time-reversal symmetry provides the most direct and controllable mechanism for quantitatively comparing SGs with and without $\Z_4$SRSIs.
Additionally, we note that the three other nonmagnetic SGs with SOC that host $\Z_4$SRSIs, $F222$ (No. 22), $I\bar{4}c2$ (No. 120), and $F23$ (No. 196), are subgroups of $F\bar{4}3c$.
While not all subgroups of $F\bar{4}3c$ necessarily support $\Z_4$SRSIs, this structural relationship provides a useful perspective for understanding their occurrence.
Furthermore, similar to $F\bar{4}3c$, we can examine whether these three SGs also lose their $\Z_4$SRSIs upon breaking time-reversal symmetry.
Indeed, explicit calculations confirm that breaking time-reversal symmetry in these SGs eliminates the $\Z_4$SRSIs, consolidating the idea that time-reversal symmetry plays a crucial role in the emergence of $\Z_4$SRSI.
Finally, as demonstrated in our analysis of magnetic SG No. 219.85, our method for computing SRSIs is not limited to nonmagnetic SGs and can be readily extended to magnetic SGs as well.

\section{Real space criteria for topology}
\label{sec:realspacecrits}
We here first establish that the SRSIs fully determine the momentum space-based symmetry indicators (SIs) in each SG, and then derive real-space criteria based on $\Z_n$SRSIs that guarantee band topology beyond SIs.
Concretely, we will show in SN~\ref{sec:stableindexdef} that $\Z$SRSIs determine the SIs.
Then, in SN~\ref{sec:realspacecriterion} we show that $\Z_n$SRSIs can diagnose topological bands arising from disconnected band representations where the topology would not be diagnosed topological from a momentum-space symmetry data perspective due to having trivial SIs.
This implies that the real-space framework of TQC provides more information than momentum-space-based SIs alone.

\subsection{SRSIs determine the symmetry indicators}
\label{sec:stableindexdef}
The SIs of a SG are linear functions of the little group irrep multiplicities at HSMs which indicate whether a given set of bands can be written as any sum (or difference) of EBRs.
They are expressed in terms of the BZ irrep multiplicity vector $B$ [SEq.~\eqref{eq:symmetrydatavec_definition}]:
\ba
\kappa = SI \cdot B \mod w,
\label{eq:SI_formula_0}
\ea
where $\kappa$ is the $r_{BR}$-dimensional vector of SIs, and $SI$ is the matrix defined in SEq.~\eqref{eq:si_and_comp} that encodes how little group irreps at HSM contribute to the SIs defined in SEq.~\eqref{eq:si_and_comp}.
Also, $w \in \N^{r_{BR}}$ is a vector made up of the elementary divisors of the $BR$ matrix and thus the ${\rm mod}$ operation is to be understood element-wise.
If $\kappa$ is nonzero, the set of bands under consideration does not correspond to any sum or difference of EBRs and therefore must be topological.
Note that the $SI$ matrix is constructed to contain only linearly independent SI formulae without redundancy.
More details about the derivation of $SI$ and SEq.~\eqref{eq:SI_formula_0} can be found in SRef.~\cite{elcoro2020application}.

While we define the SIs via SEq.~\eqref{eq:SI_formula_0} using the $SI$ matrix, it is worth noting that the SIs can also be defined using an alternative formulation involving $\td{SI}$ and $\tilde{w}$, as introduced in SRef.~\cite{Song_2018}:
\bg
\tilde{\kappa} = \td{SI} \cdot B \mod \tilde{w}.
\label{eq:SI_formula}
\eg
In this approach, the matrix $\td{SI}$ may include additional redundant SIs, in contrast to $SI$\footnote{For example, a single $\Z_8$-valued SI from $SI$ may be represented in $\td{SI}$ by a set of $\Z_2$, $\Z_4$, and $\Z_8$ indicators that are algebraically related.
Although $\td{SI}$ is redundant in this sense, it contains the same information as $SI$.}.
The possibly redundant symmetry indicators determined by $\td{SI}$ offer some physical advantages.
For instance, individual indicators in $\td{SI}$ can be associated with physical properties such as types of surface states, as studied in SRef.~\cite{Song_2018}.
For this reason, we hereafter use the SIs $\tilde{\kappa}$ defined in SEq.~\eqref{eq:SI_formula}, using the $\td{SI}$ matrix.

Nonzero SIs are a sufficient but not necessary criterion for stable band topology.
Here, we show that the $\Z$SRSIs fully determine all SIs (but the converse does not hold).
To demonstrate this, we first extend the definition of $\Z$SRSIs beyond atomic insulators to (stable or fragile) topological insulators, based on the one-to-one mapping between them and symmetry-data vector $B$.
Then, using the known relation between $B$ and SIs, we establish the mapping from $\Z$SRSIs to SIs, encompassing both atomic and topological insulators.

For this, we first explain how the $\Z$SRSIs can be defined for topological insulators beyond the atomic insulators on which we focused so far.
As we discussed in SN~\ref{subsubsec:genbandstruct_pvec}, the symmetry-representation vector $p$ takes integer values if a given band structure corresponds to atomic insulator or fragile topological insulator.
For the former case with atomic insulator, $\Z$SRSIs are determined by the relationship, $\theta_\Z = \Theta^{(0)} \cdot p$ [SEq.~\eqref{eq:SRSI_definition}] with integer-valued matrix $\Theta^{(0)}$.
Since both $\Theta^{(0)}$and $p$ are integer valued, the $\Z$SRSIs $\theta_\Z$ must be integers.
For the case of fragile topological insulators that can be considered a formal difference between atomic insulators, the $p$ vector can be expressed as the difference between the symmetry-representation vectors $p_{i=1,2}$ that corresponds to atomic insulators, i.e., $p=p_1-p_2$.
{Note that since $p_i$ represents an atomic insulator, $\theta_{\Z, i}$ is an integer-valued vector as described above.}
Thus, we can define the set of $\Z$SRSIs $\theta_\Z$ for fragile topological insulator, based on the linearity of the mapping in SEq.~\eqref{eq:SRSI_definition}; $\theta_\Z = \theta_{\Z,1} - \theta_{\Z,2}$ and takes integer values.
Thus, the $\Z$SRSIs can be used to identify symmetry-indicated fragile topological bands by noting when a given set of $\Z$SRSIs is inconsistent with any {\it non-negative} integer symmetry representation vector.
This is equivalent to the affine monoid criteria of SRef.~\cite{Song20} (which is not surprising, since the symmetry-data vector and the $\Z$SRSIs contain the same information).
Likewise, we define the $\Z$SRSIs for stable topological phases whose nontrivial topology is indicated by nonzero SIs.
In this case, since the symmetry-representation vector $p$ takes fractional values as discussed in SN~\ref{subsubsec:genbandstruct_pvec}, the integer quantization of $\Z$SRSIs $\theta_\Z$, which assumes an integer-valued $p$, no longer holds.
Consequently, $\Z$SRSIs of stable topological phases take on {\it fractional} values~\cite{ZhidaFragileTwist2,Hwang:2019aa}.

These extended $\Z$SRSIs for nonatomic insulators can be calculated from momentum-space data since the mapping between $\Z$SRSIs $\theta_\Z$, symmetry-representation vector $p$, and the symmetry-data vector $B$ are all linear.
From SEqs.~\eqref{eq:maps_B_SRSI_1} and \eqref{eq:maps_B_SRSI_2}, we can formally write the relation between SIs $\kappa$ and the $\Z$SRSIs $\theta_Z$.
First, we recall from SEqs.~\eqref{eq:maps_B_SRSI_1} and \eqref{eq:maps_B_SRSI_2} that $B = \BR \cdot [\Theta^{(0)}]^\ddagger \cdot \theta_\Z$ where $[\Theta^{(0)}]^\ddagger$ is a pseudoinverse of $\Z$SRSI matrix $\Theta^{(0)}$.
By combining this with SEq.~\eqref{eq:SI_formula}, we obtain
\ba
{\tilde{\kappa} = \td{SI}} \cdot \BR \cdot [\Theta^{(0)}]^\ddagger \cdot \theta_\Z \mod \tilde{w}.
\label{eq:stableindices_intermsof_srsis}
\ea
This implies that the full set of SIs can be determined from the $\Z$SRSIs.
In SN~\ref{sec:stitables}, we exhaustively tabulate the correspondence between $\Z$SRSIs and SIs for the 230 SGs with and without SOC, as presented in Supplementary Tables~\ref{tab: singlevaluedindices} and~\ref{tab: doublevaluedindices}, respectively.
Note that when the symmetry-data vector of a given set of bands results in trivial SIs, then we can find an integer-valued $p$ vector which has the same symmetry-data vector.
Then, from the definition of SRSIs in SEq.~\eqref{eq:SRSI_definition}, we find that we obtain integer-valued $\Z$SRSIs for this set of bands.
Conversely, suppose that the $\Z$SRSIs $\theta_\Z$ for a given set of bands are integer-valued.
From SEq.~\eqref{eq:p_from_ZSRSI}, the symmetry representation vector can be written as $p = [\Theta^{(0)}]^\ddagger \cdot \theta_\Z + p_{\theta_\Z=0}=p_1 + p_2$, where and $p_1=[\Theta^{(0)}]^\ddagger \cdot \theta_\Z$ and $p_2 = p_{\theta_\Z=0} \in {\rm ker} \, \Theta^{(0)}$.
Since $[\Theta^{(0)}]^\ddagger$ is integer-valued, $p_1$ is also integer-valued, and hence the SIs computed from the symmetry-data vector corresponding to $p_1$ must vanish.
Moreover, SIs computed from $p_2$ are also trivial: since ${\rm ker} \, \Theta^{(0)} = {\rm ker} \, BR$, the vector $p_2$ yields a zero symmetry-data vector and thus contributes no nontrivial SIs.
This analysis implies
\ba
\trm{Trivial SIs} \iff \trm{integer-valued } \theta_\Z.
\ea
This means that
\ba
\trm{Non-zero SIs} \iff \trm{fractional } \theta_\Z.
\ea
So, fractional $\Z$SRSIs reveal stable topology in the same way as non-trivial symmetry-indicators do.

\subsection{SRSI mismatch as a sufficient criterion for band topology}
\label{sec:realspacecriterion}
The combination of all valence ($V$) and conductance ($C$) bands of an insulator is always topologically trivial and can be adiabatically deformed to the bands of an unobstructed atomic insulator where all electrons are delta-localized at the sites of the atomic lattice $Lat$ (the WPs occupied by the atoms in $Lat$)~\cite{ZhidaFragileTwist2,Schindler21Noncompact}.
Formally, we can express this statement as
\ba
\mc{H}_{Lat} = V \oplus C,
\label{eq:simplebanddecomp}
\ea
where $\mc{H}_{Lat}$ refers to the full Hilbert space spanned by the lattice orbitals, while $V$ and $C$ are the Hilbert spaces spanned by the valence and conductance bands of the insulator, respectively.
For this decomposition to be well-defined (unambiguous), there must be a direct gap between valence and conductance band energies at all crystal momenta, so that $V$ indeed describes the occupied subspace of an insulator.
Importantly, even though $\mc{H}_{Lat}$ corresponds to an unobstructed atomic insulator, the same does not necessarily hold for $V$ or $C$ individually.

Let us denote the symmetry-data vector [SEq.~\eqref{eq:symmetrydatavec_definition}] of a set of bands $X$ by $B[X]$.
Since $B[X]$ is additive and adiabatically invariant -- it does not change unless there is a gap closing with the bands $X$ or the crystal symmetry is broken -- we must then have that
\ba
B[\mc{H}_{Lat}] = B[V] + B[C].
\label{eq:bvecdecomposition}
\ea
Similarly, let us denote the SRSIs of a set of bands $X$ by $\theta[X]$.
By construction, the SRSIs are always well-defined for atomic bands.
Since they follow from momentum-space symmetry data, the $\Z$SRSIs -- unlike the $\Z_n$SRSIs -- are also well-defined for topological bands.
Thus, we must have that
\ba
\theta_\Z[\mc{H}_{Lat}] = \theta_\Z[V] + \theta_\Z[C].
\label{eq:rsidecomposition_Z}
\ea
Furthermore, when both $V$ and $C$ are topologically trivial, then we can find the symmetry-representation vectors for both these sets of bands such that
\ba
p [\mc{H}_{Lat}] = p [V] + p [C].
\label{eq:band_splitting_p_vectors}
\ea
We can then obtain the $\Z_n$SRSIs for these bands as well.
As the $\Z_n$SRSIs are linear functions of $p$, SEq.~\eqref{eq:band_splitting_p_vectors} implies that
\bg
\theta_{\Z_n}[\mc{H}_{Lat}] = \theta_{\Z_n}[V] + \theta_{\Z_n}[C] \mod n.
\label{eq:rsidecomposition}
\eg
must also be satisfied.
Since the $\Z$SRSIs can be defined by momentum-space symmetry data, SEq.~\eqref{eq:rsidecomposition_Z} always follows from SEq.~\eqref{eq:bvecdecomposition} when restricting to $\Z$SRSIs.
For the $\Z_n$SRSIs, which are only well-defined for atomic bands, SEq.~\eqref{eq:rsidecomposition} poses an additional constraint, which must be satisfied together with SEq.~\eqref{eq:bvecdecomposition} in order to open a topologically trivial gap between the bands $V$ and $C$.

If SEq.~\eqref{eq:bvecdecomposition} holds with $B[V]$ and $B[C]$ that are compatible with band representations (i.e., $B[V]$ and $B[C]$ yield trivial SIs, and equivalently yield integer $\Z$SRSIs), we call SEq.~\eqref{eq:simplebanddecomp} a {\it symmetry-indicated trivial} band splitting.
We may then seek to identify the bands $V$ as equivalent to some band representation BR$_V$, and the bands in $C$ as equivalent to some band representation BR$_C$, with the same symmetry-data vectors.
However, when SEq.~\eqref{eq:rsidecomposition} is not also satisfied for the $\Z_n$SRSIs determined from BR$_V$ and BR$_C$, then $V$ and $C$ cannot actually be gapped without violating the assumption that they both form band representations/atomic insulators.
In this case, any gapped band splitting must be {\it non-symmetry-indicated topological} (NSIT), because either the valence or conductance bands (or both) must be topological.

\subsection{All split EBRs are topological}
\label{sec:ebrsplit}
We now review the central statement of SRefs.~\cite{Bradlyn17,Cano17,Cano17-2} that extends TQC beyond SIs: any split EBR must result in band topology -- even if the momentum-space symmetry data of the EBR in question, and potentially even its SRSIs, can be decomposed into that of multiple other band representations.

That is, we consider putative EBR decompositions of the form
\ba
EBR \stackrel{?}{\equiv} EBR_1 \oplus EBR_2 \oplus \dots,
\label{eq:ebr_split}
\ea
where the right-hand side contains more than one EBR and which satisfy
\ba
B[EBR] = B[EBR_1] + B[EBR_2] + \dots,
\label{eq:ebr_split_symdata}
\ea
with the symmetry-data vector $B$ defined in SEq.~\eqref{eq:symmetrydatavec_definition}.
We note that $\equiv$ in SEq.~\eqref{eq:ebr_split} denotes the adiabatic equivalence of EBRs, as defined in SRef.~\cite{Bradlyn17}.
Several previous works~\cite{Bradlyn17,Cano17,Bouhon18Fragile,BarryFragile} have shown that when the bands constituting an EBR are split by a band gap, this can results in a violation of SEq.~\eqref{eq:ebr_split_symdata}: at least one of the resulting gapped set of bands has symmetry indicated stable (with nonvanishing SIs) or symmetry-indicated fragile~\cite{Hwang:2019aa,ZhidaFragileTwist2,Song20} topology.
As we emphasize here, however, nontrivial solutions of SEq.~\eqref{eq:ebr_split_symdata} nevertheless exist in certain SGs.
(See SN~\ref{sec:tbmodel_sg75} for an explicit example and Supplementary Tables~\ref{tab: ebrdecomps_nosoc} and \ref{tab: ebrdecomps} for an exhaustive list.)

We now review the argument first presented in SRef.~\cite{Bradlyn17} that EBR splits of the form of SEq.~\eqref{eq:ebr_split} are impossible, even when SEq.~\eqref{eq:ebr_split_symdata} is satisfied.
Note that:
\begin{enumerate}
\item At least one of $EBR_1$, $EBR_2$, \dots, must all be induced from irreps at maximal a WP $W_i$ that is {\it different} from the maximal WP $W$ from which $EBR$ is induced.
Otherwise, $EBR$ would be induced from a reducible representation of the site-symmetry group at $W$, and hence would not be elementary (as defined in SN~\ref{sec:ebr_intro}).
\item $EBR$ must be induced from a site-symmetry irrep at $W$, and must not be included in the tabulated list of exceptions discussed in SNs~\ref{sec:ebr_intro} and \ref{sec:rsi_algorithm}.
This ensures that $EBR$ is elementary and cannot be adiabatically deformed to a sum of EBRs at a different maximal WP.
\end{enumerate}

As a consequence of (1) and (2), there is no adiabatic path that connects $EBR$ with the direct sum $EBR_1 \oplus EBR_2 \oplus \dots$.
Hence, splitting an EBR must {\it always} induce band topology in at least one of its split parts, even when SEq.~\eqref{eq:ebr_split_symdata} is satisfied.
We provide exhaustive tables of all EBRs splittings that are compatible with symmetry-data vector in the 230 SGs with TRS and with and without SOC in SN~\ref{sec:splitebr_tables}.
Altogether, we find 211 such cases across 51 SGs: 9 cases across 3 SGs without SOC, and 202 cases across 51 SGs with SOC.

We note for completeness that the preceding argument does not rule out the case that EBR can be rewritten as the sum $EBR_1 \oplus EBR_2 \oplus \dots$ through a {\it large} gauge transformation (i.e., a change of basis that cannot be smoothly connected to the identity transformation).
In this case the band representation $EBR$ on the left hand side of SEq.~\eqref{eq:ebr_split} would not be elementary.
We are not aware of any cases where this occurs, but defer a complete proof to future work.
Nevertheless, as we will show in SN~\ref{sec:matchingSRSIs}, the $\Z_n$SRSIs give a criteria for ruling out this possibility in many cases.

\subsection{EBR splittings with matching SRSIs}
\label{sec:matchingSRSIs}
In Supplementary Tables~\ref{tab: ebrdecomps_nosoc} and~\ref{tab: ebrdecomps} of SN~\ref{sec:splitebr_tables}, we list all linearly independent solutions to SEq.~\eqref{eq:ebr_split_symdata} in all 230 SGs with and without SOC.
By explicitly evaluating the full set of SRSIs of $EBR$ and $EBR_1$, $EBR_2$, \dots, we find that the equation
\bg
\theta[EBR] = \theta[EBR_1] + \theta[EBR_2] + \cdots,
\nn
\trm{i.e.,} \, \theta_\Z[EBR] = \theta_\Z[EBR_1] + \theta_\Z[EBR_2] + \cdots,
\quad
\theta_{\Z_n}[EBR] = \theta_{\Z_n}[EBR_1] + \theta_{\Z_n}[EBR_2] + \cdots \mod n
\label{eq:ebr_split_srsis}
\eg
is violated in almost every case, confirming that these EBR splittings are impossible by the criterion derived in SN~\ref{sec:realspacecriterion}.
Supplementary Table~\ref{table:split_exceptions} shows 8 cases of split EBRs [SEq.~\eqref{eq:ebr_split}] in 5 SGs, where SEq.~\eqref{eq:ebr_split_srsis} is satisfied in addition to SEq.~\eqref{eq:ebr_split_symdata}.
This proves that the criterion derived in SN~\ref{sec:realspacecriterion} is in fact only sufficient but not necessary for topology beyond momentum-space symmetry data.
The 8 exceptional cases all rely on SOC and arise in the SGs $Fm\bar{3}$ (No. 202), $F432$ (No. 209), $I432$ (No. 211), $Pm\bar{3}n$ (No. 223), and $Fm\bar{3}m$ (No. 225).
In these eight cases, $EBR$ and $EBR_1 \oplus EBR_2 \oplus \dots$ can be adiabatically deformed into each other {\it in the presence of auxiliary trivial bands.}
In other words, as discussed in SN~\ref{sec:stableeq}, $EBR$ and $EBR_1 \oplus EBR_2 \oplus \dots$ in Supplementary Table~\ref{table:split_exceptions} are stably equivalent due to their matching SRSIs.

\let\oldaddcontentsline\addcontentsline
\renewcommand{\addcontentsline}[3]{}
\renewcommand{\arraystretch}{1.4}
\begin{table*}
\centering
\begin{minipage}{0.85\textwidth}
\caption{List of eight cases where a split EBR and its split components, as defined in SEq.~\eqref{eq:ebr_split}, exhibit matching SRSIs.
The first column denotes the corresponding SG.
The second column presents each split EBR along with its split components.
The third column lists the auxiliary trivial bands $\mc{A}$ required for their adiabatic deformation, confirming their stable equivalence.}
\label{table:split_exceptions}
\end{minipage}
\begin{tabular*}{0.61\textwidth}{c|c|c}
\hline \hline
SG (\#) & Split EBR with matching SRSIs & Auxiliary bands $\mc{A}$ \\ \hline
\multirow{2}{*}{$Fm\bar{3}$ (No. 202)} & $({}^1\bar{F} {}^2\bar{F})_{8c} \stackrel{?}{\equiv} ({}^1\bar{F}_g{}^2\bar{F}_g)_{4a} \oplus ({}^1\bar{F}_u{}^2\bar{F}_u)_{4a}$ & $({}^1\bar{E} {}^2\bar{E})_{32f}$
\\
& $({}^1\bar{F} {}^2\bar{F})_{8c} \stackrel{?}{\equiv} ({}^1\bar{F}_g{}^2\bar{F}_g)_{4b} \oplus ({}^1\bar{F}_u{}^2\bar{F}_u)_{4b}$ & $({}^1\bar{E} {}^2\bar{E})_{32f}$
\\ \hline
\multirow{2}{*}{$F432$ (No. 209)} & $({}^1\bar{F} {}^2\bar{F})_{8c} \stackrel{?}{\equiv} (\bar{F})_{4a} \oplus (\bar{F})_{4a}$ & $({}^1\bar{E} {}^2\bar{E})_{32f}$
\\
& $({}^1\bar{F} {}^2\bar{F})_{8c} \stackrel{?}{\equiv} (\bar{F})_{4b} \oplus (\bar{F})_{4b}$ & $({}^1\bar{E} {}^2\bar{E})_{32f}$
\\ \hline
$I432$ (No. 211) & $({}^1\bar{E} {}^2\bar{E})_{8c} \stackrel{?}{\equiv} (\bar{F})_{2a} \oplus (\bar{F})_{2a}$ & $({}^1\bar{E} {}^2\bar{E})_{16f}$
\\ \hline
$Pm\bar{3}n$ (No. 223) & $({}^1\bar{E} {}^2\bar{E})_{8e} \stackrel{?}{\equiv} ({}^1\bar{F}_g{}^2\bar{F}_g)_{2a} \oplus ({}^1\bar{F}_u{}^2\bar{F}_u)_{2a}$ & $({}^1\bar{E} {}^2\bar{E})_{16i}$
\\ \hline
\multirow{2}{*}{$Fm\bar{3}m$ (No. 225)} & $(\bar{F})_{8c} \stackrel{?}{\equiv} (\bar{F}_g)_{4a} \oplus (\bar{F}_u)_{4a}$ & $(\bar{E}_1)_{32f}$
\\
& $(\bar{F})_{8c} \stackrel{?}{\equiv} (\bar{F}_g)_{4b} \oplus (\bar{F}_u)_{4b}$ & $(\bar{E}_1)_{32f}$
\\
\hline \hline
\end{tabular*}
\end{table*}
\let\addcontentsline\oldaddcontentsline

We here illustrate this phenomenon by an example of SG $I432$, which has a single symmetry-data vector decomposition with matching SRSIs given by
\ba
B[({}^1 \bar{E} {}^2 \bar{E})_{8c} \uparrow G] = B[(\bar{F})_{2a} \uparrow G] + B[(\bar{F})_{2a} \uparrow G].
\label{eq:split_EBR_example}
\ea
From Supplementary Table~\ref{tab: doublevaluedSRSIs}, we find that three $\Z$SRSIs $\theta_{1,2,3}$ and four $\Z_2$SRSIs $\theta_{4,5,6,7}$ are defined in the SG $I432$ with SOC.
For the BRs appearing on {\it both} sides of SEq.~\eqref{eq:split_EBR_example}, we find that
\ba
\theta_1[({}^1 \bar{E} {}^2 \bar{E})_{8c} \uparrow G]=\theta_1[(\bar{F})_{2a} \uparrow G]+\theta_1[(\bar{F})_{2a} \uparrow G] &= 0
\\
\theta_2[({}^1 \bar{E} {}^2 \bar{E})_{8c} \uparrow G]=\theta_2[(\bar{F})_{2a} \uparrow G]+\theta_2[(\bar{F})_{2a} \uparrow G] &= 0
\\
\theta_3[({}^1 \bar{E} {}^2 \bar{E})_{8c} \uparrow G]=\theta_3[(\bar{F})_{2a} \uparrow G]+\theta_3[(\bar{F})_{2a} \uparrow G] &= 2
\\
\theta_{4\dots 7}[({}^1 \bar{E} {}^2 \bar{E})_{8c} \uparrow G]=\theta_{4\dots 7}[(\bar{F})_{2a} \uparrow G]+\theta_{4\dots 7}[(\bar{F})_{2a} \uparrow G] &= 0 \mod 2
\ea
Thus, for the BRs appearing in the symmetry-data vector decomposition SEq.~\eqref{eq:split_EBR_example}, we have that SEq.~\eqref{eq:ebr_split_srsis} holds.
In other words, the $({}^1 \bar{E} {}^2 \bar{E})_{8c} \uparrow G$ EBR and the $[(\bar{F})_{2a} \uparrow G] \oplus [(\bar{F})_{2a} \uparrow G]$ BR have the matching $\Z$SRSIs and $\Z_n$SRSIs.
Thus, they are stably equivalent, implying their inequivalence (in the absence of auxiliary trivial bands) cannot be detected from the SRSIs.

To show the stable equivalence between $({}^1 \bar{E} {}^2 \bar{E})_{8c} \uparrow G$ and $[(\bar{F})_{2a} \uparrow G] \oplus [(\bar{F})_{2a} \uparrow G]$, we first note that
\bg
({}^1 \bar{E} {}^2 \bar{E})_{24i} \Leftrightarrow ({}^1 \bar{E} {}^2 \bar{E})_{8c} \oplus ({}^1 \bar{E} {}^2 \bar{E})_{16f}.
\label{eq:split_EBR_example2}
\eg
where the symbol $\Leftrightarrow$ means that the two orbital configurations can be connected by adiabatic processes.
This can be seen by noting that the WPs $8c$, $16f$, and $24i$ have representative positions $(1/4,1/4,1/4)$, $(x,x,x)$, and $(1/4,y,-y+1/2)$, respectively, where $x$ and $y$ are real free parameters.
Since WPs $16f$ and $24i$ are connected to WP $8c$, we can consider adiabatic deformations among their site-symmetry irreps.
In particular, we find the following adiabatic processes:
\bg
({}^1\bar{E} {}^2\bar{E})_{16f} \Leftrightarrow 2(\bar{E}_1)_{8c},
\quad ({}^1\bar{E} {}^2\bar{E})_{8c} \oplus 2(\bar{E}_1)_{8c} \Leftrightarrow ({}^1\bar{E} {}^2\bar{E})_{24i}.
\eg
Combining these, we find SEq.~\eqref{eq:split_EBR_example2}.
\bg
({}^1 \bar{E} {}^2 \bar{E})_{24i} \uparrow G
= ({}^1 \bar{E} {}^2 \bar{E})_{8c} \uparrow G \oplus ({}^1 \bar{E} {}^2 \bar{E})_{16f} \uparrow G,
\label{eq:split_EBR_example3}
\eg
The non-maximal $16f$ WP is also connected to the maximal WP $2a$.
This implies that the site-symmetry irrep $({}^1 \bar{E} {}^2 \bar{E})_{16f}$ can be adiabatically deformed to
\ba
({}^1 \bar{E} {}^2 \bar{E})_{16f} \Leftrightarrow 2(\bar{E}_1)_{2a} \oplus 2(\bar{E}_2)_{2a} \oplus 2(\bar{F})_{2a}.
\label{eq:211_adiabatic_process_1}
\ea
Moreover, the site-symmetry irrep $({}^1 \bar{E} {}^2 \bar{E})_{24i}$ can be adiabatically deformed to a linear combination of site-symmetry irreps at the WP $2a$ by considering a series of adiabatic processes involving multiple WPs.
Concretely, we have the following chain of adiabatic processes:
\bg
({}^1 \bar{E} {}^2 \bar{E})_{24i}
\Leftrightarrow 2(\bar{E})_{12d}
\Leftrightarrow ({}^1 \bar{E} {}^2 \bar{E})_{24g}
\Leftrightarrow 2(\bar{E}_1)_{6b} \oplus 2(\bar{E}_2)_{6b}
\Leftrightarrow ({}^1 \bar{E} {}^2 \bar{E})_{24h}
\Leftrightarrow 2(\bar{E}_1)_{2a} \oplus 2(\bar{E}_2)_{2a} \oplus 4(\bar{F})_{2a}.
\label{eq:211_adiabatic_process_2}
\eg
By combining SEqs.~\eqref{eq:211_adiabatic_process_1} and \eqref{eq:211_adiabatic_process_2}, we have that
\bg
({}^1 \bar{E} {}^2 \bar{E})_{24i} \uparrow G
=
2(\bar{F})_{2a} \uparrow G \oplus \mc{A} \nn
=
({}^1 \bar{E} {}^2 \bar{E})_{8c} \uparrow G \oplus \mc{A},
\label{eq:211_stable_equivalence}
\eg
where we have defined the band representation
\ba
\mc{A} = ({}^1 \bar{E} {}^2 \bar{E})_{16f} \uparrow G
= 2(\bar{E}_1)_{2a} \uparrow G \oplus 2(\bar{E}_2)_{2a} \uparrow G \oplus 2(\bar{F})_{2a} \uparrow G
\label{eq:211_A_def}
\ea
SEq.~\eqref{eq:211_stable_equivalence} shows that the EBR $({}^1 \bar{E} {}^2 \bar{E})_{8c} \uparrow G$ is {\it stably} equivalent to $(\bar{F})_{2a} \uparrow G \oplus (\bar{F})_{2a} \uparrow G$ in the presence of {\it additional trivial bands} corresponding to the band representation $\mc{A}$ in SEq.~\eqref{eq:211_A_def}.
Thus, an EBR splitting with matching SRSIs means adiabatic deformation for the relevant EBR decomposition is forbidden, however, it is possible if trivial bands are additionally considered.
Similarly, in all of the seven other split EBRs with matching SRSIs listed in Supplementary Table~\ref{table:split_exceptions}, we can show stable equivalence by adding appropriate trivial bands.
The necessary set of auxiliary bands are shown in the third column of Supplementary Table~\ref{table:split_exceptions}.
This observation calls for the development of SRSI {\it inequalities} that indicate whether a pair of site-symmetry representation vectors can be adiabatically deformed into each other {\it with} or {\it without} extra additional trivial bands.
These inequalities, which follow from treating the site-symmetry representation vectors and adiabatic processes as an {\it affine monoid}~\cite{Song20} rather than a linear space, will be derived in a future work.
For our present purposes, it is sufficient to note that site-symmetry irreps associated with EBRs cannot be adiabatically moved away from a WP W, as explained in SN~\ref{sec:ebrsplit}.
As discussed more generally for EBR decompositions for split EBRs in SN~\ref{sec:ebrsplit}, it remains a possibility, however, that for the split EBRs in Supplementary Table~\ref{table:split_exceptions} there exist gauge transformations not continuously connected to the identity that establish an equivalence between the EBR and its disconnected components.
The existence of such large gauge transformations would imply that the EBRs in Supplementary Table~\ref{table:split_exceptions} are actually composite.
We know of no cases where such large gauge transformations exist relating different band representations.
We thus conjecture that no large gauge transformations exist in the cases of Supplementary Table~\ref{table:split_exceptions}.
We defer the proof of this conjecture to future work.
Assuming this conjecture is true, we know that the eight cases of split EBRs with matching SRSIs in Supplementary Table~\ref{table:split_exceptions} cannot be adiabatically equivalent, and so must be topological band splittings.

\section{Tight-binding models}
\label{sec:tbmodels}
We now support the theoretical analysis in the previous SNs through an explicit example in a tight-binding (TB) model.
We note that SG $P41'$ has a split EBR for which the symmetry data-vectors (and therefore the $\Z$SRSIs) match.
But this SG also has $\Z_2$SRSIs, which can be used to reveal the topological nature of the split EBR as discussed in SN~\ref{sec:matchingSRSIs}.
The simple structure of SG $P41'$ (No. 75) allows us to easily construct a 4-band TB Hamiltonian with all the required symmetries.

\subsection{Topology beyond momentum-space symmetry data in SG $P41'$}
\label{sec:tbmodel_sg75}
To illustrate how SRSIs characterize NSIT band splittings, we consider a minimal example in SG $P41'$ with SOC and TRS in two spatial dimensions (2D).
We note that there are no examples of such band splittings in one dimension (1D), because there are no 1D stable or fragile topological phases protected only by crystalline symmetries and/or TRS~\cite{Shiozaki:2014aa,Bradlyn17,vanderbilt2018berry}.
This is because any set of bands in 1D can always be wannierized and is therefore topologically trivial.

Using the BCS~\cite{Aroyo2011183} for SG $P41'$ with SOC, we find that the physical EBR $({}^1\bar{E} {}^2\bar{E})_{2c} \uparrow G$ is decomposable, and so can be realized as two sets of disconnected bands, i.e.,
\ba
B [({}^1\bar{E} {}^2\bar{E})_{2c} \uparrow G]
= B [({}^1\bar{E}_1{}^2\bar{E}_1)_{1a} \uparrow G]
+ B [({}^1\bar{E}_2{}^2\bar{E}_2)_{1a} \uparrow G].
\label{eq:p75_sym_data_vecs}
\ea
This tempts us to extrapolate this equation to write
\ba
({}^1\bar{E} {}^2\bar{E})_{2c} \uparrow G 
\stackrel{?}{=} ({}^1\bar{E}_1 {}^2\bar{E}_1)_{1a} \uparrow G \oplus ({}^1\bar{E}_2 {}^2\bar{E}_2)_{1a} \uparrow G,
\label{eq:p75_decomp_candidate}
\ea
which is of the form of SEq.~\eqref{eq:simplebanddecomp} with both $V$ and $C$ forming atomic bands.
To investigate the nature of this splitting, we need to find the relevant SRSIs.
From Supplementary Table~\ref{tab: doublevaluedSRSIs}, we find the $\Z$SRSIs 
\ba
(\theta_1, \theta_2, \theta_3) [({}^1\bar{E} {}^2\bar{E})_{2c}] = (1,1,0),
\quad
(\theta_1, \theta_2, \theta_3) [({}^1\bar{E}_1 {}^2\bar{E}_1)_{1a}] = (0,1,0),
\quad
(\theta_1, \theta_2, \theta_3) [({}^1\bar{E}_2 {}^2\bar{E}_2)_{1a}] = (1,0,0),
\ea
which indeed satisfy SEq.~\eqref{eq:rsidecomposition} as expected since the $\Z$SRSIs are linear functions of the symmetry-data vector as shown in SN~\ref{sec:zsrsi_proof}.
However, the $\Z_2$SRSIs are
\ba \label{eq:sg75_srsi_mismatch}
(\theta_4, \theta_5) [({}^1\bar{E} {}^2\bar{E})_{2c}] = (0,1),
\quad 
(\theta_4, \theta_5) [({}^1\bar{E}_1 {}^2\bar{E}_1)_{1a}] = (0,0),
\quad
(\theta_4, \theta_5) [({}^1\bar{E}_2 {}^2\bar{E}_2)_{1a}] = (0,0),
\ea
which do not satisfy SEq.~\eqref{eq:rsidecomposition}.
Hence, SEq.~\eqref{eq:p75_decomp_candidate} cannot be true~\cite{AshvinFragile, Cano17,Schindler21Noncompact}, and the bands in SEq.~\eqref{eq:p75_sym_data_vecs} can only be gapped via an non-symmetry indicated topological (NSIT) band splitting as discussed in SN~\ref{sec:realspacecriterion}.

\begin{figure}[t]
\centering
\includegraphics[width=0.8\textwidth]{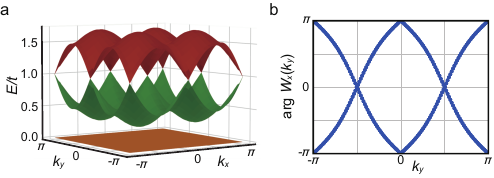}
\caption{Tight-binding model for the band splitting SEq.~\eqref{eq:p75_decomp_candidate} in SG $P41'$.
\textbf{a} Spectrum of the Bloch Hamiltonian in SEq.~\eqref{eq:bloch_hamiltonian}.
There are two zero-energy flatbands and two dispersive bands.
The latter group of bands forms the EBR $({}^1\bar{E}_1 {}^2\bar{E}_1)_{1a} \ua G$.
\textbf{b} Spectrum of the $k_x$-directed Wilson loop $W_x$ as a function of $k_y$, for the two flatbands.
The double Wilson winding is protected by $C_2 T$ symmetry and implies fragile topology.
The Wilson loop spectrum of the two dispersive bands is constant and does not wind (not shown).}
\label{fig:tb_75}
\end{figure}

We now construct a tight-binding Hamiltonian that gaps SEq.~\eqref{eq:p75_sym_data_vecs} at the expense of introducing nontrivial topology.
Our basis orbital form a lattice obtained from the EBR $({}^1\bar{E} {}^2\bar{E})_{2c} \uparrow G$.
There are two Kramers pairs of $s$-like orbitals which we denote by their spins as $\{\ket{\uparrow},\ket{\downarrow}\}$ in each unit cell located at coordinates $\hat{x}/2=(1/2,0)$ and $\hat{y}/2=(0,1/2)$, respectively, giving rise to a four-band Bloch Hamiltonian.
We write the basis states for all orbitals in the lattice as
\ba
\ket{\bb R, \alpha} = (\ket{\bb R + \hat{x}/2, \uparrow}, \ket{\bb R + \hat{x}/2, \downarrow}, \ket{\bb R + \hat{y}/2, \uparrow}, \ket{\bb R + \hat{y}/2, \downarrow})_\alpha, \quad \alpha = 1 \dots 4,
\label{eq:orbital_basis}
\ea
where $\bb R$ is a lattice vector.
In this basis, the time-reversal symmetry operator is represented by
\ba
T = \mathbb{1}_{N \times N} \otimes i \tau_0 \sg_2 \mc{K},
\label{eq:sg75_tr_lattice_basis}
\ea
where $N$ is the number of unit cells in the lattice, $\tau_i$, for $i=0,1,2,3$, are Pauli matrices which act in the space of the two sites $\hat{x}/2$ and $\hat{y}/2$ in the $2c$ WP in each unit cell, and $\sg_i$, for $i=0,1,2,3$ are Pauli matrices which act on the spin degrees of freedom at each site.
Furthermore, we abbreviated the Kronecker product $\tau_i \otimes \sg_j$ by $\tau_i \sg_j$, and $\mc{K}$ denotes complex conjugation (time-reversal is anti-unitary).
Moreover, spinful $C_4$ rotational symmetry about the maximal WP $1a$ of the unit cell at $\bb R=\bb 0$ is implemented by 
\ba
C_4 \ket{\bb R, \alpha} = \sum_\beta \ket{\bb R', \beta} C_{4,\beta \alpha}, 
\quad
\bb R' + \bb t_\beta = \bpm 0 & -1 \\ 1 & 0 \epm (\bb R + \bb t_\alpha),
\quad
C_{4,\beta \alpha} = \left[ \tau_1 e^{i \frac{\pi}{4} \sg_3} \right]_{\beta \alpha}.
\label{eq:sg75_c4_lattice_basis}
\ea
We see from SEq.~\eqref{eq:sg75_c4_lattice_basis} that $(C_4)^4 = -1$ and from SEq.~\eqref{eq:sg75_tr_lattice_basis} additionally that $[C_4, T] = 0$.

To construct a model that realizes the splitting in SEq.~\eqref{eq:p75_decomp_candidate}, we will construct localized states which transform under the $({}^1\bar{E}_1 {}^2\bar{E}_1)_{1a}$ site-symmetry representation.
We start by considering both the eigenstates of the $\sg_1$ operator localized at $\hat{x}/2$ i.e.,
\ba
\ket{\bb 0 + \hat{x}/2, \rightarrow} = \frac{1}{\sqrt{2}} (\ket{\bb 0, 1} + \ket{\bb 0, 2})
\quad
\ket{\bb 0 + \hat{x}/2, \leftarrow} = \frac{1}{\sqrt{2}} (\ket{\bb 0, 1} - \ket{\bb 0, 2}),
\ea
where $\rightarrow$ and $\leftarrow$ indicate spin aligned along the $+\hat{x}$ and $-\hat{x}$ directions, respectively.
Let us consider the following linear combinations of basis-functions generated from Kramers' pair located at $\hat{x}/2$:
\ba
\ket{\Phi^{+}_{\bb 0}} \equiv \frac{1}{2} \left[\ket{\bb 0 + \hat{x}/2, \rightarrow} + e^{-i \frac{\pi}{4}} C_4 \ket{\bb 0 + \hat{x}/2, \rightarrow} + e^{-i \frac{\pi}{2}} (C_4)^2 \ket{\bb 0 + \hat{x}/2, \rightarrow} + e^{-i \frac{3\pi}{4}} (C_4)^3 \ket{\bb 0 + \hat{x}/2, \rightarrow} \right],
\nn
\ket{\Phi^{-}_{\bb 0}} \equiv \frac{1}{2} \left[\ket{\bb 0 + \hat{x}/2, \leftarrow} + e^{+i \frac{\pi}{4}} C_4 \ket{\bb 0 + \hat{x}/2, \leftarrow} + e^{+i \frac{\pi}{2}} (C_4)^2 \ket{\bb 0 + \hat{x}/2, \leftarrow} + e^{+i \frac{3\pi}{4}} (C_4)^3 \ket{\bb 0 + \hat{x}/2, \leftarrow} \right],
\label{eq:sg75_trial_states_origin}
\ea
where the action of $C_4$ is given via SEq.~\eqref{eq:sg75_c4_lattice_basis}.
These states are a Kramers pair
\ba
T \ket{\Phi^{+}_{\bb 0}} = \ket{\Phi^{-}_{\bb 0}},
\quad T \ket{\Phi^{-}_{\bb 0}} = - \ket{\Phi^{+}_{\bb 0}},
\ea
and form a co-irrep of the $C_4$ point-group symmetry:
\ba
C_4 \ket{\Phi^{+}_{\bb 0}}
= e^{i \frac{\pi}{4}} \ket{\Phi^{+}_{\bb 0}},
\quad
C_4 \ket{\Phi^{-}_{\bb 0}} = e^{-i \frac{\pi}{4}} \ket{\Phi^{-}_{\bb 0}}.
\ea
From this, we can read off the representation matrix
\ba
\left[ \rho^{({}^1\bar{E}_1 {}^2\bar{E}_1)}_{1a}(C_4) \right]_{ab}
= \bpm e^{i \frac{\pi}{4}} & 0 \\ 0 & e^{-i \frac{\pi}{4}} \epm_{a b},
\ea
which is consistent with the representation matrix for $C_4$ in the $({}^1\bar{E}_1 {}^2\bar{E}_1)_{1a}$ site-symmetry irrep, per the BCS~\cite{elcoro2017}.
The remaining trial states $\ket{\Phi^{\pm}_{\bb R}}$ are obtained by translation from SEq.~\eqref{eq:sg75_trial_states_origin}.
The set $\{\ket{\Phi^{\pm}_{\bb R}}\}$ of basis-functions is linearly independent, but not orthonormal.
As such, although we can use them to construct a tight-binding Hamiltonian, we must be careful not to interpret the $\{\ket{\Phi^{\pm}_{\bb R}}\}$ as Wannier functions.

To construct a tight-binding model, we start by defining momentum space basis functions
\ba
\ket{\bb k, \alpha} = \frac{1}{\sqrt{V}} \sum_{\bb R} e^{i \bb k \cdot (\bb R + \bb t_\alpha)} \ket{\bb R, \alpha}
\label{eq:sg75_ebr_momentum_basis}
\ea
where $\bb t_\alpha$ is the position vector at which the $\ket{\bb R, \alpha}$ orbital is centered within the unit cell as per SEq.~\eqref{eq:orbital_basis}, and $V$ is the volume of the system.
We also define the Bloch states
\ba
u^\pm_{\bb k \alpha} \equiv e^{i \bb k \cdot \bb t_\pm} \sqrt{V} \braket{\bb k, \alpha | \Phi^\pm_{\bb 0}}
\ea
where $\bb t_\pm$ is the position vector at which $\ket{\Phi^\pm_{\bb 0}}$ is centered within the unit cell.
By construction, $\bb t_\pm = \bb 0$.
We find that the Bloch states are
\ba
u^{+}_{\bb k \alpha}
&= \frac{1}{\sqrt{2}} \left(\cos \frac{k_x}{2}, -i \sin \frac{k_x}{2}, \cos \frac{k_y}{2}, -\sin \frac{k_y}{2} \right)_\alpha,
\nn
u^{-}_{\bb k \alpha}
&= \frac{1}{\sqrt{2}} \left(-i \sin \frac{k_x}{2}, -\cos \frac{k_x}{2}, \sin \frac{k_y}{2}, -\cos \frac{k_y}{2} \right)_\alpha.
\label{eq:bloch_states}
\ea

Note that these Bloch states are not orthogonal because the trial states that we started with are not orthogonal.
However, their overlap matrix
\ba
S(\bb k) = \bpm
(u^+_\bb k)^\dagger (u^+_\bb k) & (u^+_\bb k)^\dagger (u^-_\bb k)
\\
(u^-_\bb k)^\dagger (u^+_\bb k) & (u^-_\bb k)^\dagger (u^-_\bb k)
\epm
\ea
is nonsingular, with
\ba
\det S(\bb k) = 1-\frac{1}{4}(\sin^2 k_x+\sin^2 k_y) \geq \frac{1}{2}.
\ea
This means that the Bloch states in SEq.~\eqref{eq:bloch_states} are linearly independent at every $\bb k$, and so span a two-band Hilbert space.
If desired, we could perform L\"owdin orthogonalization to obtain an orthonormal basis for this two-band subspace while preserving all the symmetries; since $S(\bb k)$ is nonsingular, this orthogonalization is an invertible transformation at every $\bb k$ and hence cannot change the symmetry data associated to the two-band Hilbert space.
We avoid performing this orthogonalization as it leads to non-compact Wannier functions forcing us to consider longer-range hoppings.
Also, this procedure leads to rather cumbersome expressions for the Bloch states compared to the ones given in SEq.~\eqref{eq:bloch_states}.
Thus, the two-band Hilbert space spanned by the $u^{\pm}_{\bb k \alpha}$ in SEq.~\eqref{eq:bloch_states} is equivalent to the EBR $({}^1\bar{E}_1 {}^2\bar{E}_1)_{1a} \uparrow G$ as desired.

By finding the orthogonal complement to the Hilbert space spanned by the $u^{\pm}_{\bb k \alpha}$ at each $\bb k$, we can also construct two other states 
{\ba
v^+_{\bb k \alpha} = \left( \frac{e^{i \pi/4} \sec k_x \, (\sin \frac{k_x + k_y}{2} + i \sin \frac{k_x - k_y}{2})}{\sqrt{2} \sqrt{1 + \sec^2 k_x}}, \frac{e^{- i \pi/4} \sec k_x \, (\cos \frac{k_x - k_y}{2} + i \cos \frac{k_x + k_y}{2})}{\sqrt{2} \sqrt{1 + \sec^2 k_x}}, 0, \sqrt{\frac{1 + \cos 2 k_x}{3 + \cos 2 k_x}} \right)_\alpha \\
v^-_{\bb k \alpha} = \left( -\frac{e^{i \pi/4} \sec k_x \, (\cos \frac{k_x - k_y}{2} + i \cos \frac{k_x + k_y}{2})}{\sqrt{2} \sqrt{1 + \sec^2 k_x}}, -\frac{e^{i \pi/4} \sec k_x \, (\sin \frac{k_x + k_y}{2} + i \sin \frac{k_x - k_y}{2})}{\sqrt{2} \sqrt{1 + \sec^2 k_x}}, 0, \sqrt{\frac{1 + \cos 2 k_x}{3 + \cos 2 k_x}} \right)_\alpha
\ea}
that span
\ba
({}^1\bar{E} {}^2\bar{E})_{2c} \uparrow G \ominus ({}^1\bar{E}_1 {}^2\bar{E}_1)_{1a} \uparrow G.
\label{eq:complementary_hilbert_space}
\ea
We will now argue that this subspace cannot be equivalent to any band representation.

We can construct a Bloch Hamiltonian whose low energy eigenspace coincides with the subspace SEq.~\eqref{eq:complementary_hilbert_space}.
To do so, we consider
\ba
H(\bb k)_{\alpha \beta}
= t(u^{+}_{\bb k \alpha} u^{+*}_{\bb k \beta} + u^{-}_{\bb k \alpha} u^{-*}_{\bb k \beta}),
\label{eq:bloch_hamiltonian}
\ea
with energy scale $t$.
Due to the orthogonality between the subspace spanned by the states in SEq.~\eqref{eq:bloch_states} and the states which span the space in SEq.~\eqref{eq:complementary_hilbert_space}, this Hamiltonian hosts two dispersive bands and two zero-energy perfectly flatbands.
Importantly, the dispersive bands are spanned by the states SEq.~\eqref{eq:bloch_states} which form the EBR $({}^1\bar{E}_1 {}^2\bar{E}_1)_{1a} \uparrow G$ by construction;
they are fully gapped from the flatbands (see SFig.~\ref{fig:tb_75}\textbf{a}), and so we have successfully gapped SEq.~\eqref{eq:p75_decomp_candidate}.
This implies that the symmetry-data vector of the flatbands is the same as the symmetry-data vector corresponding to the EBR $({}^1\bar{E}_2 {}^2\bar{E}_2)_{1a} \uparrow G$.
However, we know that the flatbands cannot carry an EBR because SEq.~\eqref{eq:rsidecomposition} is not satisfied.
Indeed, when we calculate the Wilson loop spectrum of the two flatbands, we find a double winding indicative of fragile topology (see SFig.~\ref{fig:tb_75}\textbf{b}): this winding is protected by $C_2 T$ symmetry and can only be removed by breaking a symmetry or adding a trivial band to the occupied subspace~\cite{KoreanTBG}.
In contrast, there is no Wilson loop winding for the dispersive bands, consistent with the fact that they correspond to an EBR.
Hence, we have found a non-symmetry indicated topological phase that is predicted by SRSIs.

\begin{figure}[t]
\centering
\includegraphics[width=0.9\textwidth]{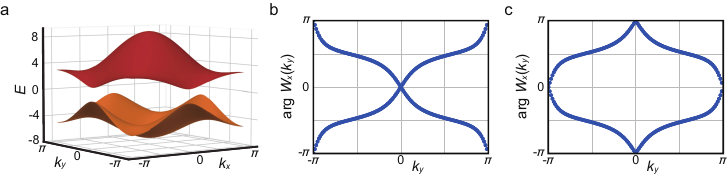}
\caption{Tight-binding model for the $\Z_2$ phase splitting of the EBR $({}^1\bar{E} {}^2\bar{E})_{2c} \uparrow G$ in SG $P41'$.
\textbf{a} Spectrum of the Bloch Hamiltonian in SEq.~\eqref{eq:z2_bloch_hamiltonian}.
There are two sets of doubly-degenerate bands.
\textbf{b}, \textbf{c} Spectra of the $k_x$-directed Wilson loop $W_x$ as a function of $k_y$, for the lower (\textbf{b}) and the upper bands (\textbf{c}), respectively.
The windings being $\pm 1$ for all the bands indicate that both sets are $\Z_2$ phase topological insulators.}
\label{fig:tb_75_z2}
\end{figure}

Note that the $({}^1\bar{E} {}^2\bar{E})_{2c} \uparrow G$ EBR can also realize gapped phases where {\it both} groups of bands are topologically nontrivial.
To see this, consider the following Bloch Hamiltonian constructed in the basis in SEq.~\eqref{eq:sg75_ebr_momentum_basis}:
\ba
H (\bb k) = 
\bpm
A (\bb k) & B (\bb k) \\
B^\dagger (\bb k) & A^\prime (\bb k)
\epm
\label{eq:z2_bloch_hamiltonian}
\ea
where we have defined the block matrices
\bg
A (\bb k) = 
\bpm
2 t_{\rm NNN} \cos k_y & 0 \\
0 & 2 t_{\rm NNN} \cos k_y
\epm,
\nn
B (\bb k) = 
\bpm
2 (t_{\rm NN} - i \lambda) \cos \frac{k_x - k_y}{2} + 2 (t_{\rm NN} + i \lambda) \cos \frac{k_x + k_y}{2} & 0 \\
0 & 2 (t_{\rm NN} + i \lambda) \cos \frac{k_x - k_y}{2} + 2 (t_{\rm NN} - i \lambda) \cos \frac{k_x + k_y}{2}
\epm,
\eg
and
\ba
A^\prime (\bb k) = 
\bpm
2 t_{\rm NNN} \cos k_x & 0 \\
0 & 2 t_{\rm NNN} \cos k_x
\epm.
\ea
We also introduce $t_{\rm NN}$ $ (\lambda)$ as the spin-independent (spin-dependent) nearest-neighbor inter-sublattice hopping parameter, and $t_{\rm NNN}$ is the next-nearest-neighbor intra-sublattice hopping.
To numerically diagonalize this model, we set $t_{\rm NN} = 2$, $t_{\rm NNN} = 1$, and $\lambda = 1$.
This model has two sets of doubly-degenerate bands separated from each other by a gap, as shown in SFig.~\ref{fig:tb_75_z2}\textbf{a}.
From our general discussion in SN~\ref{sec:ebrsplit}, we know that at least one of these groups of bands must be topologically nontrivial.
To confirm this, we show the Wilson loop spectra for both these sets of bands in SFigs.~\ref{fig:tb_75_z2}\textbf{b} and \textbf{c}.
We see that the Wilson loops for both the occupied and unoccupied band Wilson loops for this model display helical winding with odd winding number $\pm 1$, indicating that both groups of bands are topologically nontrivial.
In fact, the Wilson loops in SFigs.~\ref{fig:tb_75_z2}\textbf{b} and \textbf{c} demonstrate that the gapped bands of SEq.~\eqref{eq:z2_bloch_hamiltonian} realize a strong $\Z_2$ nontrivial quantum spin Hall insulator~\cite{Yu11}.
This is topologically distinct from the fragile topological phase realized by the flatband model.
It is possible to deform the Hamiltonian in SEq.~\eqref{eq:z2_bloch_hamiltonian} to realize a topological phase transition between the $\Z_2$ nontrivial phase and the fragile phase of the flatband mode by, for instance, constructing a Hamiltonian which continuously (though not adiabatically, as the gap must close during this process) interpolates between SEqs.~\eqref{eq:z2_bloch_hamiltonian} and \eqref{eq:bloch_hamiltonian}.
As mentioned in the main text, in well-known 4-band models of $\Z_2$ topological insulators like those in SRefs.~\cite{Kane05a,Bernevig06}, the Wilson loop spectra for both the occupied and the unoccupied bands have the same winding pattern in the $\Z_2$ nontrivial phase.
This allows band inversions to simultaneously unwind the Wilson loop for the valence and conduction bands, and so trivialize both the sets of bands resulting in the $\Z_2$ trivial phase.
In the model that we have considered, from SFigs.~\ref{fig:tb_75_z2}\textbf{b},\textbf{c}, it is clear that there is a phase shift of $\pi$ between the two Wilson loop spectra.
We speculate his phase shift is related to the obstruction to trivializing the bands via band inversions that is implied by SEq.~\eqref{eq:sg75_srsi_mismatch}.
We defer the exploration of this connection for a future study.
Nevertheless, consistent with our discussion in SN~\ref{sec:ebrsplit}, any Hamiltonian that realizes this EBR as a set of disconnected bands must have at least one topologically nontrivial group of bands; it cannot be topologically trivial.

\section{SRSI-based material diagnosis and obstructed atomic insulators}
In this SN, we explain how SRSIs can be used to diagnose obstructed atomic insulators (OAIs) in realistic materials, and clarify the logical structure underlying the construction of SRSI-based OAI indicators mentioned in the main text.
The starting point is that all SRSIs, including both $\Z$SRSIs and $\Z_n$SRSIs, are linear functions of the site-symmetry irrep multiplicity vector $p$, $\theta = \Theta \cdot p$ where $\theta$ denotes either a $\Z$SRSIs or $\Z_n$SRSIs, depending on context.
Also, $\Theta$ is the SRSI matrix introduced in SN~\ref{sec:srsi_derivation}, with $\Theta^{(0)}$ and $\Theta^{(n)}$ corresponding to $\Z$SRSIs and $\Z_n$SRSIs, respectively.

For a given material, the vector $p$ naturally decomposes as $p = (p_{\rm atom}, p_{\rm rest})$, where $p_{\rm atom}$ collects the multiplicities of site-symmetry irreps at WPs occupied by atoms in the crystal structure, and $p_{\rm rest}$ contains the remaining entries associated with unoccupied WPs.
Correspondingly, the SRSI matrix can be decomposed as $\Theta = (\Theta_{\rm atom}, \Theta_{\rm rest})$, so that
\bg
\theta = \Theta_{\rm atom} \cdot p_{\rm atom} + \Theta_{\rm rest} \cdot p_{\rm rest}.
\eg
So, we partition the columns of the SRSI matrix $\Theta$ according to whether the corresponding site-symmetry irreps associated with WPs occupied by atoms in the given material ($\Theta_{\rm atom}$) or with the remaining WPs that are not occupied ($\Theta_{\rm rest}$).
By definition, an unobstructed atomic insulator is characterized by $p_{\rm rest} = 0$, since all Wannier orbitals are centered on atomic WPs.
OAI indicators are therefore constructed by identifying linear combinations of SRSIs that necessarily vanish for all such atomic configurations.

Concretely, if there exists a row vector $\bb u$ such that $\bb u \cdot \Theta_{\rm atom} = 0$ as an identity (over $\Z$ or modulo $m$), then for any unobstructed atomic insulator one must have
\bg
\bb u \cdot \theta = \bb u \cdot \Theta_{\rm atom} \cdot p_{\rm atom} = 0
\eg
since $p_{\rm rest} = 0$.
A nonzero value of $\bb u \cdot \theta$ therefore implies $p_{\rm rest} \neq 0$, and hence diagnoses an OAI.
Accordingly, we refer to $\bb u \cdot \theta$ as a $\Z$- or $\Z_m$-valued OAI indicator, depending on whether the condition $\bb u \cdot \Theta_{\rm atom} = 0$ holds identically or modulo $m$.
The row vector $\bb u$ specifies a particular linear combination of SRSIs used to form
the OAI indicator.
Explicitly, if $\theta = (\theta_1, \theta_2, \dots)$ denotes the full set of $\Z$SRSIs or $\Z_n$SRSIs, then $\bb u \cdot \theta = u_1 \theta_1 + u_2 \theta_2 + \cdots$.
The construction of OAI indicators is purely algebraic and uses only (i) the SRSI matrix and (ii) which WPs are occupied by atoms.

Such vectors $\bb u$ can, in principle, be obtained systematically.
For example, in the case of $\Z$SRSIs, their existence is related to the Smith normal form of $\Theta_{\rm atom}$, which can give rise to both $\Z$-
and $\Z_m$-valued OAI indicators with integer $m>1$.
Analogous constructions apply to $\Z_n$SRSIs using modular algebra.
In the present work, we do not attempt a systematic construction or classification of all such indicators.
Instead, we restrict attention to OAI indicators derived from $\Z$SRSIs, which are most directly accessible from symmetry data via the one-to-one
correspondence between $\Z$SRSIs and symmetry data.
We illustrate this approach below using a simple representative material example.

As a representative example, we consider the compound Y$_3$Al$_2$, listed in the Inorganic Crystal Structure Database (ICSD No. 609643)~\cite{bergerhoff1983inorganic}, which crystallizes in SG $P4_2nm$ (No. 102).
This SG is generated by $\{ E | \bb v \in \Z^3 \}$ together with $\{ 4_{001}^+ | 1/2, 1/2, 1/2 \}$ and $\{ m_{1 \bar 1 0} | \bb 0 \}$.
It admits four WPs with representative positions, $2a \, (0,0,z)$, $4b \, (0,1/2,z)$, $4c \, (x,x,z)$, and $8d \, (x,y,z)$.
For the material Y$_3$Al$_2$, the atoms occupy the $4b$ and $4c$ WPs:
In a conventional unit cell, the crystal structure contains twelve $Y$ atoms and eight $Al$ atoms.
In the standard crystallographic description of SG $P4_2nm$, the $Y$ atoms occupy the $4b$ and $4c$ WPs, while the $Al$ atoms occupy the $4c$ WPs~\cite{bergerhoff1983inorganic}.
Thus, all atoms in this material reside on the $4b$ and $4c$ WPs, while the $2a$ and $8d$ positions remain unoccupied.

In this SG, there are four relevant site-symmetry irreps associated with these WPs, which we denote as
\bg
(\bar E)_{2a}, \quad
({}^1 \bar E {}^2 \bar E)_{4b}, \quad
({}^1 \bar E {}^2 \bar E)_{4c}, \quad
(\bar A \bar A)_{8d}.
\label{eq:sg102_kirrep_order}
\eg
The multiplicities of these irreps define one $\Z$SRSI $\theta_1$ and one $\Z_2$SRSI $\theta_2$, given by
\ba
\theta_1
&= m[(\bar{E})_{2a}] + 2 m[({}^1 \bar E {}^2 \bar E)_{4b}] + 2 m[({}^1 \bar E {}^2 \bar E)_{4c}] + 4 m[(\bar A \bar A)_{8d}],
\\
\theta_2
&= m[(\bar E)_{2a}] + m[({}^1 \bar E {}^2 \bar E)_{4b}] \pmod 2.
\ea
(The definitions of $\theta_{1,2}$ can also be found in Supplementary Table~\ref{tab: doublevaluedSRSIs}.)
Equivalently, in the notation $\theta = \Theta \cdot p$, using the basis ordering in SEq.~\eqref{eq:sg102_kirrep_order}, we have
\bg
\Theta^{(0)} = (1,2,2,4) \text{ for } \theta_1,
\quad
\Theta^{(2)} = (1,1,0,0) \text{ for } \theta_2.
\eg
Although $\theta_1$ admits a simple interpretation as one quarter of the total number of filled bands in this example, the OAI-indicator mechanism below does not rely on that interpretation, but only on the algebraic structure of $\Theta_{\rm atom}$.

Band structure and and symmetry data for this material are available through the Topological Materials Database~\cite{TopDatabase,Bradlyn17,Vergniory_2019,vergniory22}.
The momentum-space irreps of Y$_3$Al$_2$ are found to be
\bg
(39 \bar{A}_6 \oplus 39 \bar{A}_7, 39 \bar{\Gamma}_6 \oplus 39 \bar{\Gamma}_7, 39 \bar{M}_6 \bar{M}_7, 39 \bar{Z}_6 \bar{Z}_7, 39 \bar{R}_2 \bar{R}_5 \oplus 39 \bar{R}_3 \bar{R}_4, 39 \bar{X}_2 \bar{X}_3 \oplus 39 \bar{X}_4 \bar{X}_5).
\eg
From this information, and more generally from the one-to-one correspondence between $\Z$ SRSIs and the symmetry-data vector, we obtain $\theta_1 = 39$ for this material.
This result can also be understood as follows.
Since both $\bar{\Gamma}_6$ and $\bar{\Gamma}_7$ irreps are two-dimensional, the total filling is $\nu = 4 \times 39 = 156$.
Accordingly, the $\Z$SRSI $\theta_1$, which in this SG coincides with one quarter of the total filling, takes the value $\theta_1 = \nu/4 = 39$.

To construct an OAI indicator, we restrict $\Theta^{(0)}$ to the atomic WPs $(4b,4c)$, which gives
\bg
\Theta^{(0)}_{\rm atom}=(2,2).
\eg
The Smith normal form of $(2,2)$ is $(2,0)$, and the presence of a single elementary divisor equal to 2 implies the existence of a $\Z_2$-valued OAI indicator.
Equivalently, there exists an integer row vector $\bb u$ such that
\bg
\bb u \cdot \Theta^{(0)}_{\rm atom} = (0,0) \pmod 2.
\eg
For this example, one may choose $\bb u = (1)$, so the OAI indicator is simply $\theta_1 \bmod 2$.
Since $\theta_1=39$ is odd, we conclude that $p_{\rm rest} \neq 0$, and hence Y$_3$Al$_2$ is necessarily an OAI.

Moreover, from the explicit form of the indicator we obtain additional real-space information.
Indeed,
\bg
\bb u \cdot \theta_1
= (1,0,0,0)\cdot p
= m[(\bar E)_{2a}] \pmod 2,
\eg
so $\theta_1 \bmod 2 = 1$ implies that at least one Wannier orbital must be centered at the $2a$ WP, which is not occupied by atoms in this material.

We emphasize that the SG and material example discussed above were chosen for their simplicity.
The construction of OAI indicators from SRSIs is fully general and algorithmic, and applies to arbitrary material platforms.
Although the present example involves a particularly simple $\Z$SRSI, the same construction applies to general SGs that support more intricate $\Z$- and $\Z_m$-valued OAI indicators.
A more detailed and systematic study of SRSI-based OAI indicators, including their general formulation, mathematical structure, and the allowed values of $m$ for a given SG and atomic configuration, as well as large-scale materials searches, constitutes an important direction for future work.
In this way, the SRSI framework enables systematic searches for atomic, obstructed atomic, fragile, and stable topological phases in large materials databases such as the ICSD, including phases that are not detectable by SIs alone.

We note that the material considered here has also been discussed in the context of filling-enforced OAIs~\cite{xu2024filling}.
While such approaches rely on specific filling constraints, the SRSI-based diagnosis presented here does not depend on filling-enforced conditions and applies more generally.
The present example thus serves to illustrate the broader applicability of the SRSI framework beyond filling-enforced settings.
Finally, we note that this approach is complementary to previous RSI-based schemes for diagnosing obstructed atomic limits, such as the composite RSIs introduced in SRef.~\cite{xu2021threedimensional}.
While composite RSIs provide valuable insight into OAI physics, their interpretation can become subtle in situations involving nonmaximal WPs, where the identification of the relevant WPs may be ambiguous.
By contrast, the SRSI-based construction incorporates the actual atomic WPs of a given material as an explicit input, allowing OAI indicators to be defined in a systematic and unambiguous manner.
Moreover, this framework enables one to characterize the space of allowed OAI indicator values for a fixed SG and atomic configuration, providing a more complete and predictive route to diagnosing OAIs in realistic materials.

\newpage

\section{Tables for the 230 SGs with and without SOC}

\twocolumngrid

We here provide exhaustive tables that complement and support the results of the main text and supplementary information.

\listoftables

\subsection{SRSI tables}
\label{sec:srsitables}
In Supplementary Tables~\ref{tab: singlevaluedSRSIs} and~\ref{tab: doublevaluedSRSIs}, we list the SRSIs in all 230 SGs without and with SOC, respectively.
Note that we omit SGs with only one site-symmetry irrep in a unit cell, as in these cases the unique SRSI is determined by the number of bands.

\input{rsi_tables_single}

\input{rsi_tables_double}

\subsection{SI tables}
\label{sec:stitables}
In Supplementary Tables~\ref{tab: singlevaluedindices} and~\ref{tab: doublevaluedindices}, we list the SIs as expressed in terms of SRSIs in all 230 SGs without and with SOC, respectively.

\input{stableindex_tables_single}

\input{stableindex_tables_double}

\subsection{Split EBR tables}
\label{sec:splitebr_tables}
In Supplementary Tables~\ref{tab: ebrdecomps_nosoc} and~\ref{tab: ebrdecomps}, we list all putative symmetry-data-allowed EBR decompositions of the form of SEq.~\eqref{eq:ebr_split} and satisfying SEq.~\eqref{eq:ebr_split_symdata} that must result in band topology as discussed in SN~\ref{sec:ebrsplit}.

\input{ebr_tables_single}

\input{ebr_tables_double}

\onecolumngrid

\bibliography{references_supp}

%% file: stableindex_tables_single.tex
\begin{longtable}[t]{|c|l|} \caption[SIs from SRSIs without SOC]{Symmetry indicator (SI) expressions in terms of stable real-space invariants (SRSIs) in all 230 space groups without spin-orbit coupling (SOC).
We use the SI labeling conventions of SRef.~\cite{Song_2018}.
The SIs with SOC are listed in Supplementary Table~\ref{tab: doublevaluedindices}. For each space group, we list its number in the {\it International Tables for Crystallography, Vol. A}, together with the corresponding list of SIs that are expressed in terms of the SRSIs of Supplementary Table~\ref{tab: singlevaluedSRSIs}.} \label{tab: singlevaluedindices} \\ \hline SG\# & SIs without SOC \\ \hline \hline
2&$\begin{aligned}z_{2,1}=& -2\theta_{1}+2\theta_{2}+2\theta_{5} \mod 2\end{aligned}$\\ &$\begin{aligned}z_{2,2}=& -2\theta_{1}+2\theta_{2}+2\theta_{4} \mod 2\end{aligned}$\\ &$\begin{aligned}z_{2,3}=& -2\theta_{1}+2\theta_{2}+2\theta_{3} \mod 2\end{aligned}$\\ &$\begin{aligned}z_{4}=& -4\theta_{1}+4\theta_{2} \mod 4\end{aligned}$\\ \hline
3&$\begin{aligned}\alpha_2=& +4\theta_{2}+2\theta_{3}+2\theta_{4}+2\theta_{5} \mod 2\end{aligned}$\\ \hline
10&$\begin{aligned}z_{2,-}=& +2\theta_{2}-2\theta_{3}+2\theta_{5} \mod 2\end{aligned}$\\
\hline\end{longtable}

%% file: stableindex_tables_double.tex
\begin{longtable}[t]{|c|l|} \caption[SIs from SRSIs with SOC]{Symmetry indicator (SI) expressions in terms of stable real-space invariants (SRSIs) in all 230 space groups with spin-orbit coupling (SOC).
We use the SI labeling conventions of SRef.~\cite{Song_2018}.
The SIs without SOC are listed in Supplementary Table~\ref{tab: singlevaluedindices}.
For each space group, we list its number in the {\it International Tables for Crystallography, Vol. A}, together with the corresponding list of SIs that are expressed in terms of the SRSIs of Supplementary Table~\ref{tab: doublevaluedSRSIs}.} \label{tab: doublevaluedindices} \\ \hline SG\# & SIs with SOC \\ \hline \hline
2&$\begin{aligned}z_{2w,1}=& -2\theta_{1}+2\theta_{2}+2\theta_{5} \mod 2\end{aligned}$\\ &$\begin{aligned}z_{2w,2}=& -2\theta_{1}+2\theta_{2}+2\theta_{4} \mod 2\end{aligned}$\\ &$\begin{aligned}z_{2w,3}=& -2\theta_{1}+2\theta_{2}+2\theta_{3} \mod 2\end{aligned}$\\ &$\begin{aligned}z_{4}=& +4\theta_{1}-4\theta_{2} \mod 4\end{aligned}$\\ \hline
10&$\begin{aligned}z_{2w,1}=& +2\theta_{1}-2\theta_{2}-2\theta_{5} \mod 2\end{aligned}$\\ &$\begin{aligned}z_{2w,2}=& +2\theta_{1}-2\theta_{2}-2\theta_{3} \mod 2\end{aligned}$\\
\hline\end{longtable}

%% file: ebr_tables_single.tex
\begin{longtable}[t]{|c|l|} \caption[Split EBRs without SOC]{Symmetry-indicated trivial split elementary band representations (EBRs) without spin-orbit coupling (SOC).
For each space group in which such EBRs exist, we list the independent band decompositions of the form $\Lambda = V \oplus C$ that are allowed by symmetry indicators, where $\Lambda,V,C$ are all EBRs.
As discussed in SN~\ref{sec:ebrsplit}, all EBR splittings must result in band topology.
Of the 9 independent splittings listed here, all have mismatching SRSIs (indicated by a cross \xmark).} \label{tab: ebrdecomps_nosoc} \\ \hline SG\# & Split EBR \\ \hline \hline
207& $(E)_{\mathrm{c}} = (T_{1})_{\mathrm{b}} \oplus (T_{2})_{\mathrm{b}}\quad \text{\xmark} $\\ &$(E)_{\mathrm{d}} = (T_{1})_{\mathrm{a}} \oplus (T_{2})_{\mathrm{a}}\quad \text{\xmark}$ \\ \hline
211& $(E)_{\mathrm{b}} = (T_{1})_{\mathrm{a}} \oplus (T_{2})_{\mathrm{a}}\quad \text{\xmark} $\\ &$(E)_{\mathrm{c}} = (E)_{\mathrm{a}} \oplus (T_{1})_{\mathrm{a}} \oplus (T_{2})_{\mathrm{a}}\quad \text{\xmark} $\\ &$(B_{3})_{\mathrm{d}} = (T_{1})_{\mathrm{a}} \oplus (B_{2})_{\mathrm{b}}\quad \text{\xmark} $\\ &$(B_{3})_{\mathrm{d}} = (T_{2})_{\mathrm{a}} \oplus (A_{2})_{\mathrm{b}}\quad \text{\xmark} $\\ &$(B_{2})_{\mathrm{d}} = (T_{1})_{\mathrm{a}} \oplus (B_{2})_{\mathrm{b}}\quad \text{\xmark} $\\ &$(B_{2})_{\mathrm{d}} = (T_{2})_{\mathrm{a}} \oplus (A_{2})_{\mathrm{b}}\quad \text{\xmark}$ \\ \hline
222& $(E)_{\mathrm{b}} = (T_{1})_{\mathrm{a}} \oplus (T_{2})_{\mathrm{a}}\quad \text{\xmark}$ \\ \hline\end{longtable}

%% file: ebr_tables_double.tex
\begin{longtable}[t]{|c|l|} \caption[Split EBRs with SOC]{Symmetry-indicated trivial split elementary band representations (EBRs) with spin-orbit coupling (SOC).
For each space group in which such EBRs exist, we list the independent band decompositions of the form $\Lambda = V \oplus C$ that are allowed by symmetry indicators, where $\Lambda,V,C$ are all EBRs.
As discussed in SN~\ref{sec:ebrsplit}, all EBR splittings must result in band topology.
Of the 202 independent splittings listed here, 194 have mismatching SRSIs (indicated by a cross \xmark), while 8 have matching SRSIs (indicated by a checkmark \cmark).} \label{tab: ebrdecomps} \\ \hline SG\# & Split EBR \\ \hline \hline
21& $({}^1\bar{E}{}^2\bar{E})_{\mathrm{k}} = (\bar{E})_{\mathrm{a}} \oplus (\bar{E})_{\mathrm{a}}\quad \text{\xmark} $\\ &$({}^1\bar{E}{}^2\bar{E})_{\mathrm{k}} = (\bar{E})_{\mathrm{a}} \oplus (\bar{E})_{\mathrm{b}}\quad \text{\xmark} $\\ &$({}^1\bar{E}{}^2\bar{E})_{\mathrm{k}} = (\bar{E})_{\mathrm{a}} \oplus (\bar{E})_{\mathrm{c}}\quad \text{\xmark} $\\ &$({}^1\bar{E}{}^2\bar{E})_{\mathrm{k}} = (\bar{E})_{\mathrm{a}} \oplus (\bar{E})_{\mathrm{d}}\quad \text{\xmark} $\\ &$({}^1\bar{E}{}^2\bar{E})_{\mathrm{k}} = (\bar{E})_{\mathrm{b}} \oplus (\bar{E})_{\mathrm{b}}\quad \text{\xmark} $\\ &$({}^1\bar{E}{}^2\bar{E})_{\mathrm{k}} = (\bar{E})_{\mathrm{b}} \oplus (\bar{E})_{\mathrm{c}}\quad \text{\xmark} $\\
\hline\end{longtable}